\documentclass[a4paper,11pt]{article}

\usepackage[a4paper,
  left=2.5cm, right=2.5cm,
  top= 3cm, bottom=4cm]{geometry}
\usepackage{amsmath}
\usepackage{amssymb}
\usepackage{enumerate}
\usepackage[dvips]{graphicx}
\usepackage{color}
\usepackage{hyperref}

\usepackage[all]{xy}

\newcommand{\be}{\begin{equation}}  
\newcommand{\ee}{\end{equation}}

\newcommand{\nn}{\nonumber}
\newcommand{\rem}[1]{} 

\def\C{\mathbb{C}}
\def\Z{\mathbb{Z}}
\def\R{\mathbb{R}}
\def\P{\mathbb{P}}
\def\N{\mathbb{N}}
\def\Q{\mathbb{Q}}

\def\Hirz[#1]{\mathbbm{F}_{#1}}
\def\o[#1]{\overline{#1}}

\title{title}

\begin{document}

\begin{titlepage}
  
\begin{flushright}
KCL MTH-14-15 \\
IPMU14-0267
\end{flushright}
   
 \vskip 1cm
 \begin{center}
   
{\large \bf The Vertical, the Horizontal and the Rest:} \\
{\bf --- anatomy of the middle cohomology of Calabi-Yau fourfolds --- \\ 
--- and F-theory applications ---}

 \vskip 1.2cm
   
 Andreas P. Braun$^{1}$ and Taizan Watari$^{2}$

 \vskip 0.4cm
  {\it
   $^1$Department of Mathematics, King`s College, London WC2R 2LS, UK
   \\[2mm]
   
  $^2$Institute for the Physics and Mathematics of the Universe, University of Tokyo, Kashiwano-ha 5-1-5, 277-8583, Japan  
   }
\vskip 1.5cm
   
\abstract{
The four-form field strength in F-theory compactifications on Calabi-Yau fourfolds 
takes its value in the middle cohomology group $H^4$. The middle cohomology is decomposed 
into a vertical, a horizontal and a remaining component, all three of which are present in general. 
We argue that a flux along the remaining or vertical component may break some symmetry, while a purely horizontal flux does not influence the unbroken 
part of the gauge group or the net chirality of charged matter fields. This makes the decomposition  
crucial to the counting of flux vacua in the context of F-theory GUTs. 
We use mirror symmetry to derive a combinatorial formula for the dimensions of these components applicable to any toric Calabi--Yau hypersurface, and also make a
partial attempt at providing a geometric characterization of the four-cycles Poincar\'e dual to the remaining component of $H^4$.  It is also found in general 
elliptic Calabi-Yau fourfolds supporting SU(5) gauge symmetry that a remaining component can be present, for example, in a form crucial to the symmetry 
breaking ${\rm SU}(5) \longrightarrow {\rm SU}(3)_C \times {\rm SU}(2)_L \times {\rm U}(1)_Y$.
The dimension of the horizontal component is used to derive an estimate of 
the statistical distribution of the number of generations and the rank of 7-brane 
gauge groups in the landscape of F-theory flux vacua.}  
\end{center}
\end{titlepage}

\tableofcontents

\section{Introduction}

Work in string theory has traditionally focussed on the study of Calabi-Yau threefolds, as they are relevant to compactification of strings theories to four dimensions. From a mathematical point of view, it is very natural to ask about the properties of Calabi-Yau manifolds in (complex) dimensions other than three. Besides the omnipresent torus, two- and four-dimensional Calabi-Yau manifolds have subsequently acquired a central position within string theory\footnote{See \cite{Haupt:2008nu} for a discussion of M-theory compactifications on Calabi-Yau fivefolds.}.

Two-dimensional Calabi-Yau manifolds, more commonly called K3 surfaces, form a single connected family and have a long history in mathematics (see e.g. the classic \cite{hudson05}). As the simplest non-trivial Calabi-Yau manifolds, they also have long been used to compactify string and supergravity theories. Their full relevance to string theory has only been appreciated with the advent of string dualities \cite{Aspinwall:1996mn}. Besides appearing in relation to mirror symmetry \cite{Greene:1993vm}, it was also in the context of string dualities, and in particular compactifications of F- and M-theory, that Calabi-Yau fourfolds became an intense object of interest. Refs. \cite{Becker:1996gj,Sethi:1996es,Brunner:1996pk,Mayr:1996sh,Klemm:1996ts,Mohri:1997uk,Bershadsky:1997zs,Kreuzer:1997zg} form 
a partial list of early papers on the subject. 

What is common to all Calabi-Yau $n$-folds is that their K\"ahler and complex structures are measured by integrating two special harmonic differential forms, the K\"ahler form $J\in H^{1,1}$ and the holomorphic top-form $\Omega \in H^{n,0}$, over appropriate cycles. However, we may already point out a crucial difference between Calabi-Yau threefolds on one
side and K3 surfaces and Calabi-Yau fourfolds on the other side: whereas (powers of) $J$ and $\Omega^{3,0}$ live in different cohomology groups in the case of Calabi-Yau manifolds of odd complex dimensions, $\Omega^{n,0}$ and $J^{n/2}$ share the middle cohomology for Calabi-Yau manifolds of even complex dimensions.
For K3 surfaces this observation is tightly connected with the concept of polarization, where both $J$ and $\Omega^{2,0}$ are confined to lie in mutually orthogonal subspaces of $H^2$. The Torelli theorem for lattice-polarized K3 surfaces 
states that the complex structure of a K3 surface with a lattice $W \subset H^2$
generated by algebraic cycles is parametrized by the period domain 
${\cal M}_* = \P \left[\left\{ \Omega \in [W^\perp \subset H^2\otimes \C] \; | \;
  \Omega \wedge \Omega = 0, \Omega \wedge \overline{\Omega} > 0 \right\} 
\right]$. 


For Calabi-Yau fourfolds, however, there is no such convenient Torelli theorem. 
Consider a family of Calabi--Yau fourfolds $\pi: {\cal Z} \longrightarrow {\cal M}_*$ (i.e., 
$\hat{Z}_p := \pi^{-1}(p)$ for $p \in {\cal M}_*$ is a Calabi--Yau fourfold);  
${\rm Pic}(\hat{Z}_p)$ for a generic choice of $p \in {\cal M}_*$ plays 
a role similar to the lattice polarization in the case of K3 surfaces. 
For any two classes in ${\rm Pic}(\hat{Z}_p)$ with representatives $\eta_1$ and $\eta_2$, we may 
form a {\em vertical cycle} $\eta_1 \wedge \eta_2$, which defines a class 
in $H^{2,2}(\hat{Z}_p; \R) \cap H^4(\hat{Z}; \Z)$.
The subspace generated by such cycles (for $p$ not in any one of the Noether--Lefschetz 
loci within ${\cal M}_*$) is called the {\em primary vertical component} 
of $H^{2,2}$, and is denoted by $H^{2,2}_V(\hat{Z}_p)$. It is identified 
within $H^4(\hat{Z}; \Z)$ defined in {\em topology}, independently of the 
choice of the ``generic'' $p \in {\cal M}_*$. The period integral 
$\int \Omega_{\hat{Z}_p}$ takes its value in $H^4(\hat{Z}; \C)$ and satisfies 
the obvious constraints $\Omega \wedge \Omega = 0$ and 
$\Omega \wedge \overline{\Omega} > 0$---(*), 
but the image of the period map does NOT occupy all of the complement of 
the primary vertical component. That is, the period domain satisfying (*) 
may be contained in a space much smaller than 
\begin{equation}
\left[ (H^{2,2}_V \otimes \C)^\perp  \subset H^4(\hat{Z}; \C) \right].
\label{eq:non-vertical-def}
\end{equation}

References \cite{Strominger:1990pd,Greene:1993vm} introduced another 
subspace $H^4_H(\hat{Z}; \C) \subset H^4(\hat{Z}; \C)$ called the 
{\em primary horizontal} component (see 
section \ref{sec:landscape-horizontal} for a brief review). 
The period map is locally injective for Calabi--Yau fourfolds
and maps ${\cal M}_*$ into an 
$m := h^{3,1}(\hat{Z}_p) = {\rm dim}_\C {\cal M}_*$-dimensional subvariety of 
$\P[H^4(\hat{Z}; \C)]$; as we will elaborate on 
in the next section, ${\cal M}_*$ is in fact mapped into 
the projectivization of the horizontal component, 
$\P \left[ H^4_H(\hat{Z}; \C) \right]$. The middle cohomology 
$H^4(\hat{Z}; \C)$ of a Calabi--Yau fourfold is then decomposed into 
\begin{equation}
 H^4(\hat{Z}; \C) = H^4_H(\hat{Z};\C) \oplus H^{2,2}_{RM}(\hat{Z};\C) \oplus 
  H^{2,2}_V(\hat{Z}; \C),
\label{eq:H4-decomp-H-RM-V}
\end{equation}
where the decomposition is orthogonal under the intersection pairing. 
Unless the $H^4_{RM}(\hat{Z}; \C)$ component vanishes, the primary horizontal 
subspace $H^4_H(\hat{Z}; \C)$ is smaller in dimension than the non-vertical 
subspace (\ref{eq:non-vertical-def}).


There is another context---flux compactification of F-theory---where 
one is interested in the decomposition (\ref{eq:H4-decomp-H-RM-V}) 
of the middle cohomology above. An ensemble of flux vacua is specified 
by specifying a subspace of 
\begin{equation}
  \left[H^4(\hat{Z}; \Z)\right]_{\rm shift} := 
 H^4(\hat{Z}; \Z) + \frac{c_2(T\hat{Z})}{2};
\end{equation}
when this subspace is affine, the vacuum index distribution over 
the moduli space ${\cal M}_*$ is given by a concise analytic formula \cite{Ashok:2003gk,Denef:2004ze}.
As discussed already in \cite{Braun:2014ola}, and refined further 
in section \ref{sec:landscape-horizontal} in this article, we can see 
that any pair of topological four-form fluxes whose difference belongs to 
the real part of the primary horizontal subspace $H^4_H(\hat{Z}; \R)$ 
shares the same symmetry group from 7-branes in their effective theories 
below the Kaluza--Klein scale. This motivates us to choose the affine 
subspace as some form of shift of $H^4_H(\hat{Z};\R)$. Because the analytic 
formula of vacuum index density involves the dimension of the 
affine subspace, it is of interest in physics application of F-theory to 
know the dimension of $H^4_H(\hat{Z}; \R)$. 
Another question of interest is concerns how the space $H^{2,2}_{RM}(\hat{Z}; \R)$ 
arises geometrically, and which role it plays in phenomenology applications.

Mirror symmetry indicates that \cite{Strominger:1990pd, Greene:1993vm}
\begin{equation}
  \left[ H^4_H(\hat{Z}; \C) \cap H^{2,2}(\hat{Z}_p; \C) \right] =:
   H^{2,2}_H (\hat{Z}_p; \C)  \quad {\rm and} \quad  
   H^{2,2}_V(\hat{Z}_{\rm m}; \C)
\end{equation}
have the same dimensions for a mirror pair $(\hat{Z}, \hat{Z}_{m})$. 
This article shares the idea of using the dimension $h^{2,2}_V$
of the mirror $\hat{Z}_m$ to find the dimension $h^{2,2}_H$ of the original geometry 
$\hat{Z}$ with \cite{Grimm:2009ef, Bizet:2014uua}. References \cite{Grimm:2009ef, Bizet:2014uua} used the intersection ring of the vertical subspace of 
the mirror $\hat{Z}_m$ not just to determine the dimension $h^{2,2}_H(\hat{Z})=h^{2,2}_V(\hat{Z}_m)$, but also to compute 
period integrals of $\hat{Z}$.

We combine this intersection ring in the mirror $\hat{Z}_m$ with a stratification of 
$\hat{Z}_m$ and a long exact sequence of morphisms of mixed Hodge structure, 
a combination of techniques that Ref. \cite{Batyrev94dualpolyhedra} used in 
order to derive the formula for $h^{1,1}$ and $h^{k-1,1}$ of a toric-hypersurface Calabi--Yau $k$-fold. 
With this approach, we are not only able to determine the dimensions 
$h^{2,2}_V(\hat{Z}_m)=h^{2,2}_H(\hat{Z})$, $h^{2,2}_V(\hat{Z})=h^{2,2}_H(\hat{Z}_m)$ 
and $h^{2,2}_{RM}(\hat{Z})=h^{2,2}_{RM}(\hat{Z}_m)$, but also to construct the cycles representing  
the remaining component, study their geometry, and discuss their different roles in physics applications.

We start with an introductory discussion in section \ref{sec:landscape-horizontal}, which motivates the study of the space of non-vertical four-cycles (primary horizontal cycles in particular) in the context of the landscape of flux vacua in F-theory compactifications. Using mirror symmetry, we derive a combinatorial formula for $h^{2,2}_V$, $h^{2,2}_H$, and $h^{2,2}_{RM}$, the dimensions of the space of vertical, horizontal and remaining (i.e. non-vertical and non-horizontal) cycles for Calabi--Yau fourfolds obtained as hypersurfaces of toric varieties in section \ref{sect:h22hypersurface}. Already in this simple class of Calabi--Yau fourfolds, the remaining component is found to be non-zero in general. 
We provide several examples in section~\ref{sect:examples}.

A non-zero $H^{2,2}_{RM}(\hat{Z}; \C)$ already occurs for the family of Calabi--Yau fourfolds 
$\hat{Z} = {\rm K3} \times {\rm K3} = S_1 \times S_2$ with a lattice 
polarization $W_1 \subset H^2(S_1; \Z)$ and $W_2 \subset H^2(S_2; \Z)$. 
Here, 
\begin{equation}
 h^{2,2}_V = \rho_1 \rho_2+2, \quad 
 h^{2,2}_{RM} = \rho_1(22-\rho_2)+(22-\rho_1)\rho_2, \quad 
 2 + 2h^{3,1} + h^{2,2}_H = (22-\rho_1)(22-\rho_2), 
\label{eq:K3K3-rm-dim}
\end{equation}
where $\rho_i = {\rm rank}(W_i)$ \cite{Braun:2014ola}. 
In this example, the Poincar\'e duals of $H^{2,2}_{RM}(\hat{Z})$ are not represented by 
algebraic four-cycles. 

Section \ref{sect:genI5fourfolds} shows another example of a family 
$\pi: {\cal Z} \longrightarrow {\cal M}_*$ where $h^{2,2}_{RM} \neq 0$
and provides some more intuition for the geometry relevant to cycles in $h^{2,2}_{RM}$.
This family is a simple example within the class of Calabi--Yau fourfolds motivated 
by \cite{Buican:2006sn,Beasley:2008kw} for F-theory compactification 
where SU(5) unification symmetry is broken down to 
that of the Standard Model 
${\rm SU}(3)_c \times {\rm SU}(2)_L \times {\rm U}(1)_Y$ without the 
hypercharge ${\rm U}(1)_Y$ vector field becoming massive. 
The remaining component $H^{2,2}_{RM}(\hat{Z})$ in this family contains  
forms that are Poincar\'e dual to four-cycles which are non-vertical, but 
still algebraic over generic points in moduli space. This observation plays an important role in the discussion of section \ref{sec:landscape-horizontal}, which discusses the relevance of the real primary horizontal subspace 
$H^4_R(\hat{Z}; \R) \subset H^4(\hat{Z}; \R)$ in the context of the 
landscape of flux vacua in F-theory compactification.
 
In section \ref{sect:distrGNgen}, we discuss how the abundance of flux 
vacua depends on the unification group, and the number of generations. 
Under rather general assumptions, the dependence on the number of generations 
is found to factor from the distribution and to be given by a Gaussian quite 
generically for any choice of base manifold. We also develop an estimate for the
dependence of the number of flux vacua on the gauge (unification) group, generalizing earlier
attempts in \cite{Braun:2014ola}. We estimate that the abundance of flux vacua 
with e.g. gauge group $SU(5)$ is suppressed by a factor of roughly $e^{\mathcal{O}(1000)}$
when compared to models with no non-abelian gauge group. One can think of this surpression as
a fine-tuning wildly surpassing the fine-tuning of $10^{-120}$ needed to explain the smallness of the cosmological
constant. Appendix \ref{sect:dephodgegroup} contains a further elaboration of how such estimates
may be obtained. Finally, we discuss several open problems reserved for future investigation 
in section \ref{sect:outlook}.

The appendices \ref{sec:general-4fold} and \ref{sect:chirflux} mostly contain 
supplementary material and applications. Appendix \ref{sec:general-4fold} 
reviews details of the geometry of F-theory GUTs with gauge group SU(5) 
relevant for section \ref{sect:genI5fourfolds}. Appendix \ref{app:hodgeexcept}
explains how to compute the Hodge diamond of exceptional divisors in such 
a geometry by using stratification. 
Appendix \ref{sect:chirflux} reviews the construction of chirality-inducing 
four-form flux, and computation of the D3-tadpole from the geometry and this 
flux. The numerical results in the appendices \ref{sect:dephodgegroup} 
and \ref{sect:chirflux} are used as input for the discussion in 
section \ref{sect:distrGNgen}.

A letter \cite{physlett} by the same authors is focused on a subset of the
subjects discussed in this article, and is addressed to a broader spectrum  
of readers. It covers the subjects of sections \ref{sec:landscape-horizontal} 
and \ref{sect:distrGNgen} and uses the results of section \ref{sect:examplesCYna4} and 
the appendices \ref{sect:dephodgegroup} and \ref{sect:chirflux}, while omitting the
(mostly technical) material of sections \ref{sect:h22hypersurface} and \ref{sect:genI5fourfolds}.

\section{Ensembles of F-Theory flux vacua and the primary horizontal subspace}
\label{sec:landscape-horizontal}

Supersymmetric compactification of F-theory to 3+1-dimensions is 
specified by a set of data $(X, B_3, J, G^{(4)})$, where 
$\pi_X: X \longrightarrow B_3$ is an elliptic fibration with a section, 
$J \in H^{1,1}(B_3; \R)$ a K\"{a}hler form on $B_3$ and a $G_4$ a four-form flux 
in $[H^4(X; \Z)]_{\rm shift}$. Once $G^{(4)}$ is given topologically, 
the superpotential $W \propto \int_X \Omega_X \wedge G^{(4)}$ determines 
the vacuum expectation value of the complex structure of $X$, $B_3$, $\pi_X$ 
etc. For an ensemble of fluxes in $[H^4(X; \Z)]_{\rm shift}$, therefore, 
we obtain an ensemble of low-energy effective theories in 3+1-dimensions, 
called a landscape. 
 
Flux compactification not only determines the values of low-energy 
coupling constants, but also the gauge group. Once the complex structure 
of the elliptic fibration $\pi_X: X \longrightarrow B_3$ is determined by 
the mechanism above, we know the configuration of 7-branes 
(i.e. the discriminant locus of $\pi_X$). Now, remember that we usually classify low-energy effective theories in their algebraic information such as gauge group, matter representations and unbroken symmetry first, in their topological information such as the number of generations next, and then finally in their values of the coupling constants. It is thus desirable to be able to classify flux vacua in a landscape also in the same way, sorting out first in the algebraic information, secondly in the topological data and finally in the moduli data \cite{Braun:2014ola}. This requires to work out what kind of topological flux $G^{(4)}$ results in an effective theory with a given algebraic and topological information. 

In order to address this problem, it is useful to consider a family 
of elliptically fibred Calabi--Yau fourfolds characterized as 
follows \cite{Braun:2014ola}. 
First, let us choose $(B_3, [S])$ and an algebra $R$. We specify only 
the topology of an algebraic three-fold $B_3$, and let $[S]$ be a divisor class in 
${\rm Pic}(B_3)$.
$R$ is an algebra in the $A$-$D$-$E$ series, to be used for a unification group 
of one's interest.   
The family $\pi: {\cal X} \longrightarrow {\cal M}_*^R$ is that of {\it smooth} 
Calabi--Yau fourfolds $X$ with an elliptic fibration 
$\pi_X: X \longrightarrow B_3$ with a section, so that there is a locus 
of singular fibres of type\footnote{Here, `type $R$' singular fibre means 
that at a generic point in $S$, the dual graph of the fibre is given 
by the extended Dynkin diagram of $R$.} $R$ along a divisor $S$ of $B_3$ 
that belongs to the class $[S]$. The restricted moduli space ${\cal M}_*^R$ 
parametrizes the complex structure of such fourfolds $X$. 
For a generic point $p \in {\cal M}_*^R$, the corresponding fourfold 
$X_p := \pi^{-1}(p)$ has the property that 
$\left[  H^{1,1}(X_p;\Q) \cap H^2(X; \Q) \right]$ is generated by 
divisors in the base, 
the zero-section $\sigma$, and 
$\{ \hat{\mathcal{C}}_i \}_{i=1,\cdots, {\rm rank}(R)}$ called Cartan divisors.
Such a family over the restricted moduli space ${\cal M}_*^R$ is 
a useful notion to express the flux vacua distribution, as we see 
in eqs. (\ref{eq:ADD-formula-1}, \ref{eq:ADD-formula-2}).

There are a couple of physical conditions to be imposed on the fluxes. 
First of all, the flux preserves the ${\rm SO}(3,1)$ symmetry in the 
effective theory on 3+1-dimensions if and only if the flux $G^{(4)}$ does 
not have a component that has either ``two legs in the $T^2$ fibre'' or 
``no legs in the $T^2$ fibre'' \cite{Dasgupta:1999ss}. This condition is 
best paraphrased as 
\begin{equation}
 \sigma \wedge \eta \wedge G^{(4)} = 0, \qquad 
 \eta_1 \wedge \eta_2 \wedge G^{(4)} = 0 ,
\label{eq:cond-SO(3,1)}
\end{equation}
where $\sigma$ is (the differential form Poincar\'e dual to) the zero section, 
and $\eta$, $\eta_{1,2}$ are divisors on $B_3$, see \cite{Braun:2011zm,Marsano:2011hv,Grimm:2011sk,Grimm:2011fx,Cvetic:2013uta}.
The $D$-term condition (equivalently ${\cal N}=1$ supersymmetry condition, 
primitiveness condition) is that 
\begin{equation}
   J \wedge G^{(4)} = 0 \in H^6(X; \R).
\label{eq:cond-prim}
\end{equation}
The subspace of $[H^4(X; \Z)]_{\rm shift}$ satisfying the two conditions 
above, (\ref{eq:cond-SO(3,1)}, \ref{eq:cond-prim}), is denoted by 
$[H^4(X; \Z)]_{\rm shift}^{\rm Lor.prim.}$. The subspace of $H^4(X; \Z)$ without 
the shift by $c_2(TX)/2$ satisfying the condition (\ref{eq:cond-SO(3,1)}, 
\ref{eq:cond-prim}) is denoted by $[H^4(X;\Z)]^{\rm Lor.prim.}$.

For a pair of fluxes $G_1^{(4)}$ and $G_2^{(4)}$ in 
$[H^4(X; \Z)]_{\rm shift}^{\rm Lor.prim.}$ to preserve the same symmetry group 
within the 7-brane gauge group $R$, their difference needs to satisfy 
\begin{equation}
  i_{\hat{\mathcal{C}}_i}^* \left( G^{(4)}_1 - G^{(4)}_2 \right) = 0 \in H^4(\hat{\mathcal{C}}_i;\Z), 
\label{eq:cond-unbroken-sym}
\end{equation}
where $i_{\hat{\mathcal{C}}_i}: \hat{\mathcal{C}}_i \hookrightarrow X$ is the embedding of the Cartan divisors (generators) $\{ \hat{\mathcal{C}}_i \}_{i=1,\cdots, {\rm rank}(R)}$. Thus, we formulate 
an ensemble of fluxes leading to effective theories with a common unbroken 
symmetry group as 
\begin{equation}
 \left\{ G^{(4)}_{\rm tot} = G^{(4)}_{\rm scan} + G^{(4)}_{\rm fix} \; | \; 
    G^{(4)}_{\rm scan} \in H_{\rm scan} \right\}; 
\label{eq:flux-ensemble}
\end{equation}
we choose $G^{(4)}_{\rm fix}$ and $H_{\rm scan}$ such that $G^{(4)}_{\rm scan} = 0$ 
is contained in $H_{\rm scan}$. $G^{(4)}_{\rm fix}$ must be in 
$[H^4(X; \Z)]_{\rm shift}^{\rm Lor.prim.}$, while $H_{\rm scan}$ needs to be a 
sub-{\it set} of the cohomology group $H^4(X; \Z)$ satisfying all the 
conditions 
(\ref{eq:cond-SO(3,1)}, \ref{eq:cond-prim}, \ref{eq:cond-unbroken-sym}).
Because all of these conditions\footnote{The condition from D3-brane 
tadpole is treated separately from the conditions discussed above. } 
are linear in $G^{(4)}$, we always take $H_{\rm scan}$ to be a sub-{\it group} of 
$H^4(X; \Z)$. 

We now argue that the real primary horizontal subspace 
$H^4_H(X;\R) \subset H^4(X; \R)$ satisfies all of the conditions 
(\ref{eq:cond-SO(3,1)}, \ref{eq:cond-prim}, \ref{eq:cond-unbroken-sym}) 
modulo $\otimes \R$, and we can hence take $H_{\rm scan}$ to contain all 
of $H^4_H(X ; \R) \cap H^4(X; \Z)$. To see this, we take a little 
moment to provide a brief review on the definition of $H^4(X; \C)$, and to
make clear what we mean by the {\it real} primary horizontal subspace.
For any point $p \in {\cal M}_*^R$, we can define a subspace 
\begin{eqnarray}
&& {\rm Span}_\C \left\{ \Omega_{X_p}, D\Omega_{X_p}, D^2\Omega_{X_p}, D^3\Omega_{X_p}, D^4\Omega_{X_p} \right\} \label{eq:def-horizontal-space} \\
& = &
H^{4,0}(X_p,\C) \oplus H^{3,1}(X_p; \C) \oplus H^{2,2}_{H}(X_p; \C) \oplus 
  H^{1,3}(X_p; \C) \oplus H^{0,4}(X_p; \C) \subset H^4(X;\C). 
  \nonumber 
\end{eqnarray}
The cohomology group $H^4(X; \C)=\C \otimes H^4(X; \Z)$ is topological; 
there is a canonical identification between $H^4(X_p; \C)$ and 
$H^4(X_{p'}; \C)$ for $p, p'$ in a small neighbourhood in ${\cal M}_*$---called 
flat structure---and the reference to $p\in {\cal M}_*$ is suppressed in 
the expression above. The $(p,q)$-Hodge components with the canonical 
identification relative to $H^4(X; \C)$, however, vary over 
$p \in {\cal M}_*^R$. The fact that the Picard--Fuchs equations on the period integral are closed in the subspace specified above, however, implies that the space (\ref{eq:def-horizontal-space}) remains 
the same in $H^4(X; \C)$ for every $p \in {\cal M}_*$ (at least locally) under the 
canonical identification (flat structure). A reference to $p \in {\cal M}_*^R$ is hence
suppressed and the invariant subspace is called 
the primary horizontal subspace $H^4_H(X; \C)$.
The remaining component in the decomposition (\ref{eq:H4-decomp-H-RM-V}), 
$H^{2,2}(X_p; \C)$, should therefore be independent of $p \in {\cal M}_*$ 
under the canonical identification. This is why reference to 
$p \in {\cal M}_*$ is completely dropped in (\ref{eq:H4-decomp-H-RM-V}); 
the decomposition (\ref{eq:H4-decomp-H-RM-V}) is topological.

The four-form flux in M-theory/F-theory compactification is real-valued, 
while the primary horizontal subspace $H^4_H(X;\C)$ is complex-valued.
Noting, however, that the complex conjugation operation is compatible with 
the canonical identification (topological tracking) of $H^4(X_p; \C)$ for 
$p \in {\cal M}_*$, we also have a decomposition of the real part $H^4(X; \R)$: 
\begin{equation}
 H^4(X; \R) = H^4_H(X; \R) \oplus H^{2,2}_{RM}(X; \R) \oplus H^{2,2}_V(X; \R), 
\label{eq:H4-decomp-H-RM-V-real}
\end{equation}
just like in (\ref{eq:H4-decomp-H-RM-V}). 
The ``primary horizontal subspace'' component of $H^4(X; \R)$ is what we 
call the real primary horizontal subspace $H^4_H(X; \R)$.

Now, let us see that the four-forms in the real primary horizontal subspace
$H^4_R(X;\R)$ satisfy all the conditions (\ref{eq:cond-SO(3,1)}, 
\ref{eq:cond-prim}, \ref{eq:cond-unbroken-sym}) modulo $\otimes \R$.
First, noting that $\sigma \cdot \eta$ and $\eta_1 \cdot \eta_2$ in 
(\ref{eq:cond-SO(3,1)}) form only a subset of generators of the vertical 
four-cycles, and that all of the elements of the horizontal component are orthogonal 
to those in the vertical component in (\ref{eq:H4-decomp-H-RM-V}), it is 
straightforward to see that the four-forms in $H^4_H(X; \R)$ satisfy 
(\ref{eq:cond-SO(3,1)}).

In order to verify the primitiveness condition, consider the 
$(h^{3,1}+1)=(m+1)$-dimensional variety occupied by the complex line $\C[\Omega_{X_p}]$ 
(this is a $\C^\times$-cone over the period domain). Any four-form $G$ in this variety satisfies the 
condition (\ref{eq:cond-prim}), because $G$ is a $(4,0)$-form for some choice 
$p \in {\cal M}_*$ of complex structure on $X$, and $J \wedge G$ would have 
become a $(5,1)$-form under that complex structure, if it were non-zero.
The absence of such a Hodge component in a Calabi--Yau fourfold $X$ implies 
that $J \wedge G  = 0$. Since all kinds of four-forms in $H^4_H(X; \C)$ 
are obtained by taking derivatives of such $G$ with respect to the local 
coordinates in ${\cal M}_*$, we find that all the four-forms in the 
(real) horizontal primary subspace satisfy the primitiveness 
condition (\ref{eq:cond-prim}).

Finally, we can make a similar argument to show that the four-forms 
in $H^4_H(X; \R)$ satisfy the condition (\ref{eq:cond-unbroken-sym}). 
Consider again the set of four-forms realized in the form of $\Omega_{X_p}$ 
for some $p \in {\cal M}_*$. Its pull-back to any one of Cartan divisors 
$\hat{\mathcal{C}}_i$ must be a $(4,0)$-form on a complex {\it three}-fold $\hat{\mathcal{C}}_i$ 
under the complex structure of $p \in {\cal M}_*$ induced on $\hat{\mathcal{C}}_i \subset X_p$. 
Thus, the pull-back of such a four-form vanishes in $H^4(\hat{\mathcal{C}}_i ; \C)$.
Taking derivatives with respect to the local coordinates of ${\cal M}_*$, 
we find that the four-forms in $H^4_H(X; \R)$ satisfy the 
condition (\ref{eq:cond-unbroken-sym}). 

References \cite{Ashok:2003gk,Denef:2004ze,Denef:2008wq} studied how the flux vacua are 
distributed in the complex structure moduli space for ensembles of 
fluxes of the form (\ref{eq:flux-ensemble}). One of the two key ideas is 
to replace the vacuum distribution by the vacuum index distribution 
\begin{equation}\label{eq:distro}
 d\mu_I := d^{2m}z \sum_{G^{(4)}_{\rm scan}\in H_{\rm scan}} \delta^{2m}(DW, \overline{DW})
   {\rm det}\left[ \begin{array}{cc} DDW & D\overline{DW} \\ 
         \overline{D}DW & \overline{DDW} \end{array} \right]_{2m \times 2m} \;
    \Theta \left(L_* - \frac{1}{2}(G^{(4)}_{\rm tot})^2\right),
\end{equation}
so that the problem becomes easier. $(z,\bar{z})$ are local coordinates 
of ${\cal M}_*$, $W \propto \int_X \Omega_{X_p}\wedge G^{(4)}$ and 
$L_*$ is the upper bound of the D3-brane charge allowed for the fluxes 
$G^{(4)}_{\rm tot}$. The other idea is to make a continuous 
approximation to $H_{\rm scan}$, which is to treat the subgroup 
$H_{\rm scan} \subset H^4(X; \Z)$ as a vector space $H_{\rm scan} \otimes \R$, 
and to replace the sum by an integral over the vector space 
$H_{\rm scan} \otimes \R$. Under the continuous approximation, the vacuum 
index density $d\mu_I$ is of the form \cite{Ashok:2003gk,Denef:2004ze} ($L_* \gg K$)
\begin{equation}
 d\mu_I \simeq \frac{(2\pi L_*)^{K/2}}{\Gamma(K/2)} \; \rho_I, 
\qquad K := {\rm dim}_\R\left(H_{\rm scan}\otimes \R \right), 
\label{eq:ADD-formula-1}
\end{equation}
where $\rho_I$ is an $(m,m)$ form on ${\cal M}_*$. It is given as the Euler 
class of some real ${\rm rank}_\R=2m$ vector bundle on ${\cal M}_*$ 
\cite{Denef:2008wq}, and can be put in the explicit form 
\begin{equation}
 \rho_I = c_m(T{\cal M}_* \otimes {\cal L}^{-1}) = 
  {\rm det}\left[ - \frac{R}{2\pi i} + \frac{\omega}{2\pi}{\bf 1}_{m\times m}
           \right], 
\label{eq:ADD-formula-2}
\end{equation}
where $\omega$ is the K\"{a}hler form on ${\cal M}_*$ derived from 
$K = - \ln [\int \Omega \wedge \overline{\Omega} ]$, and ${\cal L}$ a 
line bundle satisfying $c_1({\cal L}^{-1} ) = \omega/(2\pi)$, 
whenever the vector space $H_{\rm scan}\otimes \R$ contains the 
real primary horizontal subspace $H^4_H(X; \R)$ 
\cite{Ashok:2003gk,Denef:2004ze,Denef:2008wq,Braun:2014ola}. 

Having reminded ourselves of how $H_{\rm scan}$ is used, let us return to 
the question of how we should take $G^{(4)}_{\rm fix}$ and $H_{\rm scan}$.
The problem we set in this article (and also in \cite{Braun:2014ola, 
physlett}) is to classify the fluxes in 
$[H^4(X; \Z)]_{\rm shift}^{\rm Lor.prim.}$ into sub-ensembles of the form 
(\ref{eq:flux-ensemble}) so that each subensemble corresponds to 
the ensemble of low-energy effective theories with a given set of 
algebraic (or algebraic and topological) information. 
$G^{(4)}_{\rm fix}$ is used as a tag of the subensemble. 
Apart from \cite{Braun:2014ola}, the $H^{2,2}_{RM}(X; \R)$ component has not been carefully discussed (at least from the perspective
of the geometry of Calabi--Yau fourfolds) in the context of F-theory phenomenology to our knowledge. 
In order to address this problem, therefore, it is necessary to reopen 
the case and understand carefully how the low-energy effective theories 
are controlled by the fluxes in each one of the components of the 
decomposition (\ref{eq:H4-decomp-H-RM-V-real}).

First of all, it is already known that $h^{2,2}_{RM}$ can be non-zero 
in a family of $X = {\rm K3} \times {\rm K3}$ \cite{Braun:2014ola}, as 
we have already reviewed in the introduction. 
We will prove in section \ref{sect:genI5fourfolds} that the class of families 
of elliptic fibred Calabi--Yau fourfolds for F-theory compactification with 
${\rm SU}(5)$ unification results in $h^{2,2}_{RM} \neq 0$ precisely in 
the cases that are well-motivated for a mechanism of ${\rm SU}(5)$ symmetry 
breaking in \cite{Buican:2006sn,Beasley:2008kw}.
Despite the presence of the $H^{2,2}_{RM}(X; \R)$ component in the middle 
cohomology, however, we have confirmed (see also the detailed 
discussion in the appendix \ref{app:detailsexcpdiv} of this article) 
that the matter surfaces for 
${\bf 10}+\overline{\bf 10}$ representations and also $\bar{\bf 5}+{\bf 5}$
representations of ${\rm SU}(5)$ belong to topological classes of 
$H^{2,2}_V(X; \Q)$, confirming \cite{Marsano:2011hv}.
This means that the net chirality of these representations is controlled 
by the flux in the $H^{2,2}_V(X; \R)$ component. 

It has also been indicated that fluxes in $H^{2,2}_{RM}(X;\R)$ may 
break the symmetry of the 7-brane gauge group $R$ based on the family 
$X = {\rm K3} \times {\rm K3}$ \cite{Braun:2014ola}.
Certainly this family is not suitable for realistic compactifications of F-theory 
in that all the matter fields in the relatively light spectrum are in 
the adjoint representation of $R$; this family may also be somehow special 
in that $h^{2,0}(X) \neq 0$.
However, we have confirmed that the flux in the $H^{2,2}_{RM}(X; \R)$ component 
may indeed break the symmetry $R$; the hypercharge flux 
of \cite{Buican:2006sn,Beasley:2008kw} turns out to be precisely 
in this category.

Given all the observations above we conclude that $H_{\rm scan} \otimes \R$ 
should be chosen such that it contains all of the horizontal component 
$H^4_H(X; \R)$ and possibly a part of $H^{2,2}_{RM}(X; \R)$. Clearly, 
$H_{\rm scan} \otimes \R$ should not contain the entire $H^{2,2}_{RM}(X; \R)$ 
for sub-ensemble of fluxes (\ref{eq:flux-ensemble}) to correspond to 
an ensemble of effective theories with a given set of algebraic and 
topological information. An example of the family of 
$X = {\rm K3} \times {\rm K3}$ suggests strongly that we should choose 
$(H_{\rm scan} \otimes \R)$ to be precisely the horizontal component, because 
any flux in $H^{2,2}_{RM}(X; \R)$ breaks some of the 7-brane gauge 
group and Higgses away vector bosons from the low-energy spectrum in this example. 
We remain inconclusive about the case of families for realistic F-theory 
compactification with $(B_3, [S], R)$ (with $R=A_4, D_5$ etc.), however, 
because the discussion in sections \ref{sect:h22hypersurface} 
and \ref{sect:genI5fourfolds} provides only a partial understanding of the 
geometry of the $H^{2,2}_{RM}(X; \R)$ component. 
This material is enough, however, to conclude that 
the formula (\ref{eq:ADD-formula-1}, \ref{eq:ADD-formula-2}) 
of \cite{Ashok:2003gk,Denef:2004ze} can be used for the ensemble of 
effective theories with a given algebraic and topological information. 
In the examples of elliptic fourfolds for F-theory compactification studied 
in section \ref{sect:elfibexamples} of this paper, the remaining component 
$H^{2,2}_{RM}(X; \R)$ is absent, so that the horizontal component can be 
identified with $H_{\rm scan} \otimes \R$.

\begin{center}
  ...................................................
\end{center}

We have so far assumed that smooth Calabi--Yau fourfolds $X$ with flat 
elliptic fibration are used for F-theory compactifications with fluxes; 
it should be rememebered, however, that it is a belief rather than a fact that 
such smooth models $X$ should be used instead of the Weierstrass models $X_s$, 
which are singular for compactifications leading to effective theories 
with unbroken non-Abelian gauge groups.
There may be alternative and/or equivalent formulations of fluxes 
that lead to the same physics end results. Even when we pursue the direction 
of using smooth models $X$,
there can be more than one choice of such a smooth model $X$. 
All of those different resolutions, however, should describe the low-energy 
physics of the same vacuum with an unbroken unified symmetry. Thus, no matter 
how fluxes in F-theory are formulated, all the observable physics consequences 
should not depend on the choice of resolutions. The dimension of the primary 
horizontal component ($(H_{\rm scan} \otimes \R)$ to be more precise) should 
be regarded as one of such physical consequences of F-theory, and hence 
$K = {\rm dim}_\R[H^4_H(X; \R)]$ must not depend on the choice 
of the smooth model $X$ for a singular $X_s$.

There is a general argument that is good enough to make us consider that 
the dimensions of $h^{2,2}_V(X; \R)$,  $h^{2,2}_H(X; \R)$ and  $h^{2,2}_{RM}(X; \R)$ 
are resolution independent. It seems reasonable to assume that, for a 
singular Calabi-Yau fourfold ($X_s$), any crepant resolution ($X$) gives rise 
to the same complex structure moduli space. If this holds, the dimension of 
the space of primary horizontal (2, 2) cycles, $h^{2,2}_H(X; \R)$, as well as 
$h^{3,1}(X)$ cannot depend on the resolution. By mirror symmetry, the same 
statement must be true also for the space of primary vertical (2, 2) cycles 
$h^{2,2}_V(X; \R) = h^{2,2}_H(X_m; \R)$, and $h^{1,1}(X) = h^{3,1}(X_m)$, 
where we assume that all the different resolutions $X$ for $X_s$ share 
some mirror geometries $X_m$ with a common complex structure moduli space.
If furthermore the Euler characteristic is invariant under which resolution 
is used, it follows that also $h^{2,2}$ cannot be resolution dependent. 
This is because $h^{2,2}$ is not an independent Hodge number but related to 
all others by \eqref{h22fromotherhnumbers}. From the independence of 
$h^{2,2}_H(X; \R)$ and $h^{2,2}_V(X; \R)$ on which resolution is used,
it then follows that also $h^{2,2}_{RM}(X; \R)$ is independent of the resolution. 

The two assumptions we have made are obvious for a Calabi-Yau manifold $X$ 
given as a hypersurface (or, more generally, complete intersection) in a 
toric ambient space: for a fixed singular Calabi-Yau manifold, different 
crepant resolutions correspond to different triangulations of the N-lattice 
polytope whereas all Hodge numbers, and, in particular the complex structure 
moduli space depend on the combinatorics of the $N$- and $M$-lattice polytopes, 
but not on triangulations. We will give a more specific proof of this for
the expressions we derive in the hypersurface case in 
section \ref{sect:triang-indep}.

More generally, our assumptions follow if different resolutions correspond 
to different cones in the extended K\"ahler moduli space, so that they are 
connected by flop transitions, which are known to leave both the complex 
structure moduli space and the Euler characteristic invariant. This state of 
affairs is realized in the geometries which are the main motivation for the 
present work: F-theory on Calabi-Yau fourfolds supporting non-abelian gauge 
groups \cite{Esole:2011sm,Hayashi:2013lra,Hayashi:2014kca,Esole:2014bka,Esole:2014hya,Braun:2014kla}.

\begin{center}
 .............................................................
\end{center}

Let us leave a few remarks at the end of this section in order to clarify 
what the vacuum index density distribution $d\mu_I$ is for, as well as what 
it is not (yet) for. These are more or less known things, and we include this 
discussion in this article only as reminder.

The first remark is that a family $\pi: {\cal X} \longrightarrow {\cal M}_*^R$
for some choice of symmetry $R$ and topology of $(B_3, [S])$ is used to describe 
the distribution of a subensemble of flux vacua that is {\it inclusive} in nature. 
Even though the scanning component of the flux, $G^{(4)}_{\rm scan}$, is chosen from 
the real primary {\it horizontal} component $H^4_H(\hat{Z}; \R) \subset 
H^4(\hat{Z}; \R)$, it may eventually end up with the Poincar\'e dual of a
algebraic cycle as a result of dynamical relaxation of the complex structure 
moduli due to the superpotential $W \propto \int \Omega \wedge G^{(4)}$. 
The delta function $\delta^{2m}(DW, \overline{DW})$ for 
a given $G^{(4)}_{\rm scan}$ eventually has a support on a point $p$ in a 
Noether--Lefschetz locus of ${\cal M}_*^R$; see \cite{Braun:2011zm,Braun:2014ola,Bizet:2014uua} and references therein. 
For some choice of $G^{(4)}_{\rm scan}$ 
(and its corresponding $p \in {\cal M}_*^R$) 
not only the rank of $H^{2,2}(X_p; \R) \cap H^4(X_p; \Z)$, but even the rank of 
$H^{1,1}(X_p; \R) \cap H^2(X_p; \Z)$ may be enhanced,\footnote{
Depending on what conditions to impose mathematically on behaviour of the family 
${\cal X} \longrightarrow {\cal M}_*^R$ at special loci, the enhancement of 
the rank of ${\rm Pic}(X_p)$ may be phrased differently. The mathematical 
conditions to impose should ultimately be determined by the (yet unknown) 
microscopic formulation of F-theory.}. For the effective theory corresponding 
to such a choice of $G^{(4)}_{\rm scan}$, this can result in an enhanced gauge symmetry 
forming a hidden sector in addition to the gauge group $R$, or the symmetry $R$ 
is enhanced to another $R'$ containing $R$. Such a subensemble of fluxes with 
$(H_{\rm scan} \otimes \R) = H^4_H(X; \R)$ for $(B_3, [S], R)$ therefore contains 
not just effective theories with the 7-brane gauge group being precisely $R$, 
but also those with an extra factor of gauge group, or those with the 7-brane 
gauge group larger than $R$. In this sense, the subensemble (and the distribution
$d\mu_I$) captured on the moduli space ${\cal M}_*^R$ is an {\it inclusive} 
ensemble. With the expectation that only a very small fraction of such an inclusive 
ensemble has a hidden sector or enhanced symmetry, however, we will refer to 
vacua in such an ensemble as those with symmetry $R$; the fraction of such 
vacua can be estimated by using the discussion in section \ref{sect:distrGNgen} 
and the appendix \ref{sect:dephodgegroup}, and we see the fraction is 
exponentially small indeed.

Secondly, let us record our understanding of the issue of potential instabilities. 
Given a topological flux $G^{(4)}_{\rm tot} = G^{(4)}_{\rm scan} + G^{(4)}_{\rm fix}$, we
may not be able to find a solution to $DW = 0$ within the restricted moduli space  
${\cal M}_*^R$. Such a flux, if there is any, has been removed
automatically from the ensemble by the time the continuous approximation 
$H_{\rm scan} \longrightarrow H_{\rm scan} \otimes \R$ is introduced and the 
distribution $d\mu_I$ is cast into the form of (\ref{eq:ADD-formula-1}). 
When the flux space integral is carried out, the delta-function picks up 
contributions only from fluxes that satisfy $DW = 0$ somewhere in ${\cal M}_*^R$.
This argument may be, however, only about the instability issue associated 
with $DW = 0$ for moduli along ${\cal M}_*^R$, i.e., $R$-singlet moduli. 
Instability associated with physical ``moduli'' transverse to ${\cal M}_*^R$
may be captured by the effective superpotential 
$W = \int_S {\rm tr}[\varphi \wedge F]$ of \cite{Donagi:2008ca, Beasley:2008dc}.

Thirdly and finally, the K\"{a}hler moduli have been (and will be)
treated in this work as if they were given by hand, but 
their stabilization of course also has to be studied separately.
The distribution (\ref{eq:ADD-formula-1}, \ref{eq:ADD-formula-2}) on the 
moduli space of complex structure ${\cal M}_*$ needs to be convoluted with 
other data, including stabilization of K\"{a}hler moduli and dynamical 
evolution of cosmology in the early universe. There is nothing to 
add in this article to this well-known zoom-out picture, aside from 
reminding ourselves that it may make sense to study the distribution over 
the complex structure moduli space separately from K\"{a}hler moduli stabilization 
when there is a separation of scales between the stabilization of two 
distinct sets of moduli. In this case the two problems can be treated separately 
and then be combined, rather than facing the mixed problem.

\section{$H^{2,2}$ of a Calabi-Yau fourfold hypersurface of a toric variety}
\label{sect:h22hypersurface}

Suppose that $\hat{Z}$ is a smooth variety and $\{ \hat{Y}_i \}$ a set of 
its divisors. We can introduce the following stratification to $\hat{Z}$:
\begin{equation}\label{stratintro}
  \hat{Z} = (\hat{Z}\setminus \cup_i \hat{Y}_i) \amalg \left( \cup_i \hat{Y}_i \setminus (\cup_{kj} \hat{Y}_k\cap\hat{Y}_j) \right) \amalg \cdots \, .
\end{equation}
The first stratum, $Z := \hat{Z} \setminus \cup_i \hat{Y}_i$, is the only 
one that has the same dimension as $\hat{Z}$; we call it the primary stratum.
For the pair $\hat{Z}$ and $Y := \hat{Z} \setminus Z$, there is the long 
exact sequence (\ref{eq:long-seq-cpt-supp}). It was with this exact sequence 
in combination with the mixed Hodge structure on these cohomology groups 
that ref. \cite{Batyrev94dualpolyhedra} derived a formula of 
$h^{1,1}(\hat{Z})$ and $h^{n-2,1}(\hat{Z})$ for a Calabi--Yau $(n-1)$-fold 
hypersurface of a toric $n$-fold and showed the beautiful mirror 
correspondence.

We find that the same combination of methods -- stratification (\ref{stratintro}) 
and mixed Hodge structure -- is also very useful in studying various 
properties of algebraic cycles in $\hat{Z}$ even when $\hat{Z}$ is not 
necessarily realized as a hypersurface of a toric variety. 
Section \ref{sect:genI5fourfolds} of this article uses this combination 
to study the space of vertical cycles of general fourfolds $\hat{Z}$, 
whereas we can derive stronger results in the case $\hat{Z}$ is a Calabi--Yau $(n-1)$-fold hypersurface of a 
toric $n$-fold. In the latter case, which is discussed in the present section, the Hodge--Deligne numbers of the cohomology 
of the primary stratum can be easily computed by following \cite{DK}.
%
%

By carefully examining the contributions to $H^{2,2}(\hat{Z})$ and using mirror symmetry, 
we are able to isolate the pieces $h^{2,2}_V(\hat{Z})=h^{2,2}_H(\hat{Z_m})$, $h^{2,2}_H(\hat{Z})=h^{2,2}_V(\hat{Z}_m)$ and 
$h^{2,2}_{RM}(\hat{Z})=h^{2,2}_{RM}(\hat{Z}_m)$. The results are given in section \ref{hyperresultVvsH}.

\subsection{General setup and known results}

Let us start by fixing our notation and reviewing some helpful facts;
subtleties associated with singularity resolution are summarized in 
section \ref{sect:tsp}. Helpful reviews on toric geometry may be found in \cite{danilov,fulton}.

A toric variety $\P^n_\Sigma$ can be constructed from a fan $\Sigma$, which is composed of strictly convex polyhedral cones in $\R^n$ for a lattice $N := \Z^{\oplus n}$. The dual lattice is denoted by $M = \Z^{\oplus n}$ and the pairing between them for $\nu \in N$ and $\tilde{\nu} \in M$ is denoted by $\langle \nu, \tilde{\nu} \rangle \in \Z$. We also use the notation $N_{\R} := N \otimes \R$ and $M_{\R} := M \otimes \R$. For a given fan $\Sigma$, $\Sigma(k)$ stands for the collection of all the $k$-dimensional cones. $\Sigma(0) = \{ \vec{0} \in N \}$. We denote the $k$-skeleton $\cup_{i=0}^k \Sigma(i)$ by $\Sigma[k]$. 

By definition, an n-dimensional toric variety contains an open algebraic torus $(\C^*)^n = \mathbb{T}^n$. From this perspective, a fan $\Sigma$ gives information on how $\mathbb{T}^n$ is compactified. In particular, the fan determines a stratification
into algebraic tori of lower dimension: simplicial cones of $\Sigma$ are in one-to-one correspondence with strata of 
$\P^n_{\Sigma}$ such that a $k$-dimensional cone $\sigma \in \Sigma(k)$ corresponds to a stratum of $\P^n_{\Sigma}$ 
isomorphic to $(\mathbb{T})^{n-k}$. For cone $\sigma$, we denote the corresponding stratum by $\mathbb{T}_{\sigma}$.

A lattice polytope is the convex hull in $N_{\R}$ (or $M_{\R}$) of a number of lattice points of the lattice $N$ (or $M$). For a lattice polytope $\Delta$ in $M$ the polar (or dual) polytope $\widetilde{\Delta}$ (in $N$) is defined as
\begin{equation}
\widetilde{\Delta} := \{v \in N_{\R}|\langle v,w \rangle \geq -1 \quad \forall w \in \Delta  \}\, .
\end{equation}
If $\widetilde{\Delta}$ is a lattice polytope as well, it follows that $\Delta$ is also the polar of $\widetilde{\Delta}$ and the two are called a reflexive pair. An elementary property which follows from reflexivity is that the origin is the only integral point which is internal to the polytope. Faces of an $n$-dimensional polytope $\Delta$ are denoted by their dimensions\footnote{Faces of a polytope $\Delta$ are often referred to in the literature by their codimensions. Codimension-1 faces of $\Delta$ are called facets, for example. So, we reserve a notation $\Theta^{(k)}=\Theta^{[n-k]}$ for codimension-$k$ faces, although we are not using this notation in the present article.} as $\Theta^{[n-k]}$.
For a pair of reflexive polytopes $\Delta$ in $M_\R$ and $\widetilde{\Delta} $ in $N_\R$, there is a one-to-one correspondence between faces 
$\Theta^{[n-k]}$ of $\Delta$ and faces $\widetilde{\Theta}^{[k-1]}$ of $\widetilde{\Delta}$.
We frequently use $(\widetilde{\Theta}^{[n-k]},\Theta^{[k-1]})$ to indicate such
a pair of dual faces. 
Note that an $n$-dimensional polytope can be considered as its own $n$-dimensional face. We indicate that a face $\Theta_a$ is on a face $\Theta_b$ by writing $\Theta_a \leq \Theta_b$. We let $\ell(\Theta)$ stand for the number of integral points on a face $\Theta$ and $\ell^*(\Theta)$ for the number of integral points interior 
to $\Theta$. The $p$-skeleton of $\Delta$, i.e. the union of all faces of $\Delta$ which have dimension $p$ or less is denoted by
$\Delta_{\leq p}$.

Given a reflexive polytope $\widetilde{\Delta}$ we can easily construct a fan $\Sigma$ by forming the cones over all faces of $\widetilde{\Delta}$. The one-to-one correspondence between $(k-1)$-dimensional faces $\{ \widetilde{\Theta}^{[k-1]} \}$ and $k$-dimensional cones $\Sigma(k)$ in $N_\R$ (resp. between $\{ \Theta^{[k-1]} \}$ and $\widetilde{\Sigma}(k)$ in $M_\R$) is described in the form of a map $\sigma: \{\widetilde{\Theta}^{[k-1]} \} \longrightarrow \Sigma(k)$ (resp. $\tilde{\sigma}: \{ \Theta^{[k-1]} \} \longrightarrow \widetilde{\Sigma}(k)$).

Well-known formulae \cite{Batyrev94dualpolyhedra,Batyrev1996901,1995alg.geom..9009B} for the Hodge numbers of a Calabi--Yau $(n-1)$-fold 
hypersurface $\hat{Z}$ are recorded here in the notation adopted above:
\begin{align}
h^{1,1}(\hat{Z})   & = \ell(\widetilde{\Delta}) - (n+1) -
 \sum_{\widetilde{\Theta}^{[n-1]}} \ell^*(\widetilde{\Theta}^{[1]}) 
+ \sum_{(\widetilde{\Theta}^{[n-2]},\Theta^{[1]})}
     \ell^*(\widetilde{\Theta}^{[n-2]})\ell^*(\Theta^{[1]})\, ,
 \label{eq:bath11} \\
h^{n-2,1}(\hat{Z}) & = \ell(\Delta) - (n+1)
  - \sum_{\Theta^{[n-1]}} \ell^*(\Theta^{[n-1]})
  + \sum_{(\Theta^{[n-2]},\widetilde{\Theta}^{[1]})}
      \ell^*(\Theta^{[n-2]})\ell^*(\widetilde{\Theta}^{[1]})\, ,
 \label{eq:bath31} \\
h^{m,1}(\hat{Z}) & = \sum_{(\widetilde{\Theta}^{[n-m-1]}, \Theta^{[m]})}
 \ell^*(\widetilde{\Theta}^{[n-m-1]}) \ell^*(\Theta^{[m]})\quad
\mbox{for} \quad n-2 > m > 1. \label{eq:bath21}
\end{align}

\subsubsection{Triangulations, smoothness and projectivity}\label{sect:tsp}

The polytope $\Delta$ in $M_\R$ is regarded as the Newton polyhedron of the 
defining equation of a Calabi--Yau hypersurface $Z_s$ in $\P^n_{\Sigma}$. 
Because $Z_s$ is not smooth, in general, we are interested in its  
projective crepant resolution; 
$\hat{Z}$ in (\ref{eq:bath11}--\ref{eq:bath21}) stands for such a resolution.  
Such resolutions may always be constructed from the data of
the polytope for $n \leq 4$ \cite{Batyrev94dualpolyhedra}. 
For Calabi-Yau hypersurfaces of complex dimension 4 (or dimensions larger than that), this is not the case, 
so that we need to explicitly check in each example.

Constructing a fan $\Sigma$ over the faces of a polytope $\widetilde{\Delta}$, the resulting cones may be non-simplicial 
or have (lattice-) volume\footnote{The ``lattice volume'' of a cone 
in $\partial \Delta$ is defined by multiplying $k!$ to the volume of 
a $k$-dimensional cone cut-off at $\partial \Delta$. The smallest 
lattice $k$-simplex has the lattice volume $1$.} greater than one. Consequently, the toric variety $\P^n_\Sigma$ has singularities. We may, however, subdivide the fan $\Sigma$ to cure such singularities. We denote the corresponding map between fans by $\phi: \Sigma' \longrightarrow \Sigma$. For $\sigma' \in \Sigma'$, $\phi(\sigma')$ is given by the $\sigma'$-containing cone $\sigma \in \Sigma$ with the smallest dimension. The map $\phi$ (resp. 
$\tilde{\phi}$) induces a toric morphism $\P^n_{\Sigma'} \longrightarrow \P^n_{\Sigma}$ of (partial) singularity resolution.  

Such a morphism will preserve the Calabi-Yau condition of a hypersurface if all of the one-dimensional cones introduced are generated by points on the polytope $\widetilde{\Delta}$ (remember that the origin is the only internal point for a reflexive polytope).
A (partial) crepant desingularization of $Z_s$ is hence equivalent to finding a triangulation of the polytope $\widetilde{\Delta}$ in which
every $n$-simplex contains the origin. This is called a star triangulation and the origin is the star point. 
A maximal desingularization of $\P^n_{\Sigma}$ keeping $Z_s$ Calabi-Yau is found by using all points\footnote{In fact, it makes sense to relax this requirement, as points which lie in the interior of facets of $\widetilde{\Delta}$ do not lead to divisors intersecting a Calabi-Yau hypersurface. This can be seen as follows: for any facet $F$ we can find a normal vector $n_F$ such that $<n_F,\nu_i>\,= 1$ for all vectors $\nu_i$ on $F$. This means that the intersection ring contains a linear relation of the form
\begin{equation}
 \sum_{\nu_i \in F} D_i + \sum_{j\,{\rm not}\,\in F} a_j D_j = 0 \, ,
\end{equation}
with includes some contribution of divisors whose corresponding primitive vectors $\nu_j$ are not in $F$ but lie on other facets. Let us now assume we have refined $\Sigma$ such that there is a point $\nu_p$ interior to the facet $F$. The associated divisor $D_p$ can only have a non-zero intersection with divisors $D_k$ for which $\nu_k$ also lies in $F$, as all others necessarily 
lie in different cones of the fan $\Sigma$. This means that the above relation implies
\begin{equation}
D_p \cdot \sum_{\nu_i \in F} D_i = 0 \, ,
\end{equation}
where we sum over all toric divisors coming from points on $F$. The Calabi-Yau hypersurface is given as the zero-locus of 
a section of $-K_{\P^n_\Sigma} = \sum_j D_j$, where we sum over all toric divisors. We now find
\begin{equation}
D_p \cdot \sum_j D_j = D_p \cdot \sum_{\nu_i \in F} D_i = 0 \, ,
\end{equation}
by using the same argument again. Hence $D_p$ does not meet a generic Calabi-Yau hypersurface. Correspondingly, a refinement of $\Sigma$ introducing $\nu_p$ does not have any influence.} on $\widetilde{\Delta}$. A triangulation using all points of a polytope is called a fine triangulation.

When subdividing cones in $\Sigma$, we want the ambient space $\P^n_{\Sigma'}$ to be a projective variety (i.e. it should be a K\"ahler manifold), which implies projectivity of the Calabi-Yau hypersurface; the K\"{a}hlerian nature combined with the trivial canonical bundle implies Ricci-flatness. For an n-dimensional toric variety $\P^n_\Sigma$ given in terms of a fan $\Sigma$, a Weil divisor $D=\sum_i a_i D_i$ is also Cartier iff we can find a support function $\psi_D$ with the following properties:
\begin{itemize}
 \item $\psi_D$ is linear on each cone
 \item For a given cone of maximal dimension, $\sigma \in \Sigma(n)$, $\psi_D$ can be described by an element $m_\sigma$ of $M$ satisfying
 \begin{equation}
  \psi_D|_{\sigma} =\, <m_{\sigma}, \nu_i >\, = -a_i  \, 
 \end{equation}
for all one-dimensional cones (determined by the primitive vectors $\nu_i$) in $\sigma$.
\end{itemize}

In this language, the cone of ample curves (or, equivalently, the K\"ahler cone) is described as the set of divisors for which $\psi_D$ is strongly convex. This means that $\psi_D|_{\sigma} > -a_j$ for all one-dimensional cones \emph{not} in $\sigma$ and implies that $m_{\sigma} \neq m_{\sigma'}$ for two different cones $\sigma$ and $\sigma'$ in $\Sigma(n)$. If this cone of ample curves is non-empty, we can find a line bundle which is very ample, i.e. it defines an embedding of the toric variety $\P^n_\Sigma$ into $\P^m$ for some $m$. Conversely, if no strongly convex support function exists, 
the corresponding toric variety cannot be projective.
        
A strongly convex support function $\psi_D$ defines a `lift' of $\tilde{\Delta}$ into $\R^{n+1}$ by assigning the value of $\psi_D$ to each point on $\tilde{\Delta}$. The triangulation can then be
seen as the upper facets of the resulting polyhedron in $\R^{n+1}$. Note that strong convexity means that no $n+2$ points of $\tilde{\Delta}$ are mapped to a hyperplane in $\R^{n+1}$, i.e. faces of $\tilde{\Delta}$ which are subdivided into more than one simplex by a triangulation are not coplanar after this lift. Conversely, any triangulation which descends from a triangulation of a lift of $\tilde{\Delta}$ (with the property that no $n+2$ points are on a common non-vertical plane) to $\R^{n+1}$ can be used to construct a projective toric variety. Triangulations with this property are called regular triangulations.

A maximal projective crepant desingularization (referred to as an \emph{MPCP} of $Z_s$ in \cite{Batyrev94dualpolyhedra}) is hence achieved by finding a fine regular star triangulation of $\widetilde{\Delta}$. If all cones of such a triangulation have lattice volume unity the ambient space, and hence the Calabi-Yau hypersurface, become completely smooth and the map $\phi:\Sigma'\rightarrow \Sigma$ can rightfully be called a resolution. Triangulations of this type are called unimodular. We will also call simplices of volume unity and cones over such simplices unimodular.

\begin{figure}
\begin{center}
\includegraphics[width=5cm]{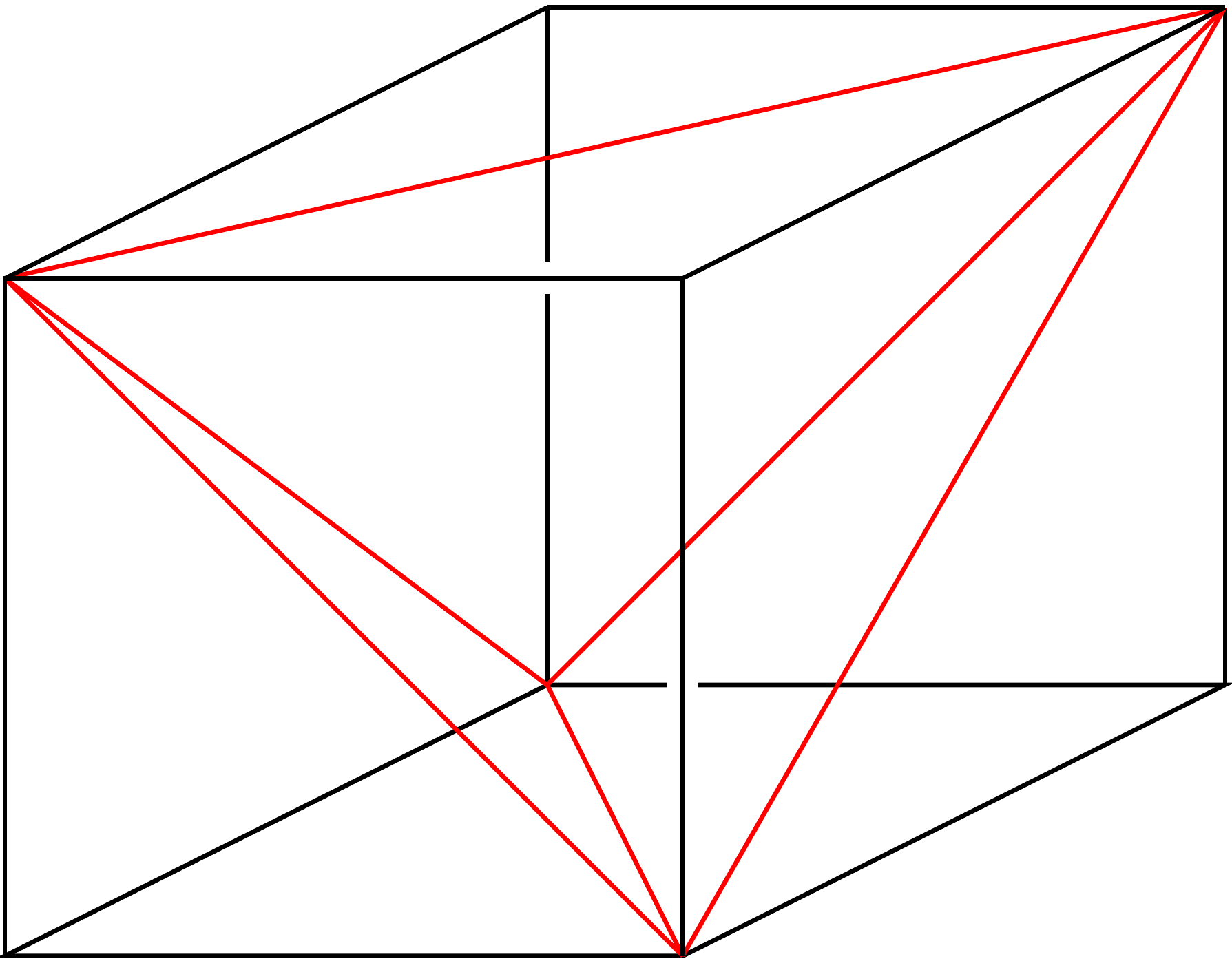}
\includegraphics[width=5cm]{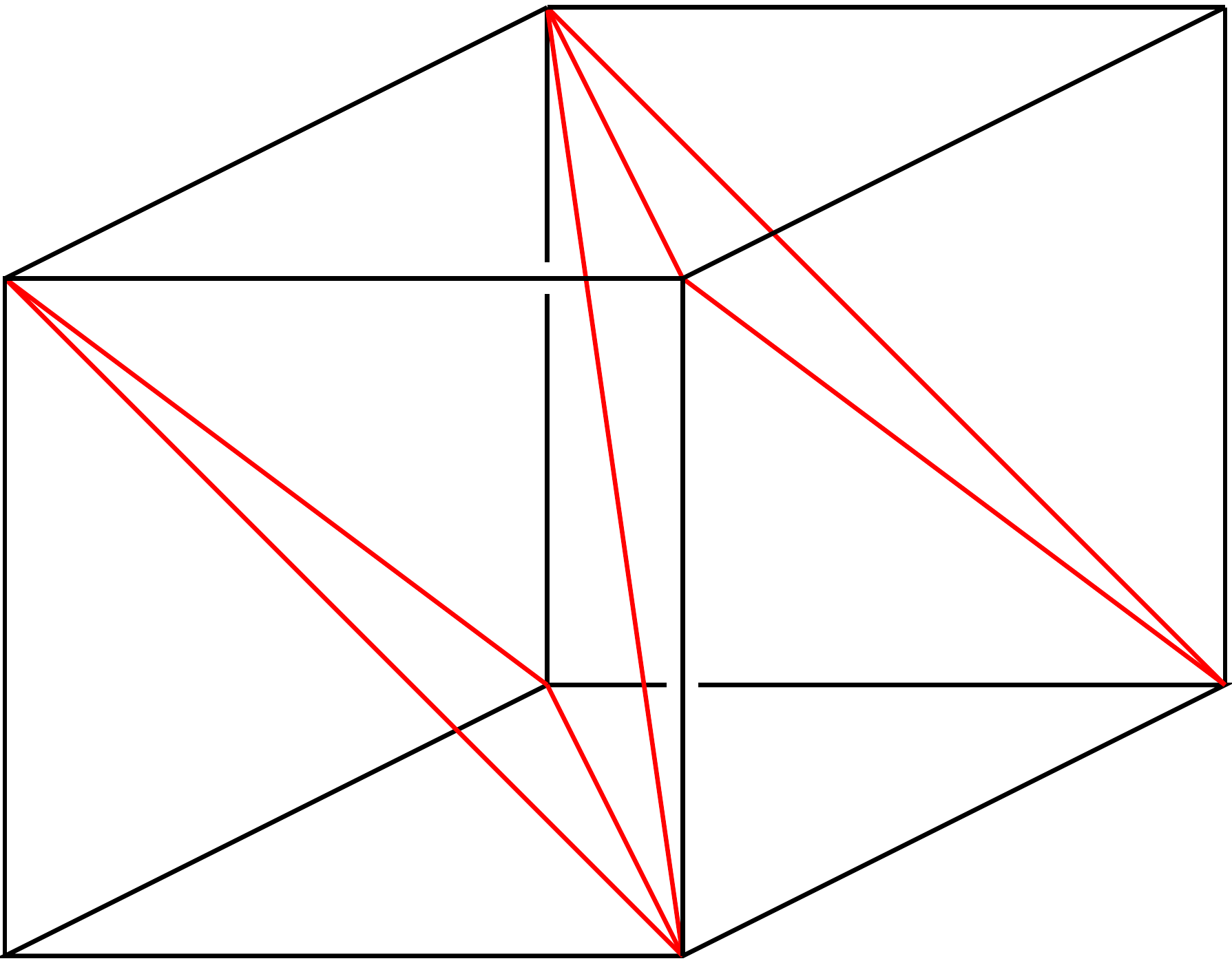}
\caption{\label{cubetr}Two fine regular triangulations of a three-dimensional cube. The `central' 3-simplex of the triangulation on the left has lattice volume 2. As there are no extra lattice points, this triangulation cannot be further refined to become unimodular. Of course the cube shown above does have a unimodular triangulation, as shown on the right. 
It may happen, however, that the central simplex of the triangulation shown on the left arises as a face of a polytope, in which case no unimodular triangulation
can exist.}
\end{center}
\end{figure}

It turns out that unimodular triangulations do not necessarily exist for polytopes of dimension $n\geq 4$. This is only relevant for Calabi-Yau manifolds of dimension $\geq 4$ though, as one may always find a triangulation which is `good enough' in the case of Calabi-Yau threefolds \cite{Batyrev94dualpolyhedra}. While any fine triangulation is also unimodular in two dimensions, a simplex can fail to be unimodular in three dimensions or more, even if it does not contain any points besides its vertices. A standard example
for this is given in figure \ref{cubetr}.
For a star triangulation of a reflexive polytope, the lattice volume\footnote{This is commonly expressed by saying that facets of reflexive polyhedra are at lattice distance one from the origin.} of any simplex is given by the volume of its `outward' face, i.e. the face which lies on a facet of $\tilde{\Delta}$. For a polytope of dimension three or less, the facets are at most two-dimensional,
so that any fine triangulation is automatically unimodular and the ambient space is smooth. For a Calabi-Yau threefold, the facets of $\tilde{\Delta}$ are three-dimensional. Hence even for a fine triangulation we are not guaranteed a smooth ambient space, as there might be a simplex $S$ with volume greater than unity. However, the singularities which are induced by such cones are point-like: they are located at the intersection of the four divisors spanning $S$. These points do not meet a generic hypersurface, so that a fine triangulation is still enough to ensure smoothness for Calabi-Yau threefolds. In the case of Calabi-Yau fourfolds, there can be singularities along curves of $\P^n_{\Sigma'}$ which are induced by faces which are not unimodular in a fine triangulation. These generically meet a hypersurface in points, so that smoothness is no longer automatic. For Calabi-Yau fourfolds, we thus need to check explicitly if a triangulation giving a resolution exists.

In the following, we will assume that we have found a fine regular star triangulation which makes $\P^n_{\Sigma'}$ a smooth projective toric variety and
leads to a smooth Calabi-Yau fourfold hypersurface $\hat{Z}$. This means we fix a fan $\Sigma'$ and a map $\phi: \Sigma'\rightarrow \Sigma$. We will 
come back to the issues discussed in this section, when we 
look at examples in section \ref{sect:examples}.

\subsection{Computing $h^{2,2}$ via decomposition}

Let $\hat{Z}$ be a non-singular compact Calabi--Yau $(n-1)$-fold 
defined as a hypersurface of a smooth toric ambient space $\P^n_{\Sigma'}$.
Let $\{\nu_i \}_{i=1,\cdots, |\Sigma'(1)|}$ be the primitive vectors 
for all the 1-dimensional cones of the toric fan $\Sigma'$.
For each one of them, there is a toric divisor $D_i$ given by 
$\{X_i=0 \} \subset \P^n_{\Sigma'}$. The restriction of $D_i$ to 
the hypersurface $\hat{Z}$ is denoted by $\hat{Y}_i$. 
This set of toric divisors $\left\{ \hat{Y}_i \right\}$ is used
to introduce a stratification (\ref{stratintro}). The primary 
stratum can be regarded as a hypersurface of $\mathbb{T}^n$, 
\begin{equation}
 Z :=  \hat{Z} \backslash Y  = 
 \hat{Z} \cap \left( \P^n_{\Sigma} \backslash \cup_i D_i \right)
 = \hat{Z} \cap \left(\mathbb{T}^n \subset \P^n_{\Sigma'} \right), 
\end{equation}
and its complement is denoted by $Y:= \cup_i \hat{Y}_i$.

For a compact non-singular irreducible algebraic variety $\hat{Z}$ 
and a divisor $Y$ with normal crossings in $\hat{Z}$ (such 
that $Z = \hat{Z} \backslash Y$ is non-singular), we can write 
a long exact sequence 
\begin{equation}
  \vcenter{\xymatrix{
 \cdots \ar[r] &
   H^{k-1}_c(Z) \ar[r] & H^{k-1}_c(\hat{Z}) \ar[r] & H^{k-1}_c(Y) 
 \ar`[rd]^<>(0.5){}`[l]`[dlll]`[d][dll] 
 & & \\
 & H^k_c(Z) \ar[r] & H^k_c(\hat{Z}) \ar[r] & H^k_c(Y) \ar[r] &  \cdots  & 
 }}
\label{eq:long-seq-cpt-supp}
\end{equation}
for the cohomology groups with compact support.
Since $\hat{Z}$ and $Y$ are both compact, $H^k(\hat{Z})$
and $H^k(Y)$ are the same as $H^k_c(\hat{Z})$ and $H_c^k(Y)$.
All the morphisms in this exact sequence are morphisms of mixed Hodge 
structure\footnote{
For mixed Hodge structure and Hodge--Deligne numbers, 
see \cite{peters2008mixed} or an easy example discussed in \cite{wikimixedHodge}.
For an algorithm of computing the Hodge--Deligne numbers, see \cite{DK}.
We do not intend to provide a systematic exposition on this (well-understood) 
issue; we just focus on deriving new results such 
as (\ref{eq:formula-H4Y-dim}) and (\ref{eq:formula-H4Z-dim}). 
The example presented in section \ref{sect:sextic}, however, is best 
suited to getting accustomed to such concepts.}  of type $(0,0)$.

In order to use the exact sequence (\ref{eq:long-seq-cpt-supp}), we first note \cite{DK} that $H^i_c(Z)$ vanishes for 
$i=0,\cdots, n-2$, and 
\begin{eqnarray} 
  h^{p,q}\left[ H^{k > (n-1)}_c(Z) \right]
& = & h^{p+1,q+1} \left[ H^{k+2}_c(\mathbb{T}^n) \right] = 
\left\{ \begin{array}{ll}
\binom{n}{k+2-n}  & p=q = k+1-n\, ,  \\
0 & {\rm otherwise}    \end{array} \right.   \, .   
     \label{eq:Hodge-Deligne-for-alg-torus}  \\
 h^{p,q}\left[H^{n-1}_c(Z) \right] & = & 0 \quad {\rm if~}p+q>n-1\, . \nn
\end{eqnarray}
More information about the non-vanishing parts of $H^{n-1}_c(Z)$ will be 
provided later.

The main focus in this article is to study $H^4(\hat{Z})$ for a
Calabi--Yau fourfold $\hat{Z}$ in a toric ambient space of dimension $n=5$.
In this case, only the weight-4 components of $H^4_c(Z)$,
$H^4_c(\hat{Z})$ and $H^4_c(Y)$ are relevant to determining $H^4(\hat{Z})$,
and we learn from \eqref{eq:long-seq-cpt-supp} that they satisfy
\begin{align}
  0 \longrightarrow & \left[H^4_c(Z) \right]^{4,0} 
    \longrightarrow \left[H^4_c(\hat{Z})\right]^{4,0} 
    \longrightarrow 0, \label{eq:H4-weight4-40}\\
 0 \longrightarrow & \left[H^4_c(Z)\right]^{3,1}
   \longrightarrow \left[H^4_c(\hat{Z})\right]^{3,1}
   \longrightarrow \left[H^4_c(Y)\right]^{3,1}
   \longrightarrow 0, \label{eq:H4-weight4-31}\\
 0 \longrightarrow & \left[H^4_c(Z)\right]^{2,2}
   \longrightarrow \left[H^4_c(\hat{Z})\right]^{2,2}
   \longrightarrow \left[H^4_c(Y)\right]^{2,2}
   \longrightarrow 0, \label{eq:H4-weight4-22} \, .
\end{align}

Let us first focus on $\left(H^4_c(Y)\right)$. 
$Y$ is not necessarily non-singular, but it can be written as
$Y= \cup_{i} \hat{Y}_i$ where each $\hat{Y}_i$ is a non-singular divisor
of $\hat{Z}$. Note that each $\hat{Y}_i$ is not necessarily irreducible. This can happen when 
$\nu_i$ is in a codimension-2 face of $\widetilde{\Delta}$.
The (co)homology groups of such geometries can be computed by using the Mayer--Vietoris spectral sequence, and one finds that 
\begin{equation}
 H^4(Y) = {\rm Kernel} \left(\left[ \oplus_i H^4(\hat{Y}_i) \right] 
  \longrightarrow
   \left[\oplus_{i<j} H^4(\hat{Y}_i \cap \hat{Y}_j) \right]\right). 
\label{eq:H4coh-Y}
\end{equation} 
As each $\hat{Y}_i$ is a smooth hypersurface of a toric variety, 
its $h^{1,0}=h^{3,1}$ vanishes (see e.g. \cite{DK}).
Hence $H^4(Y)$ contributes only to the $(H^4_c(\hat{Z}))^{2,2}$
component in (\ref{eq:H4-weight4-40}--\ref{eq:H4-weight4-22}).
Furthermore, keeping in mind that all divisors of $\hat{Z}$ occur as components of toric divisors, it is obvious from (\ref{eq:H4coh-Y}) that
all vertical $(2,2)$-forms of $\hat{Z}$ are contained in the $H^4(Y)$ quotient of $H^4(\hat{Z})$. 

We will discuss later how to compute $h^{2,2}(\hat{Y}_i) = h^{1,1}(\hat{Y}_i)$ 
and $h^{2,2}(\hat{Y}_i \cap \hat{Y}_j) = h^{0,0}(\hat{Y}_i \cap \hat{Y}_j)$  
by using toric data. In order to determine $h^4(Y) = h^{2,2}[H^4(Y)]$ 
from (\ref{eq:H4coh-Y}), we further need to know the dimension of the cokernel of the homomorphisms 
in (\ref{eq:H4coh-Y}). We claim that the homomorphism in (\ref{eq:H4coh-Y}) has a cokernel of dimension $n(n-1)/2$. 

In order to verify the claim, one needs to note that the kernel 
of (\ref{eq:H4coh-Y}) is $E_{\infty}^{0,4} = E_2^{0,4} =
{\rm Ker}[ d_1^{0,4}: E_1^{0,4} \longrightarrow E_1^{1,4} ]$ 
in the Mayer--Vietoris sequence calculation of $H^{\bullet}(Y)$.
The cokernel of the same homomorphism therefore corresponds to 
\begin{equation}
 {\rm Coker}[ d_1^{0,4}: E_1^{0,4} \longrightarrow E_1^{1,4} ] = E_2^{1,4}
   = E_{\infty}^{1,4},
\end{equation}
which gives rise to the $[H^5(Y)]^{2,2}$ component. It can be determined 
by exploiting other parts of the long exact 
sequence (\ref{eq:long-seq-cpt-supp}). 
Focusing on the weight-4 components, we find that the following is exact:
%
%
\begin{equation}
0 \longrightarrow \left[H^5_{(c)}(Y)\right]^{2,2}
  \longrightarrow \left[H^6_c(Z)\right]^{2,2} \longrightarrow 0. 
\end{equation}
Using (\ref{eq:Hodge-Deligne-for-alg-torus}), 
$h^{2,2}[H^{n+1}_c(Z)]=h^{n-3,n-3}[H^{n-3}(Z)] = n(n-1)/2$, so we find that
\begin{equation}
{\rm dim} \left({\rm Coker} \left[ \oplus_i H^4(\hat{Y}_i) \longrightarrow
   \oplus_{i<j} H^4(\hat{Y}_i \cap \hat{Y}_j) \right] \right)
=   h^{2,2}\left[H^5_{(c)}(Y)\right] 
 = \frac{n(n-1)}{2}. 
\label{eq:coker-dim}
\end{equation}
as stated before. 

\subsection{Stratification and geometry of divisors}

For the purpose of capturing $H^4_c(Y)$ and $H^4(Z)$ in terms of 
combinatorial data of the toric ambient space, we take a moment  
to digress. To start off, we describe stratifications 
of Calabi--Yau hypersurfaces $\hat{Z}$ of a toric ambient space $\P^n_{\Sigma'}$.
There are two distinct stratifications to which we pay attention:
one is associated with the toric fan $\Sigma$, and the other with its 
refinement\footnote{Remember that by assumption, $\Sigma'$ is determined by a regular star triangulation 
turning $\hat{Z}$ into a smooth hypersurface, whereas $\Sigma$ is a fan over faces of the polytope $\widetilde{\Delta}$.} $\Sigma'$.

The stratification induced by $\Sigma'$ is easier to describe, it 
descends from that on $\P^n_{\Sigma'}$ straightforwardly: each stratum of $\hat{Z}$ is of the form 
$Z_{\sigma} = \hat{Z} \cap \mathbb{T}_{\sigma}$ for $\sigma \in \Sigma'$.

The stratification corresponding to $\Sigma$ is
\begin{equation}
 \hat{Z} = (Z_{\Delta}) 
 \amalg \left( \amalg_{\Theta^{[n-1]} \leq \Delta} Z_{\Theta^{[n-1]}} \right)
 \amalg   \amalg_{2 \leq k \leq n-1} 
    \left( \amalg_{(\widetilde{\Theta}^{[k-1]},\Theta^{[n-k]}) } 
           [ E_{\widetilde{\Theta}^{[k-1]}} \times Z_{\Theta^{[n-k]}} ]
    \right); 
\label{eq:stratification-Sigma}
\end{equation}
here, the strata are labelled by the faces of $\Delta$, namely, 
$\Delta$ itself, codimension-1 faces $\Theta^{[n-1]}$'s, and all other 
faces $\Theta^{[n-k]}$'s of codimension $2 \leq k \leq n-1$.
It is understood in the expression above, and also in expressions later, 
that $(\widetilde{\Theta}^{[k-1]}, \Theta^{[n-k]})$ is a dual pair of faces.
Due to the one-to-one correspondence between faces of $\Delta$, 
the faces of $\widetilde{\Delta}$ and the cones in $\Sigma$, the strata in (\ref{eq:stratification-Sigma}) are in one-to-one 
correspondence with the cones of the fan $\Sigma$.
The primary stratum, $Z_\Delta$, corresponds to the 0-dimensional cone 
$\vec{0} \in \Sigma$, and the $Z_{\Theta^{[n-1]}}$ to the individual 
1-dimensional cones in $\Sigma(1)$.
 
The geometry of $E_{\widetilde{\Theta}^{[k-1]}}$ can be read out from 
the $(k-1)$-dimensional face $\widetilde{\Theta}^{[k-1]}$ 
of $\widetilde{\Delta}$. $E_{\widetilde{\Theta}^{[k-1]}}$ has a stratification associated with the 
maximal simplicial subdivision $\Sigma'$,
\begin{equation}
 E_{\widetilde{\Theta}^{[k-1]}} = 
      \left[ (\mathbb{T}^{k-1}\mbox{'s}) \amalg (\mathbb{T}^{k-2}\mbox{'s}) 
              \amalg \cdots \amalg {\rm points} \right],   
\label{eq:strata-Esigma}
\end{equation}
where the number of $\mathbb{T}^{k-1-p}$ in this decomposition is equal to 
the number of $p$-simplices in $\widetilde{\Theta}^{[k-1]}$ which are not 
contained in the boundary of $\widetilde{\Theta}^{[k-1]}$.

The geometries of $Z_{\Delta}$ ($k=0$) and all the other of $Z_{\Theta^{[n-k]}}$ are given by hypersurfaces of 
$\mathbb{T}^{{\rm dim}_\C(\Theta)=(n-k)}$. For each one of the $Z_{\Theta}$, only the terms in the Newton (Laurent) 
polynomial corresponding to $\Theta^{[n-k]} \cap M$ are relevant in determining $Z_{\Theta^{[n-k]}} \subset \mathbb{T}^{n-k}$
(which also means all terms originating from $\Delta \cap M$ are relevant in determining $Z_\Delta \subset \mathbb{T}^n$).
Obviously the primary stratum $Z_\Delta$ is the same as $Z := \hat{Z} \backslash Y$ introduce before.

The stratification for the fan $\Sigma'$, 
$\hat{Z} = \amalg_{\sigma \in \Sigma'} Z_\sigma$, is therefore obtained by 
decomposing the stratification (\ref{eq:stratification-Sigma}) under (\ref{eq:strata-Esigma}). 
Put differently, 
\begin{equation}
 \left[ E_{\widetilde{\Theta}^{[k-1]}} \times Z_{\Theta^{[n-k]}} \right] = 
  \amalg_{\sigma' \in \phi^{-1} \cdot \sigma (\widetilde{\Theta}^{[k-1]}) } Z_{\sigma'},  
\end{equation}
where $\phi: \Sigma' \longrightarrow \Sigma$ is the map of toric fans 
(mapping cones to cones) associated with the subdivision refining $\Sigma$ to $\Sigma'$
and $\sigma(\widetilde{\Theta}^{[k-1]})$ is the map identifying a cone of $\Sigma$ with a
face $\widetilde{\Theta}^{[k-1]}$ of $\widetilde{\Delta}$. One can think of
$E_{\widetilde{\Theta}^{[k-1]}}$ as the exceptional geometry appearing in a resolution 
of singularities of $\P^n_{\Sigma}$ associated with the $k$-dimensional cone $\sigma(\widetilde{\Theta}^{[k-1]})$.

\subsection{Combinatorial formula for $h^{2,2}[H^4_{(c)}(Y)] = 
[h^{n-3,n-3}[H^{2n-6}_c(Y)]]_{n=5}$}
\label{sect:vertial-formula-hypersurface}

Just like various Hodge numbers of a Calabi--Yau toric hypersurface
$\hat{Z}$ are computed by using the stratification and the 
Hodge--Deligne numbers of the strata \cite{DK, Batyrev94dualpolyhedra},
Hodge numbers of divisors $\hat{Y}_i$'s of a Calabi--Yau toric hypersurface $\hat{Z}$ 
can also be computed essentially with the same technique. 
Once $h^{2,2}(\hat{Y}_i)$'s are computed, it is almost straightforward 
(as already explained) to determine $h^{2,2}[H^4_c(Y)]$.

The `Euler characteristics' of Hodge--Deligne numbers of compact support cohomology groups (for some geometry $X$) are 
\begin{equation}
 e^{p,q}_c(X) := \sum_k (-)^k h^{p,q}[H^k_c(X)].
\label{eq:epqvsHodge}
\end{equation}
These numbers have the following three nice properties, which were exploited heavily in \cite{DK, Batyrev94dualpolyhedra}, and will also be in the following. 

The first two are additivity, 
\begin{equation}
 e^{p,q}_c(X_1 \amalg X_2) = e^{p,q}_c(X_1) + e^{p,q}_c(X_2),
\end{equation}
and multiplicativity,
\begin{equation}
 e^{p,q}_c(E \times Z) = \sum_{p'+p''=p} \sum_{q'+q''=q} e^{p',q'}_c(E) e^{p'',q''}_c(Z).
\end{equation}
Finally, when an algebraic variety $X$ (of complex dimension $n-1$) is compact and smooth, i.e the mixed Hodge structure on cohomology groups is pure,
\begin{equation}
 e^{p,q}_c(X) = (-)^{p+q} h^{p,q}_c(X) = (-)^{p+q}h^{p,q}(X) = 
(-)^{p+q}h^{n-1-p,n-1-q}(X)
\end{equation}
In a slight abuse of language, we will frequently refer to the $e^{p,q}$ 
as Hodge--Deligne numbers in the text.

Each divisor component $\hat{Y}_i$ (corresponding to 
$\rho_i \in \Sigma'(1)$) 
of a Calabi--Yau $(n-1)$-fold $\hat{Z}$ is compact and smooth, and hence 
$h^{n-3,n-3}(\hat{Y}_i)=h^{1,1}(\hat{Y}_i)$ is the same as\footnote{A completely parallel story holds for 
any cone $\rho \in \Sigma'(\ell)$ (not in the interior of a facet of 
$\widetilde{\Delta}$), and for any one of its Hodge numbers.} 
$e^{n-3,n-3}_c(\hat{Y}_i)$. The compact geometry $\hat{Y}_i$ has a 
stratification associated with $\Sigma'$, or to be more specific, 
\begin{equation}
 \hat{Y}_i = \left( \amalg_{\rho_i \leq \sigma \in \Sigma'} Z_\sigma \right) 
 \subset \hat{Z}.
\label{eq:collect-relevant-strata}
\end{equation}
Because of additivity, $e^{n-3,n-3}_c(\hat{Y}_i)$ is obtained by summing 
up $e^{n-3,n-3}_c(Z_\sigma)$ ($\sigma \geq \rho_i$). Using multiplicativity, this 
calculation is further boiled down to the computation of $e^{p,q}_c(Z_\Theta)$ of various faces $\Theta \leq \Delta$, for which the algorithm of \cite{DK} (in combination with (\ref{eq:Hodge-Deligne-for-alg-torus})) can be used. 

Here, we record a few crucial formulas from \cite{DK} for Hodge--Deligne 
numbers of the strata $Z_{\Theta}$. For a face of dimension $f$ there is 
the `sum rule'
\begin{equation}\label{eq:DKsumrule}
 (-1)^{f-1} \sum_q e^{p,q}(Z_{\Theta}) = (-1)^p \binom{f}{p+1} + \varphi_{f-p}(\Theta) \, ,
\end{equation}
where $f$ is the dimension of the face $\Theta$.
Here, the functions $\varphi_k$ are defined as
\begin{equation}
\varphi_k(\Theta) := \sum_{j\geq 1} (-1)^{k-j} \binom{f +1}{k-j}\ell^*(j\Theta) \, ,
\label{eq:def-varphi}
\end{equation}
where $j \Theta$ stands for the polytope which is obtained by scaling all vertices of the face $\Theta$ 
by $j$ and then taking the convex hull.
We introduce the notation 
\begin{equation}
\bar{e}^{p,q}_c(Z_{\Theta}) := e^{p,q}_c(Z_{\Theta}) - 
   \delta_{p,q} (-1)^{f-1-p} \left(\begin{array}{c} f \\ p+1 \end{array} \right),
\label{eq:def-ebar}
\end{equation}
in this article. $bar{e}^{p,q}_c(Z_{\Theta})$ is equal to $(-1)^{f-1} h^{p,q}[H^{f-1}(Z_\Theta)]$ except when $(p,q)=(0,0)$.
Obviously the sum rule (\ref{eq:DKsumrule}) can be written as 
\begin{equation}
\sum_q \bar{e}_c^{p,q}(Z_{\Theta}) = (-1)^{f-1}\varphi_{f-p}(\Theta).
\label{eq:DKsumrule-mdfd}
\end{equation}
Furthermore, the Hodge numbers obey \cite{DK}
\begin{equation}\label{eq:ep0}
e^{p,0}(Z_\Theta) = (-1)^f \sum_{\Gamma \leq \Theta, {\rm dim} \Gamma = p+1} \ell^*(\Gamma) \, ,  
\end{equation}
for any $p > 0$. For a face of dimension $f \geq 4$ we also have that
\begin{equation}\label{eq:e31}
 e^{f-2,1}(Z_\Theta) = (-1)^{f-1}\left(\varphi_2(\Theta) -  \sum_{\Gamma \leq \Theta, {\rm dim} \Gamma = f-1} \varphi_1(\Gamma) \right) .
\end{equation}
%

\subsubsection{Computation of $h^{1,1}(\hat{Y}_i)=h^{n-3,n-3}(\hat{Y}_i)$ 
for $n\geq 5$ cases}\label{sect:comph11Yhat}

We are primarily interested in $h^{2,2}(\hat{Y}_i)$, where 
$\hat{Y}_i$ is of $(n-2)=3$-dimensions in a Calabi--Yau $(n-1)=4$-fold 
$\hat{Z}$, embedded in a toric ambient space $\P^{n=5}_{\Sigma'}$. This can be regarded, 
however, as a special case of the more general problem of determining 
$h^{n-3,n-3}(\hat{Y}_i)$ of a divisor $\hat{Y}_i$ of a Calabi--Yau 
$(n-1)$-fold embedded in a toric ambient space $\P^n_{\Sigma'}$. 
We shall study the more general version of the problem for $ n \geq 5$.

For any 1-dimensional cone $\rho_i$ generated by
a primitive vector $\nu_i$ in the lattice $N$, there is a divisor $\hat{Y}_i$ of $\hat{Z}$.
Depending on which face contains $\nu_i$ in its interior, the 
computation $h^{n-3,n-3}(\hat{Y}_i)$ has to be treated separately.
The divisor $\hat{Y}_i$ is empty if $\nu_i$ is interior to an $n-1$-dimensional 
face of $\widetilde{\Delta}$.
Let $\phi(\rho_i) \in \Sigma(k_0)$, i.e., $\nu_i$ is an interior point 
of a $(k_0-1)$-dimensional face $\widetilde{\Theta}^{[k_0-1]}_i \leq 
\widetilde{\Delta}$. The cases we have to study are then
\begin{itemize}
\item $k_0=1$: i.e., $\nu_i$ is one of vertices of $\widetilde{\Delta}$.
\item $1<k_0<n-3$: (this case is absent if $n=5$.)
\item $k_0=n-3$: i.e., $\nu_i$ is an interior point of a codimension-4 face 
  (an edge, if $n=5$).
\item $k_0=n-2$: i.e., $\nu_i$ is an interior point of a codimension-3 face
  (a 2dim face, if $n=5$).
\item $k_0=n-1$: i.e., $\nu_i$ is an interior point of a codimension-2 face.
  (a 3dim face, if $n=5$)
\end{itemize}
We work on those five cases one-by-one from now.
\\
\\
{\bf The case $k_0=1$}: \\
\\
Here, $\nu_i = \widetilde{\Theta}^{[0]}_i$ is a vertex of $\widetilde{\Delta}$.
$\hat{Y}_i$ is composed of the following strata:
\begin{eqnarray}
 Z_{\Theta^{[n-1]}_i}, & \quad &  
 [{\rm pt} \times Z_{\Theta^{[n-2]}} ]_{(\widetilde{\Theta}^{[1]}, \Theta^{[n-2]})},
 \quad 
 \left[\left(\mathbb{T}\mbox{'s}+{\rm pts}\right) \times Z_{\Theta^{[n-3]}} 
   \right]_{(\widetilde{\Theta}^{[2]}, \Theta^{[n-3]})}, \nonumber \\
& & \qquad \qquad \cdots \cdots, \quad 
 \left[ \left(\mathbb{T}^{n-3}\mbox{'s}\sim {\rm points}\right) 
         \times Z_{\Theta^{[1]}} 
 \right]_{(\widetilde{\Theta}^{[n-2]}, \Theta^{[1]})}.
\label{eq:h11-Yi-case-1}
\end{eqnarray}
Note that a stratum $Z_\sigma \subset 
[E_{\widetilde{\Theta}^{[k-1]}} \times Z_{\Theta^{[n-k]}}]$ contributes to $\hat{Y}_i$ 
only when $\widetilde{\Theta}^{[k-1]}$ contains the vertex $\nu_i$, i.e. 
$\widetilde{\Theta}^{[k-1]} \geq \widetilde{\Theta}^{[0]}_i$ 
(equivalently $\Theta^{[n-k]} \leq \Theta^{[n-1]}_i$). Furthermore, 
$\mathbb{T}^{k-1-p} \subset E_{\widetilde{\Theta}^{[k-1]}}$ 
only contributes when the corresponding $p$-simplex in $\widetilde{\Theta}^{[k-1]}$ 
contains $\nu_i$ as one of its faces. This is why, for example, 
only one point $\mathbb{T}^0$ (one 1-simplex) from 
$E_{\widetilde{\Theta}^{[1]}}$ of a given dual pair 
$(\widetilde{\Theta}^{[1]}, \Theta^{[n-2]})$ contributes to $\hat{Y}_i$.
Such strata combined are denoted by $[E_{\widetilde{\Theta}^{[k-1]}} \times Z_{\Theta^{[n-k]}}]^{\rho_i \leq}$ in this article.

From the first stratum, 
\begin{equation}
 e^{n-3,n-3}_c(Z_{\Theta^{[n-1]}_i}) =   - \binom{n-1}{n-2} = -(n-1)
\end{equation}
picking up the contribution from $h^{n-3,n-3}[H^{2n-3}_c(Z_{\Theta^{[n-1]}_i})]$.

From the $k$-th group of strata ($k=2, \cdots, (n-2)$), each pair 
$(\widetilde{\Theta}^{[k-1]}, \Theta^{[n-k]})$ gives rise to  
\begin{eqnarray}
 e^{n-3,n-3}_c\left([E_{\widetilde{\Theta}^{[k-1]}} \times Z_{\Theta^{[n-k]}}]^{\rho_i \leq}
            \right) 
                   & = & \left[ {}_{\nu_i}\ell^*_1(\widetilde{\Theta}^{[k-1]}) \times
                      e^{k-2,k-2}_c(\mathbb{T}^{k-2}) \right]
    \times [e^{n-k-1,n-k-1}_c(Z_{\Theta^{[n-k]}})]    \nonumber \\ 
 & = & {}_{\nu_i}\ell^*_1(\widetilde{\Theta}^{[k-1]})\, . 
\label{eq:h11-Y-edge}
\end{eqnarray}
Here, we have introduce the notation ${}_{\nu_i}\ell^*_1(\widetilde{\Theta})$ for the number
of internal 1-simplices in the face $\widetilde{\Theta}$ which end on $\nu_i$ or, in other words,
have $\nu_i$ as a face. In the language of cones
\begin{equation}
{}_{\nu_i}\ell^*_1(\widetilde{\Theta}) := \left| \left\{ \sigma \in \Sigma'(2) \cap \phi^{-1} \cdot \sigma (\widetilde{\Theta}) 
    \; | \; \rho_i \leq \sigma \right\}\right|
\end{equation}
We also introduce ${}_{\nu_i}\ell_1(\widetilde{\Delta}_{\leq p})$ for the number of 1-simplices (again containing $\nu_i$ as a face) 
contained in the $p$-skeleton (faces of dimension $p$ or less) of $\widetilde{\Delta}$.

The contribution from the last group of strata, $k=n-1$, 
is the same as above, except that $e^{n-k-1,n-k-1}_c(Z_{\Theta^{[n-k]}})$ 
is not necessarily 1. $Z_{\Theta^{[1]}}$ consists of
\begin{equation}
e^{n-k-1,n-k-1}_c(Z_{\Theta^{[n-k]}}) = e^{0,0}_c=1+\ell^*(\Theta^{[1]}),
\end{equation}
points, which is not necessarily 1.
Hence the contribution to $e^{n-3,n-3}_c$ is $e^{0,0}_c(Z_{\Theta^{[1]}})$ times 
larger than that of (\ref{eq:h11-Y-edge}). We thus finally obtain 
\begin{eqnarray}
 h^{1,1}(\hat{Y}_i)|_{{\rm if~}k_0=1}  & = &   
 {}_{\nu_i}\ell_1(\widetilde{\Delta}_{\leq n-2}) 
 - (n-1) + \sum_{(\widetilde{\Theta}^{[n-2]},\Theta^{[1]})} 
    {}_{\nu_i}\ell^*_1(\widetilde{\Theta}^{[n-2]})
   \times  \ell^*(\Theta^{[1]}). \label{eq:formula-h11-divisor} 
\end{eqnarray}
This formula for the divisors $\hat{Y}_i$ of $\hat{Z}$ is quite like 
that of $\hat{Z}$ in (\ref{eq:bath11}). The first two terms originate from 
toric divisors (divisors of the ambient space $\P^n_{\Sigma'}$ of $\hat{Z}$ 
and the linear equivalence restricted to $\hat{Y}_i$). 
The correction term at the end of (\ref{eq:formula-h11-divisor}) also looks 
quite similar to the last term of (\ref{eq:bath11}); we will come back 
in section \ref{hyperresultVvsH} to discuss the geometric interpretation 
of this term in more detail. 
\\
\\
{\bf The three cases $1< k_0 < (n-3)$, $k_0=(n-3)$ and $k_0=(n-2)$}: 
\\
\\
The stratification of $\hat{Y}_i$ is described schematically by 
\begin{equation}
 [ E_{\widetilde{\Theta}^{[k_0-1]}_i} \times Z_{\Theta^{[n-k_0]}_i} ]^{\rho_i \leq}, \quad 
  \cdots, 
 \left[ E_{\widetilde{\Theta}^{[k-1]}} \times Z_{\Theta^{[n-k]}} 
 \right]^{\rho_i \leq}, \cdots, 
 \left[ E_{\widetilde{\Theta}^{[n-2]}} \times Z_{\Theta^{[1]}} 
 \right]^{\rho_i \leq}.
\end{equation}
In all the three cases, the contributions to $e^{n-3,n-3}_c$ from 
all but the first group of strata remain the same as in the $k_0=1$ case.
The first group of strata gives rise to 
\begin{eqnarray}
 e^{n-3,n-3}_c([E_{\widetilde{\Theta}^{[k_0-1]}_i} \times Z_{\Theta^{[n-k_0]}_i}]^{\geq \rho_i})
 & = &  e^{k_0-1,k_0-1}_c(E_{\widetilde{\Theta}^{[k_0-1]}_i}^{\geq \rho_i}) 
       e^{n-k_0-2,n-k_0-2}_c( Z_{\Theta^{[n-k_0]}_i} ) \nn \\
 & &    +\,\, e^{k_0-2,k_0-2}_c(E_{\widetilde{\Theta}^{[k_0-1]}_i}^{\geq \rho_i}) 
       e^{n-k_0-1,n-k_0-1}_c( Z_{\Theta^{[n-k_0]}_i} ).
\end{eqnarray}
In all the three cases, $1<k_0<n-3$, $k_0=n-3$ and $k_0=n-2$, we have that
\begin{itemize}
 \item $e^{k_0-1,k_0-1}_c(E_{\widetilde{\Theta}^{[k_0-1]}_i}^{\geq \rho_i}) = 1$, because 
the only contribution comes from $\mathbb{T}^{k_0-1}$ corresponding to 
the 0-plex $\nu_i$ itself 
\item $e^{n-k_0-1,n-k_0-1}_c(Z_{\Theta^{[n-k_0]}_i})=1$
\item $e^{k_0-2,k_0-2}_c(E_{\widetilde{\Theta}^{[k_0-1]}_i}^{\geq \rho_i}) = 
 -(k_0-1) + 
 {}_{\nu_i}\ell^*_1(\widetilde{\Theta}^{[k_0-1]}) $
\end{itemize}
The expression for the factor $e^{n-k_0-2,n-k_0-2}_c( Z_{\Theta^{[n-k_0]}} )$, 
however, is different in all the three cases.
If $1 < k_0 < n-3$, then $2(n-k_0-2) > (n-k_0-1)$, so that 
\begin{equation}
 e^{n-k_0-2,n-k_0-2}_c(Z_{\Theta^{[n-k_0]}_i}) = - \binom{n-k_0}{n-k_0-1}
 = -(n-k_0).
\end{equation}
For the two other cases, $k_0=(n-3)$ and $k_0=(n-2)$, however, 
there is an extra term in $e^{n-k_0-2,n-k_0-2}_c(Z_{\Theta^{[n-k_0]}})$, which 
we denote by $\bar{e}^{n-k_0-2,n-k_0-2}_c(Z_{\Theta^{[n-k_0]}})$ hereafter.

Therefore, it turns out that $h^{1,1}(\hat{Y}_i)$ remains the same as in 
(\ref{eq:formula-h11-divisor}) for $\hat{Y}_i$ with $1<k_0<(n-3)$, while
\begin{eqnarray}
h^{1,1}(\hat{Y}_i)|_{{\rm if~}k_0=(n-3),(n-2)} & = &  
%
%
{}_{\nu_i}\ell_1(\widetilde{\Delta}_{\leq n-2}) 
 - (n-1) + \sum_{(\widetilde{\Theta}^{[n-2]},\Theta^{[1]})} 
    {}_{\nu_i}\ell^*_1(\widetilde{\Theta}^{[n-2]})
   \times  \ell^*(\Theta^{[1]})   \nonumber \\   
 & & + \bar{e}^{n-k_0-2,n-k_0-2}_c(Z_{\Theta^{[n-k_0]}_i}). 
\label{eq:h11-Yi-case-middle}
\end{eqnarray}

The correction term in the last line of (\ref{eq:h11-Yi-case-middle}) is determined by using the sum rule in \cite{DK},
\eqref{eq:DKsumrule}. In the case $k_0=n-3$, $Z_{\Theta^{[n-k_0=3]}_i}$ is a surface given by a Laurent 
polynomial in $\mathbb{T}^3$. Therefore, 
\begin{eqnarray}
 \bar{e}^{n-k_0-2,n-k_0-2}_c(Z_{\Theta^{[n-k_0]}_i} ) & = & \bar{e}^{1,1}_c(Z_{\Theta^{[3]}_i})
  = \varphi_2(\Theta^{[3]}_i) - 
    \sum_{\Theta^{[2]}\leq \Theta^{[3]}_i} \varphi_1(\Theta^{[2]}), \\
 & = & \ell^*(2\Theta^{[3]}_i) - 4 \ell^*(\Theta^{[3]}_i) - 
     \sum_{\Theta^{[2]}\leq \Theta^{[3]}_i} \ell^*(\Theta^{[2]}).  
\label{eq:h11-Yi-case-middle-aux-n-3}
\end{eqnarray}
In the case $k_0=n-2$, $Z_{\Theta^{[n-k_0]}_i}$ is a curve given by a Laurent 
polynomial in $\mathbb{T}^2$. Thus, 
\begin{eqnarray}
 \bar{e}^{n-k_0-2, n-k_0-2}_c(Z_{\Theta^{[n-k_0]}_i}) & = & \bar{e}^{0,0}_c(Z_{\Theta^{[2]}_i})
 = - \varphi_2(\Theta^{[2]}_i) + \varphi_1(\Theta^{[2]}_i), \\
 & = & - \ell^*(2\Theta^{[2]}_i) + 4 \ell^*(\Theta^{[2]}_i).
\label{eq:h11-Yi-case-middle-aux-n-2}
\end{eqnarray}

Again, the first two lines of (\ref{eq:h11-Yi-case-middle}) are understood as
toric divisors of $\hat{Z}$ restricted on $\hat{Y}_i$. 
The $(n-1)$-dimensional redundancy among them is simply given by the number 
of toric linear equivalences. The geometric interpretation of the correction 
term $\bar{e}^{n-k_0-3,n-k_0-3}_c(Z_{\Theta^{[n-k_0]}})$ is discussed in detail in 
section \ref{hyperresultVvsH}. 
\\
\\
{\bf The cases $k_0=n-1$}:
\\
\\
Finally, let us work out $h^{1,1}(\hat{Y}_i)$ for $k_0=n-1$, where $\nu_i$ is in the interior of an codimension-2 face 
$\widetilde{\Theta}^{[n-2]}_i$ of $\widetilde{\Delta}$. 
The dual face of $\Delta$ is of dimension 1 and denoted by $\Theta^{[1]}_i$.
The divisor $\hat{Y}_i$ of the Calabi--Yau $(n-1)$-fold $\hat{Z}$ consists 
of $e^{0,0}_c(Z_{\Theta^{[1]}_i}) = [1 + \ell^*(\Theta^{[1]}_i)]$ irreducible 
components, each of which is a toric variety $E_{\widetilde{\Theta}^{[n-2]}_i}^{\geq \rho_i}$ given by $p$-simplices 
$\widetilde{\Theta}^{[n-2]}_i$ containing $\nu_i$ as a face.
Therefore, 
\begin{equation}
 e^{n-3,n-3}_c(E_{\widetilde{\Theta}^{[n-2]}_i}^{\geq \rho_i}) = 
  {}_{\nu_i}\ell^*_1(\widetilde{\Theta}^{[n-2]}) - (n-2)\,.
  \label{eq:h11-Yi-case-n-1}
\end{equation}
Note that the  ``number of linear equivalences'' is different from that in all the other 
cases where $k_0<(n-1)$: it is now $n-2$ instead of $n-1$. Finally, 
$h^{1,1}(\hat{Y}_i)=h^{n-3,n-3}(\hat{Y}_i)$ is obtained by multiplying the above expression with
$e^{0,0}_c(Z_{\Theta^{[1]}_i}) = [1+\ell^*(\Theta^{[1]}_i)]$.

\subsubsection{The result}
\label{sect:theresult}

Combining the results of (\ref{eq:h11-Yi-case-1}, 
\ref{eq:h11-Yi-case-middle}, \ref{eq:h11-Yi-case-middle-aux-n-3}, 
\ref{eq:h11-Yi-case-middle-aux-n-2}, \ref{eq:h11-Yi-case-n-1}), 
$h^{n-3,n-3}[H^{2n-6}_c(Y)] = h^{2,2}[H^4_c(Y)]$ is determined.  
As we have already explained, 
\begin{eqnarray}
 h^{n-3,n-3}[H^{2n-6}_c(Y)] & = & \sum_i h^{n-3,n-3}(\hat{Y}_i) 
   \label{eq:h22-H4Y-idea} \\
   & & - 
   \left[ \sum_{i<j} \left| \left\{ {\rm irr.~nonempty~components~of~}
                \hat{Y}_i \cap \hat{Y}_j \right\}\right| 
       -  \frac{n(n-1)}{2} \right]. \nonumber
\end{eqnarray}

Noting that all one-simplices on the $n-3$-skeleton of $\widetilde{\Delta}$
contribute twice in the first line of (\ref{eq:h22-H4Y-idea}) 
and by $(-1)$ in the second line, and that all one-simplices internal
to a $n-2$-dimensional face of $\widetilde{\Delta}$
do so by a factor of $[2+(-1)] \times [1+\ell^*(\Theta^{[1]})]$, we find that 
\begin{eqnarray} \label{eq:formula-H4Y-dim}
h^{n-3,n-3}[H^{2n-6}_c(Y)] & = & 
 \ell_1(\widetilde{\Delta}_{\leq n-2})
       + \frac{n(n-1)}{2} 
    \\
& - & (n-1) \left[ \ell(\widetilde{\Delta}) 
    - \sum_{\widetilde{\Theta}^{[n-1]} \leq \widetilde{\Delta}} 
         \ell^*(\widetilde{\Theta}^{[n-1]})
    - \sum_{\widetilde{\Theta}^{[n-2]} \leq \widetilde{\Delta}}
         \ell^*(\widetilde{\Theta}^{[n-2]}) -1 \right] \nonumber \\
& - & (n-2) \sum_{(\widetilde{\Theta}^{[n-2]}, \Theta^{[1]})} 
       \ell^*(\widetilde{\Theta}^{[n-2]})
      \left[ 1 + \ell^*(\Theta^{[1]}) \right], \nonumber \\
& + & \sum_{(\widetilde{\Theta}^{[n-2]}, \Theta^{[1]})} 
     \ell_1^*( \widetilde{\Theta}^{[n-2]})
        \times \ell^*( \Theta^{[1]} ), \nonumber \\
& - & \sum_{(\widetilde{\Theta}^{[n-3]},\Theta^{[2])})} 
    \ell^* ( \widetilde{\Theta}^{[n-3]} ) \times 
    \left\{ \ell^*( 2 \Theta^{[2]} ) - 4 \ell^* (\Theta^{[2]}) \right\}.
  \nonumber \\
& + & \sum_{(\widetilde{\Theta}^{[n-4]}, \Theta^{[3]})}
     \ell^*(\widetilde{\Theta}^{[n-4]}) \times     
     \left\{ \ell^*(2\Theta^{[3]})
           - 4 \ell^*(\Theta^{[3]})
           - \sum_{\Theta^{[2]}\leq \Theta^{[3]}} \ell^*(\Theta^{[2]}) \right\}. 
    \nonumber 
\end{eqnarray}
Here, $\ell_1(\widetilde{\Delta}_{\leq p})$ denotes the number of $1$-simplices on the $p$-skeleton of $\widetilde{\Delta}$ and
$\ell_1^*( \widetilde{\Theta})$ the number of internal $1$-simplices on a face $\widetilde{\Theta}$.

\subsection{Combinatorial formula for $h^{2,2}[H^4_c(Z_\Delta)] = 
[h^{n-3,2}[H^{n-1}_c(Z_{\Delta})]]_{n=5}$}
\label{sect:h22-H4Z}

Using the sum rule \eqref{eq:DKsumrule-mdfd} of \cite{DK} we immediately 
find that,
\begin{eqnarray}\label{eq:sumruleexpl}
 \sum_{q=0}^2 h^{n-3,q}\left[ H^{n-1}_c(Z_\Delta)\right] & = & \varphi_3(\Delta), \\
 \sum_{q=0}^1 h^{n-3,q}\left[ H^{n-2}_c(Z_{\Theta^{[n-1]}})\right] & = &
  \varphi_2(\Theta^{[n-1]}), \label{eq:sumruleexpl-2} \\
 h^{n-3,0}\left[ H^{n-3}_c(Z_{\Theta^{[n-2]}})\right] & = & \varphi_1(\Theta^{[n-2]}).
 \label{eq:sumruleexpl-3}
\end{eqnarray}
There are six unknowns in the left-hand sides, so that we need three more
conditions in order to determine 
$h^{n-3,2}[H^{n-1}_c(Z)]$, i.e. $h^{2,2}[H^4_c(Z)]$ ($n=5$).
They come from the fact that all of the Hodge numbers of components where
neither $p=q$, nor $p+q=n-1$ are tightly constrained for a compact smooth hypersurface of a toric variety.
For $p=0$ (or $q=0$) such Hodge numbers even vanish.
Thus, for the closure of $Z_{\Theta^{[n-1]}}$ (satisfying the condition above), 
it follows that 
\begin{equation}
 - h^{n-3,0}[H^{n-2}_c(Z_{\Theta^{[n-1]}})] + 
 \sum_{\Theta^{[n-2]}\leq \Theta^{[n-1]}} h^{n-3,0}[H^{n-3}_c(Z_{\Theta^{[n-2]}})] = 0 
\end{equation}
for any facets $\Theta^{[n-1]}$ of $\Delta$.
Similarly for the closure of $Z_\Delta = \hat{Z}$, we obtain 
\begin{eqnarray}
 0 & = & h^{n-3,0}(\hat{Z})  \\
   & = & h^{n-3,0}[H^{n-1}_c(Z_\Delta)]
 - \sum_{\Theta^{[n-1]} \leq \Delta} h^{n-3,0}[H^{n-2}_c(Z_{\Theta^{[n-1]}})]
 + \sum_{\Theta^{[n-2]}\leq \Delta} h^{n-3,0}[H^{n-3}_c(Z_{\Theta^{[n-2]}} )] \times 1, 
  \nonumber 
\end{eqnarray}
where the factor $1$ is due to 
$e^{0,0}(E_{\widetilde{\Theta}^{[2]}}) = (-)^{2+1}[\chi({\rm 2d~ball})-\chi(S_1)]=1$
(see e.g., \cite{DK}). 

The $(n-3,1)$ component (i.e. $h^{2,1}$ for the case $n=5$) does not vanish, 
but is given by the formula (\ref{eq:bath21}) (see \cite{Batyrev1996901}, 
with a correction in \cite{Klemm:1996ts,Kreuzer:1997zg}):
\begin{eqnarray}
(-) \!\!\!\!\!\!\! \!\!
   \sum_{(\widetilde{\Theta}^{[2]}, \Theta^{[n-3]})} \ell^*(\widetilde{\Theta}^{[2]}) 
   \times \ell^*(\Theta^{[n-3]})  & = & 
   (-)^{n-1} e^{n-2,2}_c(\hat{Z})
 = (-)^{n-1} e^{n-3,1}_c(\hat{Z})
 \\
& = & h^{n-3,1}[H^{n-1}_c(Z_\Delta)]
 - \sum_{\Theta^{[n-1]} \leq \Delta} h^{n-3,1}[H^{n-2}_c(Z_{\Theta^{[n-1]}})] \nonumber \\
& & 
 + \sum_{(\widetilde{\Theta}^{[1]}, \Theta^{[n-2]})} 
   h^{n-4,0}[H^{n-3}_c(Z_{\Theta^{[n-2]}})] \times
     e^{1,1}_c(E_{\widetilde{\Theta}^{[1]}})  \nonumber \\
& & - \sum_{(\widetilde{\Theta}^{[2]}, \Theta^{[n-3]})} 
   h^{n-4,0}[H^{n-4}_c(Z_{\Theta^{[n-3]}})] \times 
     e^{1,1}_c(E_{\widetilde{\Theta}^{[2]}}). \nonumber 
\end{eqnarray}
On the right hand side, 
$e^{1,1}_c(E_{\widetilde{\Theta}^{[1]}}) = \ell^*(\widetilde{\Theta}^{[1]})$, 
and $e^{1,1}_c(E_{\widetilde{\Theta}^{[2]}})$ is given by 
$ -2 \ell^*(\widetilde{\Theta}^{[2]}) + \ell_1^*(\widetilde{\Theta}^{[2]}) $.

In order to solve the six unknowns by using the six conditions above, 
we need to determine $h^{n-4,0}[H^{n-3}_c(Z_{\Theta^{[n-2]}})]$ and 
$h^{n-4,0}[H^{n-4}_c(Z_{\Theta^{[n-3]}})]$ appearing in the last condition.
As for $h^{n-4,0}[H^{n-3}_c(Z_{\Theta^{[n-2]}})]$, using a similar condition 
for the closure of $Z_{\Theta^{[n-2]}}$, 
\begin{equation}
 h^{n-4,0}[H^{n-3}_c(Z_{\Theta^{[n-2]}})] =
 \sum_{\Theta^{[n-3]} \leq \Theta^{[n-2]}} h^{n-4,0}[H^{n-4}_c(Z_{\Theta^{[n-3]}})] 
 = \sum_{\Theta^{[n-3]} \leq \Theta^{[n-2]}} \varphi_1(\Theta^{[n-3]}).
\label{eq:aux-h10-Z-Theta-codim2}
\end{equation}
For $h^{n-4,0}[H^{n-4}_c(Z_{\Theta^{[n-3]}})]$, we can simply use 
$h^{n-4,0}[H^{n-4}_c(Z_{\Theta^{[n-3]}})] = \varphi_1(\Theta^{[n-3]})$.

Now we can solve \eqref{eq:sumruleexpl}:
\begin{equation}
 h^{n-3,0}[H^{n-2}_c(Z_{\Theta^{[n-1]}})] = \sum_{\Theta^{[n-2]} \leq \Theta^{[n-1]}} 
    \varphi_1(\Theta^{[n-2]}),
\label{eq:temp1}
\end{equation}
\begin{equation}
 h^{n-3,0}[H^{n-1}_c(Z_\Delta)] = \sum_{\Theta^{[n-1]} \leq \Delta}
 \sum_{\Theta^{[n-2]} \leq \Theta^{[n-1]}} 
  \varphi_1(\Theta^{[n-2]}) - \sum_{\Theta^{[n-2]} \leq \Delta} \varphi_1(\Theta^{[n-2]}) 
 = \sum_{\Theta^{[n-2]} \leq \Delta} \varphi_1(\Theta^{[n-2]}),
\label{eq:temp2}
\end{equation}
where we used the fact that any $\Theta^{[n-2]}$ is codimension-1
in the boundary of $\Delta$, so that it is shared by precisely two faces
$\Theta^{[n-1]}$; those two results (\ref{eq:temp1}, \ref{eq:temp2}) are 
examples of (\ref{eq:ep0}).\footnote{Prop. 5.8 of \cite{DK} 
obtained the formula (\ref{eq:ep0}) by using the algorithm reviewed here.}
Let us continue with
\begin{eqnarray}
 h^{n-3,1}[H^{n-1}_c(Z_\Delta)] & = & 
   \sum_{\Theta^{[n-1]} \leq \Delta} \varphi_2(\Theta^{[n-1]})
 - \sum_{\Theta^{[n-1]} \leq \Delta} \quad \sum_{\Theta^{[n-2]} \leq \Theta^{[n-1]}} 
       \varphi_1(\Theta^{[n-2]}) \nonumber \\
&&  - \sum_{(\widetilde{\Theta}^{[1]},\Theta^{[n-2]})} 
   \ell^*(\widetilde{\Theta}^{[1]}) 
    \times 
   \sum_{\Theta^{[n-3]} \leq \Theta^{[n-2]}} \ell^*(\Theta^{[n-3]}), 
    \label{eq:formula-H4Z-21cmp}  \\
 && + \sum_{(\widetilde{\Theta}^{[2]},\Theta^{[n-3]})} 
   \left(-2 \ell^*(\widetilde{\Theta}^{[2]}) + \ell_1^*(\widetilde{\Theta}^{[2]})\right)\times \ell^*(\Theta^{[n-3]}) \nonumber \\
 & &  - \sum_{(\widetilde{\Theta}^{[2]},\Theta^{[n-3]})} 
      \ell^*(\widetilde{\Theta}^{[2]}) \ell^* (\Theta^{[n-3]}) . \nonumber 
\end{eqnarray}
Therefore, we arrive at a the formula
\begin{eqnarray}\label{eq:formula-H4Z-dim}
 h^{n-3,2}[H^{n-1}_c(Z_\Delta)] & = & 
 \varphi_3(\Delta) - \sum_{\Theta^{[n-1]} \leq \Delta} \varphi_2(\Theta^{[n-1]})
 + \sum_{\Theta^{[n-2]} \leq \Delta} \varphi_1(\Theta^{[n-2]}) \nonumber \\
  & & \qquad + 
  \sum_{(\widetilde{\Theta}^{[1]},\Theta^{[n-2]}) } \ell^*(\widetilde{\Theta}^{[1]}) \times 
       \sum_{ \Theta^{[n-3]} \leq \Theta^{[n-2]} } \ell^*(\Theta^{[n-3]}) \\
 & & \qquad + 
 \sum_{ (\widetilde{\Theta}^{[2]}, \Theta^{[n-3]})}
    \left(3 \ell^*(\widetilde{\Theta}^{[2]}) - \ell_1^*(\widetilde{\Theta}^{[2]})\right)
   \times \ell^*( \Theta^{[n-3]} ).
  \nonumber  
\end{eqnarray}
%

\subsection{Vertical, horizontal and the remaining components}
\label{hyperresultVvsH}

Due to the exact sequence \eqref{eq:H4-weight4-22}, we can now 
compute $h^{2,2}(\hat{Z})$ by summing $h^{2,2}[H^4(Y)]$ 
in \eqref{eq:formula-H4Y-dim} and $h^{2,2}(\hat{Z})$ 
in \eqref{eq:formula-H4Z-dim}. In fact, this is how 
Ref. \cite{Batyrev94dualpolyhedra} derived the formula 
(\ref{eq:bath11}, \ref{eq:bath31}) for $h^{1,1}$ and $h^{3,1}$ of 
a Calabi--Yau hypersurface fourfold $\hat{Z}$, and it is a straightforward 
generalization to use (\ref{eq:H4-weight4-22}) to determine $h^{2,2}(\hat{Z})$
in this way. The dimension of the $H^{2,2}$ component, however, does not 
have to be determined in this way: for a smooth fourfold $\hat{Z}$,
the formula \cite{Klemm:1996ts,Kreuzer:1997zg}
\begin{equation}\label{h22fromotherhnumbers}
h^{2,2}(\hat{Z}) = 44 + 4\left(h^{1,1}(\hat{Z})+h^{3,1}(\hat{Z})\right) 
    - 2 h^{2,1}(\hat{Z}) \, .
\end{equation}
makes it possible to determine $h^{2,2}(\hat{Z})$ from the three other 
Hodge numbers that are already determined 
through (\ref{eq:bath11}--\ref{eq:bath21}). 

The exact sequence (\ref{eq:H4-weight4-22}) is still very useful for the 
purpose of studying the decomposition (\ref{eq:H4-decomp-H-RM-V}). 
This is because all the independent divisors of a hypersurface Calabi--Yau 
fourfold $\hat{Z} \subset \P^5_{\Sigma'}$ appear in the form of irreducible 
components of toric divisors $\hat{Y}_i$ of $\hat{Z}$. We can therefore 
take a set of generators of the vertical component $H^{2,2}_V(\hat{Z})$ 
within $[H^4(Y)]^{2,2}$. Homological equivalence among the generators has already 
been taken care of in the study of $[H^4(Y)]^{2,2}$ in 
section \ref{sect:vertial-formula-hypersurface}.
In this section, we start off with identifying which 
subspace of $[H^4(Y)]^{2,2}$ corresponds to the vertical component 
$H^{2,2}_V(\hat{Z})$. Mirror symmetry is then used to identify the horizontal component $H^{2,2}_H(\hat{Z};\C) := H^4_H(\hat{Z}; \C) \cap H^{2,2}(\hat{Z}; \C)$. 
We will comment on the geometry associated with the remaining 
component $H^{2,2}_{RM}(\hat{Z};\C)$ at the end.

\subsubsection{The vertical component}\label{sect:vertcomp}

Let us discuss which subspace of $H^{1,1}(\hat{Y}_i)$ is generated by vertical cycles for the 
cases with $k_0=1$, $n-3=2$, $n-2=3$ and $n-1 = 4$. 
We begin with a divisor $\hat{Y}_i$ corresponding to a vertex $\nu_i$ of 
$\widetilde{\Delta}$ (i.e., $k_0=1$).
The formula (\ref{eq:formula-h11-divisor}) can be rewritten as 
\begin{equation}
h^{1,1}(\hat{Y}_i) =  {}_{\nu_i}\ell_1(\widetilde{\Delta}_{\leq n-2}) 
 - (n-1) + \sum_{(\widetilde{\Theta}^{[n-2]},\Theta^{[1]})} \!\!\!\!\!\!\!\!
    \left[ {}_{\nu_i}\ell^{*\circ}_1(\widetilde{\Theta}^{[n-2]}) + 
          {}_{\nu_i}\ell^{*\bullet}_1(\widetilde{\Theta}^{[n-2]}) \right]
   \times  \ell^*(\Theta^{[1]}), 
\label{eq:formula-h11-divisor-split}
\end{equation}
where ${}_{\nu_i} \ell^{*\circ}_1(\widetilde{\Theta}^{[n-2]})$ is the number 
of 1-simplices whose endpoints are $\nu_i$ and one of the interior points of 
a face $\widetilde{\Theta}^{[n-2]}$, and 
${}_{\nu_i} \ell^{*\bullet}_1(\widetilde{\Theta}^{[n-2]})$ is the number of
1-simplices which run through the interior of $\widetilde{\Theta}^{[n-2]}$, but 
have both end points on the boundary of $\widetilde{\Theta}^{[n-2]}$.
The first two terms account for the dimension of the space of algebraic 
cycles obtained as the intersection of $\hat{Y}_i$ with another toric divisor 
(i.e., a divisor of $\hat{Z}$ that descends from the toric ambient space 
$\P^{n=5}_{\Sigma'}$). When the intersection of a pair of toric divisors 
of the form $\hat{Y}_i \cap \hat{Y}_j$ corresponds to a 1-simplex  
counted in ${}_{\nu_i}\ell_1^{*\circ}(\widetilde{\Theta}^{[n-2]})$, however, 
the toric divisor $\hat{Y}_j$ corresponding to an interior point of a 
codimension-two face of $\widetilde{\Delta}$ consists of $\ell^*(\Theta^{[1]})+1$ 
irreducible components, each one of which are independent in $H^{1,1}(\hat{Z})$.
Therefore each one of $\hat{Y}_i \cap [\hat{Y}_j]_{\rm irr}$'s can be taken 
as an independent generator of the vertical algebraic cycles. The term   
${}_{\nu_i}\ell_1^{*\circ}(\widetilde{\Theta}^{[n-2]}) \ell^*(\Theta^{[1]})$ should 
therefore be counted as a part of the vertical component $H^{2,2}(\hat{Z})$.

1-simplices counted in ${}_{\nu_i} \ell^{*\bullet}_1(\widetilde{\Theta}^{[n-2]})$, 
however, correspond to algebraic cycles of the form $\hat{Y}_i \cap \hat{Y}_j$,
with $\hat{Y}_j$ corresponding to a point in a face of
codimension-three or higher. In such cases, $\hat{Y}_i \cap \hat{Y}_j$ 
consists of $\ell^*(\Theta^{[1]})+1$  irreducible components, but only one 
linear combination of those irreducible components, $\hat{Y}_i \cap \hat{Y}_j$, should be regarded as a vertical cycle. 
Thus, there are algebraic but non-vertical cycles left over in $h^{1,1}(\hat{Y}_i)$ if and only if 
\begin{equation}
{}_{\nu_i} \ell^{*\bullet}_1(\widetilde{\Theta}^{[n-2]}) \ell^*(\Theta^{[1]}) \neq 0.
\end{equation}

Let us move on to $H^{1,1}(\hat{Y}_i)$ for a toric divisor $\hat{Y}_i$
corresponding to an interior point of $\widetilde{\Theta}^{[1]}$ (i.e., 
$k_0=n-3=2$). The first line of (\ref{eq:h11-Yi-case-middle}) is 
rewritten just like in (\ref{eq:formula-h11-divisor-split}), and the same 
interpretation applies also in this case. The remaining 
correction term $\bar{e}^{1,1}_c(Z_{\Theta^{[3]}})$ 
in (\ref{eq:h11-Yi-case-middle}) in the $k_0 = n-3$ case is not regarded 
as part of the vertical component either. To see this, let us use 
a long exact sequence like (\ref{eq:long-seq-cpt-supp}), decomposing 
$\hat{Y}_i$ into the primary stratum 
$Z_{\rho_i}=\mathbb{T}^{k_0-1} \times Z_{\Theta^{[3]}_i}$ and the rest
$\hat{Y}_i \backslash Z_{\rho_i} = \amalg_{\rho_i < \sigma \in \Sigma'} Z_\sigma$.
\begin{equation}
  \vcenter{\xymatrix{
 \cdots \ar[r] &
   H^{2n-6}_c(Z_{\rho_i}) \ar[r] & H^{2n-6}_c(\hat{Y}_i) \ar[r] &
   H^{2n-6}_c(\hat{Y}_i \backslash Z_{\rho_i}) 
 \ar`[rd]^<>(0.5){}`[l]`[dlll]`[d][dll] 
 & & \\
 & H^{2n-5}_c(Z_{\rho_i}) \ar[r] & H^{2n-5}_c(\hat{Y}_i) \ar[r] &
   H^{2n-5}_c(\hat{Y}_i \backslash Z_{\rho_i}) \ar[r] &  \cdots\, .  & 
 }}
\label{eq:long-seq-cpt-spp-for-Yi-(n-3)}
\end{equation}
We focus on the $(n-3,n-3)$ components in this sequence in order to 
study $h^{n-3,n-3}[H^{2n-6}_c(\hat{Y}_i)]$. The cohomology groups 
$H^k_c(Z_{\rho_i})$ for $k=2n-6$ and $2n-5$ are determined by 
\begin{eqnarray}
 [H^{2n-6}_c(\mathbb{T}^{n-4} \times Z_{\Theta^{[3]}_i})]^{n-3,n-3}
 & = &
  [H^{2n-8}_c(\mathbb{T}^{n-4})]^{n-4,n-4} \otimes 
  [H^{2}_c(Z_{\Theta^{[3]}_i})]^{1,1}, \\
 { [H^{2n-5}_c(\mathbb{T}^{n-4} \times Z_{\Theta^{[3]}_i})]}^{n-3,n-3}
  & = &
  [H^{2n-9}_c(\mathbb{T}^{n-4})]^{n-5,n-5} \otimes 
  [H^4_c(Z_{\Theta^{[3]}_i})]^{2,2}  \nonumber \\
 & & \oplus 
  [H^{2n-8}_c(\mathbb{T}^{n-4})]^{n-4,n-4} \otimes 
  [H^3_c(Z_{\Theta^{[3]}_i})]^{1,1}\, ,
\end{eqnarray}
and (\ref{eq:Hodge-Deligne-for-alg-torus}). Thus we find
\begin{equation}
h^{n-3,n-3}[H^{2n-6}_c(Z_{\rho_i})] = \bar{e}^{1,1}_c(Z_{\Theta^{[3]}_i})\, , \qquad 
h^{n-3,n-3}[H^{2n-5}_c(Z_{\rho_i})] = (n-4) + 3\, .
\end{equation}
As for $H^{2n-6}_c(\hat{Y}_i \backslash Z_{\rho_i})$, it is enough to note 
that all of the irreducible components (strata) in 
$\hat{Y}_i \backslash Z_{\rho_i}$ are at most of complex dimension $(n-3)$, 
so that it suffices to count the number of such irreducible components, 
which is in turn determined by 
${}_{\nu_i}\ell_1(\widetilde{\Delta}_{\leq n-3}) + 
\sum {}_{\nu_i}\ell_1(\widetilde{\Theta}^{[n-2]})  \left (\ell(\Theta^{[1]})+1\right)$.
Therefore, the kernel of 
$[H^{2n-6}_c(\hat{Y}_i \backslash Z_{\rho_i})]^{n-3,n-3} \longrightarrow 
 [H^{2n-5}_c(Z_{\rho_i})]^{n-3,n-3}$ accounts for the first line of 
(\ref{eq:h11-Yi-case-middle}). 
The correction term---the second line of (\ref{eq:h11-Yi-case-middle})---comes 
from $[H^{2n-6}(Z_{\rho_i})]^{n-3,n-3}$. All the vertical components in 
$[H^{2n-6}_c(\hat{Y}_i)]^{n-3,n-3}$ are contained within 
$[H^{2n-6}_c(\hat{Y}_i \backslash Z_{\rho_i})]^{n-3,n-3}$, and hence 
the correction term $\bar{e}^{1,1}_c(Z_{\Theta}^{[3]})$ does not belong to the vertical component. 

In the case $k_0=n-2$, $\nu_i$ is an interior point of a two-dimensional face.
The vertical component in $H^{1,1}(\hat{Y}_i)$ 
is identified in the same way as far as the first line 
of (\ref{eq:h11-Yi-case-middle}) is concerned; the term 
${}_{\nu_i} \ell_1^{*\bullet}(\widetilde{\Theta}^{[3]}) \ell^*(\Theta^{[1]})$ 
corresponds to cycles that are algebraic, but not vertical.
The interpretation of the correction term $\bar{e}^{0,0}_c(Z_{\Theta^{[2]}_i})$ 
in (\ref{eq:h11-Yi-case-middle}), on the other hand, is quite different 
from the one in the $k_0=n-3$ case. 

Note first that in this case, $\hat{Y}_i$ is an $(n-2)$-fold which can be 
regarded as a flat family of toric $(n-3)=(k_0-1)$-dimensional varieties 
over a curve $\Sigma_{\Theta^{[2]}_i}$, which is the closure of 
$Z_{\Theta^{[2]}_i} \subset \mathbb{T}^2$. The fibre over any point in 
$Z_{\Theta^{[2]}_i}$ is given by toric data ($\nu_i$-containing simplices 
within the face $\widetilde{\Theta}^{[k_0-1]}$). The second thing to note 
is that $Z_{\Theta^{[2]}_i}$ is a Riemann surface (of genus $\ell^*(\Theta^{[2]}_i)$)
with a finite number of punctures. The number of punctures (denoted by 
$k_{\Theta^{[2]}_i}$) is given by
\begin{equation}
  k_{\Theta^{[2]}_i} - 1 = [H^1_c(Z_{\Theta^{[2]}_i})]^{0,0} = 
  2 - \bar{e}^{0,0}_c(Z_{\Theta^{[2]}_i})\, .
\label{eq:topology-relation-ebar00}
\end{equation}
The compact Riemann surface $\Sigma_{\Theta^{[2]}_i}$ is obtained by 
filling these $k_{\Theta^{[2]}_i}$ punctures in $Z_{\Theta^{[2]}_i}$.
Any one of those punctures is assigned to one of the 1-dimensional faces 
$\Theta^{[1]} \leq \Theta^{[2]}_i$ (see also footnote \ref{footnote:punctures}), and the toric $k_0-1$-dimensional fibre
geometry is determined by the $\nu_i$-containing simplices in the face 
$\widetilde{\Theta}^{[k_0]} \geq \widetilde{\Theta}^{[k_0-1]}_i$ dual to 
$\Theta^{[1]}$.
\begin{figure}[!h]
\begin{center}
\scalebox{.5}{ 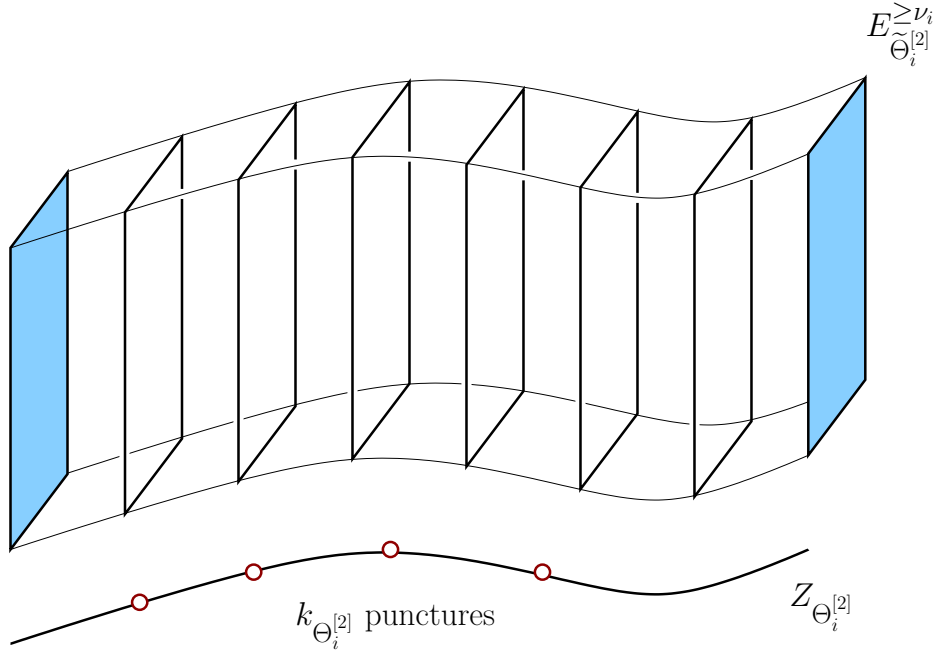 }
\caption{$\hat{Y}_i$ as a fibration of $E^{\geq \nu_i}_{\widetilde{\Theta}_i^{[2]}}$ over the
curve $Z_{\Theta_i^{[2]}}$ with $k_{\Theta_i^{[2]}}$ punctures. Divisors of a generic fibre correspond to divisors
of $\hat{Y}_i$ by sweeping them over the whole base. 
\label{fibroverZt2}}
\end{center} 
\end{figure} 
From this point of view, the first line of (\ref{eq:h11-Yi-case-middle}),   
with $-(n-1)$ replaced by $-(k_0-1)$, accounts for the number of irreducible 
divisors of $\hat{Y}_i$ modulo the toric linear equivalence acting in the 
fibre direction (see Figure \ref{fibroverZt2}). 
The remaining contribution to (\ref{eq:h11-Yi-case-middle}) is precisely 
$\bar{e}^{0,0}_c(Z_{\Theta^{[2]}_i})-(n-k_0) = 1-k_{\Theta^{[2]}_i}$.
This correction term is now understood as counting only the generic fibre 
class for an independent divisor in $\hat{Y}_i$, while removing the total fibre 
class at each one of the $k_{\Theta^{[2]}_i}$ points in $\Sigma_{\Theta^{[2]}_i}$ 
from the formula of $h^{1,1}(\hat{Y}_i)$. This subtraction is necessary 
because the total fibre class over any point in $\Sigma_{\Theta^{[2]}_i}$ is 
algebraically equivalent (see Figure \ref{fibroverZt2_2}). 
\begin{figure}[!h]
\begin{center}
\scalebox{.5}{ 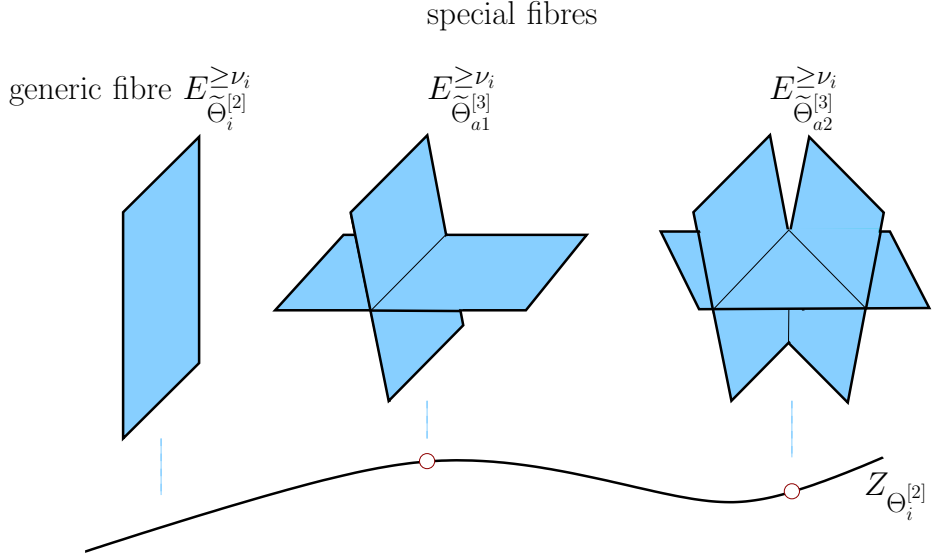 }
\caption{Over specific points, the fibre degenerates and becomes reducible. All fibres are algebraically (and hence homologically) equivalent, however. 
\label{fibroverZt2_2}}
\end{center} 
\end{figure}

The correction term $\bar{e}^{0,0}_c(Z_{\Theta^{[2]}_i}) = 3-k_{\Theta^{[2]}_i}$ 
should always be negative, because this term accounts for the algebraic 
equivalence relations among the generators of $H^{2,2}_c(\hat{Y}_i)$.
It is not hard to see this. For a dual pair of faces 
$(\widetilde{\Theta}^{[2]}_i, \Theta^{[2]}_i)$ and 
an interior point $\nu_i$ of $\widetilde{\Theta}^{[2]}_i$, 
list up all the dual pairs of faces 
$\{(\widetilde{\Theta}^{[3]}_a, \Theta^{[1]}_a) \; | \; a \in A_i\}$ labelled 
by $a \in A_i$ such that 
\begin{equation*}
\widetilde{\Theta}^{[3]}_a \geq \widetilde{\Theta}^{[2]}_i 
\qquad {\rm and} \qquad 
\Theta^{[1]}_a \leq \Theta^{[2]}_i. 
\end{equation*}
Each one of those pairs leaves 
$\ell(\Theta^{[1]}_a) = [1 + \ell^*(\Theta^{[1]}_a)]$ points in 
$\Sigma_{\Theta^{[2]}_i}$ for which the fibre of 
$\hat{Y}_i \longrightarrow \Sigma_{\Theta^{[2]}_i}$ is a (not necessarily 
irreducible) complex surface $E_{\widetilde{\Theta}^{[3]}_a}^{\geq \nu_i}$ 
rather than the generic fibre $E_{\Theta^{[2]}_i}^{\geq \nu_i}$. Therefore 
\begin{equation}
 \bar{e}^{0,0}_c(Z_{\Theta^{[2]}_i}) = 3 - k_{\Theta^{[2]}_i} =
  3 - \sum_{\Theta^{[1]}_a \leq \Theta^{[2]}_i} [1+\ell^*(\Theta^{[1]}_a)]
 = 3 - \# \{ \Theta^{[1]}_a \leq \Theta^{[2]}_i \} 
 - \sum_{\Theta^{[1]}_a \leq \Theta^{[2]}_i} \ell^*(\Theta^{[1]}_a).
\label{eq:ebar00-nmbr-punctures}
\end{equation}
The last term is obviously negative as the number of edges of a 
two-dimensional face $\Theta^{[2]}$ is always greater than or equal to 3.
Thus, $\bar{e}^{0,0}_c(Z_{\Theta^{[2]}_i}) \leq 0$.

\begin{figure}[!h]
\begin{center}
\scalebox{.5}{ 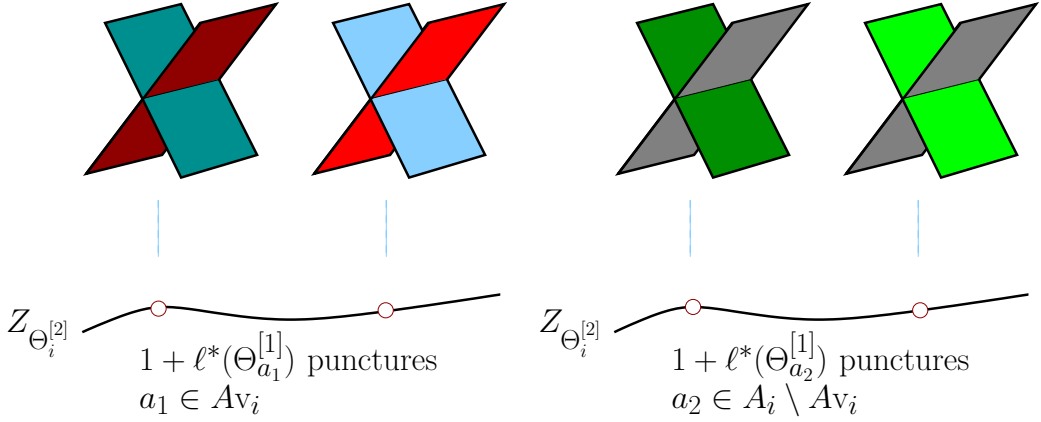 }
\caption{(colour online) Fibre components corresponding to points $\nu_i$ contributing to $\Theta^{[3]}_v$ (left hand side) and $\Theta^{[3]}_{nv}$ (right hand side). 
The fibre components drawn in red, blue and green correspond to fibre components which are obtained by intersecting appropriate divisors. These components arise from
one-simplices connecting $\nu_i$ to an interior point of the face $\widetilde{\Theta}^{[3]}_a$. When there is a one-simplex connecting $\nu_i$ to a point on the boundary of $\widetilde{\Theta}^{[3]}_a$, only the sum of several components of different fibres, drawn in grey, arises from an intersection of divisors (so that it should be considered vertical). Again, the two degenerate fibres on the left hand side are algebraically equivalent, as are the two degenerate fibres on the right hand side. 
\label{fibroverZt2_3}}
\end{center} 
\end{figure}

The algebraic equivalence relations encoded in $\bar{e}^{0,0}_c(Z_{\Theta^{[2]}_i})$ 
are sometimes among algebraic cycles in the vertical component, and
sometimes among cycles in the non-vertical component. 
Figure~\ref{fibroverZt2_3} is a schematic picture of the singular fibres of 
$\hat{Y}_i \longrightarrow \Sigma_{\Theta^{[2]}_i}$ at the $\ell^*(\Theta^{[1]}_a)+1$ 
punctures associated with a given face $\Theta^{[1]}_a \leq \Theta^{[2]}_i$.
There is always at least one combination of the total fibre classes 
that is in the vertical component; such a combination of the total fibre 
classes is in fact even in the space of vertical cycles generated by 
toric divisors. Such a fibre class is algebraically equivalent to the generic 
fibre class of $\hat{Y}_i \longrightarrow \Sigma_{\Theta^{[2]}}$, and is 
subtracted by the algebraic equivalence on $\Sigma_{\Theta^{[2]}_i}$ 
corresponding to the $(3-\# \{\Theta^{[1]}_a \leq \Theta^{[2]}_i \})$ term in 
$\bar{e}^{0,0}_c(Z_{\Theta^{[2]}_i})$ above. 

Suppose that the dual face pair $(\widetilde{\Theta}^{[3]}_a, \Theta^{[1]}_a)$ 
for an $a \in A_i$ is such that {\it all} the 1-simplices counted in 
${}_{\nu_i} \ell_1^*(\widetilde{\Theta}^{[3]}_a)$ end at interior points of 
$\widetilde{\Theta}^{[3]}_a$. 
This means that all the $[1+\ell^*(\Theta^{[1]}_a)]$ total fibre classes of 
$\hat{Y}_i \longrightarrow \Sigma_{\Theta^{[2]}_i}$ belong to the space of 
vertical cycles (see Figure~\ref{fibroverZt2_3}). 
Let $A{\rm v}_i$ be the subset of the labels $a \in A_i$ satisfying the 
condition above, and $(\widetilde{\Theta}^{[3]}_{{\rm v}}, \Theta^{[1]}_{{\rm v}})$ 
be such a dual pair of faces. The space of vertical but non-toric  
cycles originating from $H^{1,1}(\hat{Y}_i)$ is reduced by 
$- \sum_{\Theta^{[1]}_{\rm v} \leq \Theta^{[2]}}\ell^*(\Theta^{[1]}_{\rm v})$ 
due to the algebraic equivalence. 

A dual pair of faces $(\widetilde{\Theta}^{[3]}_a, \Theta^{[1]}_a)$ otherwise 
(i.e., $a \in A_i \setminus A{\rm v}_i$) has at least one 1-simplex counted in 
${}_{\nu_i} \ell_1^*(\widetilde{\Theta}^{[3]}_a)$ whose boundary other than 
$\nu_i$ is not in the interior of $\widetilde{\Theta}^{[3]}$. Such pairs of 
faces are denoted by $(\widetilde{\Theta}^{[3]}_{\rm nv}, \Theta^{[1]}_{\rm nv})$.
The remaining $- \sum_{\Theta^{[1]}_{\rm nv} \leq \Theta^{[2]}_i} \ell^*(\Theta^{[1]}_{\rm nv})$
independent algebraic equivalences reduce the dimension of the space of 
algebraic, but non-vertical four-cycles. To summarize, 
\begin{eqnarray}
 V^{\rm alg}(\hat{Z}) & = & \sum_{(\widetilde{\Theta}^{[2]}, \Theta^{[2]})} 
    \ell^*(\widetilde{\Theta}^{[2]}) 
    \left\{ - \ell^*(2\Theta^{[2]}) + 4\ell^*(\Theta^{[2]}) \right\}, \nn \\
  & = &  \sum_{(\widetilde{\Theta}^{[2]}, \Theta^{[2]})} 
    \ell^*(\widetilde{\Theta}^{[2]}) (3-k_{\Theta^{[2]}})  
  = V^{\rm alg}_{\rm tor}(\hat{Z}) + V^{\rm alg}_{\rm cor}(\hat{Z})
  + V^{\rm alg}_{\rm rm}(\hat{Z}) ;
    \label{eq:V-alg-tot}   \\
 V^{\rm alg}_{\rm tor}(\hat{Z}) & : = &   \sum_{(\widetilde{\Theta}^{[2]}, \Theta^{[2]})} 
    \ell^*(\widetilde{\Theta}^{[2]})
     \left( 3-\#\{ \Theta^{[1]} \leq \Theta^{[2]} \} \right), \nn \\
 V^{\rm alg}_{\rm cor}(\hat{Z}) & := &  -  \sum_{(\widetilde{\Theta}^{[2]}, \Theta^{[2]})} 
    \ell^*(\widetilde{\Theta}^{[2]})
     \sum_{\Theta^{[1]}_{\rm v} \leq \Theta^{[2]}} \ell^*(\Theta^{[1]}_{\rm v}), \nn \\
 V^{\rm alg}_{\rm rm}(\hat{Z}) & := &  -  \sum_{(\widetilde{\Theta}^{[2]}, \Theta^{[2]})} 
    \ell^*(\widetilde{\Theta}^{[2]})
     \sum_{\Theta^{[1]}_{\rm nv} \leq \Theta^{[2]}} \ell^*(\Theta^{[1]}_{\rm nv}). 
      \label{eq:V-alg-parts}
\end{eqnarray}
All the three terms are negative.

Finally, in the cases of $k_0 = n-1$, it is easy to see that 
all of the generators of the space 
$\oplus_{\rm irr} H^{2,2}_c([\hat{Y}_i]_{\rm irr})$ are vertical cycles.
The space of vertical cycles generated by the toric divisors, however, 
has a dimension given by (\ref{eq:h11-Yi-case-n-1}) without being 
multiplied by $e^{0,0}_c(Z_{\Theta^{[1]}_i})=[1+\ell^*(\Theta^{[1]}_i)]$.

Having seen which subspace of $H^{2,2}_c(\hat{Y}_i)$ is identified with 
a part of the vertical component $H^{2,2}(\hat{Z})$, we are now ready 
to determine the dimension of the vertical component in $H^{2,2}_V(\hat{Z})$.
Setting aside the components in $[H^4(Y)]^{2,2}$ that have turned out to be 
non-vertical, we have 
\begin{eqnarray}
h^{2,2}[H^4(Y)] & = & h^{2,2}_V(\hat{Z})
  + \left[ RM_a(\hat{Z}) +V^{\rm alg}_{\rm rm}(\hat{Z})\right] + NV_1(\hat{Z}), 
  \label{eq:H4Y-decomp-vert-RM-NV1} \\
h^{2,2}_V(\hat{Z}) & = &
   \left[ V_{\rm tor}(\hat{Z}) + V^{\rm alg}_{\rm tor}(\hat{Z}) \right]
 + \left[ V_{\rm cor}(\hat{Z}) + V^{\rm alg}_{\rm cor}(\hat{Z}) \right], 
\label{eq:V}
\end{eqnarray}
where 
\begin{eqnarray}
 V_{\rm tor}(\hat{Z}) & := & \ell_1(\widetilde{\Delta}_{\leq n-2})\nn
       + \frac{n(n-1)}{2} \nn \\ 
& - & (n-1) \left[ \ell(\widetilde{\Delta}) 
    - \sum_{\widetilde{\Theta}^{[n-1]} \leq \widetilde{\Delta}} 
         \ell^*(\widetilde{\Theta}^{[n-1]})
    - \sum_{\widetilde{\Theta}^{[n-2]} \leq \widetilde{\Delta}}
         \ell^*(\widetilde{\Theta}^{[n-2]}) -1 \right]   \nn \\
& - & (n-2) \sum_{\widetilde{\Theta}^{[n-2]}}  \ell^*(\widetilde{\Theta}^{[n-2]}) \nn \\ 
 V_{\rm cor}(\hat{Z}) & := & \sum_{(\widetilde{\Theta}^{[3]}, \Theta^{[1]})} 
    \left[ \ell^{*\circ}_1(\widetilde{\Theta}^{[3]})
         - 3 \ell^*(\widetilde{\Theta}^{[3]}) \right]
    \times \ell^*( \Theta^{[1]} ),        \nn \\
 RM_a(\hat{Z}) & = & \sum_{(\widetilde{\Theta}^{[3]}, \Theta^{[1]})}
     \ell_1^{*\bullet}(\widetilde{\Theta}^{[3]}) \ell^*(\Theta^{[1]}),  \nn \\
 NV_1(\hat{Z}) & := & \sum_{(\widetilde{\Theta}^{[1]},{\Theta}^{[3]})} 
     \ell^*(\widetilde{\Theta}^{[1]}) \times 
     \left(\ell^*(2\Theta^{[3]}) - 4 \ell^*(\Theta^{[3]}) - 
       \sum_{\Theta^{[2]}\leq \Theta^{[3]}} \ell^*(\Theta^{[2]})\right). 
\label{eq:V-parts}
\end{eqnarray}
Here, $\ell^{*\circ}_1(\widetilde{\Theta}^{[3]})$ is the number of 
1-simplices that run through the interior of a face 
$\widetilde{\Theta}^{[3]}$ and have at least one boundary point in the interior
of $\widetilde{\Theta}^{[3]}$. $\ell^{*\bullet}_1(\widetilde{\Theta}^{[3]})$ is the 
number of all other 1-simplices that run through the interior of the face 
$\widetilde{\Theta}^{[3]}$; It only counts those one-simplices which start and end
on the boundary of the face $\widetilde{\Theta}^{[3]}$.

\begin{center}
.....................................................
\end{center}

The Chow group $Ch^2(\hat{Z})$ is obtained by taking a quotient 
of the space of algebraic (complex)-two-cycles by rational equivalence, 
whereas we have also exploited algebraic equivalence in studying 
the cohomology group $H^{2,2}(\hat{Z})$. The Deligne cohomology 
$H^4_D(\hat{Z}; \Z(2))$ and the closely related Chow group $Ch^2(\hat{Z})$
not only contain information on the flux field strength 
$G^{(4)}$ for F-theory, but also on the three-form potential $C^{(3)}$. 
Truly of interest in the context of physics 
application, though, will be $H^4_D(\hat{Z}_p; \Z(2))$ for $\hat{Z}_p$ 
corresponding to a point $p\in {\cal M}_*$ in some Noether--Lefschetz locus, 
rather than that of $\hat{Z}_p$ at a generic point $p \in {\cal M}_*$. 
Fluxes which are in the primary horizontal subspace may also become algebraic 
in a Noether--Lefschetz locus, where we are expected to end up from 
the superpotential $W \propto \int_{\hat{Z}} G^{(4)} \wedge \Omega$. 
Ignoring the original context of physics applications, however, let us 
leave an interesting observation on $Ch^2(\hat{Z})$ for $\hat{Z}$ 
corresponding to a generic point $p \in {\cal M}_*$.
The relevance of the refined data contained in $Ch^2(\hat{Z})$ 
(compared with homology) to F-theory fluxes has recently been discussed 
in \cite{Bies:2014sra} (see also the literatures therein).

For $\hat{Z}_p$ at a generic point in complex structure moduli space $p \in {\cal M}_*$, 
algebraic cycles generate a subspace of $H^{2,2}(\hat{Z})$ with a dimension no less than

\begin{equation*}
V_{\rm tor}(\hat{Z}) + V_{\rm cor}(\hat{Z}) +  RM_a(\hat{Z}) + V^{\rm alg}(\hat{Z}),  
\end{equation*}
and possibly larger than this by at most $NV_1(\hat{Z})$. Only 
the first term descends from 
algebraic cycles of the toric ambient space. Apart from the last term, all the equivalence relations that have been exploited are linear (rational) 
equivalence. The last term, $V^{\rm alg}(\hat{Z})$,  
introduces algebraic equivalence relations among those cycles as 
we have already seen. They are associated with divisors $\hat{Y}_i$ of 
$\hat{Z}$ corresponding to interior points $\nu_i$ of two-dimensional faces 
$\widetilde{\Theta}^{[2]} \leq \widetilde{\Delta}$. The threefolds 
$\hat{Y}_i$ can be seen as flat fibrations of surfaces over curves 
$\Sigma_{\Theta^{[2]}}$ with genus $g = \ell^*(\Theta^{[2]})$. The fibre classes 
over any two points in $\Sigma_{\Theta^{[2]}}$ are mutually algebraically equivalent, 
and they are identified in the cohomology group. Under rational 
equivalence, however, they form a family of \emph{inequivalent} classes parametrized 
by $g = \ell^*(\Theta^{[2]})$ complex parameters. This is analogous to 
divisors (points) on the curve $\Sigma_{\Theta^{[2]}}$, which are classified under linear equivalence by ${\rm Pic}(\Sigma_{\Theta^{[2]}})$, which has  
$g = {\rm dim}_\C [{\rm Pic}^0(\Sigma_{\Theta}^{[2]})]$ complex parameters more than 
the discrete data (the first Chern class) counted in 
$H^{1,1}(\Sigma_{\Theta^{[2]}})$. Noting that 
\begin{equation}
 h^{2,1}(\hat{Z}) = \sum_{(\widetilde{\Theta}^{[2]},\Theta^{[2]})} 
   \ell^*(\widetilde{\Theta}^{[2]}) \ell^*(\Theta^{[2]}) = 
 \sum_{(\widetilde{\Theta}^{[2]},\Theta^{[2]})} 
   \ell^*(\widetilde{\Theta}^{[2]}) g(\Sigma_{\Theta^{[2]}}), 
\end{equation}
we see that the $Ch^2(\hat{Z})$ group contains $h^{2,1}(\hat{Z})$ more 
complex parameters than the cohomology group, and that this difference 
comes from divisors $\hat{Y}_i$ of $\hat{Z}$ that are regarded as flat 
surface fibration over curves with $g > 0$. 

\subsubsection{Horizontal and remaining components}

Since the vertical component $H^{2,2}_V(\hat{Z})$ has been identified within 
$[H^4(Y)]^{2,2}$, the horizontal and the remaining components in the 
decomposition (\ref{eq:H4-decomp-H-RM-V}) should live in $[H^4(Z_\Delta)]^{2,2}$ 
and the remaining space within $[H^4(Y)]^{2,2}$. 
We are going to use mirror symmetry to identify the horizontal component $H^{2,2}_H(\hat{Z})$ in 
$[H^4_c(Z_\Delta)]^{2,2}$. 

To this end, it is convenient to verify a couple of relations among 
the combinatorial data first. Let us introduce the following decomposition 
in order to facilitate the discussion:
\begin{equation}
 h^{2,2}[H^4(Z_\Delta)] := H_{\rm mon}(\hat{Z}) + H^{\rm red}(\hat{Z}) + NV_3(\hat{Z}),
\label{eq:h22-H4Z-decomp}
\end{equation}
where
\begin{eqnarray}
  H_{\rm mon}(\hat{Z}) & := & \varphi_3(\Delta)
   - \sum_{\Theta^{[4]}} \varphi_2(\Theta^{[4]})
   + \sum_{\Theta^{[3]}} \varphi_1(\Theta^{[3]}),\nn \\
  H^{\rm red}(\hat{Z}) & := & \sum_{(\widetilde{\Theta}^{[2]},\Theta^{[2]})} 
   \left( 3 \ell^*(\widetilde{\Theta}^{[2]})
            - \ell_1^*(\widetilde{\Theta}^{[2]})\right)
    \times \ell^*(\Theta^{[2]}), \nn\\
  NV_3(\hat{Z}) & := & \sum_{(\widetilde{\Theta}^{[1]},\Theta^{[3]})} 
   \sum_{\Theta^{[2]} \leq \Theta^{[3]}} 
     \ell^*(\widetilde{\Theta}^{[1]}) \times \ell^* ( \Theta^{[2]}) \, . 
  \label{eq:H-parts}
\end{eqnarray}
We claim that 
\begin{equation}
   H_{\rm mon}(\hat{Z}) = V_{\rm tor}(\hat{Z}_m), \quad 
   H^{\rm red}(\hat{Z}) = V^{\rm alg}(\hat{Z}_m), \quad 
   NV_1(\hat{Z}) + NV_3(\hat{Z}) = V_{\rm cor}(\hat{Z}_m) + RM_a(\hat{Z}_m). 
\label{eq:afew-relations}
\end{equation}
This also means that the following relation holds:
\begin{equation}
 h^{2,2}[H^4_c(Z_\Delta)] + NV_1(\hat{Z}) 
 = V_{\rm tor}(\hat{Z}_m) + V_{\rm cor}(\hat{Z}_m) + RM_a(\hat{Z}_m)
 + V^{\rm alg}(\hat{Z}_m). 
\end{equation}

\begin{figure}[tbp]
\begin{center}
\includegraphics[height=4cm]{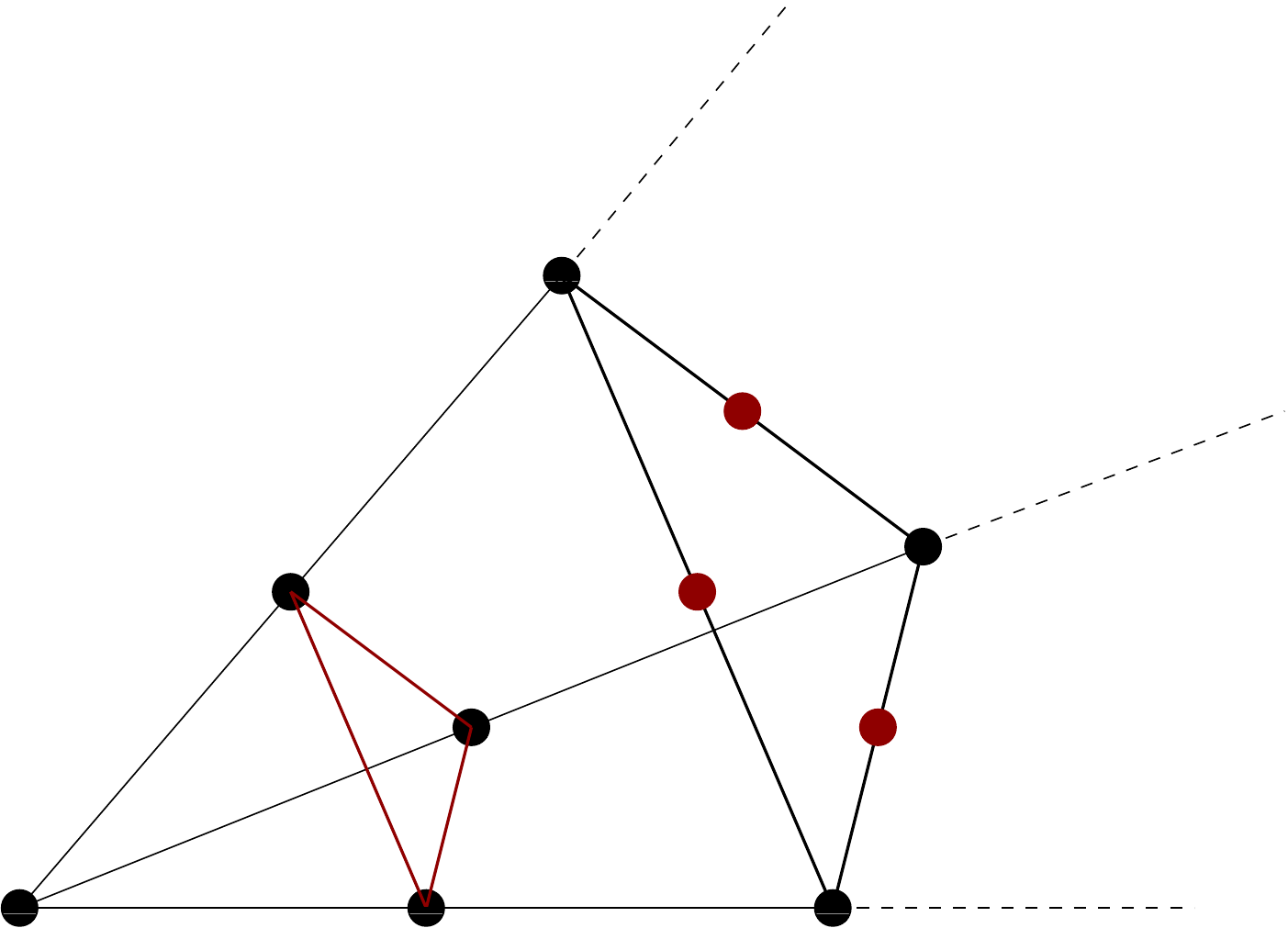}
\caption{\label{fig:distance2} (colour online) 
The lattice points at the distance-2 layer are in one-to-one correspondence 
with the lattice points and 1-simplices at the distance-1 layer.}
\end{center}
\end{figure}

Let us verify the relations (\ref{eq:afew-relations}) one by one.
As for the first one, note that 
\begin{eqnarray}
&& H_{\rm mon}(\hat{Z})\nn \\
& = & [\ell(2\Delta) - (n+1)\ell(\Delta) 
  + \frac{n(n+1)}{2}] \\
&& - \sum_{\Theta^{[n-1]} \leq \Delta} 
   [\ell^*(2\Theta^{[n-1]}) - n \ell^*(\Theta^{[n-1]})]
 + \sum_{\Theta^{[n-2]} \leq \Delta} 
     \ell^*(\Theta^{[n-2]}), \nonumber \\
& = & [\ell(2\Delta) - \ell(\Delta) 
       - \sum_{\Theta^{[n-1]}} \ell^*(2\Theta^{[n-1]}) ]
   \\
 && - n [\ell(\Delta) - \sum_{\Theta^{[n-1]}} 
             \ell^*(\Theta^{[n-1]})]  
     + \frac{n(n+1)}{2} + \sum_{\Theta^{[n-2]} \leq \Delta} 
   \ell^*(\Theta^{[n-2]}), \nonumber \\
& = &  [\ell(2\Delta) - \ell(\Delta) - \sum_{\Theta^{[n-1]}} \ell^*(2\Theta^{[n-1]}) ]
  - [\ell(\Delta) - 1 - \sum_{\Theta^{[n-1]}}
         \ell^*(\Theta^{[n-1]})], \label{eq:temp-aa} \\
& + & \left[ - n + \frac{n(n+1)}{2} \right],
   \nonumber 
   \\
& - & \left[  (n-1) \left[ \ell(\Delta) -1 
             - \sum_{\Theta^{[n-1]}} \ell^*(\Theta^{[n-1]})
             - \sum_{\Theta^{[n-2]}} \ell^*(\Theta^{[n-2]}) 
                    \right]  
          + (n-2) \sum_{\Theta^{[n-2]}} \ell^*(\Theta^{[2]}),
      \right]. \nonumber 
\end{eqnarray}
Obviously one only needs to verify that (\ref{eq:temp-aa}) is equal to 
$\ell_1(\Delta_{\leq n-2})$ in order to prove that 
$H_{\rm mon}(\hat{Z}) = V_{\rm tor}(\hat{Z}_m)$. 
Secondly, we see that the first term in (\ref{eq:temp-aa})
counts the number of lattice points in $M$ at the ``lattice-distance-2'' 
that are not in the interior of $n$-dimensional cones, while the second 
term in (\ref{eq:temp-aa}) counts lattice points 
at the ``lattice distance 1''.
Now remember that we assume existence of a fine unimodular triangulation 
of the polytope $\Delta$ (so that both $\hat{Z}_m$ and the ambient toric variety $\P^n_\Sigma$ are smooth).
For a cone whose base at the lattice-distance 1 is a minimum volume 
$k$-simplex, lattice points at the distance 1 are the $k+1$ vertices 
of the $k$-simplex, while those at the distance 2 consist of 
$\binom{k+1}{1} + \binom{k+1}{2} = \binom{k+2}{2}=:N_{k,2}$ points 
corresponding to the vertices and 1-simplices on the $k$-simplex 
(see figure~\ref{fig:distance2}).\footnote{
The slice of such a cone at the distance $h$ becomes a $k$-dimensional 
pyramid of height $h$ (see footnote \ref{fn:pyramid}). The number of 
interior points of such a pyramid is $N_{k,h-k-1} = (h-k)_k/k!$, which becomes 
positive only when $h>k$. That is, a lattice point corresponding to 
a $k$-simplex is found only at the lattice distance $h > k$. 
At the distance 2, $\partial (2\widetilde{\Delta})$, the lattice points 
correspond only to vertices or 1-simplices on $\partial \widetilde{\Delta}$.}
Because any cone of the fan $\widetilde{\Sigma}$ can be decomposed into 
such cones of the fan $\widetilde{\Sigma}'$, 
\begin{equation}
  \ell(2\Theta) = \ell(\Theta) + \ell_1(\Theta)
\label{eq:2Theta--Theta--ell1-Theta}
\end{equation}
for any faces $\Theta < \Delta$. Using recursion with respect to the faces of $\Theta$, similar relations can be derived for the number of internal one-simplices,
$\ell^*_1(\Theta)$. This proves the equality between $\ell_1(\Delta_{\leq n-2})$ 
and (\ref{eq:temp-aa}) and also the first relation 
$H_{\rm mon}(\hat{Z}) = V_{\rm tor}(\hat{Z}_m)$ in (\ref{eq:afew-relations}). 

The second one of the relations (\ref{eq:afew-relations}) 
$ H^{\rm red}(\hat{Z}) = V^{\rm alg}(\hat{Z}_m)$ can be verified for each 
one of dual pairs, $(\widetilde{\Theta}^{[2]}, \Theta^{[2]})$. 
Using the relation (\ref{eq:2Theta--Theta--ell1-Theta}) for the 2-dimensional 
face $\widetilde{\Theta}^{[2]} < \widetilde{\Delta}$, we find that\footnote{\label{footnote:punctures}
The relation (\ref{eq:topology-relation-ebar00}, 
\ref{eq:ebar00-nmbr-punctures}) between 
$\bar{e}^{0,0}_c(Z_{\widetilde{\Theta}^{[2]}}) = - \ell^*(2\widetilde{\Theta}^{[2]})
+ 4 \ell^*(\widetilde{\Theta}^{[2]})$ and the number of punctures 
$k_{\widetilde{\Theta}^{[2]}}$ can also be derived purely combinatorially, 
without looking at the geometry of the curve and punctures on it.
To see this, note first that $k_{\widetilde{\Theta}^{[2]}}=V_B$, where 
$V_B$ is the number of lattice points appearing on the boundary 
of a two-dimensional simplicial complex $\widetilde{\Theta}^{[2]}$.
Now, let $T$ be the number of 2-simplices in $\widetilde{\Theta}^{[2]}$, 
$E_I$ and $E_B$ the number of 1-simplices in the interior and boundary of 
$\widetilde{\Theta}^{[2]}$, and $V_I$ and $V_B$ the 
number of points in the interior and boundary of $\widetilde{\Theta}^{[2]}$.
The topology of $\widetilde{\Theta}^{[2]}$ and 
$\partial \widetilde{\Theta}^{[2]}$ indicates that 
\begin{eqnarray}
 0 & = & V_B - E_B, \\
 1 & = & (V_B+V_I)-(E_B+E_I) + T, \\
 3T & = & 2E_I + E_B.
\end{eqnarray}
From this, we find that 
\begin{equation}
3V_I-E_I = 3-V_B = 3-k_{\Theta^{[2]}}.
\end{equation}
The left-hand side is precisely the right-hand side 
of (\ref{eq:Valg=Hred-proof}), and is hence equal to 
$\bar{e}^{0,0}_c(Z_{\widetilde{\Theta}^{[2]}})$ combinatorially. 
This completes a combinatorial proof of the relation 
(\ref{eq:topology-relation-ebar00}, 
\ref{eq:ebar00-nmbr-punctures}), which also follows from the geometry of 
the curve $Z_{\widetilde{\Theta}^{[2]}})$. } 
\begin{equation}
 \bar{e}^{0,0}_c(Z_{\widetilde{\Theta}^{[2]}}) =
 - \ell^*(2\widetilde{\Theta}^{[2]}) + 4 \ell^*(\widetilde{\Theta}^{[2]})
 = \left(3 \ell^*(\widetilde{\Theta}^{[2]})
  - \ell_1^*(\widetilde{\Theta}^{[2]})\right) .  
\label{eq:Valg=Hred-proof}
\end{equation}

The third relation in (\ref{eq:afew-relations}) is equivalent to
\begin{equation}
\ell^*(2\Theta^{[3]})-4\ell^*(\Theta^{[3]}) =
 \ell_1^*(\Theta^{[3]})-3\ell^*(\Theta^{[3]})
\end{equation}
for each one of the faces $\Theta^{[3]} < \Delta$. This relation also 
follows from (\ref{eq:2Theta--Theta--ell1-Theta}).

\begin{center}
 .............................................
\end{center}

Toric divisors on $\hat{Z}_m$ are mirror to monomial deformations of the 
defining equation of the hypersurface $\hat{Z}$. It is thus reasonable to 
consider that the space spanned by mutual intersections of toric divisors --- a 
$V_{\rm tor}(\hat{Z}_m)$-dimensional subspace  of the primary vertical subspace 
of $\oplus_p H^{p,p}(\hat{Z}_m)$ --- is mirror to second order monomial deformations 
of $\hat{Z}$, which must be a subspace of the primary horizontal 
subspace of $\oplus_p H^{p,n-1-p}(\hat{Z})$. This subspace should have 
a dimension $H_{\rm mon}(\hat{Z}) = V_{\rm tor}(\hat{Z}_m)$ because of 
mirror symmetry. This reasoning seems to work very well:
the first line of the expression (\ref{eq:temp-aa}) of $H_{\rm mon}(\hat{Z})$
counts the number of lattice points $2\Delta \cap M$ of the form 
$\tilde{\nu}_\alpha + \tilde{\nu}_\beta$ where 
$\tilde{\nu}_\alpha$ and $\tilde{\nu}_\beta$ have corresponding 
monomial deformations $D_\alpha \Omega_{\hat{Z}}$ and $D_\beta \Omega_{\hat{Z}}$, 
and there is a 1-simplex joining the lattice points $\tilde{\nu}_\alpha$ 
and $\tilde{\nu}_\beta$. It is quite reasonable to identify 
the point $\tilde{\nu}_\alpha + \tilde{\nu}_\beta$ in $2\Delta \cap M$ with the 
quadratic complex structure deformation $D_\alpha D_\beta \Omega_{\hat{Z}}$ (or
$(\Omega^{-1}D_\alpha \Omega)\cdot (\Omega^{-1}D_\beta \Omega)$ in 
$H^2(\hat{Z}; \wedge^2 T\hat{Z})$ of \cite{Greene:1993vm}). 

According to mirror symmetry, the subspace of non-vertical components with the
dimension $NV_1(\hat{Z})+NV_3(\hat{Z})$ must be decomposed into 
\begin{eqnarray}
 H_{\rm cor}(\hat{Z}) & := & V_{\rm cor}(\hat{Z}_m) = 
  \sum_{(\widetilde{\Theta}^{[1]}, \Theta^{[3]})} 
     \ell^*( \widetilde{\Theta}^{[1]} ) \times 
    \left[ \ell^{*\circ}_1(\Theta^{[3]})
         - 3 \ell^*(\Theta^{[3]}) \right], \nn \\
 RM^m_a(\hat{Z}) & := & RM_a(\hat{Z}_m) =
  \sum_{(\widetilde{\Theta}^{[1]}, \Theta^{[3]})}
     \ell^*(\widetilde{\Theta}^{[1]}) \ell_1^{*\bullet}(\Theta^{[3]})\, .
 \label{eq:H-parts-2}
\end{eqnarray}
The space generated by $H_{\rm cor}$ 
generators must also be part of the primary horizontal subspace. 
The $H_{\rm cor}$-dimensional subspace of the primary horizontal component 
must correspond to $D^2 \Omega_{\hat{Z}}$ which involve at least one deformation 
that is not represented by a monomial, i.e. the last term of (\ref{eq:bath31}).
It is reasonable that the expression of $H_{\rm cor}(\hat{Z})$ above vanishes 
when there is no pair of dual faces where 
$\ell^*(\widetilde{\Theta}^{[1]}) \ell^*(\Theta^{[3]}) \neq 0$, 
because there should be no non-monomial deformation of complex structure 
in that case.  

The correction term $H^{\rm red}(\hat{Z})$ in (\ref{eq:h22-H4Z-decomp}) 
is mirror to the $V^{\rm alg}(\hat{Z}_m)$ algebraic equivalences, and hence 
will represent some redundancy in the description of the quadratic 
deformations of complex structure by $H_{\rm mon}(\hat{Z})+H_{\rm cor}(\hat{Z})$ 
generators and $RM^m_a(\hat{Z})$ for the non-horizontal non-vertical 
components. $H^{\rm red}(\hat{Z})$ can be split into three, just like 
we did for $V^{\rm alg}(\hat{Z})$. The dimensions of those three pieces 
are denoted by $H^{\rm red}_{\rm mon}(\hat{Z})$, $H^{\rm red}_{\rm cor}(\hat{Z})$ 
and $H^{\rm red}_{\rm rm}(\hat{Z})$. 

By using mirror symmetry, we finally arrive at the following 
formula for the vertical, horizontal and remaining components of $H^{2,2}(\hat{Z})$ 
of a Calabi--Yau fourfold $\hat{Z}$ obtained as a hypersurface of a 
toric variety $\P^5_{\Sigma'}$.
\begin{eqnarray}
  h^{2,2}(\hat{Z}) & = &
    h^{2,2}_V(\hat{Z}) + h^{2,2}_{RM}(\hat{Z}) + h^{2,2}_H(\hat{Z}), \\
  h^{2,2}_H(\hat{Z}) & = &
    \left[ H_{\rm mon}(\hat{Z}) + H^{\rm red}_{\rm mon}(\hat{Z}) \right]
  + \left[ H_{\rm cor}(\hat{Z}) + H^{\rm red}_{\rm cor}(\hat{Z}) \right], 
     \label{eq:H} \\
  h^{2,2}_{RM}(\hat{Z}) & = & \left[ RM_a(\hat{Z}) + V^{\rm alg}_{\rm rm}(\hat{Z})\right]
                + \left[RM_a^m(\hat{Z}) + H^{\rm red}_{\rm rm}(\hat{Z})\right].
     \label{eq:RM}
\end{eqnarray}
This result shows under which circumstances the remaining component is present.
It is quite reasonable from the perspective of mirror symmetry, though, 
that the remaining component $H^{2,2}_{RM}(\hat{Z})$ 
has a dimension that is symmetric under the exchange of 
$\Delta$ and $\widetilde{\Delta}$. 

The term $RM_a(\hat{Z})$ describes the space of algebraic cycles 
on divisors $\hat{Y}_i$ of $\hat{Z}$ that are not obtained by restriction 
of divisors of $\hat{Z}$; that is, they come from 
\begin{equation}
{\rm Coker} \left[ i^*_{\hat{Y}_i \hookrightarrow \hat{Z}}: 
   NS(\hat{Z}) \longrightarrow NS(\hat{Y}_i) \right].
\label{eq:rm-from-coker}
\end{equation}
We do not have a robust theory on the component with the 
dimension $RM^m_a(\hat{Z})$ at this moment. Some of this component 
come from $NV_1(\hat{Z})$-dimensional subspace of $[H^4(Y)]^{2,2}$, but 
some may also come from the $NV_3(\hat{Z})$-dimensional subspace of 
$[H^4_c(Z_{\Delta})]^{2,2}$. There are some examples where (a part of) 
the $RM^m_a(\hat{Z})$-dimensional space also represents algebraic 
but non-vertical cycles characterized by (\ref{eq:rm-from-coker}), 
as we will see in section \ref{sect:correction-term}.
Not all of the components in this $RM^m_a(\hat{Z})$-dimensional space 
may be of this form for general choices of $\Delta$ and $\widetilde{\Delta}$, 
however. We just simply do not know at this moment.

\subsubsection{Triangulation (resolution) independence}
\label{sect:triang-indep}

As we have already explained in section \ref{sec:landscape-horizontal}, 
formulating F-theory compactification on resolved fourfolds only makes sense
if the dimensions of the vertical, horizontal and the remaining components (i.e., $h^{2,2}_V$, $h^{2,2}_H$ and 
$h^{2,2}_{RM}$) of a Calabi--Yau fourfold $\hat{Z}$ are independent of 
the choice of crepant resolution of singularities of $Z_s$ (fine regular unimodular 
triangulation of $\widetilde{\Delta}$).\footnote{To be more precise, 
the formulation of F-theory suggests this resolution independence only for 
Calabi--Yau fourfolds where $Z_s$ is given by a Weierstrass-model 
elliptic fibration over $B_3$, and $\hat{Z}$ is a crepant resolution 
of $Z_s$ such that $\hat{Z} \longrightarrow B_3$ remains a flat fibration. 
Thus, the statement here is stretching the ``suggestion'' a bit too far 
by not demanding a flat elliptic fibration, and also restricting the range 
of validity by focusing on $\hat{Z}$ which are obtained as hypersurfaces 
of toric fivefolds. 
Thus, an attempt of formulating flux in F-theory using resolved models 
$\hat{Z}$ will still survive, even when the dimensions $h^{2,2}_V$, 
$h^{2,2}_H$ and $h^{2,2}_{RM}$ may turn out to depend on resolutions for some 
Calabi--Yau fourfolds $Z_s$ which do not admit elliptic fibrations.} 
As we have discussed, this follows from the independence of the complex
structure moduli space on which resolution is chosen.

In this section we supplement this general argument 
with a more specific discussion of triangulation independence for the construction
discussed in this section, i.e. for resolutions $\hat{Z}$ of $Z_s$ obtained by fine unimodular 
triangulations of the polytope $\widetilde{\Delta}$.

It is easy to see, first, that $H_{\rm mon}(\hat{Z})$ and 
$V^{\rm alg}_{\rm tor}(\hat{Z})$ do not depend on the triangulation of 
$\Delta$ (or $\widetilde{\Delta}$), because their expressions only involve 
the numbers of lattice points in polytopes. Since $V_{\rm tor}(\hat{Z})$ and 
$H^{\rm red}_{\rm mon}(\hat{Z})$ are mirror to $H_{\rm mon}(\hat{Z}_m)$ and 
$V^{\rm alg}_{\rm tor}(\hat{Z}_m)$, they are also independent of the triangulation 
of $\widetilde{\Delta}$ (or $\Delta$); although the expression of 
$V_{\rm tor}(\hat{Z})$ involves a number of 1-simplices explicitly, we have 
seen by using (\ref{eq:2Theta--Theta--ell1-Theta}) that 
$V_{\rm tor}(\hat{Z})$ is equal to $H_{\rm mon}(\hat{Z}_m)$, and the number of 
1-simplices used in $V_{\rm tor}(\hat{Z})$ does not depend on the choice of 
triangulation. This means that both 
\begin{equation*}
  \left[ V_{\rm tor}(\hat{Z}) + V^{\rm alg}_{\rm tor}(\hat{Z}) \right]
 \quad {\rm and} \quad 
  \left[ H_{\rm mon}(\hat{Z}) + H^{\rm red}_{\rm mon}(\hat{Z}) \right]
\end{equation*}
are independent of triangulations. 

The dimensions of other components such as $V_{\rm cor}(\hat{Z})$, 
$V^{\rm alg}_{\rm cor}(\hat{Z})$, $RM_a(\hat{Z})$, etc., however, involve 
counting the number of 1-simplices with much more specific restrictions, and it is 
not obvious at first sight how we see triangulation-independence. 
Let us look at (\ref{eq:RM}), however, where four terms are grouped 
into two. The first two terms do not depend on the triangulation of 
$\Delta$, but they may depend on the triangulation of $\widetilde{\Delta}$; 
the last two terms, on the other hand, do not depend on the triangulation of 
$\widetilde{\Delta}$, but they may depend on the triangulation of $\Delta$.
The dimension of $h^{2,2}_{RM}(\hat{Z})$, however, 
has no chance of depending on the choice of triangulation of $\Delta$ by construction. 
This means that the last two terms of (\ref{eq:RM}) combined---$[RM^m_a(\hat{Z}) + 
H^{\rm red}_{\rm rm}(\hat{Z})]$---should not depend on the choice of a
triangulation of $\Delta$, and not just on the triangulation of $\widetilde{\Delta}$.
Taking its mirror, we see that the combination of the first two terms,  
\begin{equation*}
[RM_a(\hat{Z}) + V^{\rm alg}_{\rm rm}(\hat{Z})], 
\end{equation*}
also does not depend on the triangulation of $\widetilde{\Delta}$. 
This proves that $h^{2,2}_{RM}(\hat{Z})$ is independent of which (fine, regular, unimodular)
triangulation is chosen.

In order to prove that $h^{2,2}_V(\hat{Z})$ is also independent of the
triangulation of $\widetilde{\Delta}$, note that $V^{\rm alg}(\hat{Z})$ and 
$NV_1(\hat{Z}_m)+NV_3(\hat{Z}_m)$ are independent of triangulation; they depend 
only on numbers of lattice points, not on 1-simplices. This means that 
the combination $V^{\rm alg}_{\rm cor}(\hat{Z})+V^{\rm alg}_{\rm rm}(\hat{Z})$ is 
also independent of triangulation, because $V^{\rm alg}_{\rm tor}(\hat{Z})$ is,  
and so is the combination $[V_{\rm cor}(\hat{Z}) + RM_a(\hat{Z})]$ 
because of the relation (\ref{eq:afew-relations}). From all above, 
we see that the combination
\begin{equation*}
  \left[ V_{\rm cor}(\hat{Z}) + V^{\rm alg}_{\rm cor}(\hat{Z}) \right],
\end{equation*}
the second group of terms in (\ref{eq:V}), is also independent of 
triangulation. 
Obviously the independence of 
$[H_{\rm cor}(\hat{Z}) + H^{\rm red}_{\rm cor}(\hat{Z})]$ also follows from 
mirror symmetry.

We have therefore seen that the six groups of terms in (\ref{eq:V}, 
\ref{eq:H}, \ref{eq:RM}) are separately independent of the choice of 
triangulations of $\widetilde{\Delta}$ and $\Delta$ in a 
toric-{\it hypersurface} realization of a smooth $\hat{Z}$ and singular $Z_s$. 
This statement is almost the same as the similar statement in section 
\ref{sec:landscape-horizontal}, although the argument in 
section \ref{sec:landscape-horizontal} is about arbitrary crepant resolutions 
$\hat{Z}$ of $Z_s$, i.e. $\hat{Z}$ does not have to be a toric hypersurface.
When a Calabi--Yau fourfold hypersurface $\hat{Z}$ of a toric fivefold 
is also realized as a complete intersection in an ambient space of higher 
dimensions, the separation between $[V_{\rm tor}+V^{\rm alg}_{\rm tor}]$
and $[V_{\rm cor}+V^{\rm alg}_{\rm cor}]$ may not remain the same, in general.
One can also see that the argument for triangulation-independence given here 
exploits some combinatorics of toric data, (\ref{eq:2Theta--Theta--ell1-Theta}), 
but still relies partially on mirror symmetry. This means that there must be 
some triangulation independent relations involving such numbers as 
$\ell_1^{*\bullet}(\Theta^{[3]})$, $\ell_1^*(\widetilde{\Theta}^{[1]})$ etc.

\section{Examples}\label{sect:examples}

A couple of examples of toric-hypersurface Calabi-Yau fourfolds are presented
in this section. We begin with the pair 
of sextic $(6) \subset \P^5$ and its mirror in section \ref{sect:sextic}, 
where the geometry is so simple that we can compute everything by hand.
It serves well for the purpose of digesting such notions as 
stratification and mixed Hodge structure. We will see how things work together 
nicely so that the long exact sequence (\ref{eq:long-seq-cpt-supp}) holds.
For more complicated toric-hypersurface Calabi--Yau fourfolds, however, 
we need to use the computer packages {\verb TOPCOM } \cite{TOPCOM} and 
{\verb sage } \cite{sage} partially in the computation (see 
section \ref{sect:howtocompute}).
Examples in section \ref{sect:correction-term} bring the 
formulae (\ref{eq:V}, \ref{eq:H}, \ref{eq:RM}) and 
(\ref{eq:V-alg-parts}, \ref{eq:V-parts}, \ref{eq:H-parts}, 
\ref{eq:H-parts-2}) to life. We chose examples where various terms have 
non-zero contributions, so that we can test our geometric interpretation 
developed in the previous section.
In section \ref{sect:elfibexamples}, we work on examples to be used 
in F-theory compactification for unified theories, and compute 
$h^{2,2}_V$, $h^{2,2}_H$ and $h^{2,2}_{RM}$. 
The results in this section are used as an input 
in section \ref{sect:distrGNgen}, along with additional results 
from appendices \ref{sect:chirflux} and \ref{sect:dephodgegroup}, 
to study how the number of flux vacua depends on the number of 
generations $N_{\rm gen}$ or on the choice of the unification group of 
low-energy effective theories.

\subsection{The sextic and its mirror}
\label{sect:sextic}

As a first canonical example, let us discuss the sextic fourfold 
$\hat{Z}_6$, a degree-six hypersurface of $\P^5$, and its mirror manifold 
denoted by $\hat{Z}_{6,m}$. With this definition, it is easy to find 
(using index theorems and the Lefschetz hyperplane theorem) that the
Hodge numbers of $\hat{Z}_6$ are
\begin{align}
h^{p,q}(\hat{Z}_6) = 
\begin{array}{|ccccc}
 1 &  & & & 1 \\
  & 426 & 0 & 1 & \\
 & 0 & 1752 & 0 &\\
 & 1 & 0 & 426&\\
1 & & & & 1 \\
\hline
\end{array}.
\label{eq:diamond-sextic}
\end{align} 
In this presentation of the Hodge numbers, $p$ starts from 0 and increases 
to the right, while $q$ begins with 0 and increases upward. 
We will use the same presentation style in this article, when we write down 
the numbers $e^{p,q}_c$. 
For the sextic, $h^{1,1}(\hat{Z}_6) = 1$ is generated by $H|_{\hat{Z}_6}$, 
restriction of the hyperplane class of $\P^5$. The vertical component 
$H^{2,2}_V(\hat{Z})$ is generated by $H^2|_{\hat{Z}_6}$, and we 
expect $h^{2,2}_V(\hat{Z})=1$. Mirror symmetry also indicates that 
$h^{2,2}_H(\hat{Z}_{6,m})=1$. 
We compute $h^{2,2}_H(\hat{Z}_6) = h^{2,2}_V(\hat{Z}_{6,m})$ and 
$h^{2,2}_{RM}(\hat{Z}_{6}) = h^{2,2}_{RM}(\hat{Z}_{6,m})$ 
in section \ref{sect:Z6(m)-V-H-RM}. 
Sections~\ref{sect:Z6} and \ref{sect:Z6m} are only meant to be warming 
up exercise for readers unfamiliar with such notions as 
mixed Hodge structure or toric stratification.

As a toric variety, $\P^5$ can be described by a fan over the faces of 
a polytope $\tilde{P}_6$, whose six vertices in $\mathbb{Z}^{\oplus 5}$
are given by 
\begin{align}
 \left(\begin{array}{rrrrrr}
-1 & 0 & 0 & 0 & 0 & 1 \\
-1 & 0 & 0 & 0 & 1 & 0 \\
-1 & 0 & 0 & 1 & 0 & 0 \\
-1 & 0 & 1 & 0 & 0 & 0 \\
-1 & 1 & 0 & 0 & 0 & 0
\end{array}\right).
\end{align}
The dual polytope $P_6$ has six vertices in $\mathbb{Z}^{\oplus 5}$:
\begin{equation}
 \left(\begin{array}{rrrrrr}
-1 & -1 & -1 & -1 & -1 & 5 \\
-1 & -1 & -1 & -1 & 5 & -1 \\
-1 & -1 & -1 & 5 & -1 & -1 \\
-1 & -1 & 5 & -1 & -1 & -1 \\
-1 & 5 & -1 & -1 & -1 & -1
\end{array}\right), 
\label{eq:toric-data-sextic-M}
\end{equation}
and one quickly recognizes that the fan $\Sigma$ over the faces of $P_6$ 
gives rise to an orbifold of $\P^5$. 

\subsubsection{Geometry of the sextic: mixed Hodge structure of its 
subvarieties}
\label{sect:Z6}

The sextic $\hat{Z}_6$ has six toric divisors $\hat{Y}_i$, $i=1,\cdots, 6$, 
corresponding to the six vertices of $\tilde{P}_6$.
These toric divisors $\hat{Y}_i$ are all $(6) \subset \P^4$. Similarly, 
$\hat{Y}_i \cap \hat{Y}_j$ are surfaces $(6) \subset \P^3$, while 
$\hat{Y}_i \cap \hat{Y}_j \cap \hat{Y}_k$ are curves $(6) \subset \P^2$ and 
$\hat{Y}_i \cap \hat{Y}_j \cap \hat{Y}_k \cap \hat{Y}_l$ are six points in $\P^1$.
These facts can be read out from the fact that the $k$-dimensional faces 
$\Theta^{[k]}$ of the polytope $P_6$ are $k$-dimensional pyramids of 
height-6 (7 points in one edge),\footnote{\label{fn:pyramid}
A $k$-dimensional pyramid of height-$h$ is (anything lattice-isomorphic to) 
the minimal $k$-dimensional simplex in a lattice $\Z^{\oplus k}$ enlarged by a 
positive integer $h$. The number of lattice points on such a pyramid is 
$N_{k,h} := [(h+1)(h+2)\cdots(h+k)]/k!$, while the number of interior points 
is given by $N_{k,h-k-1}$.} which are regarded as the complete linear 
system of the divisor $6H$ of $\P^k$. Exploiting e.g. index theorems in combination with the Picard-Lefschetz hyperplane theorem 
one easily finds that 
\begin{align}
 h^{p,q}(\hat{Y}_i) = \begin{array}{|cccc}
 5 &  &  & 1 \\
  & 255 & 1 &  \\
  & 1 & 255 &  \\
 1 &  &  & 5 \\
 \hline 
 \end{array}\, \quad \quad 
 h^{p,q}(\hat{Y}_i\cap \hat{Y}_j) = \begin{array}{|ccc}
  10 &  & 1  \\
  & 86 &   \\
 1 &  & 10 \\
 \hline 
 \end{array}\,  \nn \\ 
 &&&&\nn\\
  h^{p,q}(\hat{Y}_i\cap \hat{Y}_j\cap \hat{Y}_k) = \begin{array}{|cc}
 10 & 1 \\
 1 & 10  \\
 \hline 
 \end{array}\,  \quad\quad
  h^{p,q}(\hat{Y}_i\cap \hat{Y}_j\cap \hat{Y}_k\cap \hat{Y}_l) = \begin{array}{|cc}
  &    \\
 6 &   \\
 \hline 
 \end{array}\,  .
\label{eq:diamond-subvar-sextic}
\end{align}

We begin with the direct computation of $H^k(Y)$ using the explicit results 
(\ref{eq:diamond-subvar-sextic}) and the Mayer--Vietoris spectral sequence, 
where $Y = \cup_i \hat{Y}_i$ is 
the complement of the primary stratum $Z_{P_6}$ in $\hat{Z}_6$. 
The compact support cohomology groups of $Z_{P_6}$, on the other hand, are 
obtained by following the algorithm of \cite{DK} (which we have reviewed partially 
in the previous section). Those computations allow us to see that the 
long exact sequence (\ref{eq:long-seq-cpt-supp}) nicely reproduces the 
Hodge diamond (\ref{eq:diamond-sextic}); 
see (\ref{eq:long-exact-seq-cpt-supp-sextic}).

It is important in the sum rule (\ref{eq:DKsumrule}, \ref{eq:DKsumrule-mdfd}) 
and the algorithm of \cite{DK} that they can be applied to polytopes 
that do not necessarily correspond to the complete linear system defining a family 
of Calabi--Yau hypersurfaces; the sum rule (\ref{eq:DKsumrule-mdfd}) has been 
used in (\ref{eq:sumruleexpl-2}, \ref{eq:sumruleexpl-3}), for example. 
Thus, we will compute $e^{p,q}_c(Z_{\Theta^{[k]}})$ from the Hodge diamonds of the 
subvarieties (\ref{eq:diamond-subvar-sextic}), and confirm that the sum rule (\ref{eq:DKsumrule}, 
\ref{eq:DKsumrule-mdfd}) is indeed satisfied at the end of this 
section. 

The cohomology group of the compact (but singular) geometry 
$Y = \cup_i \hat{Y}_i$ is computed by the Mayer--Vietoris spectral sequence.
At the stage of $d_1^{pq}: E^{p,q}_1 \longrightarrow E_1^{p+1,q}$, we have 
\begin{equation}
 \begin{array}{|c|cc|cc|cc}
 6\times (1) \\
 0 \\
 \hline
 6\times (0,1,0) & \longrightarrow & 15 \times (1) \\
 6\times (5,255,255,5) & \longrightarrow & 0 \\
 \hline
 6\times (0,1,0) & \longrightarrow & 15 \times (10,86,10) & \longrightarrow &
  20\times (1) & & \\
 0 & \longrightarrow & 0 & \longrightarrow & 20 \times (10,10) & & \\
\hline
 6\times (1) & \longrightarrow & 15\times(1) & \longrightarrow &
 20\times (1) & \longrightarrow & 15\times (6) \\
\hline
 \end{array},
\end{equation}
where ${\rm dim}(E_1^{p,q})$ is shown in the $(p+1)$-th column from the left 
and the $(q+1)$-th row from the bottom, in a form maintaining the information 
of the Hodge filtration. 
To proceed to the stage $E_2^{p,q}$ in the spectral sequence calculation, 
we need to know the morphisms $d_1^{pq}: E_1^{p,q} \longrightarrow E_1^{p+1,q}$.
In the case of the sextic, the morphism  
$d_1: E_1^{0,q} \longrightarrow E_1^{1,q}$ has a 5-dimensional image 
for $q=0,2,4$, while $d_1: E_1^{1,q} \longrightarrow E_1^{2,q}$
has a 10-dimensional image for $q=0,2$. The morphism 
$d_1:E_1^{2,0} \longrightarrow E_1^{3,0}$ has a 10-dimensional image.
Combining all of the above, we have that 
\begin{equation}
 {\rm dim}(E_2^{p,q}) = 
 \begin{array}{|c|c|c|c}
   (6) \\
       \\
  \hline
   (0,1,0) & (10) \\
 (30,1530,1530,30) & \\
  \hline 
  (0,1,0) & (150,1275,150) & (10) \\ 
          &                & (200,200) \\
  \hline 
   (1) & (0) & (0) & (80) \\
\hline
 \end{array}.
\end{equation}
We conclude from this that 
\begin{equation}
  h^{3,3}[H^6(Y)] = 6, \quad 
  h^{2,2}[H^5(Y)] = 10, \quad 
  h^{0,0}[H^0(Y)] = 1, \quad
  h^{1,1}[H^2(Y)] = 1,
\label{eq:HDnumber-HotherY}
\end{equation}
and all other Hodge--Deligne numbers of the cohomology groups $H^{k}(Y)$ with 
$k=0,1,2,5,6$ vanish. As for $H^k(Y)$ with $k=3,4$, 
\begin{equation}
 h^{p,q}[H^3(Y)] = \begin{array}{|cccc}
    30 & & & \\
    150 & 1530 & \\
    200 & 1275 & 1530 & \\
    80 & 200 & 150 & 30 \\
   \hline \end{array}, \qquad 
 h^{p,q}[H^4(Y)] = \begin{array}{|cccc}
    & 0 & & \\
    & 0 & 1 & \\
    & 10 & 0 & 0 \\
    &  &  & \\
    \hline \end{array}.
\label{eq:HDnumber-H34Y}
\end{equation}

Let us now move on to the computation the Hodge--Deligne numbers of the 
cohomology group of the top (primary) stratum 
$Z_6 = \hat{Z}_6 \setminus Y = Z_{P_6}$. They are determined by 
the algorithm of \S 5 in \cite{DK}, which is precisely the one we 
adopted in section \ref{sect:h22-H4Z}. We only need the values of the 
functions $\varphi_i$ for the polytope $P_6$, 
\begin{equation}
\varphi_1(P_6) = 1, \quad \varphi_2(P_6)= 456, \quad 
\varphi_3(P_6) = 3431, \quad \varphi_4(P_6)= 3431, \quad 
\varphi_5(P_6) = 456,
\label{eq:value-varphi-P6}
\end{equation}
in order to use the algorithm.\footnote{This task has already 
been carried out partially in section \ref{sect:h22-H4Z}.
$h^{2,1}$ of $H^4_c(Z_{P_6})$ is given by \eqref{eq:formula-H4Z-21cmp}, 
while \eqref{eq:formula-H4Z-dim} determines $h^{2,2}$.  
The formulae (\ref{eq:ep0}, \ref{eq:temp1}, \ref{eq:temp2}) can be used 
to determine $h^{p,q}$'s of $H^4_c(Z_{P_6})$ with $p=1,2,3,4$, 
while $h^{3,1}$ is determined by (\ref{eq:e31}, \ref{eq:bath31}).
The sum rule (\ref{eq:DKsumrule}, \ref{eq:DKsumrule-mdfd}) still 
has to be used along with (\ref{eq:value-varphi-P6}), however, 
to determine two more Hodge--Deligne numbers 
$h^{1,1}$ and $h^{0,0}$ of $H^4_c(Z_{P_6})$.} It turns out that 
\begin{equation}
  h^{p,q}[H^4_c(Z_{P_6})] = \begin{array}{|ccccc}
    1 \\
    30 & 426 \\
    150 & 1530 & 1751 \\
    200 & 1275 & 1530 & 426 \\
    80 & 200 & 150 & 30 & 1\\
   \hline \end{array} .
\label{eq:HDnumber-H4Z}
\end{equation}
Using the Hodge--Deligne numbers of other cohomology groups 
$H^k_c(Z_{P_6})$ given in \eqref{eq:Hodge-Deligne-for-alg-torus}, 
we also see that 
\begin{equation}
  e^{p,q}_c(Z_{P_6}) = 
   \begin{array}{|ccccc}
   1 &&&& 1\\
 30 &  426 &  & -5& \\
  150 & 1530 & 1761 & & \\
  200 & 1265 & 1530 & 426 & \\
 80 & 200 & 150 & 30 &1 \\
 \hline 
 \end{array}\, .
\end{equation}

With all the Hodge--Deligne numbers of the cohomology groups 
$H^k_c(Z_{P_6})$ and $H^k(Y)$ determined, we are now ready to see 
that the long exact sequence (\ref{eq:long-seq-cpt-supp}) reproduces 
the Hodge diamond of the sextic (\ref{eq:diamond-sextic}):
\begin{equation}
\begin{array}{ccccccccc}
  [H^4_c(Z_{P_6})] & \rightarrow & [H^4(\hat{Z})] & \rightarrow & [H^4(Y)] &
  \rightarrow & [H^5_c(Z_{P_6})] & \rightarrow & [H^5(\hat{Z})] \\
  h^{2,2}=1751 & \rightarrow & h^{2,2}=1752 & \rightarrow & h^{2,2}=1 &
  \rightarrow & 0 & &  \\
  h^{3,1}=426  & \rightarrow & h^{3,1}=426  & \rightarrow & h^{3,1}=0 &
  &   & &  \\
  h^{4,0}=1    & \rightarrow & h^{4,0}=1    & \rightarrow & h^{4,0}=0 & 
  &    & &  \\
              & & 0 & \rightarrow & h^{1,1}=10 & \rightarrow & h^{1,1}=10 & 
  \rightarrow & 0 
\end{array};
\label{eq:long-exact-seq-cpt-supp-sextic}
\end{equation}
$h^{p,q}[H^4_c(Z_{P_6})]$ for $(p,q)$ with $p+q < 4$ are irrelevant here, 
because $h^{p,q}[H^3(Y)] = h^{p,q}[H^4_c(Z_{P_6})]$ for $p+q < 4$, as one 
can see from (\ref{eq:HDnumber-H34Y}) and (\ref{eq:HDnumber-H4Z}).

The study above used the Mayer--Vietoris spectral sequence and the Hodge 
numbers of subvarieties (\ref{eq:diamond-subvar-sextic}) to determine 
the Hodge--Deligne numbers of $H^k(Y)$. This is doable by hand only for 
such a simple geometry as the sextic, however. The Hodge--Deligne numbers of 
$H^k(Y)$ in more complicated geometries are dealt with much more 
systematically  
under the approach using the toric stratification (\ref{stratintro}) and 
the algorithm and sum rule of \cite{DK}.
We therefore confirm from here, toward the end of this section, 
that the sum rule (\ref{eq:DKsumrule}) is satisfied indeed by 
$e^{p,q}_c(Z_{\Theta^{[k]}})$'s of various faces $\Theta^{[k]}$ of the polytope $P_6$. 

The polytope $P_6$ has six vertices $\Theta^{[0]}$ 
in (\ref{eq:toric-data-sextic-M}), $\binom{6}{2}=15$ faces $\Theta^{[1]}$ of 
dimension-1, $\binom{6}{3} = 20$ faces $\Theta^{[2]}$ of dimension-2, 
$\binom{6}{4} =15$ faces $\Theta^{[3]}$ of dimension-3, and 
$\binom{6}{5} = 6$ facets $\Theta^{[4]}$. All the faces of a given dimension $k$ 
are lattice-isomorphic, thanks to the high symmetry of the 
sextic--mirror-sextic. Each one of the $k$-dimensional faces 
has $\binom{k+1}{1}$ facets $\Theta^{[k-1]}$ of dimension-$(k-1)$, 
$\binom{k+1}{2}$ faces $\Theta^{[k-2]}$ of dimension-$(k-2)$, etc., all the 
way to $\binom{k+1}{k}=(k+1)$ faces $\Theta^{[0]}$, i.e., vertices.   

Let us begin with the face $\Theta^{[1]}$. The hypersurface 
$Z_{\Theta^{[1]}} \subset \mathbb{T}^1$ consists of six points, 
just like in the last entry of (\ref{eq:diamond-subvar-sextic}).
This means that $e^{0,0}_c(Z_{\Theta^{[1]}})=6$, and 
$\bar{e}^{0,0}_c(Z_{\Theta^{[1]}})=5$. Because $\Theta^{[1]}$ is a 1-dimensional 
pyramid of height 6, $\ell^*(\Theta^{[1]}) = N_{1,6-2} = 5$ 
(see footnote \ref{fn:pyramid}), 
the sum rule (\ref{eq:DKsumrule-mdfd}) --- which states that $\bar{e}^{0,0}_c(Z_{\Theta^{[1]}})
=\varphi_1(\Theta^{[1]}) = \ell^*(\Theta^{[1]})=5$ in this case --- is 
satisfied indeed.

The next step is to verify the sum rule for $Z_{\Theta^{[2]}}$. 
It is a degree-6 curve of $\P^2=\P^2_{\Theta^{[2]}}$ (i.e., $g=10$ curve, 
$\hat{Y}_i \cap \hat{Y}_j \cap \hat{Y}_k$ in (\ref{eq:diamond-subvar-sextic})), 
with 3 set of six points (i.e., $Z_{\Theta^{[1]}}$) removed. Using the 
additivity of $e^{p,q}_c$'s, we find that 
\begin{equation}
 e^{p,q}_c(Z_{\Theta^{[2]}}) = \begin{array}{|cc}
 -10 & 1 \\
 1 & -10  \\
 \hline 
 \end{array}\, 
 - 3\,\, \begin{array}{|cc}
  &    \\
 6 &   \\
 \hline 
 \end{array}\,  
 =
 \begin{array}{|cc}
 -10 & 1    \\
 -17 &  -10 \\
 \hline 
 \end{array}\,  , \qquad 
\bar{e}^{p,q}_c(Z_{\Theta^{[2]}}) = \begin{array}{|cc}
 -10 & \\ -15 & -10 \\ \hline \end{array}.
\end{equation}
The sum $\sum_q \bar{e}^{p,q}_c$ is indeed equal to $-\varphi_{2-p}(\Theta^{[2]})$ 
given by $\varphi_1(\Theta^{[2]})=\ell^*(\Theta^{[2]})=N_{2,6-3}=10$ and 
$\varphi_2(\Theta^{[2]})=\ell^*(2\Theta^{[2]})-3\ell^*(\Theta^{[2]})=
N_{2,12-3}-3\cdot N_{2,6-3}= 25$.

A similar computation for $\Theta^{[3]}$ and $\Theta^{[4]}$ proceeds as 
follows:
\begin{equation}
  e^{p,q}_c(Z_{\Theta^{[3]}}) = \begin{array}{|ccc}
  10 &  & 1  \\
  & 86 &   \\
 1 &  & 10 \\
 \hline 
 \end{array}\, - 4 \,\, \begin{array}{|cc}
 -10 & 1    \\
 -17 &  -10 \\
 \hline \end{array}
 -6\,\, \begin{array}{|cc}
  &    \\
 6 &   \\
 \hline 
 \end{array}\,  
 =\,\, \begin{array}{|ccc}
  10 &  & 1  \\
  40 & 82 &   \\
 33 & 40  & 10 \\
 \hline 
 \end{array}\, ,
\end{equation}
\begin{equation}
 e^{p,q}_c(Z_{\Theta^{[4]}}) = 
 \begin{array}{|cccc}
 -5 &  &  & 1 \\
  & -255 & 1 &  \\
  & 1 & -255 &  \\
 1 &  &  & -5 \\
 \hline 
 \end{array}\, 
 - 5\,\, \begin{array}{|ccc}
  10 &  & 1  \\
  40 & 82 &   \\
 33 & 40  & 10 \\
 \hline 
 \end{array}\,
 - 10 \,\,  \begin{array}{|cc}
 -10 & 1    \\
 -17 &  -10 \\
 \hline 
 \end{array}\,
  -10 \,\, \begin{array}{|cc}
   &   \\
 6 &   \\
 \hline 
 \end{array} . 
\end{equation}
We thus find that 
\begin{equation}
\bar{e}^{p,q}_c(Z_{\Theta^{[3]}}) = \begin{array}{|ccc}
  10 & & \\ 40 & 85 & \\ 30 & 40 & 10 \\ \hline \end{array}, \qquad 
\bar{e}^{p,q}_c(Z_{\Theta^{[4]}}) = - \,\, \begin{array}{|cccc}
  5 & & & \\ 50 & 255 & & \\ 100 & 425 & 255 & \\ 50 & 100 & 50 & 5 \\
   \hline \end{array},
\end{equation}
which are consistent with the sum rule on $\sum_q \bar{e}^{p,q}_c(\Theta^{[k]})$
using the value of $\varphi_{k-p}(\Theta^{[k]})$ given by 
\begin{eqnarray}
 &&\varphi_1(\Theta^{[3]}) = 10, \quad \varphi_2(\Theta^{[3]}) = 125, \quad 
 \varphi_3(\Theta^{[3]}) = 80, \\
 &&\varphi_1(\Theta^{[4]}) = 5, \quad \varphi_2(\Theta^{[4]}) = 305, \quad 
 \varphi_3(\Theta^{[4]}) = 780, \quad \varphi_4(\Theta^{[4]}) = 205.
\end{eqnarray}

As a final consistency check, one can compute $e^{p,q}_c(Y)$ by using 
the additivity of $e^{p,q}_c$ and the stratification of $Y$ into non-compact 
and smooth subsets:
\begin{align}
 e^{p,q}_c(Y) &= \binom{6}{2}e^{p,q}_c(Z_{\Theta^{[1]}})+ \binom{6}{3}e^{p,q}_c(Z_{\Theta^{[2]}})+ \binom{6}{4}e^{p,q}_c(Z_{\Theta^{[3]}}) +
 \binom{6}{5} e^{p,q}_c(Z_{\Theta^{[4]}}) \nn\\
 &=
 \begin{array}{|cccc}
 -30 &  &  & 6 \\
  -150 & -1530 & -9 &  \\
  -200 & -1264 & -1530 &  \\
 -79 & -200 & -150 & -30 \\
 \hline 
 \end{array}\, .
\end{align}
One can see that this result is consistent with $e^{p,q}_c(Y)$ given directly 
from the Hodge--Deligne numbers of $H^k(Y)$ in (\ref{eq:HDnumber-HotherY}, 
\ref{eq:HDnumber-H34Y}). Also, $e^{p,q}_c(Z_{P_6})+e^{p,q}_c(Y)$ reproduces 
the Hodge diamond (\ref{eq:diamond-sextic}), as it should be.

\subsubsection{Geometry of the mirror-sextic: toric stratification}
\label{sect:Z6m}

Let us now use the mirror-to-sextic $\hat{Z}_{6,m}$ fourfold
to have a close look at the stratification associated with the toric 
divisors $\left\{ \hat{Y}_i \right\}$. The stratification of $\hat{Z}_6$ 
was very simple---$Z_6 + \binom{6}{1} Z_{\Theta^{[4]}} + \binom{6}{2} Z_{\Theta^{[3]}} 
+ \cdots + \binom{6}{4} Z_{\Theta^{[1]}}$---because all the faces 
$\widetilde{\Theta}^{[k]}$ of the polytope $\tilde{P}_6$ are 
lattice-isomorphic to the minimal simplex of $k$-dimension, so that 
any one of $E_{\widetilde{\Theta}^{[k]}}$'s consists of a single point.
The faces $\Theta^{[k]}$ ($k < n=5$) of the polytope $P_6$ are not, however.

To the primary stratum of $\hat{Z}_{6,m}$, $\binom{6}{1} \times 
Z_{\widetilde{\Theta}^{[4]}}$ is added first. 
Each one of the $Z_{\widetilde{\Theta}^{[4]}}$'s is a $\P^3$ with five hyperplanes 
removed.

Coming next are the strata  
$\binom{6}{2} \times E_{\Theta^{[1]}} \times Z_{\widetilde{\Theta}^{[3]}}$.
The geometry of $Z_{\Theta^{[3]}}$ is a $\P^2$ with four hyperplanes removed, 
while $E_{\Theta^{[1]}}$ is the exceptional locus of an $A_4$ singularity 
resolution. This exceptional locus $E_{\Theta^{[1]}}$ has a stratification 
that consists of five $\mathbb{T}^1$'s and six points, because 
$\ell^*(\Theta^{[1]})=5$ and $\ell_1^*(\Theta^{[1]})=6$.

\begin{figure}[tbp]
\begin{center}
\scalebox{.4}{ 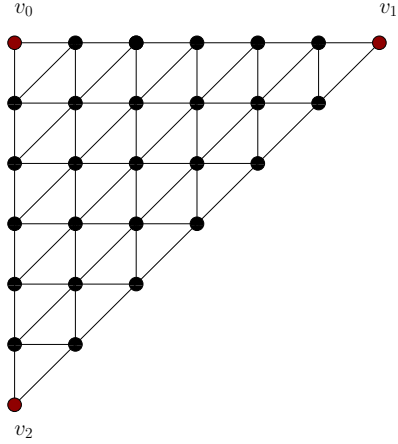 }
\caption{A 2-dimensional face of $P_6$, $\Theta^{[2]}$. 
We have shown vertices, integral points, and a simple triangulation.
\label{sextic2dface}}
\end{center} 
\end{figure}

Similarly, there are strata $\binom{6}{3} \times E_{\Theta^{[2]}} \times 
Z_{\widetilde{\Theta}^{[2]}}$. Each one of $Z_{\widetilde{\Theta}^{[2]}}$ is $\P^1$ 
with three points removed, while $E_{\Theta^{[2]}}$ has a stratification 
comprised of ten $\mathbb{T}^2$'s, $3 \times 15=45$ $\mathbb{T}^1$'s and 
thirty-six points. If a polytope $\Theta^{[2]}$ is given the triangulation 
shown in figure \ref{sextic2dface}, then $E_{\Theta^{[2]}}$ contains ten compact 
complex surfaces isomorphic to $dP_3$. 

The only remaining strata are $\binom{6}{4} \times 
Z_{\widetilde{\Theta}^{[1]}} \times E_{\Theta^{[3]}}$; each 
$Z_{\widetilde{\Theta}^{[1]}}$ consists of a single point. 
It is not too difficult to find a unimodular fine triangulation of 
$\Theta^{[3]}$ by hand. For this triangulation, 
it turns out that $\ell_1^*(\Theta^{[3]}) = 155$. 
This number satisfies 
$\ell^*(2\Theta^{[3]}) - \ell^*(\Theta^{[3]})=N_{3,8}-N_{3,2}=155 = 
\ell_1^*(\Theta^{[3]})$, a special case of \eqref{eq:2Theta--Theta--ell1-Theta}. Because of this relation, $\ell_1^*(\Theta^{[3]})$ does not depend on the triangulation we choose.

For the mirror sectic, we can also compute the numbers $e^{p,q}_c$'s for $E_{\Theta^{[k]}}$ and $Z_{\widetilde{\Theta}^{[4-k]}}$ 
as well the primary stratum, and we can then use their multiplicativity 
and additivity to compute $e^{p,q}(Y)$, just as we did for the case of the sextic. All these details, however, 
are not recorded here. 

\subsubsection{Evaluation of $h^{2,2}_V$, $h^{2,2}_H$ and $h^{2,2}_{RM}$}
\label{sect:Z6(m)-V-H-RM}

Let us now evaluate the dimension of the vertical, horizontal and 
the remaining components in $H^{2,2}$ given 
by (\ref{eq:V}, \ref{eq:H}, \ref{eq:RM}), respectively; we do this 
both for the sextic $\hat{Z}_6$ and its mirror $\hat{Z}_{6,m}$. 

Let us begin with the vertical component $H^{2,2}_V$ of the sextic $\hat{Z}_6$.
Because all the faces of the polytope $\widetilde{P}_6$ do not have 
an interior point (except $\widetilde{P}_6$ itself), both 
$V_{\rm cor}(\hat{Z}_6)$ and $V^{\rm alg}(\hat{Z}_6)$ vanish. Therefore 
the vertical component comes purely from intersections of toric divisors,
\begin{equation}
 h^{2,2}_V(\hat{Z}_6) =  V_{\rm tor}(\hat{Z}_6) =
      \binom{6}{2} + 10 - 4 \cdot 6 = 1 \, , 
\end{equation}
where we have used (\ref{eq:V-parts}). The vertical component 
$H^{2,2}_V(\hat{Z})$ is generated by $H^2|_{\hat{Z}_6}$, as we discussed 
already at the beginning of this section.

The same formula (\ref{eq:V}) can be used also to determine the 
dimension of the vertical (2,2)-forms on the mirror fourfold $\hat{Z}_{6,m}$, 
$h^{2,2}_V(\hat{Z}_{6,m})$.
We first note that $V_{\rm cor}(\hat{Z}_{6,m})$ and $V^{\rm alg}(\hat{Z}_{6,m})$ 
vanish, as there are no interior points in $\widetilde{\Theta}^{[1]}$, 
$\widetilde{\Theta}^{[2]}$ and $2 \widetilde{\Theta}^{[2]}$. 
In order to evaluate $V_{\rm tor}(\hat{Z}_{6,m})$, we need to count the 
number of 1-simplices in the polytope $P_6$. We find that 
\begin{equation}
\ell_1((P_{6})_{\leq 3}) = \binom{6}{4} \cdot \ell_1(\Theta^{[3]})
 + \binom{6}{3} \cdot \ell_1^*(\Theta^{[2]})
 + \binom{6}{2} \cdot \ell_1(\Theta^{[1]}) = 3315 , 
\end{equation}
using the 1-simplex counting presented in section \ref{sect:Z6m}.
To determine the dimensions subtracted due to the toric rational equivalence, 
the second and third line of $V_{\rm tor}(\hat{Z}_{6,m})$, we also need 
$\sum_{\Theta^{[4]} \leq P_6} \ell^*(\Theta^{[4]}) = \binom{6}{5} \cdot N_{4,1} = 30$ 
and 
$\sum_{\Theta^{[3]} \leq P_6} \ell^*(\Theta^{[3]}) = \binom{6}{4} \cdot N_{3,2} = 150$. 
Hence
\begin{align}
h^{2,2}_V(\hat{Z}_{6,m}) = V_{\rm tor}({\hat{Z}_{6,m}}) = 
   3315 + 10 - 4(462 - 30 - 150 - 1) - 3 \cdot 150  =  1751.
\end{align}

This result already indicates that $h^{2,2}_V(\hat{Z}_{6,m})=1751$ and 
$h^{2,2}_H(\hat{Z}_{6,m})=h^{2,2}_V(\hat{Z}_6)=1$ add up to yield all of 
$h^{2,2}(\hat{Z}_{6,m})=1752$. Thus, the remaining component should be 
absent in $\hat{Z}_{6,m}$ as well as in $\hat{Z}_6$. It is not difficult 
to see this directly. First, $RM_a(\hat{Z}_6) =RM^m_a(\hat{Z}_{6,m})$
vanishes because all the faces $\widetilde{\Theta}^{[3]}$ of 
$\widetilde{P}_6$ are minimal three-dimensional simplices and no 1-simplices 
are introduced upon triangulation; $\ell_1^{*\bullet}(\widetilde{\Theta}^{[3]})=0$.
Secondly, $RM^m_a(\hat{Z}_6)=RM_a(\hat{Z}_{6,m})$ also vanishes because 
all of the one-dimensional faces $\widetilde{\Theta}^{[1]}$ of $\widetilde{P}_6$ 
do not have an interior point. 

Similarly, one can evaluate the dimension of the horizontal component 
$h^{2,2}_H(\hat{Z}_6)$ and $h^{2,2}_H(\hat{Z}_{6,m})$ directly from the formulae 
in the previous section, although we already know their results.
As in the discussion for the vertical component above, one can see that both
$H^{\rm red}$ and $H_{\rm cor}$ vanish for both $\hat{Z}_6$ and $\hat{Z}_{6,m}$.
Thus, the horizontal component only comes from the monomial deformation
$H_{\rm mon}$; 
the results are
\begin{equation}
h^{2,2}_H(\hat{Z}_6) = H_{\rm mon}(\hat{Z}_6) = 3431 - 6 \cdot 305 + 15 \cdot 10
 = 1751\, ,
\end{equation}
and 
\begin{equation}
h^{2,2}_{H}(\hat{Z}_{6,m}) = H_{\rm mon}(\hat{Z}_6) = 1 - 0 + 0 = 1\, 
\end{equation}
by using $\varphi_3(\widetilde{P}_6)=1$. 

\subsection{Computations in practice}\label{sect:howtocompute}

For practical applications to fourfolds more complicated than the sextic, 
we clearly do not want to evaluate (\ref{eq:V}, \ref{eq:H}, \ref{eq:RM}) 
by hand. What is even worse, it is a non-trivial task 
to find an appropriate triangulation of $\widetilde{\Delta}$. Certainly 
the value of $h^{2,2}_V$, $h^{2,2}_H$ and $h^{2,2}_{RM}$ do not depend 
on the choice of triangulation, as we have seen at the end 
of section \ref{hyperresultVvsH}, but at least we have to make sure that 
there is at least one triangulation that is fine, regular and unimodular 
(see section \ref{sect:tsp}), 
or otherwise the formulae in the previous section cannot be applied. 

In this section we hence explain how to evaluate 
(\ref{eq:V}, \ref{eq:H}, \ref{eq:RM}) in practice using existing computer 
software. The computation of Hodge numbers is most efficiently
carried out using the package PALP \cite{Kreuzer:2002uu} (described in some more detail in \cite{Braun:2012vh}). 
In the present context, PALP is also useful to construct a reflexive polytope 
from a combined weight system.

In order to obtain triangulation we use the package {\verb TOPCOM } \cite{TOPCOM}.
It is able to perform fine regular triangulations, which are, however, not necessarily star. Furthermore, as we have explained in section \ref{sect:tsp}, a fine triangulation does not 
always give rise to a smooth Calabi-Yau hypersurface, let alone a smooth ambient space.

Even though a fine star triangulation is naturally related to a maximally subdivided fan, we may also
use a fine regular triangulation of $\widetilde{\Delta}$ which is not star for practical purposes. 
Let us hence assume we have found a regular, fine (non-star) triangulation of $\widetilde{\Delta}$. This clearly gives rise to triangulations of all faces of $\widetilde{\Delta}$ which are mutually consistent, i.e. the triangulations of any two neighbouring faces induce the same triangulation on their intersection. 
We may then construct a fan $\Sigma'$ (or, if we like a star triangulation) over all simplices on
$\partial \widetilde{\Delta}$ obtained this way. As discussed e.g. in \cite{Batyrev94dualpolyhedra} (see also \cite{rambau}), regularity of the original triangulation implies existence of a strongly convex support function on the simplices on $\partial \widetilde{\Delta}$, which in turn can be lifted to a strongly convex
support function on the cones of $\Sigma'$. 

We hence feed the configuration of integral points on $\partial \widetilde{\Delta}$ into {\verb TOPCOM } to generate a fine regular triangulation, given in terms of five-simplices. As explained above, this can be cast into the data of a fan $\Sigma'$, or, equivalently, a star triangulation. 
Such manipulations can be conveniently carried out using the computer algebra system {\verb sage } \cite{sage}.
In particular, {\verb sage } already contains many routines to construct and analyse lattice polytopes. 

Having obtained a regular star triangulation, we only need to check unimodularity. Again, this can be easily
done by checking that the lattice volumes of all five-dimensional cones are unity. 

With a smooth ambient space at hand, we are ready to evaluate the formulae (\ref{eq:V}, \ref{eq:H}, \ref{eq:RM}). This can again efficiently be done using {\verb sage }. The computation of the whole procedure outlined above (without any optimization) can be done in a few hours (for polytopes $\widetilde{\Delta}$ with $\mathcal{O}(100)$ points such as the mirror sextic) to a few days (for polytopes with several 1000 points such as the mirrors of the cases discussed in section \ref{sect:elfibexamples}) using an off-the-shelf PC at the time of writing.

Note that a straightforward evaluation of eqs. (\ref{eq:V}, \ref{eq:H}, \ref{eq:RM}) requires a triangulation of both $\widetilde{\Delta}$ and $\Delta$, even though we expect the final result to be independent of any triangulation.

\subsection{Correction terms at work}
\label{sect:correction-term}

The sextic and its mirror are clearly very degenerate examples for the evaluation 
of (\ref{eq:V}, \ref{eq:H}, \ref{eq:RM}), as the polytope $\widetilde{P}_6$
for $\P^5$ does not have any interior points of any of its faces and furthermore does not 
require triangulation. In this section \ref{sect:correction-term}, 
we present some examples for which various correction terms 
in (\ref{eq:V}, \ref{eq:H}, \ref{eq:RM}) give a non-zero contribution.

Our first example is given by the following pair of reflexive polytopes
\begin{equation}
\widetilde{P}_B =  \left(\begin{array}{rrrrrrrrrrrrrrrr}
-2 & -2 & 0 & 0 & 0 & 0 & 0 & 0 & 0
& 0 & 0 & 0 & 0 & 0 & 2 & 2 \\
-2 & 0 & -2 & -2 & 0 & 0 & 0 & 0 & 0
& 0 & 0 & 0 & 2 & 2 & 0 & 2 \\
0 & 0 & -2 & 0 & -2 & -2 & 0 & 0 & 0
& 0 & 2 & 2 & 0 & 2 & 0 & 0 \\
0 & 0 & 0 & 0 & -2 & 0 & -2 & 0 & 0
& 2 & 0 & 2 & 0 & 0 & 0 & 0 \\
1 & 1 & 1 & 1 & 1 & 1 & 1 & -1 & 1 &
-1 & -1 & -1 & -1 & -1 & -1 & -1
\end{array}\right) \, ,
\end{equation}
and
\begin{equation}
P_B =\left(\begin{array}{rrrrrrrrrrrrrrrr}
-1 & -1 & -1 & 0 & 0 & 0 & 0 & 0 & 0
& 0 & 0 & 0 & 0 & 1 & 1 & 1 \\
0 & 0 & 0 & -1 & -1 & 0 & 0 & 0 & 0
& 0 & 0 & 1 & 1 & 0 & 0 & 0 \\
-1 & 0 & 0 & 0 & 0 & -1 & 0 & 0 & 0
& 0 & 1 & 0 & 0 & 0 & 0 & 1 \\
0 & -1 & 0 & -1 & 0 & 0 & -1 & 0 & 0
& 1 & 0 & 0 & 1 & 0 & 1 & 0 \\
-1 & -1 & -1 & -1 & -1 & -1 & -1 & -1 &
1 & 1 & 1 & 1 & 1 & 1 & 1 & 1
\end{array}\right) \, .
\end{equation}
As before, we have simply given the polytopes in terms of a matrix 
containing their vertices. We have confirmed, however, that there is 
a fine unimodular regular triangulation for both of the polytopes 
$\widetilde{P}_B$ and $P_B$, following the procedure described 
in section \ref{sect:howtocompute}. 
The Calabi-Yau fourfold $\hat{Z}_B \subset \P^5_{\Sigma'_B}$ 
has the Hodge numbers
\begin{equation}
h^{p,q}(\hat{Z}_{B}) = 
\begin{array}{|ccccc}
 1 &  & & & 1 \\
  & 11 & 0 & 97 & \\
 & 0 & 476 & 0 &\\
 & 97 & 0 & 11&\\
1 & & & & 1 \\
\hline
\end{array}. 
\end{equation}
The Hodge diamond of the mirror $\hat{Z}_{B,m} \subset 
\P^5_{\widetilde{\Sigma}'_B}$ is the left-right (or top-bottom) 
flip of the Hodge diamond above. 

Evaluation of (\ref{eq:V}, \ref{eq:H}, \ref{eq:RM}) gives that 
$h^{2,2}(\hat{Z}_B) = h^{2,2}(\hat{Z}_{B,m})$ should be decomposed as 
$h^{2,2}_V(\hat{Z}_B)=440$, $h^{2,2}_H(\hat{Z}_B)=32$ and 
$h^{2,2}_{RM}(\hat{Z}_B)=4$. 
In particular 
\begin{equation}\label{eq:tablePB}
 \begin{array}{c|cc|cc|cc}
 & V_{\rm tor}+V^{\rm alg}_{\rm tor} & V_{\rm cor}+V^{\rm alg}_{\rm cor} & 
 RM_a + V^{\rm alg}_{\rm rm} & RM^m_a+H^{\rm red}_{\rm rm} & 
 H_{\rm cor}+H^{\rm re}_{\rm cor} & H_{\rm mon}+H^{\rm red}_{\rm mon} \\
		\hline
\hat{Z}_B &
 442-2 & 0-0 &    0-0 & 4-0 &    0-0 & 32-0 \\
\hat{Z}_{B,m} & 
 32 -0 & 0-0 & 4-0 & 0-0 & 0-0 & 442 - 2 \\
 \end{array}\, .
\end{equation}

\begin{figure}[h!]
\begin{center}
a) \includegraphics[height=3cm]{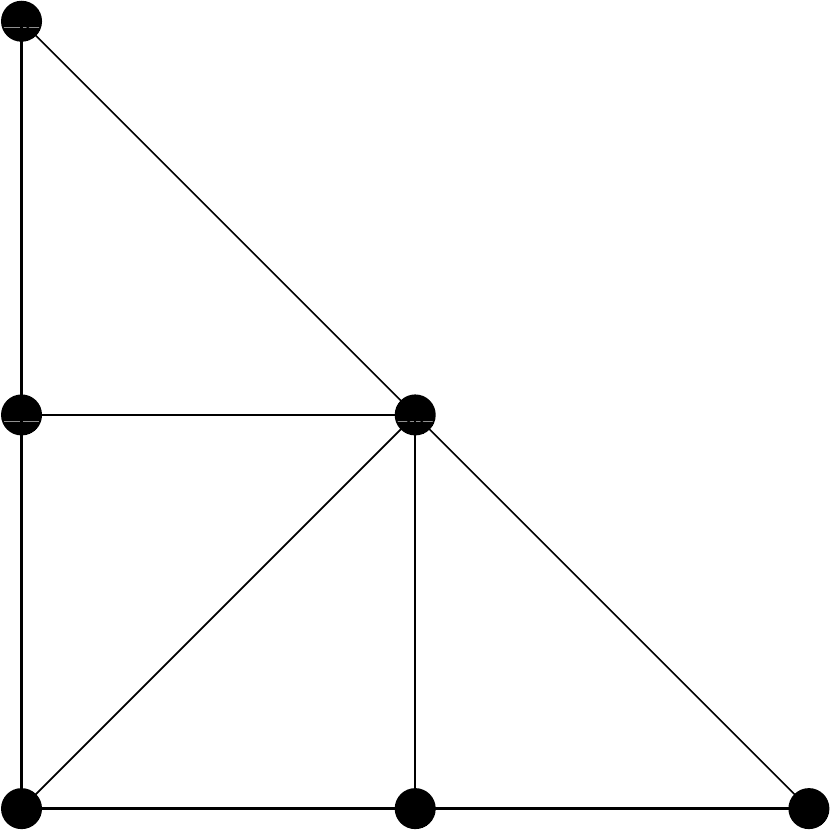}
\hspace{1cm}
b) \includegraphics[height=3cm]{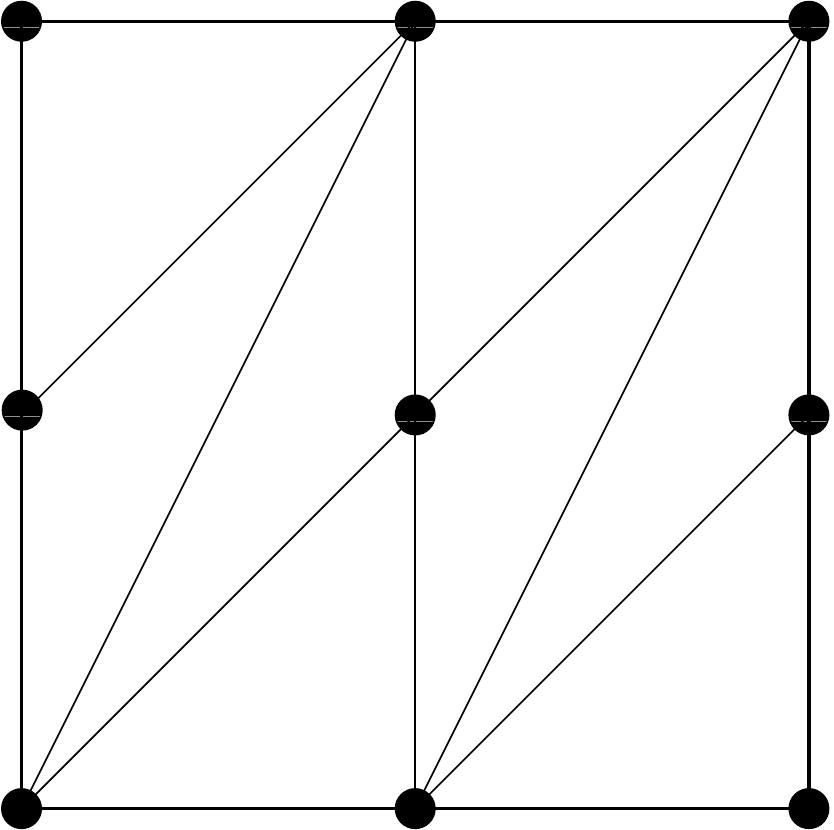}
\hspace{1cm}
c) \includegraphics[height=1.5cm]{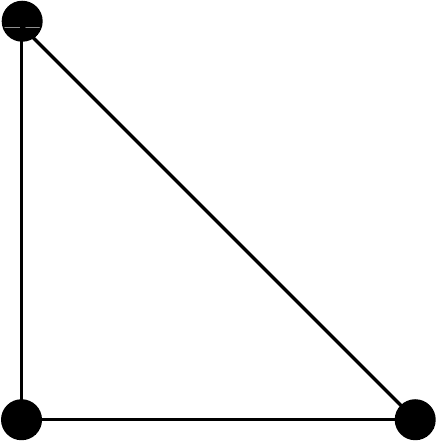}
\hspace{1cm}
d) \includegraphics[height=1.5cm]{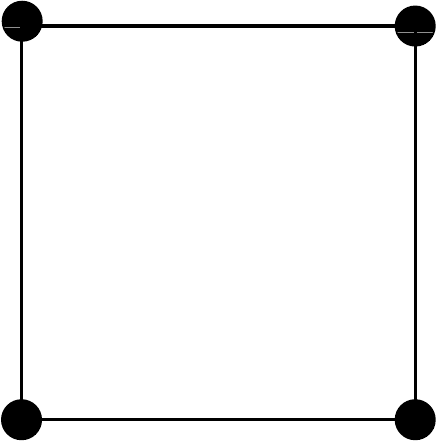}
 
 \caption{\label{fig:facestT2} 
The two isomorphism classes of 2-dimensional faces of the polytope 
$\widetilde{P}_B$, $\widetilde{\Theta}^{[2]}_T$ and $\widetilde{\Theta}^{[2]}_S$
are shown in (a) and (b), whereas those of the polytope $P_B$, 
$\Theta^{[2]}_t$ and $\Theta^{[2]}_s$ are in (c) and (d), respectively.
A specific fine unimodular triangulation of those faces is  
also shown in (a) and (b).
}
\end{center}
\end{figure}

As can already be seen from the lists of vertices above, we find 
$\ell^*(\widetilde{\Theta}^{[1]})=1$ for all 1-dimensional faces of 
$\widetilde{P}_B$ and $\ell^*(\Theta^{[1]})=0$ for all 1-dimensional faces 
of $P_B$. This already explains why 
$[V_{\rm cor}(\hat{Z}_B)+RM_a(\hat{Z}_B)]=0$, 
$[NV_1(\hat{Z}_{B,m})+NV_3(\hat{Z}_{B,m})]=0$.
Furthermore, none of the 2-dimensional faces of $P_B$ has any interior 
points (there are $80$ triangles $\Theta^{[2]}_t$ and $18$ squares 
$\Theta^{[2]}_s$ with $3$, resp. $4$ integral points; 
see figure~\ref{fig:facestT2}~(c,d)), so that also 
$V^{\rm alg}(\hat{Z}_{B,m})=0$, $H^{\rm red}(\hat{Z}_B)=0$ and     
$NV_3(\hat{Z}_B) = 0$ follows.   

The correction term $V^{\rm alg}(\hat{Z}_B)=-2$ is understood as follows. 
There are 98 dual pairs of faces $(\widetilde{\Theta}^{[2]}, \Theta^{[2]})$
in the pair of polytopes $(\widetilde{P}_B, P_B)$; among them, 
64 are of the type $(\widetilde{\Theta}^{[2]}_T, \Theta^{[2]}_t)$, 
16 are of the type $(\widetilde{\Theta}^{[2]}_T, \Theta^{[2]}_s)$,
16 are of the type $(\widetilde{\Theta}^{[2]}_S, \Theta^{[2]}_t)$ and 
2 are of the type $(\widetilde{\Theta}^{[2]}_S, \Theta^{[2]}_s)$, 
modulo lattice-isomorphism; see figure~\ref{fig:facestT2}~(a--d) for 
the notation of $\widetilde{\Theta}^{[2]}_{T,S}$ and $\Theta^{[2]}_{t,s}$.
The correction term $V^{\rm alg}(\hat{Z}_B) = -2$ originates 
from the 2 dual pairs $(\widetilde{\Theta}^{[2]}_S, \Theta^{[2]}_s)$. 
There is a divisor $\hat{Y}_i$ of $\hat{Z}_B$ for each pair,
which is regarded as a surface fibration over a curve $\Sigma_{\Theta^{[2]}_s}$.
The generic fibre is a surface, which is 
$\P^1 \times \P^1 \subset E_{\widetilde{\Theta}^{[2]}_S}$ in the case the 
triangulation of $\widetilde{\Theta}^{[2]}_S$ is the one in 
figure~\ref{fig:facestT2}~(b). The base $\Sigma_{\Theta^{[2]}_s}$ 
is a curve of genus $g = \ell^*(\Theta^{[2]}_s)=0$,
because $\Theta^{[2]}_s$ corresponds to the complete linear system 
of $H_1+H_2$ of $\P^1 \times \P^1$. he fibre may degenerate 
at $k_{\Theta^{[2]}_s}=4$ points $\Sigma_{\Theta^{[2]}_s} \setminus Z_{\Theta^{[2]}_s}$.
A naive toric calculation counts the fibre class $k_{\Theta^{[2]}_s}-2$ times, 
and the correction term $\Delta V^{\rm alg}(\hat{Z}_B) = 
\bar{e}^{0,0}_c(\Theta^{[2]}_s)=-1$ removes the overcounting of the 
fibre class; $k_{\Theta^{[2]}_s}-2+\bar{e}^{0,0}_c=1$
because of (\ref{eq:topology-relation-ebar00}). Because all the four 
faces $\Theta^{[1]}$ of $\Theta^{[2]}_s$ do not have an interior point 
(see figure~\ref{fig:facestT2}~(d)), this contributes to 
$V^{\rm alg}_{\rm tor}(\hat{Z}_B)$ and not to $V^{\rm alg}_{{\rm cor}+{\rm rm}}$.

Although we do not have a particular geometric interpretation for 
the correction term $H^{\rm red}(\hat{Z}_{B,m})$, certainly the contribution 
$H^{\rm red}(\hat{Z}_{B,m})=-2$ comes from the same two dual pairs of 
the faces isomorphic to $(\widetilde{\Theta}^{[2]}_S, \Theta^{[2]}_s)$.

\begin{figure}[h!]
\begin{center}
 \includegraphics[height=3cm]{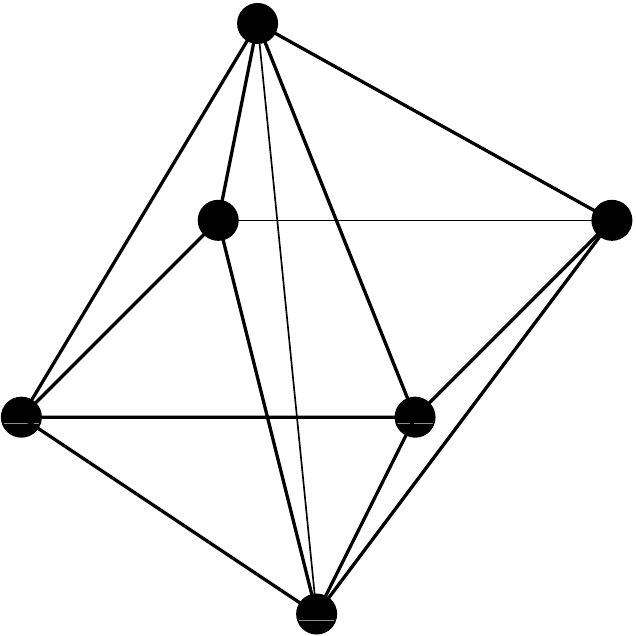}
 \caption{\label{fig:facdiamond}The 64 faces of 3-dimensions in $P_B$ form 
four lattice-isomorphism classes. One of them looks like the one 
in this figure, and is denoted by $\Theta^{[3]}_{\rm dia}$. Four out of 64
belong to this class. 
The face $\Theta^{[3]}_{\rm dia}$ contains a single internal one-simplex 
(also shown in the figure) after triangulation. One example of such a face 
has $\left(-1,\,0,\,-1,\,0,\,-1\right)$,
$\left(-1,\,0,\,0,\,-1,\,-1\right)$, $\left(-1,\,0,\,0,\,0,\,-1\right)$,
$\left(0,\,0,\,0,\,0,\,1\right)$, $\left(0,\,0,\,0,\,1,\,1\right)$, and
$\left(0,\,0,\,1,\,0,\,1\right)$ as its vertices.}
\end{center}
\end{figure}

The correction term $RM_a(\hat{Z}_{B,m})=4$ is understood as follows. 
The 64 dual pairs of faces $(\widetilde{\Theta}^{[1]}, \Theta^{[3]})$ are 
classified (modulo lattice isomorphisms) into four types. All the 
$\widetilde{\Theta}^{[1]}$'s are isomorphic, and $\Theta^{[3]}$'s have one of 
the four shapes, a tetrahedron, a pyramid with a rectangular base, 
a 3-dimensional prism, and a diamond, shown in figure \ref{fig:facdiamond}. The correction term $RM_a(\hat{Z}_{B,m})$ 
comes from the pairs $(\widetilde{\Theta}^{[1]}, \Theta^{[3]}_{\rm dia})$.
The one-simplex in $\Theta^{[3]}_{\rm dia}$ in figure \ref{fig:facdiamond} 
is surrounded by four triangles and four tetrahedra, realizing 
a geometry $\mathbb{T}^1 \times \P^1 \times \P^1$ in the toric ambient 
space $\P^5_{\widetilde{\Sigma}'_B}=\P^5_{\widetilde{P}_B}$ for $\hat{Z}_{B,m}$. 
The hypersurface equation selects 
$[1+\ell^*(\widetilde{\Theta}^{[1]})]=\ell(\widetilde{\Theta}^{[1]})-1=2$ points 
out of $\mathbb{T}^1$. Thus, this one-simplex does not represent 
a single irreducible algebraic cycle, but consists of two irreducible ones, 
each of which is isomorphic to $\P^1 \times \P^1$.
These two irreducible algebraic cycles combined are vertical and can be obtained as 
the intersection of divisors corresponding to the two boundary points of 
the 1-simplex in $\Theta^{[3]}_{\rm dia} \leq P_B$. This piece is counted 
as part of $V_{\rm tor}(\hat{Z}_{B,m})$. 
The other $\ell^*(\widetilde{\Theta}^{[1]})=1$ combination of the 
irreducible algebraic cycles, however, is not realized as a vertical cycle, 
and is counted as a part of $RM_a(\hat{Z}_{B,m})$. There are four 
such dual pairs of faces $(\widetilde{\Theta}^{[1]}, \Theta^{[3]}_{\rm dia})$, 
and this is how $RM_a(\hat{Z}_{B,m})=4$ is obtained. 

Finally, we obtain $RM^m_a(\hat{Z}_B)= 4$ in the following way. 
It comes from the same four dual pairs of type  
$(\widetilde{\Theta}^{[1]}, \Theta^{[3]}_{\rm dia})$.
As noted already, $\ell^*(\widetilde{\Theta}^{[1]})=1$ for all 
1-dimensional faces of the polytope $\widetilde{P}_B$, and 
the 1-simplex in figure~\ref{fig:facdiamond} contributes to 
$\ell_1^{*\bullet}(\Theta^{[3]}_{\rm dia})$ {\it combinatorially};   
each pair of dual faces $(\widetilde{\Theta}^{[1]}, \Theta^{[3]}_{\rm dia})$ 
gives rise to a $1 \times 1$ contribution to $RM^m_a(\hat{Z}_B)$, and there 
are four such pairs. The {\it geometric} interpretation of this component, 
however, is not as directly related to the 1-simplex in 
$\Theta^{[3]}_{\rm dia}$ as the interpretation of $RM_a(\hat{Z}_{B,m})$ is.  
Remember that $NV_3(\hat{Z}_B)=0$, and note that
$V_{\rm cor}(\hat{Z}_{B,m})=H_{\rm cor}(\hat{Z}_B)$ vanishes in this example.
The face $\Theta^{[3]}_{\rm dia}$ does not have an interior point or 
1-simplex ending on such an interior point. This means 
that $RM^m_a(\hat{Z}_B)$ originates purely from $NV_1(\hat{Z}_B)$ 
in this example. Let us now focus on one of the four pairs of 
$(\widetilde{\Theta}^{[1]}, \Theta^{[3]}_{\rm dia})$, and 
let $\nu_i$ and $\hat{Y}_i$ be the interior point of $\widetilde{\Theta}^{[1]}$
and the corresponding divisor of $\hat{Z}_B$, respectively. This divisor 
$\hat{Y}_i$ contains a group of strata forming a $\P^1$ fibration over 
a surface $Z_{\Theta^{[3]}_{\rm dia}}$. This surface has 
$\bar{e}^{1,1}_c(Z_{\Theta^{[3]}_{\rm dia}})=1$, which is the origin of 
$RM^m_a(\hat{Z}_B) \neq 0$. The Hodge diamond of this divisor $\hat{Y}_i$ 
is reconstructed by collecting $e^{p,q}_c$ of all the relevant 
strata (see (\ref{eq:collect-relevant-strata})), and we found that it is 
\begin{equation}
  h^{p,q}(\hat{Y}_i) = \begin{array}{|cccc}
   & & & 1 \\ & & 11 & \\  & 11 & & \\ 1 & & & \\ \hline \end{array}; 
\end{equation}
$h^{2,0}(\hat{Y}_i)$ vanishes because $\ell^*(\Theta^{[3]}_{\rm dia})=0$.
All of the components in $H^{1,1}_c(\hat{Y}_i)$ and $H^{2,2}_c(\hat{Y}_i)$ are 
therefore algebraic, but the exact 
sequence (\ref{eq:long-seq-cpt-spp-for-Yi-(n-3)}) shows that only a
10-dimensional subspace of the $h^{2,2}_c(\hat{Y}_i)=11$-dimensional space of
algebraic cycles is realized in the form of vertical cycles.
The remaining $\bar{e}^{1,1}_c =1$-dimensional contribution to $H^{2,2}_c(\hat{Y}_i)$ 
is regarded as part of the $RM^m_a(\hat{Z}_B)$ component, which is 
algebraic (and hence non-horizontal), but not vertical. 
In this example, we see that the $RM^m_a(\hat{Z}_B)$-component is 
also characterized by (\ref{eq:rm-from-coker}), 
not just the elements in $RM_a(\hat{Z})$ are.

In this example, the correction terms for $H^{1,1}(\hat{Z}_B)$ and $H^{1,1}(\hat{Z}_{B, m})$, i.e. the last term in \eqref{eq:bath11}, vanish
and all divisor classes are generated by toric divisors for both fourfolds. This means we can easily compute the dimensions of $H^{2,2}_V(\hat{Z}_B)$
and $H^{2,2}_V(\hat{Z}_{B , m})$ by using the intersection ring of the ambient toric space restricted to $\hat{Z}_B$ and $\hat{Z}_{B, m}$. This computation
reproduces the dimensions of the vertical components in \eqref{eq:tablePB}, $h^{2,2}_V(\hat{Z}_B)= 440 $ and $h^{2,2}_V (\hat{Z}_{B, m}) = 32$. 
Using mirror symmetry, we can then also recover the dimensions of the horizontal components. From $h^{2,2}= 476$ it follows again that 
the remaining component must be 4-dimensional.  
\\
\\

A second example is defined by the dual pair of polyhedra
\begin{equation}
\widetilde{P}_S = \left(\begin{array}{rrrrrr}
-1 & 0 & 0 & 0 & 0 & 1 \\
-1 & 0 & 0 & 0 & 1 & 0 \\
-15 & -1 & 5 & 11 & 0 & 0 \\
-3 & 1 & 1 & 1 & 0 & 0 \\
-9 & -1 & 5 & 5 & 0 & 0
\end{array}\right)\, , \, P_S = 
 \left(\begin{array}{rrrrrr}
-1 & -1 & -1 & -1 & -1 & 5 \\
-1 & -1 & -1 & -1 & 5 & -1 \\
-1 & 0 & 0 & 1 & 0 & 0 \\
0 & -1 & 4 & -1 & -1 & -1 \\
2 & 0 & -1 & -1 & 0 & 0
\end{array}\right) \, ,
\end{equation}
where we have displayed the vertices in the form of a matrix. It turns out 
that these two polyhedra are equivalent, so that they define the same 
Calabi--Yau fourfold, which is hence self-mirror. We have found fine 
regular and unimodular triangulations for both $\widetilde{P}_S$ and $P_S$, 
which are used for the computations below. The hodge numbers are
\begin{equation}
h^{p,q}(\hat{Z}_{S}) = 
\begin{array}{|ccccc}
 1 &  & & & 1 \\
  & 30 & 101 & 30 & \\
 & 101 & 82 & 101 &\\
 & 30 & 101 & 30&\\
1 & & & & 1 \\
\hline
\end{array} \, .
\end{equation}
and we find
\begin{equation}
 \begin{array}{c|cc|cc|cc}
 & V_{\rm tor}+V^{\rm alg}_{\rm tor} & V_{\rm cor}+V^{\rm alg}_{\rm cor} & 
 RM_a + V^{\rm alg}_{\rm rm} & RM^m_a+H^{\rm red}_{\rm rm} & 
 H_{\rm cor}+H^{\rm re}_{\rm cor} & H_{\rm mon}+H^{\rm red}_{\rm mon} \\
		\hline
\hat{Z}_S &
 41 - 0 & 0-0 &    150-150 & 150-150 &    0-0 & 41-0 
 \end{array}\, .
\end{equation}
As expected for a self-mirror fourfold with $h^{2,2}_{RM} = 0$, the result is $h^{2,2}_V = \tfrac12 h^{2,2}$.

Let us comment on the computation of $NV_3(\widetilde{P}_S)$, which was zero in the example before.
There are three 1-dimensional faces with $\ell^*(\widetilde{\Theta}^{[1]})=5$. 
In $\widetilde{P}_S$ they have vertices at $\{(0, 0, -1, 1, -1), (0, 0, 5, 1, 5)\}$, $\{(0, 0, -1, 1, -1), (0, 0, 11, 1, 5)\}$ and $\{(0, 0, 5, 1, 5), (0, 0, 11, 1, 5)\}$. They are dual to 3-dimensional faces each of which contains 10 points interior to their 2-dimensional faces. As all other 1-dimensional faces have $\ell^* = 0$, we hence find $NV_3(\widetilde{P}_S) = 150$. 

Just like in the example $P_B$, the correction term for $H^{1,1}(\hat{Z}_S)$ vanishes and we can independently verify that 
$h^{2,2}_V(\hat{Z}_S)= h^{2,2}_V(\hat{Z}_{S ,m})= 41$ by using the intersection ring of the ambient space.

We now discuss an example for which $h^{2,2}_{RM} \neq 0$ and not all toric divisors are irreducible. The vertices of the polytopes defining this example and its mirror are
\begin{equation}
\widetilde{P}_3 = \left(\begin{array}{rrrrrrrrrrrrr}
-1 & 0 & 0 & 0 & 0 & 0 & 0 & 0 & 0 &
0 & 0 & 0 & 1 \\
-3 & -2 & -2 & -1 & -1 & 0 & 0 & 0 & 2
& 2 & 4 & 4 & 0 \\
-5 & -3 & -2 & -2 & -2 & 0 & 0 & 4 & 1
& 1 & 3 & 4 & 0 \\
-8 & -5 & -4 & -3 & -2 & 0 & 2 & 2 & 0
& 1 & 1 & 2 & 0 \\
-7 & -5 & -5 & -2 & -2 & 1 & 1 & -3 & 1
& 1 & 1 & 1 & 0
\end{array}\right)
\end{equation}
and
\begin{equation}
P_3 = \left(\begin{array}{rrrrrrrrrrrrr}
-1 & -1 & -1 & -1 & -1 & -1 & -1 & -1 &
0 & 1 & 1 & 8 & 10 \\
-1 & -1 & -1 & 0 & 1 & 1 & 3 & 4 & -1
& -1 & 4 & 0 & 1 \\
1 & 1 & 2 & 0 & -2 & -1 & 0 & -1 & 1
& 2 & -1 & 0 & -1 \\
-1 & 0 & 0 & 0 & 2 & 0 & 0 & 0 & -1
& 0 & 0 & 0 & 0 \\
1 & 0 & -1 & -1 & -1 & -1 & -1 & -1 & 1
& -1 & -1 & -1 & -1
\end{array}\right) \, .
\end{equation}
We have found a fine, regular and unimodular triangulation for both polytopes. The hodge numbers are
\begin{equation}
h^{p,q}(\hat{Z}_{3}) = 
\begin{array}{|ccccc}
 1 &  & & & 1 \\
  & 76 & 11 & 67 & \\
 & 11 & 594 & 11 &\\
 & 67 & 11 & 76 &\\
1 & & & & 1 \\
\hline
\end{array} \, .
\end{equation}
Evaluation of (\ref{eq:V}, \ref{eq:H}) and (\ref{eq:RM}) gives
\begin{equation}
 \begin{array}{c|cc|cc|cc}
 & V_{\rm tor}+V^{\rm alg}_{\rm tor} & V_{\rm cor}+V^{\rm alg}_{\rm cor} & 
 RM_a + V^{\rm alg}_{\rm rm} & RM^m_a+H^{\rm red}_{\rm rm} & 
 H_{\rm cor}+H^{\rm re}_{\rm cor} & H_{\rm mon}+H^{\rm red}_{\rm mon} \\
		\hline
\hat{Z}_3 & 212 -6 & 0 -0 & 177 -74 &  118 - 61 & 29 - 0 & 199 - 0 \\
\hat{Z}_{3, m} & 199 -0 & 29 -0 & 118 -61 & 177 -74 & 0 - 0 & 212 - 6 
 \end{array}\, ,
\end{equation}
so that
\begin{equation}
 h^{2,2}_V(\hat{Z}_3) = 206 \quad\quad  h^{2,2}_{RM}(\hat{Z}_3) = 160 \quad\quad  h^{2,2}_H(\hat{Z}_3) = 228 
\end{equation}

In this example, the correction term to $h^{1,1}(\hat{Z}_{3, m})$ is equal to $10$ and some of the toric divisors are reducible, giving
rise to a non-zero contribution to $V_{\rm cor}(\hat{Z}_{3, m})$. One can check that the intersection ring of the ambient toric space restricted
to $\hat{Z}_{3, m}$ gives rise to $199$ four-cycles in this case, as expected from $V_{\rm tor}(\hat{Z}_{3, m})+V^{\rm alg}_{\rm tor}(\hat{Z}_{3, m}) = 199$. For  $V_{\rm cor}(\hat{Z}_{3})$, the correction term to $h^{1,1}$ vanishes and all 206 vertical four-cycles are obtained by the restriction of intersections of toric divisors.
 
\subsection{Elliptic fourfolds}\label{sect:elfibexamples}

In this section we come to our objects of interest, which are elliptic fourfolds. With the machinery to treat toric hypersurfaces in place,
let us discuss elliptic fibrations over toric base manifolds. A smooth elliptic fourfold over a toric base $B_{\, 3}$
not supporting any gauge group can be obtained by taking a generic Calabi-Yau hypersurface in an appropriate fibration of 
$\P^2_{123}$ over $B_{\, 3}$. The elliptic fibration morphism $\pi$ then descends from the toric morphism projecting down to $B_{\, 3}$. Torically, we may realize such a situation by setting up a polytope $\widetilde{\Delta}^{\pi}_{B_{\, 3}}$ with vertices
\begin{align}\label{eq:eftoricambient}
\left(\begin{array}{c}
       -1 \\ 0 \\ 0 
      \end{array}
\right)  \, ,
\left(\begin{array}{c}
       0 \\ -1 \\  0
      \end{array}
\right)  \, ,
\left(\begin{array}{c}
       3 \\ 2 \\  0
      \end{array}
\right)  \, ,
\left(\begin{array}{c}
       3 \\ 2 \\ \vec{v}_1  
      \end{array}
\right)  \, ,
\cdots  \, ,
\left(\begin{array}{c}
       3 \\ 2 \\ \vec{v}_n   
      \end{array}
\right)  \, .
\end{align}
Here, the $\vec{v}_i$ are the generators of the 1-dimensional cones in the fan of the toric variety $B_{\, 3}$. The above assignment ensures that $X_1$ (the coordinate associated with $(-1,0,0,0,0)$) is a section of $-3 K_{B_{\, 3}}$ and $X_2$ (the coordinate associated with $(0,-1,0,0,0)$)
is a section of $-2 K_{B_{\, 3}}$.

For appropriate base manifolds $B_{\, 3}$, a generic hypersurface $\hat{Z}^\pi_{B_{\, 3}}$ in the toric variety defined by a fan over faces of the polytope $\widetilde{\Delta}^{\pi}_{B_{\, 3}}$ is a smooth elliptic Calabi-Yau manifold with $h^{1,1}(\hat{Z}^\pi_{B_{\, 3}})=h^{1,1}(B_{\,3})+1$. A compactification of F-theory on such a manifold will not give rise to any gauge symmetry in the low energy effective action.

The (combinatorially) simplest example is $B_{\, 3}=\P^3$, where we choose
\begin{equation}
\vec{v}_1 = \left(\begin{array}{c}
                    1 \\ 0 \\ 0
                   \end{array}\right) \, , \,
\vec{v}_2 = \left(\begin{array}{c}
                    0 \\ 1 \\ 0
                   \end{array}\right) \,  , \,
\vec{v}_3 = \left(\begin{array}{c}
                    0 \\ 0 \\ 1
                   \end{array}\right) \,  , \,
\vec{v}_4 = \left(\begin{array}{c}
                    -1 \\ -1 \\ -1
                   \end{array}\right) \,   .                
\end{equation}
Another class of examples we will discuss in the following is given by taking a toric base defined by
\begin{equation}\label{eq:hirz3folds}
\vec{v}_1 = \left(\begin{array}{c}
                    1 \\ 0 \\ 0
                   \end{array}\right) \, , \,
\vec{v}_2 = \left(\begin{array}{c}
                    -1 \\ 0 \\ 0
                   \end{array}\right) \, , \,
\vec{v}_3 = \left(\begin{array}{c}
                    n \\ 1 \\ 1
                   \end{array}\right) \, , \,
\vec{v}_4 = \left(\begin{array}{c}
                    0 \\ -1 \\ 0
                   \end{array}\right) \,  ,  \,
\vec{v}_5 = \left(\begin{array}{c}
                    0 \\ 0 \\ -1
                   \end{array}\right) \,   .                
\end{equation}
Let us denote the resulting threefolds by $B_3^{\, (n)}$. 
We can also characterize them as $\mathbb{P} \left[{\cal O}_{\mathbb{P}^2} \oplus {\cal O}_{\mathbb{P}^2}(n) \right]$,
i.e. they are themselves fibrations of $\P^1$ over $\P^2$. Hence the resulting fourfolds are K3 fibred and for F-theory compactifications there is a heterotic dual. One may think of $[z_1:z_2]$ as coordinates on the $\P^1$ fibre (they also define two sections of the $\P^1$ fibration) and $[z_3:z_4:z_5]$ as coordinates on the $\P^2$ base.

Using the methods described earlier in this paper, we may now easily compute the Hodge numbers and the dimension 
of $H^{2,2}_V$, $H^{2,2}_{H}$ and $H^{2,2}_{RM}$ for these examples. The results are given in Table \ref{tab:hodge-nogauge}.
\begin{table}[htbp]
\begin{center}
 \begin{tabular}{c|cccccccc}
  $B_{\, 3}$ & $\P^3$ & $B_3^{\, (-3)}$ & $B_3^{\, (-2)}$ & $B_3^{\, (-1)}$ & $B_3^{\, (0)}$ & $B_3^{\, (1)}$ & $B_3^{\, (2)}$ & $B_3^{\, (3)}$ \\
  \hline
  $h^{1,1}$ & 2 & 3 & 3 & 3 & 3 & 3 & 3 & 3\\
  $h^{2,1}$ & 0 & 1 & 0 & 0 & 0 & 0 & 0 & 1\\
  $h^{3,1}$ & 3878 & 4358 & 3757 & 3397 & 3277 & 3397 & 3757 & 4358 \\
  $h^{2,2}$ & 15564 &17486&15084&13644&13164&13644&15084&17486\\
  $h^{2,2}_V$ & 2 &4&4&4&4&4&4&4\\
  $h^{2,2}_H$ & 15562 &17482&15080&13640&13160&13640&15080&17482 \\
  $h^{2,2}_{RM}$ & 0 & 0 & 0 & 0 & 0 & 0 & 0 & 0   \\
  $K$ & 23320 &  26200 &22596&20436&19716&20436&22596&  26200
 \end{tabular}
\caption{\label{tab:hodge-nogauge} Hodge numbers and $h^{2,2}_V$, $h^{2,2}_H$ and $h^{2,2}_{RM}$ 
for generic elliptic fibrations, based on the polytopes with vertices \eqref{eq:eftoricambient},
over base spaces $\P^3$, and $B_3^{\, (n)}$.}
\end{center}
\end{table}

For a fourfold given by toric data such as \eqref{eq:eftoricambient}, we may engineer models with a prescribed gauge group $G$ over a divisor $S$ in the base by choosing the monomials appearing in the defining equation of the hypersurface $\hat{Z}^\pi_{\, B_{\,3}}$ appropriately. If $S$ is given by a toric divisor of the ambient space, this is equivalent to deleting points from the dual polytope $\Delta_{B_{\, 3},\,G}^\pi$ in the $M$-lattice. If this process results in another reflexive polytope $\Delta_{B_{\, 3},\,G}^\pi$, we may construct the dual $\widetilde{\Delta}_{B_{\, 3},\,G}^\pi$, which contains more points than $\widetilde{\Delta}_{B_{\, 3}}^\pi$. If the resulting hypersurface $\hat{Z}_G$ is smooth, these extra points can be interpreted as the exceptional divisors of a resolution.

We may also reverse this process and define our models with an appropriate $\widetilde{\Delta}_{B_{\, 3}}^\pi$, which is what we will do in the following. This anticipates the resolution process (which is another way to find the new vertices to add)
and only allows those monomials in $\Delta$ leading to the desired fibre structure. This approach is similar to the one used in
\cite{Blumenhagen:2009yv} and the tops of \cite{Candelas:1996su,Candelas:1997eh,Perevalov:1997vw}. Equivalently, the polytopes used may also be constructed from the combined weight systems of appropriately resolved fourfolds.

In the following we will present a few examples for gauge groups SU(5) and SO(10).

\subsubsection{SU(5) along $\P^2$}\label{sect:examplesCYna4}

For a fourfold given as a hypersurface in a toric ambient space via a reflexive polytope such as
\eqref{eq:eftoricambient}, we may engineer fibres of type $I_5$ in Kodaira's classification along a divisor $[z_k]$ in the base 
(leading to SU(5) gauge symmetry in a compactification of F-theory) by
adding the vertices\footnote{To be more precise, we take the convex hull of these points and the polytope \eqref{eq:eftoricambient}.
In the resulting polytope, $ \vec{v}_{e_1} $ is not a vertex, but lies on an edge between $\vec{v}_{k}$ and $\vec{v}_{e_2}$.}:
\begin{equation}\label{eq:extraptssu5}
 \vec{v}_{e_1} = \left(\begin{array}{c}
                        2 \\ 1 \\ \vec{v}_{k}
                       \end{array}\right) \, , \, 
 \vec{v}_{e_2} = \left(\begin{array}{c}
                        1 \\ 0 \\ \vec{v}_{k}
                       \end{array}\right) \, , \, 
 \vec{v}_{e_3} = \left(\begin{array}{c}
                        0 \\ 0 \\ \vec{v}_{k}
                       \end{array}\right) \, , \,                   
 \vec{v}_{e_4} = \left(\begin{array}{c}
                        1 \\ 1 \\ \vec{v}_{k}
                       \end{array}\right) \,  
\end{equation}
to the polytope \eqref{eq:eftoricambient}. For the examples of $B_{\, 3}$ discussed in the last section, this leads to reflexive polytopes $\widetilde{\Delta}_{\P^3,\,SU(5)}^\pi$ and $\widetilde{\Delta}_{B_3^{\, (n)},\,SU(5)}^\pi$ (for $n=-3 \cdots 3$). For the bases $B_3^{\, (n)}$, we place
$S$ along the divisor $[z_1]$ (the divisor corresponding to $\vec{v}_1$ in \eqref{eq:hirz3folds}). This means that $S$ is a $\P^2$ in the base of the elliptic fibration. 

Evaluating \eqref{eq:V}, \eqref{eq:H} and \eqref{eq:RM} gives the results in 
Table \ref{tab:hodge-SU(5)}.
\begin{table}[htbp]
\begin{center}
 \begin{tabular}{c|cccccccc}
  $B_{\, 3}$ & $\P^3$ & $B_3^{\, (-3)}$ & $B_3^{\, (-2)}$ & $B_3^{\, (-1)}$ & $B_3^{\, (0)}$ & $B_3^{\, (1)}$ & $B_3^{\, (2)}$ & $B_3^{\, (3)}$ \\
  \hline
  $h^{1,1}$ & 6 & 7 & 7 & 7 & 7 & 7 & 7 & 7\\
  $h^{2,1}$ & 0 & 1 & 0 & 0 & 0 & 0 & 0 & 0\\
  $h^{3,1}$ & 2204 & 1249 & 1423 & 1723 & 2148 & 2698 & 3373 & 4173 \\
  $h^{2,2}$ & 8884 &5066 &5764&6964&8664&10864&13564&16764\\
  $h^{2,2}_V$ & 7 &9&9&9&9&9&9&8\\
  $h^{2,2}_H$ & 8877 &5057 &5755&6955&8655&10855&13555&16756 \\
  $h^{2,2}_{RM}$ & 0 & 0 & 0 & 0 & 0 & 0 & 0 & 0 \\
  $K$ & 13287 & 7557 & 8603 & 10403 & 12953 & 16253 & 20303 & 25104 
 \end{tabular}
\caption{\label{tab:hodge-SU(5)}Hodge numbers, decomposition of $H^{2,2}$ and $K$ for Calabi--Yau
fourfolds with SU(5) gauge group based on the reflexive polytopes $\widetilde{\Delta}_{\P^3,\,SU(5)}^\pi$ and $\widetilde{\Delta}_{B_3^{\, (n)},\,SU(5)}^\pi$.}
\end{center}
\end{table}

Note that in each case, $h^{1,1}$ has increased by four, corresponding to the exceptional divisors. The value of $h^{2,2}_V(\hat{Z}^\pi_{B_3^{\, (0)},\, SU(5)}) = 9$ has already been independently computed in \cite{Grimm:2011fx}. In the last case, $h^{2,2}_V(\hat{Z}^\pi_{B_3^{\, (3)},\, SU(5)}) = 8$ because the ${\bf 10}$ matter curve is absent there.

\subsubsection{SO(10) along $\P^2$}\label{sect:examplesCYnd5}

In a similar fashion as done in the last section we may construct the polytopes
$\widetilde{\Delta}^\pi_{B_{\, 3}, \, SO(10)}$. As (generic) models with gauge 
group SO(10) can be obtained by a further degeneration of SU(5) models, the
polytope $\widetilde{\Delta}^\pi_{B_{\, 3}, \, SO(10)}$ contains $\widetilde{\Delta}^\pi_{B_{\, 3}, \, SU(5)}$. The vertices which have to be added to $\widetilde{\Delta}^\pi_{B_{\, 3}, \, SU(5)}$ to
achieve gauge group SO(10) along $z_1$ are:
\begin{equation}
 \vec{v}_{e_5} = \left(\begin{array}{c}
                        2 \\ 1 \\ 2 \vec{v}_{k}
                       \end{array}\right) \, , \, 
 \vec{v}_{e_6} = \left(\begin{array}{c}
                        3 \\ 2 \\ 2 \vec{v}_{k}
                       \end{array}\right) \, .
\end{equation}

For these cases we find the hodge numbers given 
in Table \ref{tab:hodge-SO(10)}.
\begin{table}
\begin{center}
 \begin{tabular}{c|cccccccc}
  $B_{\, 3}$&$\P^3$& $B_3^{\, (-3)}$ & $B_3^{\, (-2)}$ & $B_3^{\, (-1)}$ & $B_3^{\, (0)}$ & $B_3^{\, (1)}$ & $B_3^{\, (2)}$ & $B_3^{\, (3)}$ \\
  \hline
  $h^{1,1}$ & 7   & 8       & 8    & 8    & 8    & 8    & 8    & 8\\
  $h^{2,1}$ & 0   & 0       & 0    & 0    & 0    & 0    & 0    & 0\\
  $h^{3,1}$ & 2189    & 1221    & 1402 & 1708 & 2138 & 2692 & 3370 & 4172 \\
  $h^{2,2}$ &  8828   &4958     & 5684 &6908  &8628  &10844 &13556 &16764 \\
  $h^{2,2}_V$ & 8  &10       &10     &10    &10     &10     &10     &10 \\
  $h^{2,2}_H$ &  8820   &4948     & 5674 &6898  &8618  &10834 &13546 &16754 \\
  $h^{2,2}_{RM}$ & 0 & 0 & 0 & 0 & 0 & 0 & 0 & 0 \\
  $K$ & 13200 & 7392 & 8480 & 10316 & 12896 & 16220 & 20288 & 25100
 \end{tabular}
\caption{\label{tab:hodge-SO(10)}Hodge numbers, decomposition of $H^{2,2}$ and $K$ for Calabi--Yau
fourfolds with $SU(5)$ gauge group based on the reflexive polytopes $\widetilde{\Delta}_{\P^3,\,SO(10)}^\pi$ and $\widetilde{\Delta}_{B_3^{\, (n)},\,SO(10)}^\pi$.}
\end{center}
\end{table}

\section{Fourfolds which are not hypersurfaces of toric varieties}
\label{sect:genI5fourfolds}

In the last section, we have shown how $H^{2,2}(\hat{Z},\Q)$ is
decomposed into vertical cycles, horizontal cycles and the remaining part 
$H^{2,2}_{RM}(\hat{Z},\Q)$ for the case of hypersurfaces in toric varieties.
As we have seen, the remaining part can be non-zero when divisors of $\hat{Z}$
have algebraic cycles which are not obtained by intersection with other divisors,
cf. \eqref{eq:rm-from-coker}. In this section we discuss another class 
of Calabi--Yau fourfolds which can have a non-vanishing remaining part and are motivated 
in the study of F-theory compactifications: elliptic fibrations
with singular fibres of type $I_5$ along a divisor $S$.

Let $B_3$ be a smooth Fano threefold, and let 
$Z_s$ be a fourfold given by imposing a Weierstrass equation in 
the ambient space 
$\P[ {\cal O}_{B_3} \oplus {\cal O}_{B_3}(-2K_{B_2}) \oplus 
     {\cal O}_{B_3}(-3K_{B_3}) ]$. We may choose the Weierstrass equation 
to be 
\begin{equation}
 X_1^2 = X_2^3 + a_5 X_1 X_2 X_3 + a_4 X_2^2 X_3^2 s  + a_3 X_1 X_3^3 s^2  
 + a_2 X_2 X_3^4 s^3 + a_0 X_3^6 s^5,
\label{eq:generalized-Weierstrass-Tate-SU5}
\end{equation}
where $X_{3,2,1}$ are the homogeneous coordinates of an
ambient space containing the fibre
$\P[{\cal O}\oplus {\cal O}(-2K_{B_3}) \oplus {\cal O}(-3K_{B_3}) ]$.
A morphism $\bar{\pi}: Z_s \longrightarrow B_3$ defining an elliptic fibration  naturally follows from this. 
Let $[S]$ be some divisor class of $B_3$ and $s$ a global section of 
${\cal O}_{B_3}(S)$, so that the zero locus of $s$ is $S$.
$a_{r=5,4,3,2,0}$'s are global sections of appropriate line bundles
${\cal O}_{B_3}(-(6-r)K_{B_3}  -(5-r)S)$ making \eqref{eq:generalized-Weierstrass-Tate-SU5}
a Calabi-Yau manifold. 
These conditions leave moduli for $s$ and the $a_{r=5,4,3,2,0}$, and we assume that they are chosen generic (in the sense that they are not in a Noether--Lefschetz locus). Fourfolds $Z_s$ obtained in this 
way have an $A_4$ singularity along the subvariety $s = X_2 = X_1 = 0$. 
A compact and non-singular fourfold $\hat{Z}$ can be obtained by carrying out 
the canonical resolution of the $A_4$ singularity in $Z_s$ first, and then 
going through a small resolution. The elliptic fibration 
$\pi: \hat{Z} \longrightarrow B_3$ remains a flat morphism \cite{Esole:2011sm}. 
Some more details about this resolution procedure are reviewed in 
appendices \ref{ssec:construction} and \ref{ssec:fib-degen} so that 
there is no ambiguity in the notation used in this article.

This class of geometries generalizes the examples discussed in section \ref{sect:elfibexamples}, 
where $\hat{Z}$ is realized as a hypersurface of a toric variety. As a canonical example,
we can think of $B_3 = \P^3$, and $S$ a quadratic or cubic 
(or possibly degree $d=4$) hypersurface of $B_3$. It is known that 
$S = \P^1 \times \P^1$ in the $d=2$ case, $S = dP_6$ when $d =3$, and 
$S = {\rm K3}$ for $d=4$. In these cases, 
$i^* : H^{1,1}(B_3; \C) \longrightarrow H^{1,1}(S; \C)$ is not 
surjective, which is a welcome feature in F-theory compactifications
\cite{Buican:2006sn,Beasley:2008kw}.

In this section we study the difference between the algebraic and vertical components 
in $H^{2,2}(\hat{Z}; \Q)$ for this class of geometries. This is 
done by dividing $\hat{Z}$ into a collection of (transversally intersecting) 
divisors $Y = \cup_i \hat{Y}_i \subset \hat{Z}$ and its complement 
$Z := \hat{Z} \backslash Y$, just like we did in section \ref{sect:h22hypersurface}
when studying the case where $\hat{Z}$ is a toric hypersurface. 
Here, we choose a collection of divisors $Y = \cup_i \hat{Y}_i$ exploiting 
the fibration structure:
the first five $\hat{Y}_i$, $i=0,1,2,3,4$, are reserved for the five 
irreducible components of the family of $I_5$ Kodaira fibres over the 
divisor $S$ in $B_3$. Apart from those, the collection $\cup_i \hat{Y}_i$
consists of the zero section $\sigma$ of the elliptic fibration 
$\pi: \hat{Z} \longrightarrow B_3$ and the divisors $\pi^*(D_i)$'s, where 
$\{ D_i \}$ is some basis of ${\rm Pic}(B_3)$.

We use the long exact sequence (\ref{eq:long-seq-cpt-supp}) with 
the separation $\hat{Z} = Y \amalg Z$ described above. 
The $H^4_c(Y)$-based algebraic cycles are studied by looking at 
\begin{equation}
 {\rm Ker} \left[ [H^4_c(Y)]^{2,2} \longrightarrow [H^5_c(Z)]^{2,2} \right].
\end{equation}
There are two things to be worked out:
the first is to determine the (dimension of the) algebraic components in 
$[H^4_c(Y)]^{2,2}$ for cases where $\hat{Z}$ is not necessarily a 
hypersurface of a toric variety, and the second is to 
determine $h^{2,2}[H^5_c(Z)] = h^{2,2}[H^3(Z)]$. It is known that 
$h^{2,2}[H^5_c(\hat{Z})] = 0$ when $\hat{Z}$ is a hypersurface of a toric 
variety, but this is not necessarily true for more general cases
(note that $2+2 \geq 3$, and $Z$ is smooth and non-compact).

Let us now work out $[H^4_c(Y)]^{2,2}$. It is determined by 
the Mayer--Vietoris spectral sequence, and by (\ref{eq:H4coh-Y}), 
in particular. This task consists of determining $H^4(\hat{Y}_i)$, 
$H^4(\hat{Y}_i \cap \hat{Y}_j)$ and the morphism between them.

\subsection{The dimension of $h^{1,1}(\hat{Y}_i)$'s}\label{sect:h11excpdiv}

We begin by computing the dimension of $h^2(\hat{Y}_i) = h^4(\hat{Y}_i)$ 
for the compact and generically smooth divisors 
$\hat{Y}_i \in \left\{ \sigma, \pi^*(D_i), \hat{Y}_{0,1,2,3,4} \right\}$.

For $\hat{Y}_i = \sigma \simeq B_3$, 
$H^{2,2}_c(B_3)=H^{1,1}(B_3) = NS(B_3)$. 

For $\hat{Y}_i = \pi^*(D_i)$ for one of the basis elements of 
$NS(B_3)$, $H^{1,1}(\pi^*(D))$ is generated by restrictions of 
the divisors $\sigma$, $\hat{Y}_{0,1,2,3,4}$ and divisors of $B_3$, 
where we assume that $D_i \cdot S$ is not empty in $B_3$ for now.
To verify this, the Hodge diamond of the threefold $\pi^*(D) \subset \hat{Z}$
can be determined by decomposing $\hat{Z}$ into strata first, and 
then by collecting strata that belong to $\pi^*(D)$.
It is convenient to start from the following stratification of the 
threefold $\pi^*(D)$: the surface $D \subset B_3$ is decomposed into 
$D \backslash (D \cap S)$, 
$D \cap (S \backslash (\Sigma_{({\bf 10})} \cup \Sigma_{({\bf 5})}))$ 
and $D \cap \Sigma_{({\bf 10})}$ and $D \cap \Sigma_{({\bf 5})}$, and 
the elliptically fibred threefold $\pi^*(D) \subset \hat{Z}$ is also 
decomposed accordingly to the stratification in the base. 
The Hodge number of the compact and smooth geometry $\pi^*(D)$ is obtained 
by summing up the Hodge--Deligne numbers of the strata. 
To $h^{1,1}(\pi^*(D)) = h^{2,2}(\pi^*(D))$, only the first two strata contribute. 
\begin{eqnarray}
 h^{2,2}(\pi^*(D)) & = &
\left[  e^{0,0}_c(T^2) e^{2,2}_c(D \backslash (D \cap S)) 
+ e^{1,1}_c(T^2) e^{1,1}_c(D \backslash (D \cap S)) \right] \nonumber \\
& + & e^{1,1}_c(I_5) e^{1,1}_c(D \cap S)  \nonumber \\
& = & 1 + [h^{1,1}_{\rm alg}(D \subset B_3) -1] + 5 = 1 + h^{1,1}_{\rm alg}(D) + 4.
\end{eqnarray}
The claim is now verified \footnote{
To be more precise, the term $e^{1,1}_c(T^2)e^{1,1}_c(D\backslash (D \cap S))$
should be decomposed into the region where the fibre is a smooth $T^2$  
and type II fibre. However, $e^{1,1}_c=1$ for the type II fibre, so that there is no difference after all.}.
If $D_i \cdot S = 0$ in $B_3$, however, $H^{1,1}(\pi^*(D_i))$ is 
generated by $\pi^*(D_j|_{D_i})$ for $D_j$'s in $NS(B_3)$ and 
$\sigma|_{\pi^*(D_i)}$.

We assume that a basis $\{ D_i \}$ of $NS(B_3)$ is chosen 
such that the number of $D_i$'s with $D_i \cdot S \neq \phi$ 
is the same as the dimension $\tilde{\rho}_B$ of the image of 
\begin{equation}
 i^* : H^{1,1}(B_3) \longrightarrow H^{1,1}(S).
\label{eq:B3-S-H11}
\end{equation}

In order to determine the Hodge numbers of the five other divisors  
$\hat{Y}_{0,1,2,3,4}$ comprising the $I_5$ fibre over the divisor 
$S \subset B_3$, we introduce the following stratification of $S$: 
\begin{equation}
 S = S^{\circ} \amalg \Sigma_{({\bf 10})}^{\circ} \amalg 
  \widetilde{\Sigma}_{({\bf 5})}^{\circ} \amalg P_{E6} \amalg P_{D6}\, .
\end{equation}
Here, $\Sigma_{({\bf 10})}$ is the curve in $S$ given by 
$a_5|_{S} = 0$, and $\Sigma_{({\bf 5})}$ the curve in $S$ given by 
$(a_0 a_5^2 - a_2 a_5 a_3 + a_4 a_3^2)|_{S} = 0$.
Those two curves intersect in $S$, and 
$P_{E6}$ and $P_{D6}$ denote the collection of such intersection points of 
two different kinds; $P_{E6}$ are the points where $a_5 = a_4 = 0$, and 
$P_{D6}$ where $a_5 = a_3 = 0$.
$\Sigma_{({\bf 10})}^\circ $ is defined as $ \Sigma_{({\bf 10})} \backslash (P_{E6} \cup P_{D6})$.
The curve $\Sigma_{({\bf 5})}$ forms a double point singularity at each 
one of the points in $P_{D6}$; $\widetilde{\Sigma}_{({\bf 5})}^\circ$ is obtained 
by resolving the double point singularities (to obtain a curve 
$\widetilde{\Sigma}_{({\bf 5})}$), and then removing the lift of the 
points of $P_{E6}$ and $P_{D6}$.
Finally, $S^\circ := S \backslash (\Sigma_{({\bf 10})} \cup \Sigma_{({\bf 5})})$
(see \cite{Andreas:1999ng, Hayashi:2008ba}).

The Hodge numbers of the five divisors can be calculated by using 
the stratification of the geometry of $S$ as described above, 
and the additivity and multiplicativity of the Euler characteristics of the Hodge--Deligne numbers $e^{p,q}$. Appendix \ref{app:hodgeexcept} 
explains how to carry out the computation in practice, taking 
the Hodge numbers of $\hat{Y}_{4}$ as an example. The result is 
\begin{equation}
 e^{p,q}_c(\hat{Y}_{4}) = \begin{array}{|cccc}
 0 & h^{2,0}(S) & 0 & 1 \\
 h^{2,0}(S) & -2g_{10} & [h^{1,1}(S)+3] & 0 \\
 0 & [h^{1,1}(S)+3] & -2g_{10} & h^{2,0}(S) \\
 1 & 0 & h^{2,0}(S) & 0 \\
 \hline 
 \end{array} = (-)^{p+q} h^{3-p,3-q}(\hat{Y}_4) \, , 
\end{equation}
where the Hodge numbers are given in terms of $h^{2,0}(S)$ and $h^{1,1}(S)$  and the genera $g_{10}$ and $\tilde{g}_{5}$ 
of the (resolved) matter curves; we assume that $h^{1,0}(S)=0$ here. 
The appendix \ref{sec:general-4fold} contains more information on the 
generators of $H^{1,1}(\hat{Y}_{4})$. 

Similar computations can be carried out for the four other divisors 
in the $I_5$ fibre. The results are:
\begin{equation}
 e^{p,q}_c(\hat{Y}_0) = e^{p,q}_c(\hat{Y}_1) = \begin{array}{|cccc}
 0 & h^{2,0}(S) & 0 & 1 \\
 h^{2,0}(S) & 0 & [h^{1,1}(S)+1] & 0 \\
 0 & [h^{1,1}(S)+1] & 0 & h^{2,0}(S) \\
 1 & 0 & h^{2,0}(S) & 0 \\
 \hline 
 \end{array}\, , 
\end{equation}
\begin{equation}
 e^{p,q}_c(\hat{Y}_{2}) = \begin{array}{|cccc}
 0 & h^{2,0}(S) & 0 & 1 \\
 h^{2,0}(S) & -g_{10} & [h^{1,1}(S)+2] & 0 \\
 0 & [h^{1,1}(S)+2] & -g_{10} & h^{2,0}(S) \\
 1 & 0 & h^{2,0}(S) & 0 \\
 \hline 
 \end{array}\, ,
\end{equation}
\begin{equation}
 e^{p,q}_c(\hat{Y}_{3}) = \begin{array}{|cccc}
 0 & h^{2,0}(S) & 0 & 1 \\
 h^{2,0}(S) & -\tilde{g}_5 & [h^{1,1}(S)+2] & 0 \\
 0 & [h^{1,1}(S)+2] & -\tilde{g}_5 & h^{2,0}(S) \\
 1 & 0 & h^{2,0}(S) & 0 \\
 \hline 
 \end{array}\, . 
\end{equation}
%

\subsection{Determination of $H^4(Y)$}

Next, $H^4(Y)$ is determined as the kernel of (\ref{eq:H4coh-Y}). 
Since we have chosen all of the $\hat{Y}_i$'s to be irreducible, 
$h^4(\hat{Y}_i \cdot \hat{Y}_j)=1$ for any pair of those divisors 
($i \neq j$) with a non-empty intersection $\hat{Y}_i \cdot \hat{Y}_j$. 
We switch to the dual (homology group) language in this subsection and 
the next. $H_4(Y)$ is the cokernel of the morphism 
\begin{equation}
d^1: \quad 
  \left[ \oplus_{i<j} H_4( \hat{Y}_i \cdot \hat{Y}_j ) \right] = E^1_{1,4}
\longrightarrow  E^1_{0,4} = \left[ \oplus_i H_4( \hat{Y}_i ) \right], 
\label{eq:H4hom-Y}
\end{equation}
and the contribution to $H_4(\hat{Z})$ from $H_4(Y)$ is 
characterized as the cokernel of the boundary map 
\begin{equation}
\delta:  \left[ H_5^{\rm BM}(Z) = H_5(\hat{Z}, Y) \right] \longrightarrow H_4(Y).
\label{eq:hom-boundary-5to4}
\end{equation}

We have already learned that 
\begin{eqnarray}\label{h4divisors}
 {\rm dim}[ H_4(\sigma \simeq B_3)] & = & \rho_B,\nn \\
 {\rm dim}[ H_4(\pi^*(D_i) ] & = & h^{1,1}(D_i) + 1, 
    \qquad \qquad ({\rm if~} D_i \cdot S = \phi) \nn\\
 {\rm dim}[ H_4(\pi^*(D_i) ] & = & h^{1,1}(D_i) + 1 + 4, 
    \qquad \quad ({\rm if~} D_i \cdot S \neq \phi) \nn\\
 {\rm dim}[ H_4(\hat{Y}_0) ] & = & \rho_S + 1, \nn\\
 {\rm dim}[ H_4(\hat{Y}_1) ] & = & \rho_S + 1, \nn\\
 {\rm dim}[ H_4(\hat{Y}_{2}) ] & = & \rho_S + 2, \nn\\
 {\rm dim}[ H_4(\hat{Y}_{3}) ] & = & \rho_S + 2, \nn\\
 {\rm dim}[ H_4(\hat{Y}_{4}) ] & = & \rho_S + 3, 
\end{eqnarray}
where $\rho_S := h^{1,1}(S)$ and $\rho_B = h^{1,1}(B_3)$. 
On the other hand,  
\begin{equation}
 {\rm dim} \left[ \oplus_{i<j} H_4(\hat{Y}_i \cap \hat{Y}_j) \right] 
   = \rho_B + \frac{\rho_B(\rho_B-1)}{2} + 5 \tilde{\rho}_B + 8\, , 
\end{equation}
and $\oplus_{i<j} H_4(\hat{Y}_i \cap \hat{Y}_j)$ is generated by four-cycles of the form 
\begin{enumerate}[i)]
 \item $\sigma \cdot \pi^*(D_i)$
 \item $\pi^*(D_i \cdot D_j)$ with $i \neq j \leq \rho_B$
 \item  $\pi^*(D_i) \cdot \hat{Y}_{0,1,2,3,4}$ for $D_i \cdot S \neq \phi$
 \item  intersections among $\hat{Y}_{0,1,2,3,4}$, neglecting self-intersections
\end{enumerate}

There are $\rho_B$ four-cycles in the group i) above.
In $[\oplus_i H_4(\hat{Y}_i)]$, a given cycle of the form $(\hat{Y}_i \cdot \hat{Y}_j) = \sigma \cdot D_i$
appears once in $H_4(\sigma)$ and once in $H_4(\pi^*(D_i))$. Thus, there are $\rho_B$ independent four-cycles $[\sigma \cdot D_i]$ in $H_4(Y)$ remaining in the Cokernel of the morphism (\ref{eq:H4hom-Y}). 

The second group ii) of four-cycles above are mapped into 
$\oplus_{i} H_4(\pi^*(D_i))$, and also leave a cokernel of dimension 
$\rho_B$. These cycles are represented by $\pi^*({\rm curves} \subset B_3)$.

A given four-cycle of the form $\pi^*(D_i) \cdot \hat{Y}_{1,2,3,4}$  (with $D_i \cdot S \neq \phi$) among the group iii) of four-cycles appears once in $H_4(\hat{Y}_{1,2,3,4})$ and once in $H_4(\pi^*(D_i))$. 
Thus, all of the $4\rho_S$ generators of $H_4(\hat{Y}_{1,2,3,4})$ are in the cokernel. The remaining four-cycles in the third group are of the form $\pi^*(D_i) \cdot \hat{Y}_S$. As only a $\tilde{\rho}_B$-dimensional subspace of $H^{1,1}(S)$ descends from
$H^{1,1}(B_3)$, the images of these cycles in $H_4(\pi^*(D_i))$ are mapped to a $\tilde{\rho}_B$-dimensional subspace of $H_4(\hat{Y}_S)$, so that their contribution to the cokernel is $\rho_S - \tilde{\rho}_B$-dimensional.

So far, we have identified $2\rho_B + 4 \rho_S + \rho_S-\tilde{\rho}_B$ independent generators of the cokernel of \eqref{eq:H4hom-Y} in $H_4(Y)$. Besides the summands already covered in the discussion above, there are $(1+1+2+2+3)=9$-dimensions remaining in $\oplus_i H_4(\hat{Y}_i)$, and eight dimensions remaining  in $\oplus_{i<j} H_4(\hat{Y}_i \cdot \hat{Y}_j)$. We choose $\hat{Y}_{2} \cdot \hat{Y}_{4}$ as the remaining representative of thecokernel and may conclude that 
\begin{equation}
 {\rm dim}[H_4(Y)] = 2\rho_B + 5\rho_S - \tilde{\rho}_B + 1.
\end{equation}
%

\subsection{The vertical components in $H_4(Y)$}

The $(2\rho_B + 1)$-dimensional subspace of $H_4(Y)$ which may be represented by 
$\sigma\cdot D_i$ and $D_i\cdot D_j$ is clearly composed of vertical four-cycles. 
However, not all of $H_4(Y)$ arises in this way.

The remaining $4 \rho_S + (\rho_S-\tilde{\rho}_B)$-dimensional subspace 
of $H_4(Y)$ is generated by four-cycles of the form 
$\hat{Y}_{1,2,3,4}\cdot \pi^*(\omega)$ or $\hat{Y}_S \cdot \pi^*(\omega)$ 
for some $\omega \in H^{1,1}(S)$ (Poincar\'e duality is implicit everywhere).
If $\omega$ is in the image of (\ref{eq:B3-S-H11}), then such a 
four-cycle is vertical. Thus, at least a subspace of dimension
$4\tilde{\rho}_B$ is also vertical. 

Intersections of the form $\hat{Y}_{i} \cdot \hat{Y}_j$ for 
$i, j \in \{0,1,2,3,4\}$ and $i \neq j$ are all vertical 
four-cycles as well. They are already contained in the 
$(2\rho_B+1)$-dimensional subspace of $H_4(Y)$ referred 
to at the beginning of this subsection, however. 

The last remaining group of vertical four-cycles are of the form 
$\hat{Y}_i \cdot \hat{Y}_i$ for $i=0,1,2,3,4$. Such self-intersections can be computed by using such relations as 
\begin{equation}
 \hat{Y}_1 \cdot \hat{Y}_1 \sim  \hat{Y}_1 \cdot \left[
  \nu_{\rm tot}^*(D_S) - \hat{Y}_0 - \hat{Y}_{2} - \hat{Y}_{3} - \hat{Y}_{4} 
    \right].
\end{equation}
Noting that both $c_1(N_{S|B_3})$ and $c_1(B_3)|_S$ are in the image of 
(\ref{eq:B3-S-H11}), and using the relations 
(\ref{eq:rat-equiv-except-divdiv-a}--\ref{eq:rat-equiv-except-divdiv-f}),
one finds that these vertical four-cycles are not independent from those 
that we have discussed already. Therefore, we conclude that the vertical 
four-cycles form a subspace of dimension
\begin{equation}\label{eq:vertellfib}
 2\rho_B + 4 \tilde{\rho}_B + 1
\end{equation}
in $H_4(Y)$.

The remaining $5(\rho_S - \tilde{\rho}_B)$-dimensional 
subspace of $H_4(Y)$ cannot be represented by vertical four-cycles. 
This subspace has a clear geometric interpretation: for any cycle in $S$
which does not descend from $B_3$, we can form five vertical four-cycles
by taking the fibre components of the $I_5$ fibre over $S$.

\subsection{The remaining component}

A four-cycle in $H_4(Y)$ represents a topological four-cycle 
in $\hat{Z}$ only if it is in the cokernel of the boundary map, 
\begin{equation}\label{bdry5to4}
  {\rm Coker} \left[ H_5(\hat{Z},Y) \longrightarrow H_4(Y) \right]
   \subset H_4(\hat{Z}).
\end{equation}
For the general class of geometries $\hat{Z}$ described at the beginning 
of this section, we do not have a clear method of computation for 
$H_5(\hat{Z},Y)$. Although it is difficult to find {\it all} the generators 
of $H_5(\hat{Z},Y)$ and their images in $H_4(Y)$, it is easier to identify  
{\it some} generators of $H_5(\hat{Z}, Y)$ and study their images in $H_4(Y)$.
This way, we can at least establish an upper bound on the cokernel. 

For every two-cycle $\Sigma$ in $H_2(S)$ whose Poincar\'e dual is in the 
cokernel of (\ref{eq:B3-S-H11}), there exists a three-chain $\gamma$ 
in $B_3$ with $\Sigma = \partial \gamma$.
This three-chain $\gamma$ and the boundary morphism 
$H_3(B_3, S) \longrightarrow H_2(S)$ are lifted under the elliptic fibration. 
Clearly, $\pi^{-1}(\gamma)$ is in $H_5(\hat{Z},Y)$ and its boundary 
in $H_5(\hat{Z},Y) \longrightarrow H_4(Y)$ is represented by 
$\pi_{\hat{Y}_0}^{-1}(\Sigma)+\pi_{\hat{Y}_1}^{-1}(\Sigma) + \cdots +
 \pi_{\hat{Y}_{4}}^{-1}(\Sigma)$; there are $(\rho_S - \tilde{\rho}_B)$ independent 
choices of $\Sigma$ and consequently $(\rho_S - \tilde{\rho}_B)$ independent 
choices of $\pi^{-1}(\gamma)$. This means that only 
$4(\rho_S - \tilde{\rho}_B)$ out of the $5(\rho_S - \tilde{\rho}_B)$ 
non-vertical cycles in $H_4(Y)$ can become non-trivial four-cycles in 
$H_4(\hat{Z})$. If there are more cycles in $H_5(\hat{Z},Y)$ than 
the ones considered, this number may decrease.

As the five-chains $\pi^{-1}(\gamma)$ constructed above map only to 
non-vertical cycles in $H_4(Y)$ under the boundary map \eqref{bdry5to4}, 
we find that there is at most a $(2\rho_B+4\tilde{\rho}_B+1)$-dimensional 
subspace of $H_4(\hat{Z})$ that is represented by vertical four-cycles. 
If there are more five-chains in $H_5(\hat{Z}, Y)$ than those we discussed 
above, this subspace may be smaller. Since all the independent 
divisors are included in the collection $\cup_i \hat{Y}_i$, all 
the vertical four-cycles have already been listed up. 

The case $(\rho_S-\tilde{\rho}_B) > 0$ is not without phenomenological 
motivation \cite{Buican:2006sn,Beasley:2008kw}. The non-vertical cycles 
spanning at most $4(\rho_S-\tilde{\rho}_B)$-dimensions in $H_4(\hat{Z})$, 
are precisely the Poincar\'e duals of four-forms 
that can break ${\rm SU}(5)$ unification symmetry down to the gauge group of 
the Standard Model, while keeping the vector field of ${\rm U}(1)_Y$ massless.
In this scenario, those four-cycles of $H_4(Y)$ need to represent 
topologically non-trivial four-cycles in $H_4(\hat{Z})$.
If $h^{2,0}(S)=0$, those four-cycles are automatically algebraic, 
so they are not horizontal either. 
The four-forms (four-cycles) of this type in the class of geometries
studied in this section are another class of examples of the 
remaining component $H^{2,2}_{RM}(\hat{Z})$, and they are also characterized 
by the property (\ref{eq:rm-from-coker}). See \cite{Braun:2014pva} for an
explicit construction of such cycles.

Certainly, all of the components of $H^{2,2}_V(\hat{Z})$ are contained
in (\ref{bdry5to4}), and some cycles from $H^{2,2}_{RM}(\hat{Z})$ also are, but
we cannot say we have exhausted all the components of $H^{2,2}_{RM}(\hat{Z})$.
Therefore we should use mirror symmetry, $h^{2,2}_H(\hat{Z})=h^{2,2}_V(\hat{Z}_m)$, 
in order to determine $h^{2,2}_H(\hat{Z})$. That will be doable once we can
construct a smooth model $\hat{Z}$ as a complete intersection Calabi--Yau of
a toric ambient space \cite{Borisov93, BatyrevBorisov}.
%
%
Such an analysis will be a generalization of what we have done in section \ref{sect:h22hypersurface}.
A crucial step in constructing such a smooth model is to carry out the resolution
of the singularities in the Weierstrass model $Z_s$ in toric language.
This process was to add the points \eqref{eq:extraptssu5} in section \ref{sect:examplesCYna4}. The essence of this procedure lies in the fact that
the divisor $S$, the component of the discriminant locus for an $I_5$ fibre, is a toric
divisor of $B_3$ (represented by a vertex in the polytope). This procedure works not
just in the case when $B_3$ itself is toric (as in section \ref{sect:examplesCYna4}), but also in cases
where $B_3$ is regarded as a complete intersection in a toric ambient space $\P^k_{\Sigma'_*}$, as long as this property --- $S$ is regarded as a toric divisor of
$P^k_{\Sigma'_*}$ restricted on $B_3$ --- is maintained, as discussed in \cite{Blumenhagen:2009yv}.\footnote{
In case $B_3$ is a complete intersection of a toric ambient space $\P^k_{\Sigma'_*}$
and $S$ is given by $s|_{B_3} = 0$ of some section
$s \in \Gamma(\P^k_{\Sigma'_*}; {\cal O}(D_S))$ on the toric ambient space,
($D_S$ is a Cartier divisor of the fine unimodular $\Sigma'_*$)
it is possible to embed the original toric ambient space $\P^k_{\Sigma'_*}$ into
another toric variety, the total space of ${\cal O}(D_S)$, by $s$, and so are $B_3$ and $S$.
$B_3$ is still a complete intersection of the new toric ambient space, and now $S$ is also regarded as a toric divisor, satisfying the condition in the main text.
This case, however, lacks generality, in that the divisor class of $S$ needs to be in the
image of $i^*: {\rm Pic}(\P^k_{\Sigma'_*}) \longrightarrow {\rm Pic}(B_3)$.
Such a situation can be improved by choosing a different embedding in many cases.
An alternative approach exists in the case the divisor $S$ of $B_3$ is very ample.
One can use the projective embedding of $B_3$ using the complete
linear system of $S$; the divisor $S$ is regarded as the hyperplane divisor of the new
ambient space $\P^{{\rm dim}|S|-1}$ restricted on the image of $B_3$, one of the conditions
assumed in the main text. We do not know, however, what kind of conditions have to be imposed on $B_3$ 
so that the image of $B_3$ in $\P^{{\rm dim}|S|-1}$ is regarded as a complete
intersection. 
}
It is an extra step to verify that a further small resolution exists, which is equivalent
to finding an appropriate triangulation of the polytope; a smooth model $\hat{Z}$ needs
to come out as a complete intersection Calabi--Yau, while satisfying the flat-fibration
condition \cite{Esole:2011sm}. 
It is beyond the scope of this article, however, to study all of these issues.

The examples discussed in section \ref{sect:elfibexamples} indeed satisfy \eqref{eq:vertellfib}.
In section \ref{sect:elfibexamples}, we furthermore found that $H^{2,2}_{RM} = 0$ for all examples.
This of course fits with the facts that $S$ was a toric divisor of the (toric) bases $B_{3}$
of the elliptic fibrations in all cases, so that all of $H^{1,1}(S)$ is obtained by restricting divisors of the base; at least there is no component in 
$H^{2,2}_{RM}$ that is characterized by (\ref{bdry5to4}) because of 
$\rho_S = \tilde{\rho}_B$.

\section{Distribution of rank of unification group and number 
of generations}
\label{sect:distrGNgen}

There is mounting evidence that the decomposition of $H^4(\hat{Z};\R)$ of 
a Calabi--Yau fourfold $\hat{Z}$ in a family 
$\pi: {\cal Z} \longrightarrow {\cal M}_*$ has a non-trivial $H^{2,2}_{RM}(\hat{Z}; \R)$
component. In addition to the families of $\hat{Z} = {\rm K3} \times {\rm K3}$ 
in \cite{Braun:2014ola} (see also (\ref{eq:K3K3-rm-dim})), we have seen 
in sections \ref{sect:h22hypersurface}--\ref{sect:genI5fourfolds} that 
the four-forms in $H^{2,2}_{RM}(\hat{Z}; \R)$ are often Poincar\'e dual to topologically non-trivial
cycles on divisors. Flux in such a four-cycle violates the condition 
(\ref{eq:cond-unbroken-sym}), and hence we are led to take 
$H_{\rm scan} \otimes \R$ to be the primary horizontal subspace $H^4_H(\hat{Z}; \R)$
of a family ${\cal Z} \longrightarrow {\cal M}_*$ specified by $(B_3, [S], R)$, for an ensemble
with the same unbroken symmetry in the effective theories (as stated already in section \ref{sec:landscape-horizontal}).

The vacuum index density distribution $d\mu_I$ of such a subensemble of 
F-theory flux vacua becomes a product of a prefactor and 
a distribution $\rho_I$ (top-dimensional differential form) on ${\cal M}_*$, 
as in (\ref{eq:ADD-formula-1}, \ref{eq:ADD-formula-2}). 
Since the integral of $\rho_I$ over the fundamental domain of ${\cal M}_*$
often returns a number of order unity (see \cite{Ashok:2003gk}; so called ``D-limit'' 
regions may have to be removed from ${\cal M}_*$), 
the prefactor therefore gives an estimate of the number of flux 
vacua with a given set of gauge group, matter representations,
multiplicities and choice of $(B_3, [S])$. The distribution of the value of 
coupling constants of such a class of low-energy effective theories is 
given by the distribution $\rho_I$ on ${\cal M}_*$ without being integrated.
In this article and ref. \cite{physlett}, we only discuss the physics consequences 
coming from the prefactor. We learn how the number of vacua depends on the 
choice of the algebraic information (7-brane gauge group = unification group 
$R$) and topological information (number of generations $N_{\rm gen}$) of 
the low-energy effective theories, starting with concrete examples and 
extracting the essence later on. 

Let us begin with the choice $R={\rm SU}(5)$ for the unification group and use the 
$(B_3, [S])$ we studied explicitly already in section \ref{sect:examplesCYna4} 
(except the one with $B_3 = \P^3$).
At the beginning, we pay attention to the sub-ensembles for individual 
$N_{\rm gen}$, the net chirality in the ${\bf 10}+\bar{\bf 5}$ vs 
$\overline{\bf 10}+{\bf 5}$ representations of SU(5).
This is done by taking $H_{\rm scan} \otimes \R$ to be that of the 
real primary horizontal subspace of the families in 
section \ref{sect:examplesCYna4}, and $G^{(4)}_{\rm fix}$ to be the 
flux generating chirality. As this class of F-theory compactifications 
has a dual description in Heterotic string theory, it is well-known 
how chirality is generated. For the Heterotic string compactified on 
an elliptically fibred Calabi--Yau threefold with the base $B_2 = S=\P^2$, 
$\pi_{\rm Het}: Z_{\rm Het} \longrightarrow B_2 = S$.
Reference \cite{Friedman:1997yq} introduced the origin of chirality in the form of 
\begin{equation}
\label{eq:FMW-flux}
 \gamma_{FMW} = \lambda_{FMW} \left[ 
    5 j_* (\Sigma_{({\bf 10})}) - (\pi_{\rm Het} \cdot i_C)^*(\Sigma_{\bf (10)}) \right],
  \qquad 
  \lambda_{FMW} \in \frac{1}{2} + \Z,
\end{equation}
where $\gamma$ is a divisor with (possibly half) integral coefficient 
on a spectral surface $C \subset Z_{\rm Het}$, $i_C: C \hookrightarrow Z_{\rm Het}$ 
and $j: \Sigma_{\bf (10)} \hookrightarrow C$ are the embedding maps.
It is known from \cite{Curio:1998vu,Diaconescu:1998kg} that 
\begin{equation}
\label{eq:Het-chirality}
  N_{\rm gen} = -(18-n)(3-n) \lambda_{FMW}.
\end{equation}
The F-theory dual description of $\gamma_{FMW}$ has also been known 
in the literature \cite{Marsano:2011hv}; necessary details are reviewed briefly in the 
appendix \ref{sect:chirflux}; see (\ref{fluxquantgenform}) for 
the explicit form of the flux. The maximum contribution to the D3-brane 
charges from $G^{(4)}_{\rm scan}$, $L_*$, is given by (\ref{eq:tadpole-bound}),
where $G^{(4)}_{\rm fix}$ is assumed to be orthogonal to the scanning space 
$H_{\rm scan}$.

The flux scanning space $H_{\rm scan}\otimes \R$ is chosen to be the real primary 
horizontal subspace $H^4_H(\hat{Z};\R)$, for the reason discussed in 
section \ref{sec:landscape-horizontal}. The dimension of this subspace, $K$, 
and the value of $L_*$ above, are computed in section \ref{sect:examplesCYna4}
and appendix~\ref{sect:chirflux}, respectively. These numbers can be 
fed into the prefactor $(2\pi L_*)^{K/2}/[(K/2)!]$ 
in (\ref{eq:ADD-formula-1}). 

We should note by looking at the values of $L_*$ and $K$ summarized in 
table~\ref{tab:tadpole-value-SU(5)}, however, that the condition 
$K \ll L_*$ is not satisfied. It is thus expected that the formula 
(\ref{eq:ADD-formula-1}) is not going to be very precise. In fact, it is 
not surprising that $L_* \ll K$ holds in the examples of our interest, 
rather than $L_* \gg K$. As we have considered base spaces
with only small $h^{1,1}$, the resulting fourfolds $\hat{Z}$ also 
only have a rather small number of divisors whereas $h^{3,1}(\hat{Z})$ is 
rather large. More generally, one could try to argue that taking $B_3$ to 
be Fano leads to $h^{3,1}(\hat{Z}) \gg h^{1,1}(\hat{Z})$. 
In any case, $h^{2,2}_H \gg h^{2,2}_V$ in table \ref{tab:hodge-SU(5)} should 
be regarded as a natural consequence of $h^{3,1} \gg h^{1,1}$.
Although we do not have a general feeling about when $h^{2,2}_{RM}$ becomes large 
or small relatively to $h^{2,2}_{H}$ or $h^{2,2}_V$ for general Calabi--Yau 
fourfolds, at least the remaining component is much smaller than the horizontal 
component in dimension, $h^{2,2}_{RM} \ll h^{2,2}_H$, 
in table \ref{tab:hodge-SU(5)}. It seems reasonable to even make a guess 
from this experience and the result (\ref{eq:K3K3-rm-dim}) that 
$h^{2,2}_{RM} \ll {\rm max}[h^{2,2}_H, h^{2,2}_V]$ whenever 
$h^{3,1} \gg h^{1,1}$ or $h^{3,1} \ll h^{1,1}$. 

In cases where 
\begin{equation}
 h^{3,1} \gg h^{1,1}, \qquad h^{2,2}_H \gg h^{2,2}_V, \qquad 
 h^{2,2}_H \gg h^{2,2}_{RM}, \quad {\rm and} \quad 
 h^{2,1} \approx {\cal O}(1) \, ,
\label{eq:hodge-inequality-cpx-rich}
\end{equation}
we find a relation 
\begin{equation}
  \chi(\hat{Z}) \approx K, \qquad 
  L_*|_{G^{(4)}_{\rm fix}=0} \sim \frac{\chi(\hat{Z})}{24} \sim \frac{K}{24}, 
    \label{K24L}
\end{equation}
which implies $L_* \ll K$. In such cases, one can even use (\ref{h22fromotherhnumbers}) 
to derive a relation 
\begin{equation}
  K \approx 6 h^{3,1} + {\rm const}., 
\label{eq:K-h31}
\end{equation}
where the constant term comes from ``44'' in (\ref{h22fromotherhnumbers}); 
the relation (\ref{h22fromotherhnumbers}) implies that $h^{2,2}_H$ 
does not increase quadratically in $h^{3,1}$, but only linearly. 

Let us now try to find a better estimate for the number of vacua when $L_* \ll K$ holds. 
The reason \eqref{eq:ADD-formula-1} predicts essentially no vacua in this case is that it approximates
the number of lattice points at a specific distance $L_*$ from the origin in a $K$ dimensional lattice by the 
surface area of a sphere. We hence count the number of lattice points at 'generic' positions on such a sphere.
If we increase the dimension $K$ while holding $L_*$ fixed it is hence no surprise that no such points remain.
This does not mean, however, that there are no more solutions; this is easily seen by restricting the flux quanta 
to an $n$-dimensional subspace. Applying \eqref{eq:ADD-formula-1}, we then count the points at generic positions of 
an $n$-dimensional sphere. 

Following this logic, we may approximate the number of flux vacua in the case $L_* \ll K$ by\footnote{A similar approach to tackle
cases with $L_* \ll K$ is proposed in \cite{Ashok:2003gk}.}
\begin{equation}
 \sum_{n=1}^{L_*} \binom{K}{n} \frac{(2 \pi L_*)^{n/2}}{(n/2)!} \, .
\end{equation}
The reason we only sum contributions up to $n = L_*$ is that we expect no more solutions with
$ n > L_* $ non-zero flux quanta switched on. We may further approximate the above sum by 
only taking the dominant contribution, corresponding to the last term with $ n = L_*$.
This gives the estimate
\begin{align}
 \binom{K}{L_*} \frac{(2 \pi L_*)^{L_*/2}}{(L_*/2)!} \, \approx \, e^{L_* \ln\left(\sqrt{2\pi} K/L_*\right)}
\end{align}
for the number of flux vacua in the case $L_* \ll K$.

$L_*$ is always an upper convex quadratic function of $N_{\rm gen}$, 
\begin{equation}
L_* =  L_*|_{G^{(4)}_{\rm fix}=0} - c N_{\rm gen}^2
\label{eq:L*-c}
\end{equation}
for some $c > 0$ as in (\ref{eq:Ngen-dep-L*}), and we can expand the 
exponent in terms of $N_{\rm gen}$ for relatively small number of generations. 
Retaining only the next-to-leading term and using \eqref{K24L} we obtain 
\begin{align}
& \exp[L_*|_{G^{(4)}_{\rm fix}=0} \ln(K/ L_*|_{G^{(4)}_{\rm fix}=0})-c N_{\rm gen}^2(1+ \ln(K/ L_*|_{G^{(4)}_{\rm fix}=0}))] \\
\simeq & \,\, e^{K/6}  \exp \left[ - 5 c N_{\rm gen}^2 \right] .
\label{eq:Gaussian-distr}
\end{align}
for the distribution of vacua as a function of $N_{\rm gen}$.

It is interesting to note that the distribution 
function (\ref{eq:Gaussian-distr}) has a factorized form. The first factor 
$e^{K/6}$ depends on the algebraic information such as $R$ as well as 
$(B_3, [S])$, but it does not depend on $N_{\rm gen}$. The second factor
contains the $N_{\rm gen}$-dependence. The distribution function on the value of 
coupling constants of the low-energy effective theories, $\rho_I$, can be 
multiplied to (\ref{eq:Gaussian-distr}), as it was in (\ref{eq:ADD-formula-1}), 
if one wishes. One should not expect that the 
formula (\ref{eq:ADD-formula-2}) remains to be a very good approximation 
at the quantitative level because the condition $K \ll L_*$ does not hold, 
but still it is not terribly bad to expect that $\rho_I$ still retains
qualitative aspects of the distribution of the coupling constants 
after smearing over local regions in the moduli space ${\cal M}_*$.

The distribution of the number of generations in the low-energy 
effective theories is given by the second factor, which is the 
Gaussian distribution with the variance $1/(10 c)$, for relatively 
small $N_{\rm gen}$. This should be regarded as a very robust prediction 
of landscapes based on F-theory flux compactification, as long as 
the condition (\ref{eq:hodge-inequality-cpx-rich}) is satisfied 
in the relevant family of fourfolds 
$\pi: {\cal Z} \longrightarrow {\cal M}_*^R$.
As for the coefficient $c$ in (\ref{eq:L*-c}), we can read it out from 
(\ref{eq:Ngen-dep-L*}),
\begin{equation}
c = \frac{5}{(18-n)(3-n)} ,
\label{eq:coeff-Ngen2}
\end{equation}
for the examples we studied in section \ref{sect:examplesCYna4}.
Also for many other choices of $(B_3, [S])$, the coefficient $c$ is 
determined by the intersection ring associated with only a few divisors in 
$\hat{Z}$, and is expected to be more or less of order unity, 
as in (\ref{eq:coeff-Ngen2}).
Of course the Gaussian distribution is not a good approximation for 
$N_{\rm gen}^2$ large enough to be comparable to the absolute limit 
$\chi(\hat{Z})/24$; the distribution absolutely vanishes, since it is 
impossible to satisfy the D3-tadpole condition without supersymmetry breaking. 
Since the value of $\chi(\hat{Z})/24$ often comes at the order of hundreds 
to thousands \cite{Klemm:1996ts,Kreuzer:1997zg,Gray:2013mja}, however, the 
distribution of $N_{\rm gen}$ is approximately Gaussian at least for the range 
$N_{\rm gen} \lesssim 10$. The distribution for this range 
(covering $N_{\rm gen}=3$) will be sufficient information for our practical 
interest. The variance $\langle N_{\rm gen}^2\rangle$ is of order
$1/(10 c)\sim 10^{-1}$.

Let us now focus on the first factor of the distribution 
function (\ref{eq:Gaussian-distr}), $e^{K/6}$, from which we learn 
the statistical cost of unified symmetry (7-brane gauge group) $R$ with 
higher rank. For the purpose of taking ratios of the number of vacua with 
different unified symmetries $R_1$ and $R_2$, the ratio $e^{(K_1-K_2)/6}$
may be studied instead by $e^{h^{3,1}_1-h^{3,1}_2}$, when the condition 
(\ref{eq:hodge-inequality-cpx-rich}) is satisfied. This makes things 
much easier, since it is much easier to compute/estimate $h^{3,1}$ than 
$h^{2,2}_H$. Appendix~\ref{sect:dephodgegroup} exploits this 
$\Delta K \approx 6 \Delta h^{3,1}$ relation and develops the discussion further.
In the rest of this section, however, we stick to the dimension $K$ of 
the primary horizontal space.
It is also worth mentioning that $e^{K/6}$ is regarded as a 
refinement of the popular ``$10^{500}$'', a crude estimate of the number of 
flux vacua of Type IIB Calabi--Yau orientifold compactifications.

It has been widely accepted at the intuitive level that more general flux 
leads to geometry stabilized at more general values of the complex structure, and hence 
with less unbroken symmetry (fewer independent divisors).  
This intuition has been made quantitative by the factor $e^{K/6}$; 
Ref.~\cite{Braun:2014ola} used a family of Calabi--Yau fourfolds that are 
topologically $\hat{Z} = {\rm K3} \times {\rm K3}$, and the computation 
of $K$ for other families in this article adds more examples. The results 
in the family $\hat{Z} = {\rm K3} \times {\rm K3}$ and in the cases 
in section \ref{sect:elfibexamples} remain similar, though there is 
difference in detail. In the family of $\hat{Z} = {\rm K3} \times {\rm K3}$, 
$K$ is linear in the rank of 7-brane gauge group, and 
$K = 21 \times (20 -{\rm rank}_7)$ (where vanishing cosmological 
constant is not required; see \cite{Braun:2014ola} for more).
The value of $K$ becomes smaller for a subensemble with higher-rank 
unification group $R$. This remains true in the cases we studied in 
section \ref{sect:elfibexamples} (see table \ref{tab:K-dep-on-R}).

At the quantitative level, the difference in the dimension of the flux 
scanning space $K$ for SU(5) unification and SO(10) unification, 
$K_{R=A4} - K_{R=D5}$, remains more or less around order ${\cal O}(10)$ for 
all the geometries listed in table \ref{tab:K-dep-on-R}, and the ratio of 
the number of vacua with a stack of SU(5) 7-branes to that with a stack of 
SO(10) 7-branes remains of order $e^{{\cal O}(10)}$. 
The fraction of vacua with a stack of SU(5) 7-branes in $[S]$ in all the 
flux vacua on $B_3$ is given by $e^{-\Delta K_{0\mbox{-}4}/2}$; the value of 
$\Delta K_{0\mbox{-}4} = K_{R = \phi} - K_{R=A4}$ can be $1000$--$10000$ for 
$B_3 = \P^3$ and $B_3^{(n)}$ studied in section \ref{sect:elfibexamples}, 
as opposed to the value $\Delta K_{0\mbox{-}4} \sim 100$ in the family 
$\hat{Z} = {\rm K3} \times {\rm K3}$. As we have discussed, we may estimate $\Delta K_{R_1 R_2}$ 
by (six times) the number of complex structure moduli which are fixed when enhancing $R_1$ to $R_2$. The result for 
$\hat{Z} = {\rm K3} \times {\rm K3}$ is certainly very special. For more general fourfolds, 
where the number of moduli is much larger, it seems plausible that the number of complex structure moduli 
we need to fix to achieve a gauge group of rank 4 or 5 is typically $\mathcal{O}(100)$ to $\mathcal{O}(1000)$, 
irrespective of the specific $(B_3, [S])$ used. Besides our concrete results (summarized in table \ref{tab:K-dep-on-R}), 
such numbers are typical for the dimensions of the middle cohomology of Calabi-Yau fourfolds.

Note that we have assumed a fixed choice of $[S]$ so far.
This way of thinking is perfectly reasonable when comparing the statistical cost of enhancing a group of low
rank, which we assume is given on $[S]$, to a group of higher rank. When we want to tackle the physically more
interesting question of comparing the abundance of models with no gauge group to a model with gauge group $R$
for \emph{some} $[S]$, we need to be able to sum over various choices of $[S]$. While we do not expect this
to make a relevant contribution in cases where $h^{1,1}(B_3)$ is small, an exhaustive discussion is beyond the scope of the present work.

\begin{table}[tbp]
\begin{center}
\begin{tabular}{|c|cccccccc|c|}
\hline 
  $B_{\, 3}$&$\P^3$& $B_3^{\, (-3)}$ & $B_3^{\, (-2)}$ & $B_3^{\, (-1)}$ & $B_3^{\, (0)}$ & $B_3^{\, (1)}$ & $B_3^{\, (2)}$ & $B_3^{\, (3)}$ & $\P^1 \times {\rm K3}$ \\
  \hline
$K_{R = \phi}$ & 23320 & 26200 & 22596 & 20436 & 19716 & 20436 & 22596 & 26200& \\
$K_{R = A4}$ & 13287 & 7557 & 8603 & 10403 & 12953 & 16253 & 20303 & 25104 & \\
$K_{R = D5}$ & 13200 & 7392 & 8480 & 10316 & 12896 & 16220 & 20288 & 25100 & \\
   \hline
$\Delta K_{0\mbox{-}4}$ & 10033 & 18663 & 13993 & 10033 & 7123 & 4183 & 2293
    & 1096 & 84 \\
$\Delta K_{4\mbox{-}5}$ & 87 & 165 & 123 & 87 & 57 & 33 & 15 & 4 & 21 \\
\hline 
\end{tabular}
\caption{ \label{tab:K-dep-on-R} 
The value of $K_{R=\phi}$, $K_{R=A4}$ and $K_{R=D5}$ 
in the first three rows are extracted from tables \ref{tab:hodge-nogauge}, 
\ref{tab:hodge-SU(5)} and \ref{tab:hodge-SO(10)}, respectively. 
In the case $B_3 = \P^1 \times {\rm K3}$, the value of 
$\Delta K_{0\mbox{-}4} = K_{R=\phi}-K_{R=A4}$ and 
$\Delta K_{4\mbox{-}5}=K_{R=A4}-K_{R=D5}$ is determined by using 
$21 \times \Delta {\rm rank}_7$. 
}
\end{center}
\end{table}

\section{Outlook}
\label{sect:outlook}

Let us leave a few remarks on open problems at the end of the main text. We start off with an obvious open problem of mathematical flavour. 
The study we carried out in section \ref{sect:h22hypersurface} for 
toric-hypersurface Calabi--Yau fourfolds has a clear generalization to Calabi--Yau fourfolds obtained 
as complete intersection in toric ambient spaces \cite{Borisov93,BatyrevBorisov}. This makes it possible to determine the dimension of the primary horizontal 
subspace $H^4_H(\hat{Z}; \R)$ and also to elucidate the geometric origin of the cycles in the remaining component $H^{2,2}_{RM}(\hat{Z})$ for more general cases. We 
stopped at the level of finding evidence that $H^{2,2}_{RM}(\hat{Z})$
is non-empty in geometries that are motivated for phenomenological 
applications, and provided their geometric 
characterization (\ref{eq:rm-from-coker}) for the cycles we identified; 
not all of the cycles in the remaining component may have been found, however. 
More detail will be learned about this geometry by this generalization, as 
we already stated at the end of section \ref{sect:genI5fourfolds}.

One can come up with various physics applications of the idea 
developed in \cite{Ashok:2003gk,Denef:2004ze,DeWolfe:2004ns,Braun:2014ola} and this article.
For example, it is an interesting question whether SU(5) unification due 
to the presence of a stack of SU(5) 7-brane is more ``natural'' than 
accidental gauge coupling unification in supersymmetric Standard Models 
with stacks of U(3), U(2) and U(1) D7-branes wrapped on different topological 
cycles, with one U(1) remaining anomaly free.
At least by comparing the number of flux vacua in those two categories, 
we will have a partial information that is needed in answer to 
this ``naturalness'' question. Although a very crude baby-version study was carried out 
in \cite{Braun:2014ola}, this question deserves much more serious study.

This article did not try to include the symmetry breaking 
${\rm SU}(5)_{\rm GUT} \longrightarrow {\rm SU}(3)_C \times {\rm SU}(2)_L 
\times {\rm U}(1)_Y$. If one is to use the geometry with non-surjective 
$H^{1,1}(B_3) \longrightarrow H^{1,1}(S)$ as motivated 
in \cite{Buican:2006sn,Beasley:2008kw}, the flux $G^{(4)}_{\rm fix}$ needs 
to have non-zero component both 
in $H^{2,2}_V(\hat{Z})$ for $N_{\rm gen} \neq 0$, and in $H^{2,2}_{RM}(\hat{Z})$ 
for the SU(5)$_{\rm GUT}$ symmetry breaking. One then needs to think 
more carefully about the integral structure (flux quantization) of 
the space $H^{2,2}_V(\hat{Z}) \oplus H^{2,2}_{RM}(\hat{Z}) \oplus 
H^{2,2}_H(\hat{Z})$; a study of the integral structure 
in the $H_{\rm mon}$ component is found in \cite{Grimm:2009ef, Bizet:2014uua}.
Note also that if we set $G^{(4)}_{\rm fix}$ so that 
$G^{(4)}_{\rm scan}=0$ is contained in $H_{\rm scan}$, then $G^{(4)}_{\rm fix}$ may 
not always be orthogonal to the elements of $H_{\rm scan}$.

The statistical cost of requiring an extra global U(1) symmetry can also 
be studied by computing the dimension $K={\rm dim}_\R[H_{\rm scan} \otimes \R]$.
One needs to be careful in how we take the scanning space of the flux 
$H_{\rm scan}$. The U(1) vector field does not have to remain in the massless 
spectrum (indeed, we do not want it to be), but its symmetry should not 
be broken spontaneously (possibly apart from non-perturbative symmetry 
breaking exponentially suppressed to the level harmless in phenomenology). 
Eventually this statistical cost needs to be compared against the cost 
of alternatives.  In the application to the dimension-4 proton decay problem, 
an alternative will be a discrete symmetry, while in the application to 
the ``approximately rank-1'' problem of Yukawa matrices 
(cf. \cite{Hayashi:2009bt,Cordova:2009fg}), the alternative is to tune a moduli parameter. 
See also \cite{Bizet:2014uua}.

Obviously one can also exploit the distribution $\rho_I$ 
in (\ref{eq:ADD-formula-2}) to study distribution of the values of the 
coupling constants in a low-energy effective theory. Certainly 
the formula\footnote{See \cite{Bizet:2014uua} and references therein for the period integral computation to be used in
formula \eqref{eq:ADD-formula-2}. In the presence of corrections to the K\"{a}hler 
potential $- \ln [\int_{\hat{Z}} \Omega \wedge \overline{\Omega}]$, 
$\rho_I$ may not be given simply by the formula (\ref{eq:ADD-formula-2}) 
with the Weil--Peterson metric of ${\cal M}_*$. 
The authors thank Y.~Nakayama and Y.~Sumitomo for raising this issue.}
(\ref{eq:ADD-formula-2}) is not expected to be very precise, because 
we expect $L_* \ll K$ in many cases. 
Experience in explicit numerical studies suggests (see \cite{Denef:2004ze} 
and also Fig. 5 of \cite{Braun:2014ola}) that qualitative 
aspects of the actual distribution are still captured by the distribution function $\rho_I$
even in cases where $L_* \ll K$.

It should also be noted that the scanning component of the flux, 
$G^{(4)}_{\rm scan}$, may play another role in addition to determining 
the coupling constants of the low-energy effective theories.
All the fluxes in the form of $G^{(4)}_{\rm fix}+G^{(4)}_{\rm scan}$ with 
$G^{(4)}_{\rm scan} \in H^4_H(\hat{Z}; \R)$ give rise to effective theories 
with the same $N_{\rm gen}$, but the number of extra vector-like pairs of matter 
in the ${\rm SU}(5)_{\rm GUT}$ $\bar{\bf 5}+{\bf 5}$ representations may vary 
among such an ensemble of vacua. It has been known widely since the 
work of \cite{Donagi:2004ia} that there tend to be many vector-like pairs that do 
not seem to be present in supersymmetric Standard Models that work 
phenomenologically well. It took an enormous effort to find a topological 
choice that leads to small number of vector like pairs. Such studies as 
\cite{Donagi:2004ia,Hayashi:2009bt}, however, are equivalent to only use  
the flux $G^{(4)}_{\rm fix}$ in the form of (\ref{fluxquantgenform}), 
or $\gamma_{FMW}$ in (\ref{eq:FMW-flux}), and set $G^{(4)}_{\rm scan} = 0$. 
With the freedom for $G^{(4)}_{\rm scan}$, however, there may be a new 
insight on the issue of the number of vector like pairs.

As already mentioned in section \ref{sect:vertcomp}, specifying fluxes in F-theory
(or, more generally M-theory) backgrounds via a four-form $G_4$ in (co)homology 
is not sufficient information to properly characterize all degrees of 
freedom \cite{Curio:1998bva,Diaconescu:2003bm}. In particular, we need to specify
the three form $C_3$. It is an open problem how to include this data in the 
vacuum counting problem, i.e. the setup of \cite{Ashok:2003gk,Denef:2004ze}.

\section*{Acknowledgements}

We thank Andr\'es Collinucci, Roberto Valandro, Ryuichiro Kitano and 
Sakura Sch\"afer-Nameki for discussions.
The work of A.~P.~B. was supported by the STFC under grant ST/J002798/1
and that of T.~W. by WPI Initiative, MEXT, Japan 
and a Grant-in-Aid for Scientific Research on Innovative Areas 2303.

\appendix

\section{Geometry of Elliptic-fibred Calabi--Yau fourfold for general F-theory 
SU(5) models}
\label{sec:general-4fold}

\subsection{Construction}
\label{ssec:construction}

The geometry considered in section \ref{sect:genI5fourfolds}, smooth elliptically fibred Calabi--Yau fourfolds 
for general F-theory SU(5) models, has been studied in \cite{Esole:2011sm,Marsano:2011hv,Krause:2011xj}, 
see also \cite{Blumenhagen:2009yv,Grimm:2009yu}. We briefly review the construction of the geometry here, 
so that no ambiguity remains in the notation used. 

The construction begins with an ambient space 
\begin{equation}
 A_0 := \P \left[ {\cal O}_{B_3} \oplus {\cal O}_{B_3}(-2K_{B_3}) \oplus 
  {\cal O}_{B_3}(-3K_{B_3}) \right],
\end{equation}
where $B_3$ is a complex 3-dimensional (Fano) variety.
A fourfold $Z_s$ is defined as a hypersurface of $A_0$ by 
the equation (\ref{eq:generalized-Weierstrass-Tate-SU5}).
The projection $\pi_{A_0}:A_0 \longrightarrow B_3$ defines 
an elliptic fibration morphism $\pi_{Z_s}: Z_s \longrightarrow B_3$.

Let $X_3$, $X_2$ and $X_3$ be the homogeneous coordinates corresponding 
to the $W\P_{1:2:3}^2$ fibre of the ambient space 
$\pi_{A_0}: A_0 \longrightarrow B_3$, and $\sigma$ be the zero section 
defined by $X_3=0$. $D_{X_1}$, $D_{X_2}$ and $D_{X_3}$ denotes 
the zero locus of $X_1$, $X_2$ and $X_3$ in $A_0$, respectively.
It follows that 
\begin{equation}
 D_{X_1} \sim 3 \left( \sigma + c_1(B_3) \right), \qquad 
 D_{X_2} \sim 2 \left( \sigma + c_1(B_3) \right).
\end{equation}

Consider a line bundle on $B_3$, and let $s$ be a global holomorphic 
section of this line bundle. The divisor defined by the zero locus 
of $s$ is denoted by $S$, i.e. $s \in \Gamma(B_3; {\cal O}_{B_3}(S))$. 
The section $s$ is used in the hypersurface equation of $Z_s$ 
in (\ref{eq:generalized-Weierstrass-Tate-SU5}), which implements the 
condition for an $I_5$ Kodaira fibre over $S$ \cite{Bershadsky:1996nh}.
The divisor $\pi_{A_0}^*(S)$ in $A_0$ is denoted by $D_S$.

{\bf First blow-up}
The fourfold $Z_s$ has $A_4$ singularity along a subvariety of 
$A_0$ given by
\begin{equation}
 Y_1 := D_S \cdot D_{X_1} \cdot D_{X_2}. 
\end{equation}
Thus, we blow up at this locus and let $A_1 := {\rm Bl}_{Y_1}A_0$. The blow-up morphism 
is denoted by $\nu_1: A_1 \longrightarrow A_0$, and the exceptional 
divisor by ${\cal E}_1$. The center of the blowup $Y_1$ is contained in $Z_s$, and 
$Z_s$ is of multiplicity $k=2$ along $Y_1$. Thus, 
\begin{equation}
 \nu_1^*(Z_s) = Z_s^{(1)} + 2 {\cal E}_1,
\end{equation}
where $Z_s^{(1)}$ is the proper transform of $Z_s$.
The proper transforms of $D_S$, $D_{X_1}$ and $D_{X_2}$ are denoted by 
$D_S^{(1)}$, $D_{X_1}^{(1)}$ and $D_{X_2}^{(1)}$, respectively.

{\bf Second blow-up} The fourfold $Z_s^{(1)}$ still has a singularity of type $A_2$
along the subvariety 
\begin{equation}
 Y_2 := {\cal E}_1 \cdot D_{X_1}^{(1)} \cdot D_{X_2}^{(1)}
\end{equation}
in $A_1$. Thus, let $A_2 := {\rm Bl}_{Y_2} A_1$. The blow-up morphism 
is denoted by $\nu_2: A_2 \longrightarrow A_1$, and the exceptional 
divisor ${\cal E}_2$. The hypersurface $Z_s^{(1)}$ of $A_1$ contains 
$Y_2$ with the multiplicity 2, and hence 
\begin{equation}
 \nu_2^*(Z_s^{(1)}) = Z_s^{(2)} + 2 {\cal E}_2, 
\end{equation}
where $Z_s^{(2)}$ is the proper transform of $Z_s^{(1)}$ 
under this blow-up morphism.
The proper transforms of ${\cal E}_1$, $D_{X_1}^{(1)}$ and $D_{X_2}^{(1)}$ are 
denoted by ${\cal E}_1^{(2)}$, $D_{X_1}^{(2)}$ and $D_{X_2}^{(2)}$, respectively. 
$\nu_2^*(D_{S}^{(1)})$ is denoted by $D_S^{(2)}$.

{\bf Small resolution} The fourfold $Z_s^{(2)}$ still has singularities
at loci with codimension higher than two. These singularities can be resolved  
while the proper transform of $Z_s^{(2)}$ remains a flat family of curves 
over $B_3$ \cite{Esole:2011sm}. We provide a description of two such small resolutions
in the following. The two small resolutions correspond, when $B_3$ and $[S]$ 
are the ones studied in section \ref{sect:examplesCYna4}, to having 
$\langle \vec{v}_{e_1}, \vec{v}_{e_3} \rangle$ in the SR ideal, 
or having $\langle \vec{v}_{e_2}, \vec{v}_{e_4} \rangle$ in the SR ideal.
We only describe the first resolution in detail and then add 
a brief comment concerning the second type. 

{\bf Small resolution [a]}: 
Let $Y^{[a]}_3 := {\cal E}_1^{(2)} \cdot D_{X_1}^{(2)}$, and 
$A_3^{[a]} : = {\rm Bl}_{Y_3^{[a]}} A_2$. The blow-up morphism 
is $\nu_3^{[a]}: A_3^{[a]} \longrightarrow A_2$, and the exceptional 
divisor is ${\cal E}_3^{[a]}$. The hypersurface $Z_s^{(2)}$ of $A_2$ 
contains $Y_3^{[3]}$ with multiplicity 1, and hence 
\begin{equation}
 (\nu_3^{[a]})^*(Z_s^{(2)}) = Z_s^{(3)[a]} + {\cal E}_3^{[a]},
\end{equation}
where $Z_s^{(3)[a]}$ is the proper transform of $Z_s^{(2)}$ under 
the morphism $\nu_3^{[a]}$. Proper transforms of other divisors in $A_2$ 
are denoted by 
\begin{equation}
 (\nu_3^{[a]})^*({\cal E}_1^{(2)}) = {\cal E}_1^{(3)[a]} + {\cal E}_3^{[a]}, \quad 
  (\nu_3^{[a]})^*(D_{X_1}^{(2)}) = D_{X_1}^{(3)[a]} + {\cal E}_3^{[a]}  
\end{equation}
for divisors involved in the blow-up, and 
$D_{X_2}^{(3)[a]} := (\nu_3^{[a]})^*(D_{X_2}^{(2)})$, 
$D_S^{(3)[a]} := (\nu_3^{[a]})^*(D_S^{(2)})$ and 
${\cal E}_2^{(3)[a]} := (\nu_3^{[a]})^*({\cal E}_2)$ for the others.

The fourfold $Z_s^{(3)[a]}$ only becomes non-singular after one more step 
of blow-up in the ambient space. Let us take 
$Y_4^{[a]} := {\cal E}_2^{(3)[a]} \cdot D_{X_1}^{(3)[a]}$ as the centre of 
the blow-up; $A_4^{[a]}:= {\rm Bl}_{Y_4^{[a]}}A_3^{[a]}$. The blow-up morphism 
is denoted by $\nu_4^{[a]}: A_4^{[a]} \longrightarrow A_3^{[a]}$, and the 
exceptional divisor by ${\cal E}_4^{[a]}$. The hypersurface $Z_s^{(3)[a]}$ 
of $A_3^{[a]}$ contains $Y_4^{[a]}$ with the multiplicity 1, and hence 
\begin{equation}
 (\nu_4^{[a]})^* (Z_s^{(3)[a]}) = \hat{Z}^{[a]} + {\cal E}_4^{[a]}.
\end{equation}
The proper transforms of other divisors of $A_3$ are 
\begin{equation}
 (\nu_4^{[a]})^*({\cal E}_2^{(3)[a]}) = {\cal E}_2^{(4)[a]} + {\cal E}_4^{(4)[a]}, 
\qquad 
 (\nu_4^{[a]})^*(D_{X_1}^{(3)[a]}) = \hat{D}_{X_1}^{[a]} + {\cal E}_4^{(4)[a]},
\end{equation}
and 
\begin{eqnarray}
 (\nu_4^{[a]})^*(D_{X_2}^{(3)[a]}) = \hat{D}_{X_2}^{[a]}, & \qquad & 
 (\nu_4^{[a]})^*(D_{S}^{(3)[a]}) = \hat{D}_{S}^{[a]}, \\
 (\nu_4^{[a]})^*({\cal E}_1^{(3)[a]}) = {\cal E}_1^{(4)[a]}, & \qquad & 
 (\nu_4^{[a]})^*({\cal E}_3^{[a]}) ={\cal E}_3^{(4)[a]}. 
\end{eqnarray}
In order to simplify the notation, we now relabel the exceptional divisors according to their intersections. Over a generic point on $S$, the four exceptional divisors ${\cal E}_{1,2,3,4}^{(4)[a]}$ together with $D_S^{(4)[a]}$ meet according to the extended Dynkin diagram of $A_4$ (see the first picture 
in figure~\ref{fig:SU(5)-singl-fibr}). 
Namely, 
\begin{align}
D_S^{(4)[a]} \equiv \hat{E}_0 & & \\
{\cal E}_1^{(4)[a]} \equiv \hat{E}_1 &&  {\cal E}_2^{(4)[a]} \equiv \hat{E}_2  \\
{\cal E}_3^{(4)[a]} \equiv \hat{E}_4 &&  {\cal E}_4^{(4)[a]} \equiv \hat{E}_3 \, .
\end{align}

As a consequence of four successive morphisms, 
$(\nu_{\rm tot}^{[a]} := \nu_1 \cdot \nu_2 \cdot \nu_3^{[a]} \cdot \nu_4^{[a]}):
A_4 \longrightarrow A_0$, 
\begin{equation}
 \nu_{\rm tot}^*(Z_s) = \hat{Z}^{[a]}
   +2 \hat{E}_1 +4 \hat{E}_2 +5\hat{E}_3 + 3\hat{E}_4.
\end{equation}
One also finds that 
\begin{eqnarray}
 \nu_{\rm tot}^*(D_S) & = & \hat{D}_S
 + \hat{E}_{1} + \hat{E}_{2} + \hat{E}_{3} + \hat{E}_{4}, 
     \\
 \nu_{\rm tot}^*(D_{X_2}) & = & \hat{D}_{X_2}^{[a]}
 + \hat{E}_1 + 2 \hat{E}_{2} + 2 \hat{E}_{3} + \hat{E}_{4},
     \\
 \nu_{\rm tot}^*(D_{X_1}) & = & \hat{D}_{X_1}^{[a]} 
 + \hat{E}_1 + 2 \hat{E}_{2} + 3 \hat{E}_{3} + 2 \hat{E}_{4}.
\end{eqnarray}

\begin{figure}[!h]
\begin{center}
\scalebox{.5}{ 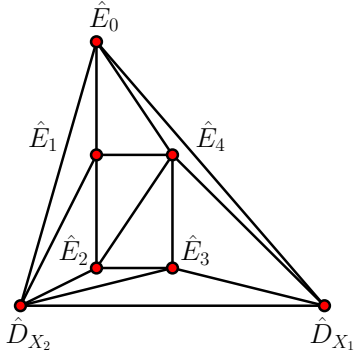 }
\caption{A schematic diagram showing how (some of) the divisors of the 
ambient space $A_4^{[a]}$ intersect; if a set of points in this diagram 
does not share a simplex, the corresponding divisors has empty intersection.
This diagram shows the history of successive blow-ups starting from the 
original ambient space $A_0$. This diagram for $A_4^{[b]}$ is such that 
the 1-simplex joining $\hat{E}_2$ and $\hat{E}_4$ is replaced by another 
joining $\hat{E}_1$ and $\hat{E}_3$. In the examples we used 
in section \ref{sect:examplesCYna4}, this diagram appears in the form of 
2-dimensional faces of the polytope $\widetilde{\Delta}$.
\label{A4figure}}
\end{center} 
\end{figure}

\begin{figure}[tbp]
\begin{center}
\begin{tabular}{ccc}
  $S^\circ$ & $\Sigma^{\circ}_{({\bf 10})}$ & $\widetilde{\Sigma}^\circ_{({\bf 5})}$ \\
&&  \\
  \includegraphics[scale=0.3]{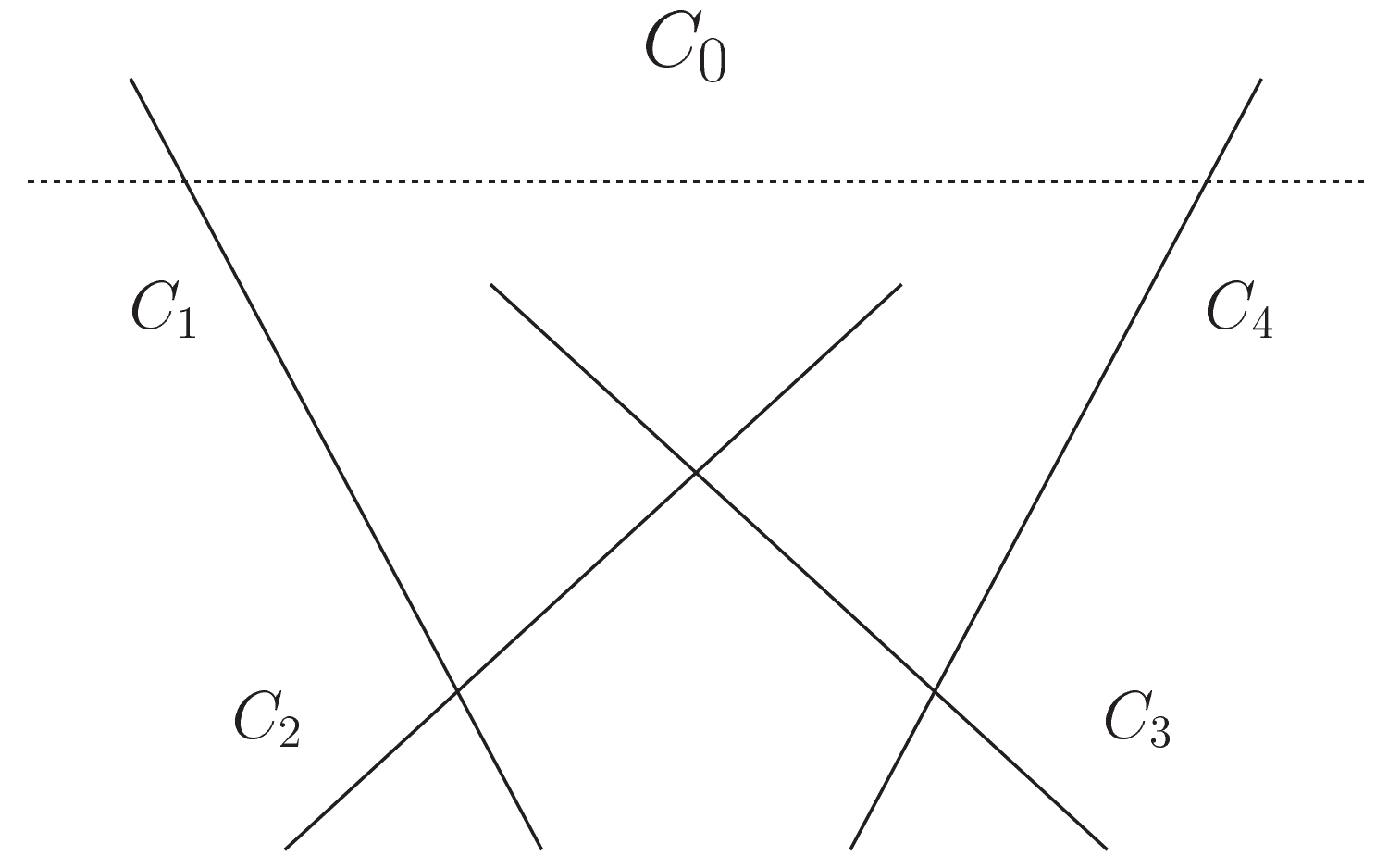} &
  \includegraphics[scale=0.3]{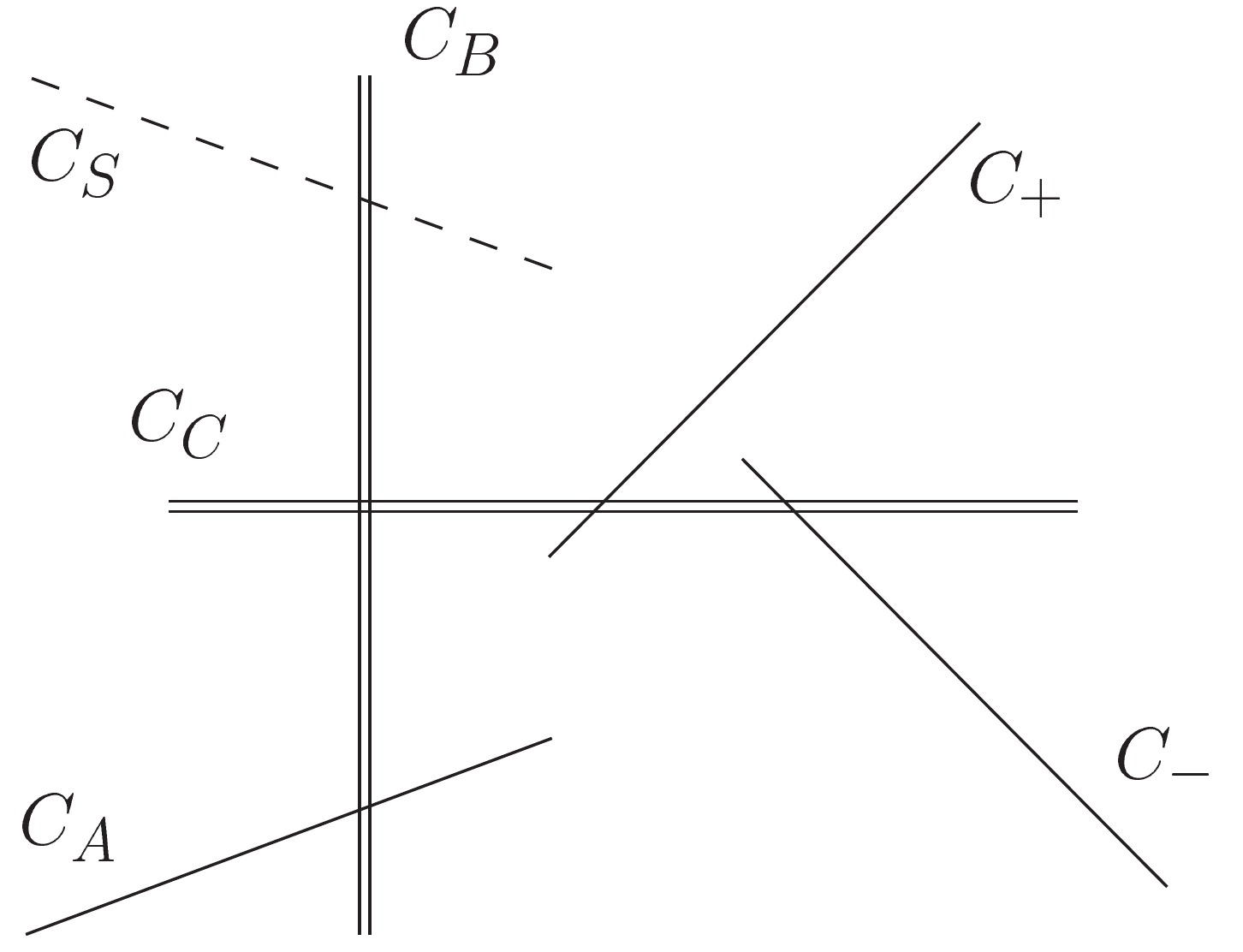} & 
  \includegraphics[scale=0.3]{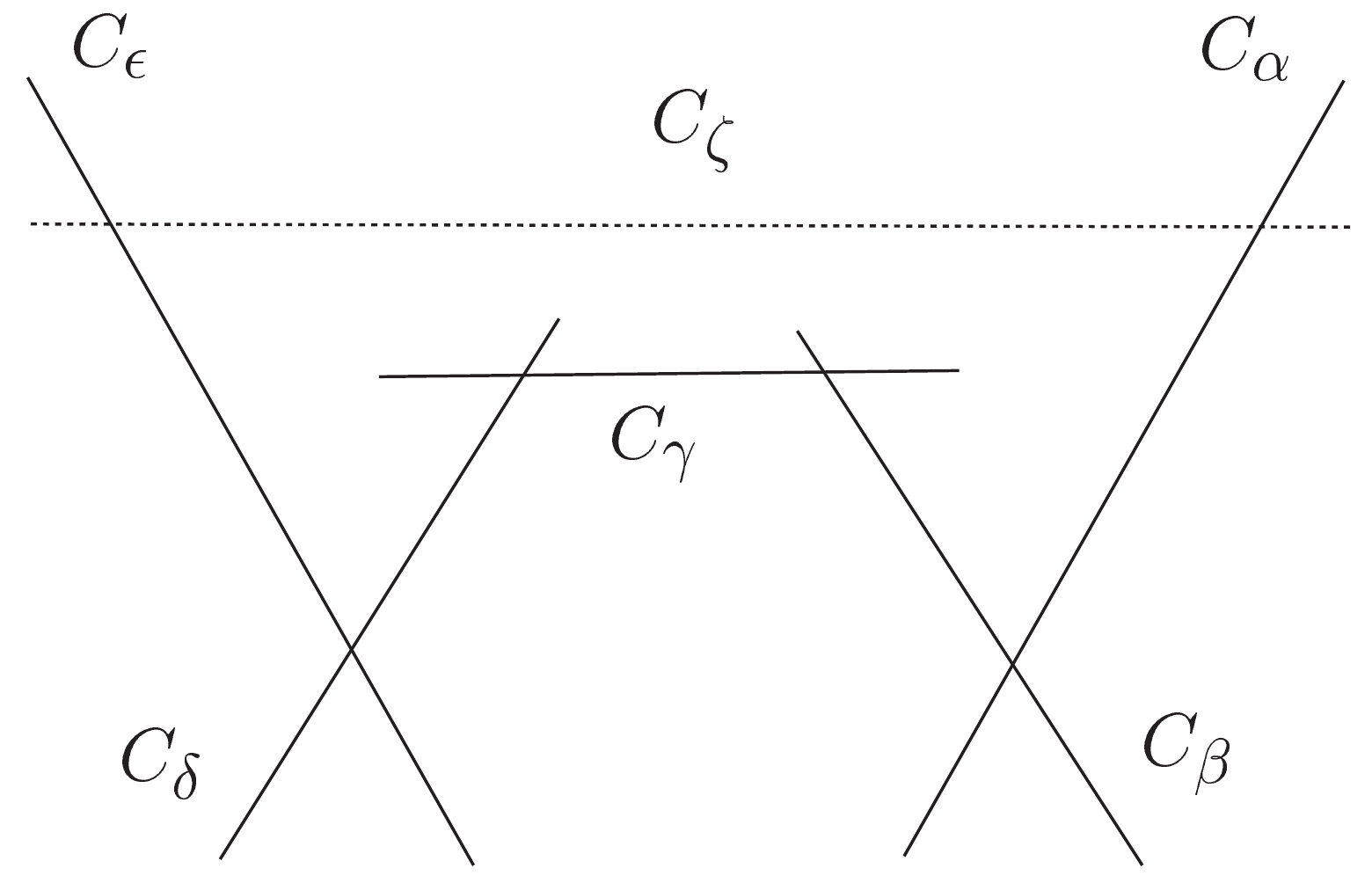} \\ 
 && \\
 \hline
&& \\
    &
  \includegraphics[scale=0.3]{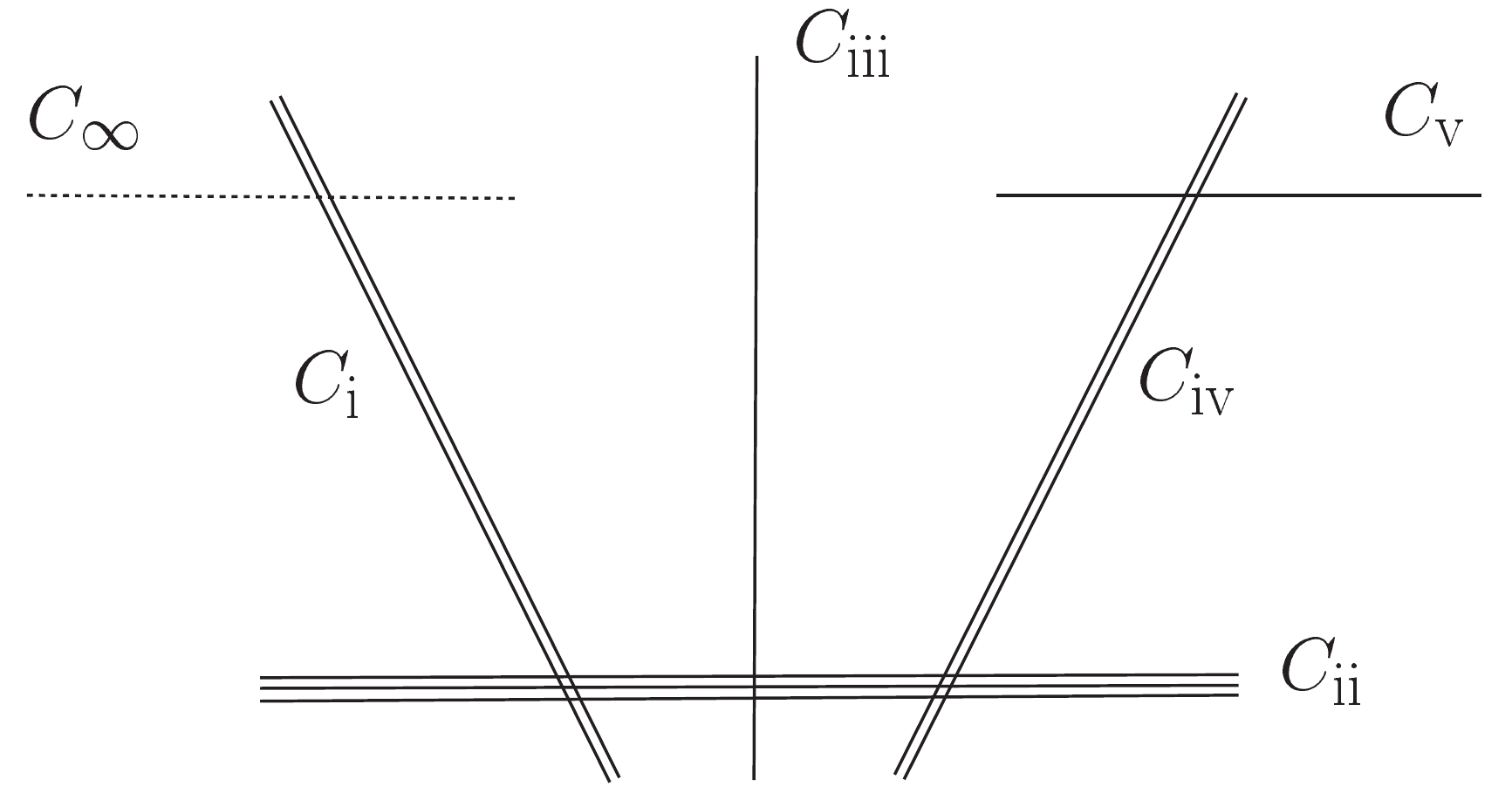} &
  \includegraphics[scale=0.3]{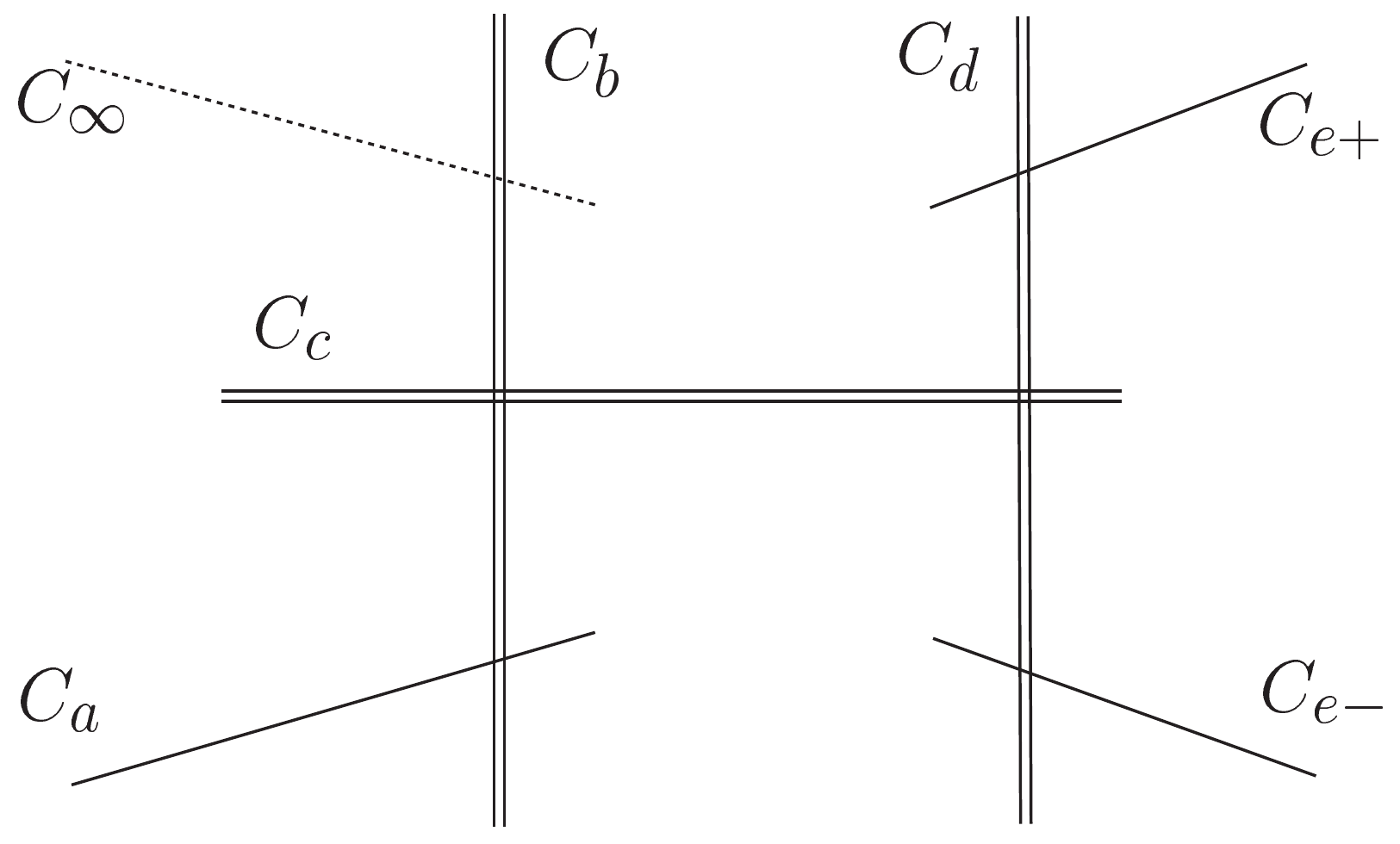} \\
&&\\
  & $P_{E6}$ & $P_{D6}$
\end{tabular}
\caption{\label{fig:SU(5)-singl-fibr}
Schematic picture of how irreducible curves in singular fibre 
share points. This information is used in computing the Hodge diagrams 
of the divisors $\hat{Y}_{0,1,2,3,4}$ in section \ref{app:hodgeexcept}. }
\end{center}
\end{figure}

A simple way to represent intersections among divisors of the ambient space is by a diagram such as figure \ref{A4figure}. It
organizes the information in the same way as a fan over faces of a polytope does in the context of toric geometry. In fact, figure~\ref{A4figure} is a two-dimensional projection of a $A_4$ `top' with a triangulation corresponding to the resolution $[a]$.

The divisors $\hat{D}_{X_1}$, $\hat{D}_{X_2}$, $\hat{E}_{1,2,3,4}$ and 
$\hat{E}_{0}$ of $A_4^{[a]}$ define divisors of $\hat{Z}^{[a]}$. They 
are denoted by $\hat{Y}_{X_1}$, $\hat{Y}_{X_2}$, $\hat{Y}_{1,2,3,4}$ 
and $\hat{Y}_{0}$, respectively. 

{\bf Small Resolution [b] }  In an alternative small resolution, 
one can take $Y_3^{[b]} := {\cal E}_2 \cdot D_{X_1}^{(2)}$ in $A_2$, so that 
$A_3^{[b]} := {\rm Bl}_{Y_3^{[b]}}A_2$. In the next step, 
the subvariety $Y_4^{[b]} := {\cal E}_1^{(3)[b]} \cdot D_{X_1}^{(3)[b]}$ of 
$A_3^{[b]}$ is chosen as the centre of the blow-up; it will be clear what 
the divisors ${\cal E}_1^{(3)[b]}$ and $D_{X_1}^{(3)[b]}$ stand for. 
The resulting non-singular fourfold is denoted by $\hat{Z}^{[b]}$.

Note that in the small resolution [b] ${\cal E}_4^{[b]}$ corresponds to the
fourth fibre component and ${\cal E}_3^{(4)[b]}$ to the third, instead of the other way around. By computing which intersections between the exceptional divisors vanish, one finds that this resolution furthermore corresponds to a triangulation where the line connection between $\hat{E}_2$ and $\hat{E}_4$ has been replaced by a line from $\hat{E}_1$ to $\hat{E}_3$


It is obvious from this property of the intersection ring that 
$\hat{Z}^{[a]}$ and $\hat{Z}^{[b]}$ are not the same geometries. 
%
%
In fact, we can turn $\hat{Z}^{[a]}$ into $\hat{Z}^{[b]}$ by a flop 
transition, which can already be anticipated from the fact that 
they only differ by a small resolution. The rich net of phases connected 
by flop transitions in F-theory compactifications with non-abelian 
gauge groups and the connection to group- and gauge theory has been 
explored in \cite{Esole:2011sm,Hayashi:2013lra,Hayashi:2014kca,Esole:2014bka,Esole:2014hya,Braun:2014kla}.

Any physics consequences in an SU(5) symmetric vacuum should remain 
the same whether the resolution [a] or [b] is used in formulating the 
flux background, as we have remarked in section \ref{sec:landscape-horizontal}. 
It is quite likely (see
section \ref{sec:landscape-horizontal}) that $h^{2,2}_V$, $h^{2,2}_H$ and 
$h^{2,2}_{RM}$ will indeed turn out to be the same for both [a] and [b].
For this reason, we pay attention only to the small resolution [a] 
described above, and use in section \ref{sect:genI5fourfolds} in the 
main text, as well as in the rest of this appendix. 
The superscript [a] will hence be dropped completely in the following. 

\subsection{Degeneration of singular fibres in higher codimension}
\label{ssec:fib-degen}

An elliptic fibration $\pi_{\hat{Z}}: \hat{Z} \longrightarrow B_3$ is obtained 
by restricting 
$(\pi_{A_4} := \pi_{A_0} \cdot \nu_{\rm tot}): A_4 \longrightarrow B_3$ on 
$\hat{Z} \subset A_4$. In this notation, the total fibre class over 
$S \subset B_3$ is 
$\pi_{\hat{Z}}^*(S) = \hat{Y}_0+\hat{Y}_1+\hat{Y}_2+\hat{Y}_3+\hat{Y}_4$.
Over a generic point in $S$, the fibres of this projection become curves $C_0+C_1+C_{2}+C_{3} + C_{4}$, which are all $\P^1$'s. A further restriction is 
given by only considering the divisors $\hat{Y}_i$, 
$\pi_{\hat{Y}_i}: \hat{Y}_i \longrightarrow S$.

Let us record known results on degeneration of singular fibre components, 
as we need them in the computation in the appendix \ref{app:hodgeexcept}.
The degeneration over the matter curve $\Sigma_{({\bf 10})} \subset S$ is:
\begin{eqnarray}
\pi_{\hat{Y}_0}^*(\Sigma_{({\bf 10})}) = S_{\infty}, & &
\pi_{\hat{Y}_1}^*(\Sigma_{({\bf 10})}) = S_B, \qquad 
\pi_{\hat{Y}_2}^*(\Sigma_{({\bf 10})}) = S_C + S_+,  \nonumber \\ 
\pi_{\hat{Y}_3}^*(\Sigma_{({\bf 10})}) = S_-, && 
\pi_{\hat{Y}_4}^*(\Sigma_{({\bf 10})}) = S_A + S_B + S_C.
\end{eqnarray}
Within $\hat{Z}$, 
$\hat{Y}_1 \cdot \hat{Y}_{4} = S_B$, and 
$\hat{Y}_{2} \cdot \hat{Y}_{4} = S_C$.
At a generic point in $\Sigma_{({\bf 10})}$, the fibre becomes 
a collection of curves, $C_S+C_A + 2C_B + 2C_C + C_+ + C_-$, all $\P^1$'s.

The degeneration over the matter curve $\Sigma_{({\bf 5})}$ is:
\begin{eqnarray}
\pi_{\hat{Y}_0}^*(\Sigma_{({\bf 5})}) = S_{\zeta}, \qquad 
\pi_{\hat{Y}_1}^*(\Sigma_{({\bf 5})}) = S_{\epsilon}, & & 
\pi_{\hat{Y}_2}^*(\Sigma_{({\bf 5})}) = S_{\delta}, \nonumber \\
\pi_{\hat{Y}_3}^*(\Sigma_{({\bf 5})}) = S_{\gamma} + S_{\beta}, & &
\pi_{\hat{Y}_4}^*(\Sigma_{({\bf 5})}) = S_{\alpha}, 
\end{eqnarray}
%
Over a generic point in $\Sigma_{({\bf 5})}$ these surfaces
become curves that are denoted by $C_{\alpha, \beta, \gamma, \delta, \epsilon}$ 
and $C_{\zeta}$, respectively.

At any one of $E_6$ points, 
\begin{eqnarray}
 C_0 \longrightarrow C_\infty, & & 
 C_1 \longrightarrow C_{\rm i},  \qquad 
 C_{2} \longrightarrow C_{\rm ii}+C_{\rm iii}, \nonumber \\
 C_{3} \longrightarrow C_{\rm iv}+C_{\rm v}, & & 
 C_{4} \longrightarrow C_{\rm i} + 2C_{\rm ii} + C_{\rm iv}, 
\end{eqnarray}

At any one of $D_6$ points, 
\begin{eqnarray}
 C_0 \longrightarrow  C_\infty, & & C_1 \longrightarrow C_{b}, \quad
 C_{2} \longrightarrow C_c +C_d, \nonumber \\
 C_{3} \longrightarrow C_d + C_{e+} + C_{e-}, & & 
 C_{4} \longrightarrow C_a + C_b + C_c
\end{eqnarray}
%


\subsection{Computation of Hodge numbers of the exceptional divisors 
$\hat{Y}_i = \hat{E}_{i}|_{\hat{Z}}$}\label{app:hodgeexcept}

Using the result reviewed in the appendix \ref{ssec:fib-degen} 
and the stratification described in section \ref{sect:h11excpdiv}, 
we compute the Hodge numbers of the divisors 
$\hat{Y}_{0,1,2,3,4}$ in $\hat{Z}$. The computation of $\chi(\hat{Z})$ 
in \cite{Andreas:1999ng} is similar in spirit. 

Let us first work out the Hodge--Deligne numbers of the
various strata of the surface $S \subset B_3$.
Using the divisor $\eta := c_1(N_{S|B_3}) + 6 c_1(S)$ on $S$
the number of $E_6$ and $D_6$ points is given by 
\begin{equation}
 N_E := (5K_S+\eta) \cdot (4K_S + \eta), \qquad 
 N_D := (5K_S + \eta) \cdot (3K_S + \eta).
\end{equation}
Thus the only non-vanishing $e^{p,q}_c$ of $P_{E6}$ and $P_{D6}$ are
\begin{equation}
 e^{0,0}_c(P_{E6}) = N_E, \qquad e^{0,0}_c(P_{D6}) = N_D\, .
\end{equation}
The genus $g_{10}$ of the matter curve $\Sigma_{({\bf 10})}$ is determined by 
\begin{equation}
 2g_{10}-2 = (5K_S + \eta) \cdot (6K_S + \eta). 
\end{equation}
The complement of $P_{E6}$ and $P_{D6}$ in $\Sigma_{({\bf 10})}$, denoted by 
$\Sigma_{({\bf 10})}^\circ$, has the following Hodge--Deligne numbers:
\begin{equation}
 e^{p,q}_c(\Sigma_{({\bf 10})}^\circ ) = \begin{array}{|cc}
    -g_{10} & 1 \\ (1-N_E-N_D) & -g_{10} \\ \hline \end{array}.
\end{equation}

The genus $g_5$ of the matter curve $\Sigma_{({\bf 5})}$ is formally given by 
\begin{equation}
2g_5-2 = (10K_S + 3 \eta) \cdot (11K_S + 3 \eta)\, . 
\end{equation}
However, there is a double point singularity at each of the points 
in $P_{D6}$. The resolved curve $\widetilde{\Sigma}_{({\bf 5})}$
has the genus 
\begin{equation}
  \tilde{g}_5 = g_5 - N_D.
\end{equation}
The curve $\widetilde{\Sigma}_{({\bf 5})}^\circ$ is obtained by removing 
the points $P_{E6}$ and the lift of the points $P_{D6}$ from 
$\widetilde{\Sigma}_{({\bf 5})}$. Thus we find that 
\begin{equation}
 e^{p,q}_c(\widetilde{\Sigma}_{({\bf 5})}^\circ) = \begin{array}{|cc}
  -\tilde{g}_5 & 1 \\ (1-N_E-2N_D) & - \tilde{g}_5 \\ \hline 
  \end{array}.
\end{equation}

Finally, the Hodge--Deligne numbers of $S^\circ := S \backslash 
(\Sigma_{({\bf 10})} \cup \Sigma_{({\bf 5})})$ are obtained by using 
the additivity of $e^{p,q}_c$. Assuming that $h^{1,0}(S)=0$, we find that 
\begin{eqnarray}
 e^{p,q}_c(S^\circ) & = & (-)^{p+q} h^{p,q}(S) - e^{p,q}_c(\Sigma_{({\bf 10})}^\circ)
  - e^{p,q}_c(\widetilde{\Sigma}_{({\bf 5})}^\circ) - e^{p,q}_c(P_{E6}) - e^{p,q}_c(P_{D6})
    \nonumber \\
 & = & \begin{array}{|ccc}
   h^{2,0}(S) & 0 & 1 \\ 
  (g_{10}+\tilde{g}_5) & (h^{1,1}(S)-2) & 0 \\
  (-1+N_E+2N_D) & (g_{10}+\tilde{g}_5) & h^{2,0}(S) \\
  \hline 
 \end{array}.
\end{eqnarray}

We are now ready to compute the Hodge numbers of the exceptional 
divisors $\hat{Y}_{0,1,2,3,4} = \hat{E}_{0,1,2,3,4}|_{\hat{Z}}$. 
Let us take $\hat{Y}_{4}$ as an example; the Hodge numbers of the other $\hat{Y}_{i}$ are determined analogously. The stratification of $\hat{Y}_{4}$ is:
\begin{eqnarray}
 \hat{Y}_{4} & = & 
 \left[ C_{4} \times S^\circ\right] \;\; \amalg \;\; 
 \left[ (C_A+C_B+C_C) \times \Sigma_{({\bf 10})}^\circ \right] \;\; \amalg \;\; 
 \left[ C_\alpha \times \widetilde{\Sigma}_{({\bf 5})}^\circ \right] \nonumber \\
 & & \amalg \;\;\;
 \left[ (C_{\rm i}+C_{\rm ii}+C_{\rm iv}) \times P_{E6} \right] \;\;\; \amalg \;\;\;
 \left[ (C_a+C_b+C_c) \times P_{D6} \right].
\end{eqnarray}
Using the multiplicativity and additivity of $e^{p,q}_c$, we find that 
\begin{eqnarray}
e^{2,2}_c(\hat{Y}_{4}) & = &
 e^{1,1}_c(C_{4}) e^{1,1}_c(S^\circ) + e^{0,0}_c(C_{4}) e^{2,2}_c(S^\circ) \nonumber \\
&& + e^{1,1}_c(C_A+C_B+C_C)e^{1,1}_c(\Sigma_{({\bf 10})}^\circ)
 + e^{1,1}_c(C_\alpha)e^{1,1}_c(\widetilde{\Sigma}_{({\bf 5})}^\circ) \nonumber \\
& = & (h^{1,1}(S)-2)+1+3+1 = h^{1,1}(S)+3, \\
e^{2,1}_c(\hat{Y}_{4}) & = & 
 e^{1,1}_c(C_{4})e^{1,0}_c(S^\circ) + e^{0,0}_c(C_{4})e^{2,1}_c(S^\circ) \nonumber \\
 & & + e^{1,1}_c(C_A+C_B+C_C)e^{1,0}_c(\Sigma_{({\bf 10})}^\circ) 
     + e^{1,1}_c(C_\alpha)e^{1,0}_c(\widetilde{\Sigma}_{({\bf 5})}^\circ) \nonumber \\
 & = & (g_{10}+\tilde{g}_5) + 0 - 3g_{10} - \tilde{g}_5 = -2g_{10}. 
\end{eqnarray}
$h^{2,2}(\hat{Y}_{4})=h^{1,1}(\hat{Y}_{4})$ is given by 
$e^{2,2}_c(\hat{Y}_{4})$, and $h^{2,1}(\hat{Y}_4)$ by $- e^{2,1}_c(\hat{Y}_{4})$.

\subsection{More on the geometry of the Cartan divisors $\hat{Y}_{1,2,3,4}$}
\label{app:detailsexcpdiv}

$H^{1,1}(\hat{Y}_0)$ is generated by $\pi_{\hat{Y}_0}^*(H^{1,1}(S))$ 
and $\hat{Y}_1|_{\hat{Y}_0}$. Although $\sigma$ defines a divisor 
on $\hat{Y}_0$ as well, it can be written in terms of the generators above. To see this, note that 
\begin{equation}
 \nu_{\rm tot}^*[-c_1(B_3)] = \sigma + \nu_{\rm tot}^*(D_{X_1}-2D_{X_2})
  = \sigma + \hat{D}_{X_1}-2\hat{D}_{X_2} -\hat{E}_1 - 2 \hat{E}_{2} - \hat{E}_{3}.
\end{equation}
From this, we obtain 
\begin{equation}
  0= \left[\sigma - \hat{E}_1 + \nu_{\rm tot}^*(c_1(B_3))\right]|_{\hat{E}_0},
\end{equation}
because $\hat{E}_0 \cdot \hat{E}_{2} = \hat{E}_0 \cdot \hat{E}_{3}=0$
in the ambient space $A_4$ (see figure~\ref{A4figure}). Using the relations  
$\hat{Z} \sim (3\hat{D}_{X_2} + \hat{E}_1 + 2\hat{E}_{2} + \hat{E}_{3}) \sim  
(2\hat{D}_{X_1} + \hat{E}_{3}+\hat{E}_{4})$ and the intersection ring 
information of the ambient space $A_4$ in figure \ref{A4figure}, 
it also follows that 
$(\hat{Y}_1-\hat{Y}_{4})|_{\hat{Y}_0} = (\hat{E}_1 - \hat{E}_{4}) \cdot 
\hat{E}_0 \cdot \hat{Z} \sim 0$.

$H^{1,1}(\hat{Y}_1)$ is generated by $\pi_{\hat{Y}_1}^*(H^{1,1}(S))$ 
and $\hat{Y}_{2}|_{\hat{Y}_1}$. 
A few more comments are in order here. First, 
\begin{equation}
  \left. \left(\hat{Y}_0 - \hat{Y}_{2} \right)\right|_{\hat{Y}_1} \sim 
    \left. \left(\nu_{\rm tot}^* (D_S) - \nu_{\rm tot}^*(D_{X_2}) \right)
    \right|_{\hat{Y}_1} \sim  
  \pi_{\hat{Y}_1}^* \left[ c_1(N_{S|B_3}) - 2 c_1(B_3)|_S \right].
\end{equation}
That is, both $\hat{Y}_0|_{\hat{Y}_1}$ and $\hat{Y}_{2}|_{\hat{Y}_1}$ 
are sections of the flat fibration 
$\pi_{\hat{Y}_1}: \hat{Y}_1 \longrightarrow S$, but they are different 
only modulo $\pi_{\hat{Y}_1}^*(H^{1,1}(S))$.
In the derivation above, we have used
\begin{equation}
 \hat{D}_{X_2}|_{\hat{Y}_1} = \hat{D}_{X_2} \cdot \hat{E}_1 \cdot \hat{Z} 
  \sim \hat{D}_{X_2} \cdot \hat{E}_1 \cdot (2\hat{D}_{X_1}+\hat{E}_3+\hat{E}_4)
 = 0 \, ,
\end{equation}
because none of $\hat{D}_{X_1}$, $\hat{E}_3$ and $\hat{E}_4$ share a face 
together with both $\hat{D}_{X_2}$ and $\hat{E}_1$.
This is a generalization of the toric statement that 
the 1-simplex $\langle \hat{D}_{X_2}, \hat{E}_1 \rangle$
($\langle \nu_2, \nu_9 \rangle$ in the notation of \cite{Grimm:2011fx})
is inside a facet of the polytope and hence does not give rise to 
a non-trivial divisor in $\hat{Y}_1$. Finally, one can see that 
$\hat{Y}_1 \cdot \hat{Y}_{4} = S_B$ is the total fibre component over 
the matter curve $\Sigma_{({\bf 10})}$ 
in $\pi_{\hat{Y}_1}: \hat{Y}_1 \longrightarrow S$, by observing that
\begin{eqnarray}
 \nu_{\rm tot}^*(D_{X_1} - D_{X_2})|_{\hat{Y}_1} 
 & = & \left[\hat{D}_{X_1} - \hat{D}_{X_2} + \hat{E}_{3} + \hat{E}_{4}
   \right]|_{\hat{Y}_1} = \hat{E}_{4}|_{\hat{Y}_1}, \\
 \nu_{\rm tot}^*(\sigma + c_1(B_3))|_{\hat{Y}_1} & = & \nu_{\rm tot}^*(c_1(B_3))
  = \pi_{\hat{Y}_1}^*(c_1(B_3)|_S)\, .
\end{eqnarray}
The matter curve $\Sigma_{({\bf 10})}$ is in the class 
$[5K_S + \eta] = c_1(N_{S|B_3})+c_1(S)=c_1(B_3)|_S$.

$H^{1,1}(\hat{Y}_{2})$ is generated by 
$\pi_{\hat{Y}_{2}}^*(H^{1,1}(S))$, $\hat{Y}_{3}|_{\hat{Y}_{2}}$ and 
$\hat{Y}_{4}|_{\hat{Y}_{2}}$. Both 
$\hat{Y}_1|_{\hat{Y}_{2}}$ and $\hat{Y}_{3}|_{\hat{Y}_{2}}$ are sections 
of the flat fibration $\pi_{\hat{Y}_{2}}: \hat{Y}_{2} \longrightarrow S$, 
but they are different only by the pullback of $H^{1,1}(S)$:  
\begin{equation}
 \left. \left( \hat{Y}_1 - \hat{Y}_{3} \right)\right|_{\hat{Y}_{2}} =  
  \nu_{\rm tot}^* \left( 2D_S - D_{X_1} \right)|_{\hat{Y}_{2}} \sim  
  \pi_{\hat{Y}_{2}}^*( 2c_1(N_{S|B_3})-3c_1(B_3)|_S )\, .
\end{equation}
The other generator $\hat{Y}_{4}|_{\hat{Y}_{2}} = S_C$ is one of the two singular 
fibre components in the flat family $\pi_{\hat{Y}_2}: \hat{Y}_2 \longrightarrow S$ along with the matter curve $\Sigma_{({\bf 10})}$. This four-cycle 
can be chosen as the matter surface for the hypermultiplets 
in the ${\rm SU}(5)$-$({\bf 10}+\overline{\bf 10})$ representations.

$H^{1,1}(\hat{Y}_{3})$ is generated by $\pi_{\hat{Y}_{3}}^*(H^{1,1}(S))$ 
and two other independent generators, which can be chosen as  
$\hat{Y}_{4}|_{\hat{Y}_{3}}$ and $\hat{Y}_{2}|_{\hat{Y}_{3}}$. $\hat{Y}_{3}$ is also regarded as a flat family 
of curves over $S$, $\pi_{\hat{Y}_{3}}: \hat{Y}_{3} \longrightarrow S$, 
of which both $\hat{Y}_{2}|_{\hat{Y}_{3}}$ and $\hat{Y}_{4}|_{\hat{Y}_{3}}$ define a section. 
Divisors $\hat{D}_{X_1}|_{\hat{Y}_{3}}$ and $\hat{D}_{X_2}|_{\hat{Y}_{3}}$
are linearly equivalent to the generators above; using 
\begin{equation}
 \nu_{\rm tot}^*(D_{X_2} - 2D_S) = 
  \hat{D}_{X_2} - 2\hat{D}_S - \hat{E}_1 - \hat{E}_{4}, 
\end{equation}
for example, one obtains a relation 
\begin{equation}
 \hat{Y}_{4}|_{\hat{Y}_{3}} \sim \hat{D}_{X_2}|_{\hat{Y}_{3}} + \pi_{\hat{Y}_3}^*(2K_S).
\end{equation}
Also, by using $\nu_{\rm tot}^*(D_{X_1}-3D_S)$ instead of 
$\nu_{\rm tot}^*(D_{X_2}-2D_S)$, we find that 
\begin{equation}
 \hat{D}_{X_1}|_{\hat{Y}_{3}} \sim 
   \hat{Y}_4|_{\hat{Y}_3} + \hat{Y}_2 |_{\hat{Y}_3} - \pi_{_hat{Y}_3}^*(3K_S) \, .
\end{equation}

The matter surface for the 
${\rm SU}(5)$-$(\bar{\bf 5}+{\bf 5})$-representations is given by 
a difference between the two divisors on $\hat{Y}_{3}$:
${\rm div}(a_3 X_{\hat{E}_{4}} + a_5 X_{\hat{D}_2})|_{\hat{Y}_{3}} - 
 \hat{E}_{2}|_{\hat{Y}_{3}}$, where $X_{\hat{E}_4}$ and $X_{\hat{D}_{X_2}}$ are the 
homogeneous coordinates of $A_4^{[a]}$ corresponding to the divisors 
$\hat{E}_4$ and $\hat{D}_{X_2}$, respectively. 
The divisor $\{ a_3 X_{\hat{E}_4}+a_5 X_{\hat{D}_{X_2}}=0 \}$ on 
$\hat{Y}_3$ consists of two irreducible components. One component is 
$\hat{E}_2|_{\hat{Y}_3}$, to be subtracted away, and the other one is an 
irreducible component of the singular fibre over the matter curve 
$\Sigma_{({\bf 5})}$, the matter surface $S_\gamma$. 
Thus, it is in the divisor class 
$[ \hat{E}_{4}+ \nu_{\rm tot}^*(c_1(N_{S|B_3})+3c_1(B_3)|_S) ]-\hat{E}_{2}$ on 
$\hat{Y}_{3}$. From this, we can conclude that the matter surface for 
the $(\bar{\bf 5}+{\bf 5})$ representations also belongs to the 
class of a vertical cycles.

$H^{1,1}(\hat{Y}_{4})$ is generated by $\pi_{\hat{Y}_{4}}^*(H^{1,1}(S))$ and three more divisors $\hat{Y}_{3}|_{\hat{Y}_{4}}$, $\hat{Y}_1|_{\hat{Y}_{4}}=S_B$ and $\hat{Y}_{2}|_{\hat{Y}_{4}}=S_C$. The last two of these form two out of the three singular fibre components over the matter curve $\Sigma_{({\bf 10})}$. Although $\hat{Y}_0|_{\hat{Y}_{4}}$ also defines a section of the flat fibration $\pi_{\hat{Y}_{4}}: \hat{Y}_{4} \longrightarrow S$, there is a relation 
\begin{eqnarray}
 \left. \left(\hat{Y}_{3} - \hat{Y}_0\right)\right|_{\hat{Y}_{4}} 
 & = & \nu_{\rm tot}^*(D_{X_2}-D_S)|_{\hat{Y}_{4}}
  - (\hat{E}_{2}+\hat{D}_{X_2})|_{\hat{Y}_{4}} \nonumber \\
 & = & 
 - \hat{Y}_{2}|_{\hat{Y}_{4}} + \pi_{\hat{Y}_{4}}^*(2c_1(B_3)|_S - c_1(N_{S|B_3}))\, .
\end{eqnarray}

Using all the relations above, one can derive the following 
rational equivalence relations:
\begin{eqnarray}
 \hat{Y}_0 \cdot \hat{Y}_1 & \sim & \pi^*(D_S) \cdot \sigma 
    + c_1(B_3)\cdot \hat{Y}_0, \label{eq:rat-equiv-except-divdiv-a}\nn \\
 \hat{Y}_1 \cdot \hat{Y}_{2} & \sim & \pi^*(D_S) \cdot \sigma 
    + c_1(B_3) \cdot (\hat{Y}_0 + 2\hat{Y}_1)
    - c_1(N_{S|B_3}) \cdot \hat{Y}_1, \nn\\
 \hat{Y}_{2} \cdot \hat{Y}_{3} & \sim & \pi^*(D_S) \cdot \sigma
    + c_1(B_3) \cdot (\hat{Y}_0 + 2\hat{Y}_1+3\hat{Y}_{2})
    - c_1(N_{S|B_3}) \cdot (\hat{Y}_1+2\hat{Y}_{2}),\nn \\
 \hat{Y}_{3} \cdot \hat{Y}_{4} & \sim & \pi^*(D_S) \cdot \sigma
    - (\hat{Y}_{2} \cdot \hat{Y}_{4}) 
    + c_1(B_3) \cdot (\hat{Y}_0 + 2\hat{Y}_{4}) 
    - c_1(N_{S|B_3}) \cdot \hat{Y}_{4}, \nn \\
  \hat{Y}_{4} \cdot \hat{Y}_0 & \sim & \pi^*(D_S) \cdot \sigma
     + c_1(B_3) \cdot \hat{Y}_0,
\end{eqnarray}
and 
\begin{equation}
  \hat{Y}_1 \cdot \hat{Y}_{4} \sim c_1(B_3) \cdot \hat{Y}_1.
\label{eq:rat-equiv-except-divdiv-f}
\end{equation}

The matter surfaces for the ${\bf 10}$ and $\bar{\bf 5}$ 
representations are different under rational equivalence only by
\begin{eqnarray}
[S_\gamma] - [-S_C] & \sim & \left[ \hat{Y}_{3} \cdot 
    \left\{ \hat{Y}_{4} + c_1(N)+3c_1(B_3) - \hat{Y}_{2} \right\} \right]- 
  \left[ -  \hat{Y}_{2} \cdot \hat{Y}_{4} \right]  \\
& \sim & c_1(B_3) \cdot (-2\hat{Y}_1-3\hat{Y}_{2} + 3 \hat{Y}_{3} + 2\hat{Y}_{4})
+ c_1(N_{S|B_3}) \cdot (\hat{Y}_1 + 2\hat{Y}_{2} + \hat{Y}_{3} - \hat{Y}_{4}).
   \nonumber 
\end{eqnarray}
If we turn on a flux which does not break SU(5) gauge invariance, this guarantees that the chirality in 
the ${\bf 10}$ representation is the same as that in $\bar{\bf 5}$ representation, as also observed in \cite{Marsano:2011hv}.

\section{Dependence of Hodge numbers on the rank of 7-brane symmetry}
\label{sect:dephodgegroup}

The number of flux vacua that have a stack of 7-branes with symmetry $R$ 
scales as $e^{K/6}$, $K={\rm dim}[H^4_H(X;\R)]$, as we have seen in 
section \ref{sect:distrGNgen}. In this appendix, 
we use the approximate relation (\ref{eq:K-h31}), which holds in 
the cases satisfying the condition (\ref{eq:hodge-inequality-cpx-rich}), 
to study the statistics of the unification symmetry $R$. 
Instead of studying how the dimension of the primary horizontal subspace $K$ 
depends on the symmetry $R$, we will estimate how $h^{3,1}$ changes for 
different choices of $R$.

Let us start from a family for $(B_3, [S], R_1)$, and suppose that the 
enhancement of the 7-brane symmetry $R_1$ to $R_2$ occurs when a section 
$f \in \Gamma(B_3; {\cal O}(D))$ of some line bundle ${\cal O}_{B_3}(D)$ 
vanishes along $S$; the section $f$ is used in a defining equation of 
the Weierstrass model, and the geometry gets more singular for $f|_S=0$.
The line bundle ${\cal O}_{B_3}(D)$ here is determined by Tate's 
algorithm, as we will discuss more explicitly later on. Requiring 
$f$ to vanish along $S$, the number of independent complex structure moduli 
is reduced roughly by 
\begin{equation}
  \Delta h^{3,1} = h^0(B_3; {\cal O}_{B_3}(D)) - h^0(B_3; {\cal O}_{B_3}(D-S)).
\end{equation}
One can then use the exact sequence 
\begin{equation}
 0 \longrightarrow {\cal O}_{B_3}(D-S) \longrightarrow {\cal O}_{B_3}(D) 
   \longrightarrow i_* \left[ {\cal O}_S(D|_S) \right] \longrightarrow 0  
\end{equation}
to set an upper bound 
\begin{equation}
 \Delta h^{3,1} \leq h^0(S; {\cal O}_S(D|_S)).
\end{equation}
This inequality is saturated when $h^1(B_3; {\cal O}_{B_3}(D-S))$ vanishes.
From this, we can derive the crude estimate
\begin{equation}
e^{h^0(S; {\cal O}_S(D|_S))}
\end{equation}
on the statistical cost of unification symmetry $R_2$ relatively to $R_1$. 

Let us use the examples we studied in section \ref{sect:elfibexamples} to 
see how this works in practice. We focus on the case 
$B_3^{(n)}=\P[{\cal O}_{\P^2} \oplus {\cal O}_{\P^2}(n)]$ and study the enhancement 
from $R_1 = A_4$ to $R_2 = D_5$ along $S = \P^2$. 
The condition for this $A_4 \rightarrow D_5$ enhancement is 
$(f=a_5)|_S = 0$; note also that $a_5 \in \Gamma(B_3; {\cal O}_{B_3}(-K_{B_3}))$ 
and hence $D|_S = S - K_S = (3-n)H_{\P^2}$. The estimated upper bound  
on $\Delta h^{3,1}$ therefore becomes 
\begin{equation}
  h^0 \left( S; {\cal O}_S(D|_S) \right) =
  h^0 \left(\P^2; {\cal O}_{\P^2}((3-n)H) \right) = 
 \frac{(5-n)(4-n)}{2}.
\end{equation}
The value of $\Delta K_{4\mbox{-}5}$ in table \ref{tab:K-dep-on-R} is 
estimated reasonably well 
by $6 \times h^0(S; {\cal O}_S(D|_S)) = 3(5-n)(4-n)$ indeed.\footnote{
It is a little misleading to use these examples to emphasize that this 
estimate is good. The $\P^1$-fibration structure over $S$ in $B_3^{(n)}$ 
defines a normal coordinate of $S$ that remains well-defined globally 
on $B_3^{(n)}$. We do not intend to claim that $6 h^0(S; {\cal O}_S(D|_S))$ 
is a good estimate of $\Delta K$ rather than a good estimate of the 
upper bound on $\Delta K$.}

Let us now use this argument to study how the statistical cost varies 
for different enhancement of symmetries $R_1 \rightarrow R_2$.
Suppose that the Weierstrass model $Z_s$ is parametrized by 
the generalized Weierstrass equation (Tate's form),
\begin{equation}
\label{eq:tateform}
 y^2 + b_1 xy + b_3 y = x^3 + x^2 b_2 + x b_4 + b_6 \, .
\end{equation}
When $b_i$ vanishes along $S$ with the order of vanishing $n_i$, 
let $b_i = s^{n_i} b_{i|n_i}$.
The dictionary between the order of vanishing $n_i$'s and the 7-brane 
symmetry $R$ is known \cite{Bershadsky:1996nh}, and the necessary information 
is recorded in table \ref{tab:Tate-condition}.
\begin{table}[tbp]
\begin{center}
\begin{tabular}{ccc}
\begin{tabular}{|cc|cccc|}
\hline 
 & & $b_1$ & $b_3$ & $b_4$ & $b_6$ \\
\hline 
$A_m$ & $I_{m+1}^s$ & 0 & $m/2$ & $m/2+1$ & $m+1$ \\ 
$A_{m+1}$ & $I_{m+2}^s$ & 0 & $m/2+1$ & $m/2+1$ & $m+2$ \\ 
$A_{m+2}$ & $I_{m+3}^s$ & 0 & $m/2+1$ & $m/2+2$ & $m+3$ \\ 
\hline
$D_{m+1}$ & $I_{m-3}^{*s}$ & 1 & $m/2$ & $m/2+1$ & $m+1$ \\ 
$D_{m+2}$ & $I_{m-2}^{*s}$ & 1 & $m/2+1$ & $m/2+1$ & $m+1$ \\ 
$D_{m+3}$ & $I_{m-1}^{*s}$ & 1 & $m/2+1$ & $m/2+2$ & $m+3$ \\ 
\hline 
\end{tabular}
& $\quad$ & 
\begin{tabular}{|cc|ccccc|}
\hline
 & & $b_1$ & $b_2$ & $b_3$ & $b_4$ & $b_6$ \\
\hline
$A_4$ & $I_5^{s}$ & 0 & 1 & 2 & 3 & 5 \\ 
$D_5$ & $I_1^{*s}$ & 1 & 1 & 2 & 3 & 5 \\ 
$E_6$ & ${\rm IV}^{*s}$ & 1 & 2 & 2 & 3 & 5 \\ 
$E_7$ & ${\rm III}^{*}$ & 1 & 2 & 3 & 3 & 5 \\ 
$E_8$ & ${\rm II}^*$ & 1 & 2 & 3 & 4 & 5 \\ 
\hline  
\end{tabular}
\end{tabular}
\caption{\label{tab:Tate-condition}
Order of vanishing $n_i$ of $b_i$'s required 
for various types of singular fibre (7-brane gauge group); information 
relevant to the discussion in this section is extracted from a table 
in \cite{Bershadsky:1996nh}. $m \in 2\N$ is assumed in this table. 
The order of vanishing for $b_2$ is 1 for any one in the $I_k^s$ and $I_k^{*s}$ 
series.}
\end{center}
\end{table}

An immediate generalization of the $A_4 \rightarrow D_5$ enhancement is 
the enhancement $A_m \rightarrow D_{m+1}$, $m \in 2\N$. In these cases, 
$f = b_{1|0}$ and $D = -K_{B_3}$ for any $m \in 2\N$, not just for $m=4$. 
Thus, the same value 
$h^0(S; {\cal O}_S(-K_{B_3}|_S)) = h^0(S; {\cal O}_S(S-K_S))$ provides an 
approximate upper bound on $\Delta h^{3,1}$ for any $m \in 2\N$. 

One can also think of two separate chains of symmetry enhancement, 
$A_m \rightarrow A_{m+1} \rightarrow A_{m+2}$ and 
$D_{m+1} \rightarrow D_{m+2} \rightarrow D_{m+3}$. The statistical cost 
increases at the same pace in these two chains; as the rank increases by 
two from $A_m$ to $A_{m+2}$, or from $D_{m+1}$ to $D_{m+3}$, we need to set 
the sections $b_{3|m/2}|_S$, $b_{4|m/2+1}|_S$, $b_{6|m+1}|_S$ and $b_{6|m+2}|_S$ 
to zero in any one of those two chains (see table \ref{tab:Tate-condition}). 
This is consistent with the result above that the statistical cost 
for the enhancement $A_m \rightarrow D_{m+1}$ remains much the same 
for any $m \in 2\N$. Noting that $b_{i|n_i} \in 
\Gamma(B_3; {\cal O}_{B_3}(-i K_{B_3} -n_i S))$, and hence 
\begin{align}
& b_{3|m/2}|_S \in \Gamma(S; {\cal O}_{S}(-3K_{B_3}|_S -m/2 S)), \quad 
  b_{6|m+1}|_S \in \Gamma(S; {\cal O}_S(-6K_{B_3}|_S-(m+1)S)), \\
& b_{4|m/2+1}|_S \in \Gamma \left(S; 
        {\cal O}_S(-4K_{B_3}|_S -\frac{m+2}{2}S) \right), \quad 
  b_{6|m+2}|_S \in \Gamma(S; {\cal O}_S(-6K_{B_3}|_S-(m+2)S)), \nonumber 
\end{align}
one finds that the statistical cost for enhancement by rank-one, 
measured by $\Delta K/\Delta m \approx 6 \Delta h^{3,1}/\Delta m$, 
becomes larger for higher rank $m$ in the case $c_1(N_{S|B_3}) < 0$; 
if $c_1(N_{S|B_3}) > 0$, on the other hand, the cost for one-rank 
enhancement decreases for higher rank $m$, because $\Delta h^{3,1}$ is bounded 
from above by smaller value. One should be careful in interpreting this 
phenomenon for $c_1(N_{S|B_3}) > 0$; it looks as if higher rank gauge groups 
become just as ``natural'' as lower rank gauge groups at first sight, 
but it may also be that the choice of complex structure for such a high 
rank enhancement has already become impossible. 

Let us finally look at the chain of symmetry enhancements 
$E_m \rightarrow E_{m+1}$; $A_4 \rightarrow D_5$ is also regarded as a part 
of this chain. In this chain, we need to set the section 
\begin{equation}
 \left.  \left[ a_{9-m|0} \propto b_{m-3|m-4} \right]\right|_S \in 
  \Gamma(S; {\cal O}_S(S-(m-3)K_S)) 
\end{equation}
to zero for the enhancement $E_m \rightarrow E_{m+1}$. Thus, the statistical 
cost for one-rank enhancement, $\Delta K/\Delta m$, becomes increasingly 
large in higher rank (larger $m$), if $K_S < 0$. [Note that there are 
chiral multiplets in the adjoint representation of $E_m$ in the spectrum 
below the Kaluza--Klein scale, if $K_S >0$ instead.]\footnote{In the 
case of $\hat{Z} = {\rm K3} \times {\rm K3}$ \cite{Braun:2014ola}, 
the dimension of the primary horizontal subspace, $K$, and $h^{3,1}$ are 
linear in the rank of the 7-brane gauge group ${\rm rank}_7$, and 
$\Delta K/\Delta {\rm rank}_7$ remains constant, in particular. 
This simple result for ${\rm K3} \times {\rm K3}$ compactifications of 
F-theory should be understood as an artefact of trivial bundles $K_S$ and 
$N_{S|B_3}$.} Interestingly, the behaviour of $\Delta K/\Delta m$ is 
controlled by the normal bundle $N_{S|B_3}$ in the (IIB-like) $A_m$ type and 
$D_m$ type chains of symmetry enhancement, and by the canonical bundle 
$K_S$ in the $E_m$ type chain available in F-theory.

\section{Flux controlling the net chirality}\label{sect:chirflux}

In order to consider an ensemble of fluxes leading to effective theories 
with a given number of generations (net chirality) $N_{\rm gen}$, 
$G^{(4)}_{\rm fix}$ needs to be chosen so that 
it generates the net chirality $N_{\rm gen}$. If we take the scanning space 
$H_{\rm scan}$ to be the real primary horizontal subspace $H^4_H(X;\R)$, 
then all the flux vacua end up with effective theories with one and the 
same value of $N_{\rm gen}$ in such an ensemble.
This appendix begins with writing down 
the four-form flux generating the net chirality, which is already 
a well-understood subject in the literature.  We then move on to compute the 
D3-tadpole bound 
\begin{equation}
 \frac{\chi(\hat{Z})}{24} - \frac{1}{2} (G^{(4)}_{\rm fix})^2 = L_*
   \geq \frac{1}{2} (G^{(4)}_{\rm scan})^2.
\label{eq:tadpole-bound}
\end{equation}
We implicitly used that $G^{(4)}_{\rm fix} \wedge G^{(4)}_{\rm scan} = 0$, which follows
because the chirality generating flux $G^{(4)}_{\rm fix}$ is chosen 
within the vertical component $H^{2,2}_V(\hat{Z}; \R)$ (because the 
matter surface belongs to the space of vertical cycles), and the primary 
horizontal subspace is orthogonal to the vertical component. 

The choice of $(B_3, [S])$ in section \ref{sect:examplesCYna4} is a simple 
generalization of \cite{Grimm:2011fx}; 
$B_3 = \P[{\cal O}_{\P^2} \oplus {\cal O}_{\P^2}(n)]$ instead of 
$B_3 = \P[{\cal O}_{\P^2} \oplus {\cal O}_{\P^2}] = \P^1 \times \P^2$.
In order to determine the flux $G^{(4)}_{\rm fix}$ generating the chirality 
of SU(5) unification models, the conditions for Lorentz SO(3,1) symmetry 
(\ref{eq:cond-SO(3,1)}) and unbroken SU(5) 
symmetry (\ref{eq:cond-unbroken-sym}) were imposed on the space of 
vertical four-forms $H^{2,2}_V(\hat{Z}; \R)$. As we can think of the K\"ahler form
as being expanded in a basis consisting of divisors of the base, the section, the
generic fibre class and the exceptional fibre components, these constraints also
automatically make the $D$-term \eqref{eq:cond-prim} vanish.
It is legitimate, as we stated above, to search 
the chirality generating flux $G^{(4)}_{\rm fix}$ from $H^{2,2}_V(\hat{Z}; \R)$, 
because the matter surface belongs to the space of vertical four-cycles. 
We have seen in section \ref{sect:examplesCYna4} that the space 
$H^{2,2}_V(\hat{Z}; \R)$ has nine dimensions, while the conditions of 
unbroken SO(3,1) Lorentz and SU(5) unified symmetries result in 
eight independent constraints. This is true for all the cases with 
$-3 \leq n \leq 2$, not just the case $n=0$ in \cite{Grimm:2011fx}. 
After a straightforward computation, it turns out that 
\begin{equation}
\label{fluxquantgenform}
G^{(4)}_{\rm fix} =  \lambda 
 \left(  5 \hat{Y}_4 \cdot \hat{Y}_2
      + (2S+(3+n)H_{\P^2}) \cdot (2\hat{Y}_1 - 2\hat{Y}_4-\hat{Y}_2 + \hat{Y}_3)
 \right),
\end{equation}
where $\lambda \in \R$ and $H_{\P^2}$ is the hyperplane divisor of the base 
$\P^2$. We have confirmed that this flux is equivalent to the one given in \cite{Marsano:2011hv} and, 
in the case of $n=0$, to the choice of \cite{Grimm:2011fx} with $\lambda = - 3 \beta$, 
exploiting rational equivalence. The net chirality for the SU(5) 
${\bf 10}$--$\overline{\bf 10}$ representations is given by 
\begin{equation}
 N_{\rm gen} = \chi_{\bf 10} = - \lambda (18-n)(3-n).
\end{equation}

The choice of such chirality generating flux is quantized due to the 
condition $G^{(4)}_{\rm fix} \in [H^4(\hat{Z}; \Z)]_{\rm shift}$. It is not 
an easy task, to say the least, to determine the integral basis of 
$[H^4(\hat{Z}; \Z)]_{\rm shift} \cap H^{2,2}_V(\hat{Z}; \Q)$, but fortunately 
this task can be detoured in the cases we are dealing with. Here,
a dual description in Heterotic string theory exists, and it is known that (\ref{eq:FMW-flux}) gives rise to 
the net chirality (\ref{eq:Het-chirality}).
We should thus identify $\lambda_{FMW} = \lambda$, and the quantization 
\begin{equation}
   \lambda = \frac{1}{2} (1+2a), \qquad a \in \Z
\end{equation}
follows from that of that of $\lambda_{FMW}$ \cite{Marsano:2011hv}.
As discussed in \cite{Marsano:2011hv}, $G^{(4)}_{\rm fix}$ with this quantization 
condition satisfies $G^{(4)}_{\rm fix} \in [H^4(\hat{Z}; \Z)]_{\rm shift}$.

Now that $G^{(4)}_{\rm fix}$ is given explicitly, we are ready to compute 
the D3-tadpole upper bound $L_*$. The standard computation techniques 
of toric geometry allow us to compute 
$\int_{\hat{Z}} G^{(4)}_{\rm fix} \wedge G^{(4)}_{\rm fix}$. We find that 
\begin{eqnarray}
  L_* & = & \frac{2163}{4} + \frac{125}{8} n(n+7) -
  \frac{5 (18-n)(3-n)}{2} \lambda^2 \\
   & = &  \frac{2163}{4} + \frac{125}{8}n(n+7) -
  \frac{5 N_{\rm gen}^2}{2(18-n)(3-n)}.
\label{eq:Ngen-dep-L*}
\end{eqnarray}
It is not obvious from the above equation
whether $L_*$ becomes integer or not, but $L_*$ is indeed; see 
table \ref{tab:tadpole-value-SU(5)}.
\begin{table}[tbp]
\begin{center}
 \begin{tabular}{c|cccccc}
 $n$					& $-3$ & $-2$ & $-1 $& $0$ & $1$ & $2$ \\
 \hline
 $\chi(\hat{Z})/24$			&$1263/4$& $719/2$& $869/2$ & $2163/4$ & $2713/4$ & $847$ \\
 $\frac{1}{2} (G^{(4)}_{\rm fix})^2$     &$315/4$& $125/2$& $95/2$ & $135/4$ & $85/4$ & $10$ \\
 $L_*^{\rm max}$ & 237 & 297 & 387 & 507 & 657 & 837 \\
 $K$ & 7557 & 8603 & 10403 & 12953 & 16253 & 20303
 \end{tabular}
\caption{\label{tab:tadpole-value-SU(5)}
The value of $\chi(\hat{Z})/24$ as well as 
$(G^{(4)}_{\rm fix})^2/2$ and $L_*$ for $\lambda = \pm 1/2$, when 
$L_*$ becomes maximal for a given $n$. 
The result of $K := {\rm dim}_\R [H^4_H(\hat{Z};\R)]$ is copied 
from table~\ref{tab:hodge-SU(5)}.
}
\end{center}
\end{table}
The computation here is essentially that of \cite{Andreas:1999ng}, 
although a K3-fibre is used instead of the stable degeneration limit 
$dP_9 \cup dP_9$, and a specific resolution of singularity is employed.
The value of the D3-tadpole upper bound $L_*$ depends on the number of 
generation $N_{\rm gen}$. This result is used in section \ref{sect:distrGNgen} 
(and also \cite{physlett}) to derive the distribution of 
$N_{\rm gen}$ in the landscape of F-theory flux vacua.


\begin{thebibliography}{99}
 
\bibitem{Haupt:2008nu}
A.~S. Haupt, A.~Lukas, and K.~Stelle, ``{M-theory on Calabi-Yau Five-Folds},''
  \href{http://dx.doi.org/10.1088/1126-6708/2009/05/069}{{\em JHEP} {\bf 0905}
  (2009)  069},
\href{http://arxiv.org/abs/0810.2685}{{\tt arXiv:0810.2685 [hep-th]}}.

\bibitem{hudson05}
R.~Hudson, {\em Kummer's Quartic Surface}.
\newblock Cambridge, 1905.

\bibitem{Aspinwall:1996mn}
P.~S. Aspinwall, ``{K3 surfaces and string duality},''
\href{http://arxiv.org/abs/hep-th/9611137}{{\tt arXiv:hep-th/9611137
  [hep-th]}}.

\bibitem{Greene:1993vm}
B.~R. Greene, D.~R. Morrison, and M.~R. Plesser, ``{Mirror manifolds in higher
  dimension},'' \href{http://dx.doi.org/10.1007/BF02101657}{{\em
  Commun.Math.Phys.} {\bf 173} (1995)  559--598},
\href{http://arxiv.org/abs/hep-th/9402119}{{\tt arXiv:hep-th/9402119
  [hep-th]}}.

\bibitem{Becker:1996gj}
K.~Becker and M.~Becker, ``{M theory on eight manifolds},''
  \href{http://dx.doi.org/10.1016/0550-3213(96)00367-7}{{\em Nucl.Phys.} {\bf
  B477} (1996)  155--167},
\href{http://arxiv.org/abs/hep-th/9605053}{{\tt arXiv:hep-th/9605053
  [hep-th]}}.

\bibitem{Sethi:1996es}
S.~Sethi, C.~Vafa, and E.~Witten, ``{Constraints on low dimensional string
  compactifications},''
  \href{http://dx.doi.org/10.1016/S0550-3213(96)00483-X}{{\em Nucl.Phys.} {\bf
  B480} (1996)  213--224},
\href{http://arxiv.org/abs/hep-th/9606122}{{\tt arXiv:hep-th/9606122
  [hep-th]}}.

\bibitem{Brunner:1996pk}
I.~Brunner and R.~Schimmrigk, ``{F theory on Calabi-Yau fourfolds},''
  \href{http://dx.doi.org/10.1016/0370-2693(96)01100-8}{{\em Phys.Lett.} {\bf
  B387} (1996)  750--758},
\href{http://arxiv.org/abs/hep-th/9606148}{{\tt arXiv:hep-th/9606148
  [hep-th]}}.

\bibitem{Mayr:1996sh}
P.~Mayr, ``{Mirror symmetry, N=1 superpotentials and tensionless strings on
  Calabi-Yau four folds},''
  \href{http://dx.doi.org/10.1016/S0550-3213(97)00196-X}{{\em Nucl.Phys.} {\bf
  B494} (1997)  489--545},
\href{http://arxiv.org/abs/hep-th/9610162}{{\tt arXiv:hep-th/9610162
  [hep-th]}}.

\bibitem{Klemm:1996ts}
A.~Klemm, B.~Lian, S.~Roan, and S.-T. Yau, ``{Calabi-Yau fourfolds for M theory
  and F theory compactifications},''
  \href{http://dx.doi.org/10.1016/S0550-3213(97)00798-0}{{\em Nucl.Phys.} {\bf
  B518} (1998)  515--574},
\href{http://arxiv.org/abs/hep-th/9701023}{{\tt arXiv:hep-th/9701023
  [hep-th]}}.

\bibitem{Mohri:1997uk}
K.~Mohri, ``{F theory vacua in four-dimensions and toric threefolds},''
  \href{http://dx.doi.org/10.1142/S0217751X99000415}{{\em Int.J.Mod.Phys.} {\bf
  A14} (1999)  845--874},
\href{http://arxiv.org/abs/hep-th/9701147}{{\tt arXiv:hep-th/9701147
  [hep-th]}}.

\bibitem{Bershadsky:1997zs}
M.~Bershadsky, A.~Johansen, T.~Pantev, and V.~Sadov, ``{On four-dimensional
  compactifications of F theory},''
  \href{http://dx.doi.org/10.1016/S0550-3213(97)00393-3}{{\em Nucl.Phys.} {\bf
  B505} (1997)  165--201},
\href{http://arxiv.org/abs/hep-th/9701165}{{\tt arXiv:hep-th/9701165
  [hep-th]}}.

\bibitem{Kreuzer:1997zg}
M.~Kreuzer and H.~Skarke, ``{Calabi-Yau four folds and toric fibrations},''
  \href{http://dx.doi.org/10.1016/S0393-0440(97)00059-4}{{\em J.Geom.Phys.}
  {\bf 26} (1998)  272--290},
\href{http://arxiv.org/abs/hep-th/9701175}{{\tt arXiv:hep-th/9701175
  [hep-th]}}.

\bibitem{Strominger:1990pd}
A.~Strominger, ``{Special Geometry},''
\href{http://dx.doi.org/10.1007/BF02096559}{{\em Commun.Math.Phys.} {\bf 133}
  (1990)  163--180}.

\bibitem{Ashok:2003gk}
S.~Ashok and M.~R. Douglas, ``{Counting flux vacua},''
  \href{http://dx.doi.org/10.1088/1126-6708/2004/01/060}{{\em JHEP} {\bf 0401}
  (2004)  060},
\href{http://arxiv.org/abs/hep-th/0307049}{{\tt arXiv:hep-th/0307049
  [hep-th]}}.

\bibitem{Denef:2004ze}
F.~Denef and M.~R. Douglas, ``{Distributions of flux vacua},''
  \href{http://dx.doi.org/10.1088/1126-6708/2004/05/072}{{\em JHEP} {\bf 0405}
  (2004)  072},
\href{http://arxiv.org/abs/hep-th/0404116}{{\tt arXiv:hep-th/0404116
  [hep-th]}}.

\bibitem{Braun:2014ola}
A.~P. Braun, Y.~Kimura, and T.~Watari, ``{The Noether-Lefschetz problem and
  gauge-group-resolved landscapes: F-theory on K3 $\times$ K3 as a test
  case},'' \href{http://dx.doi.org/10.1007/JHEP04(2014)050}{{\em JHEP} {\bf
  1404} (2014)  050},
\href{http://arxiv.org/abs/1401.5908}{{\tt arXiv:1401.5908 [hep-th]}}.

\bibitem{Grimm:2009ef}
T.~W. Grimm, T.-W. Ha, A.~Klemm, and D.~Klevers, ``{Computing Brane and Flux
  Superpotentials in F-theory Compactifications},''
  \href{http://dx.doi.org/10.1007/JHEP04(2010)015}{{\em JHEP} {\bf 1004} (2010)
   015},
\href{http://arxiv.org/abs/0909.2025}{{\tt arXiv:0909.2025 [hep-th]}}.

\bibitem{Bizet:2014uua}
N.~C. Bizet, A.~Klemm, and D.~V. Lopes, ``{Landscaping with fluxes and the E8
  Yukawa Point in F-theory},''
\href{http://arxiv.org/abs/1404.7645}{{\tt arXiv:1404.7645 [hep-th]}}.

\bibitem{Batyrev94dualpolyhedra}
V.~V. Batyrev, ``{Dual polyhedra and mirror symmetry for Calabi-Yau
  hypersurfaces in toric varieties},'' {\em J. Alg. Geom} (1994)  493--535,
  \href{http://arxiv.org/abs/alg-geom/9310003}{{\tt arXiv:alg-geom/9310003}}.

\bibitem{Buican:2006sn}
M.~Buican, D.~Malyshev, D.~R. Morrison, H.~Verlinde, and M.~Wijnholt,
  ``{D-branes at Singularities, Compactification, and Hypercharge},''
  \href{http://dx.doi.org/10.1088/1126-6708/2007/01/107}{{\em JHEP} {\bf 0701}
  (2007)  107},
\href{http://arxiv.org/abs/hep-th/0610007}{{\tt arXiv:hep-th/0610007
  [hep-th]}}.

\bibitem{Beasley:2008kw}
C.~Beasley, J.~J. Heckman, and C.~Vafa, ``{GUTs and Exceptional Branes in
  F-theory - II: Experimental Predictions},''
  \href{http://dx.doi.org/10.1088/1126-6708/2009/01/059}{{\em JHEP} {\bf 0901}
  (2009)  059},
\href{http://arxiv.org/abs/0806.0102}{{\tt arXiv:0806.0102 [hep-th]}}.

\bibitem{physlett}
A.~P. Braun and T.~Watari, ``{Distribution of Number of Generations in Flux
  Compactifications},'' {\em to appear}  .

\bibitem{Dasgupta:1999ss}
K.~Dasgupta, G.~Rajesh, and S.~Sethi, ``{M theory, orientifolds and G -
  flux},'' \href{http://dx.doi.org/10.1088/1126-6708/1999/08/023}{{\em JHEP}
  {\bf 9908} (1999)  023},
\href{http://arxiv.org/abs/hep-th/9908088}{{\tt arXiv:hep-th/9908088
  [hep-th]}}.

\bibitem{Braun:2011zm}
A.~P. Braun, A.~Collinucci, and R.~Valandro, ``{G-flux in F-theory and
  algebraic cycles},''
  \href{http://dx.doi.org/10.1016/j.nuclphysb.2011.10.034}{{\em Nucl.Phys.}
  {\bf B856} (2012)  129--179},
\href{http://arxiv.org/abs/1107.5337}{{\tt arXiv:1107.5337 [hep-th]}}.

\bibitem{Marsano:2011hv}
J.~Marsano and S.~Schafer-Nameki, ``{Yukawas, G-flux, and Spectral Covers from
  Resolved Calabi-Yau's},''
  \href{http://dx.doi.org/10.1007/JHEP11(2011)098}{{\em JHEP} {\bf 1111} (2011)
   098},
\href{http://arxiv.org/abs/1108.1794}{{\tt arXiv:1108.1794 [hep-th]}}.

\bibitem{Grimm:2011sk}
T.~W. Grimm and R.~Savelli, ``{Gravitational Instantons and Fluxes from
  M/F-theory on Calabi-Yau fourfolds},''
  \href{http://dx.doi.org/10.1103/PhysRevD.85.026003}{{\em Phys.Rev.} {\bf D85}
  (2012)  026003},
\href{http://arxiv.org/abs/1109.3191}{{\tt arXiv:1109.3191 [hep-th]}}.

\bibitem{Grimm:2011fx}
T.~W. Grimm and H.~Hayashi, ``{F-theory fluxes, Chirality and Chern-Simons
  theories},'' \href{http://dx.doi.org/10.1007/JHEP03(2012)027}{{\em JHEP} {\bf
  1203} (2012)  027},
\href{http://arxiv.org/abs/1111.1232}{{\tt arXiv:1111.1232 [hep-th]}}.

\bibitem{Cvetic:2013uta}
M.~Cvetič, A.~Grassi, D.~Klevers, and H.~Piragua, ``{Chiral Four-Dimensional
  F-Theory Compactifications With SU(5) and Multiple U(1)-Factors},''
  \href{http://dx.doi.org/10.1007/JHEP04(2014)010}{{\em JHEP} {\bf 1404} (2014)
   010},
\href{http://arxiv.org/abs/1306.3987}{{\tt arXiv:1306.3987 [hep-th]}}.

\bibitem{Denef:2008wq}
F.~Denef, ``{Les Houches Lectures on Constructing String Vacua},''
\href{http://arxiv.org/abs/0803.1194}{{\tt arXiv:0803.1194 [hep-th]}}.

\bibitem{Esole:2011sm}
M.~Esole and S.-T. Yau, ``{Small resolutions of SU(5)-models in F-theory},''
  \href{http://dx.doi.org/10.4310/ATMP.2013.v17.n6.a1}{{\em
  Adv.Theor.Math.Phys.} {\bf 17} (2013)  1195--1253},
\href{http://arxiv.org/abs/1107.0733}{{\tt arXiv:1107.0733 [hep-th]}}.

\bibitem{Hayashi:2013lra}
H.~Hayashi, C.~Lawrie, and S.~Schafer-Nameki, ``{Phases, Flops and F-theory:
  SU(5) Gauge Theories},''
  \href{http://dx.doi.org/10.1007/JHEP10(2013)046}{{\em JHEP} {\bf 1310} (2013)
   046},
\href{http://arxiv.org/abs/1304.1678}{{\tt arXiv:1304.1678 [hep-th]}}.

\bibitem{Hayashi:2014kca}
H.~Hayashi, C.~Lawrie, D.~R. Morrison, and S.~Schafer-Nameki, ``{Box Graphs and
  Singular Fibers},'' \href{http://dx.doi.org/10.1007/JHEP05(2014)048}{{\em
  JHEP} {\bf 1405} (2014)  048},
\href{http://arxiv.org/abs/1402.2653}{{\tt arXiv:1402.2653 [hep-th]}}.

\bibitem{Esole:2014bka}
M.~Esole, S.-H. Shao, and S.-T. Yau, ``{Singularities and Gauge Theory
  Phases},''
\href{http://arxiv.org/abs/1402.6331}{{\tt arXiv:1402.6331 [hep-th]}}.

\bibitem{Esole:2014hya}
M.~Esole, S.-H. Shao, and S.-T. Yau, ``{Singularities and Gauge Theory Phases
  II},''
\href{http://arxiv.org/abs/1407.1867}{{\tt arXiv:1407.1867 [hep-th]}}.

\bibitem{Braun:2014kla}
A.~P. Braun and S.~Schafer-Nameki, ``{Box Graphs and Resolutions I},''
\href{http://arxiv.org/abs/1407.3520}{{\tt arXiv:1407.3520 [hep-th]}}.

\bibitem{Donagi:2008ca}
R.~Donagi and M.~Wijnholt, ``{Model Building with F-Theory},''
  \href{http://dx.doi.org/10.4310/ATMP.2011.v15.n5.a2}{{\em
  Adv.Theor.Math.Phys.} {\bf 15} (2011)  1237--1318},
\href{http://arxiv.org/abs/0802.2969}{{\tt arXiv:0802.2969 [hep-th]}}.

\bibitem{Beasley:2008dc}
C.~Beasley, J.~J. Heckman, and C.~Vafa, ``{GUTs and Exceptional Branes in
  F-theory - I},'' \href{http://dx.doi.org/10.1088/1126-6708/2009/01/058}{{\em
  JHEP} {\bf 0901} (2009)  058},
\href{http://arxiv.org/abs/0802.3391}{{\tt arXiv:0802.3391 [hep-th]}}.

\bibitem{DK}
V.~I. Danilov and A.~G. Khovanskii, ``{Newton Polyhedra and an Algorithm for
  Computing Hodge--Deligne Numbers},'' {\em Math. USSR. Izv.} {\bf 29} (1987)
  279--298.

\bibitem{danilov}
V.~I. Danilov, ``The geometry of toric varieties,'' {\em Russian Mathematical
  Surveys} {\bf 33} (1978) no.~2, 97.

\bibitem{fulton}
W.~Fulton, {\em {Introduction to toric varieties}}.
\newblock Princeton University Press, Princeton, 1993.

\bibitem{Batyrev1996901}
V.~V. Batyrev and D.~I. Dais, ``Strong mckay correspondence, string-theoretic
  hodge numbers and mirror symmetry,'' {\em Topology} {\bf 35} (1996) no.~4,
  901 -- 929, \href{http://arxiv.org/abs/arXiv:alg-geom/9410001}{{\tt
  arXiv:alg-geom/9410001}}.

\bibitem{1995alg.geom..9009B}
V.~V. {Batyrev} and L.~A. {Borisov}, ``{Mirror duality and string-theoretic
  Hodge numbers},'' {\em Inventiones mathematicae} {\bf 126 (1)} (1995)
  183--203, \href{http://arxiv.org/abs/arXiv:alg-geom/9509009}{{\tt
  arXiv:alg-geom/9509009}}.

\bibitem{peters2008mixed}
C.~Peters and J.~Steenbrink, {\em Mixed Hodge Structures}.
\newblock Ergebnisse der Mathematik und ihrer Grenzgebiete. 3. Folge / A Series
  of Modern Surveys in Mathematics. Springer Berlin Heidelberg, 2008.

\bibitem{wikimixedHodge}
``Hodge structure.''
  \url{http://en.wikipedia.org/wiki/Hodge_structure#Example_of_curves}.
\newblock Accessed: 2014-08-12.

\bibitem{Bies:2014sra}
M.~Bies, C.~Mayrhofer, C.~Pehle, and T.~Weigand, ``{Chow groups, Deligne
  cohomology and massless matter in F-theory},''
\href{http://arxiv.org/abs/1402.5144}{{\tt arXiv:1402.5144 [hep-th]}}.

\bibitem{TOPCOM}
J.~Rambau, ``{TOPCOM}: Triangulations of point configurations and oriented
  matroids,'' in {\em Mathematical Software---ICMS 2002}, A.~M. Cohen, X.-S.
  Gao, and N.~Takayama, eds., pp.~330--340.
\newblock World Scientific, 2002.

\bibitem{sage}
W.~Stein {\em et al.}, {\em {S}age {M}athematics {S}oftware ({V}ersion 6.1.1)}.
\newblock The Sage Development Team, 2014.
\newblock {\tt http://www.sagemath.org}.

\bibitem{Kreuzer:2002uu}
M.~Kreuzer and H.~Skarke, ``{PALP: A Package for analyzing lattice polytopes
  with applications to toric geometry},''
  \href{http://dx.doi.org/10.1016/S0010-4655(03)00491-0}{{\em
  Comput.Phys.Commun.} {\bf 157} (2004)  87--106},
\href{http://arxiv.org/abs/math/0204356}{{\tt arXiv:math/0204356 [math-sc]}}.

\bibitem{Braun:2012vh}
A.~P. Braun, J.~Knapp, E.~Scheidegger, H.~Skarke, and N.-O. Walliser, ``{PALP -
  a User Manual},''
\href{http://arxiv.org/abs/1205.4147}{{\tt arXiv:1205.4147 [math.AG]}}.

\bibitem{rambau}
J.~De~Loera, J.~Rambau, and F.~Santos, {\em Kummer's Quartic Surface}.
\newblock Springer, Heidelberg Dordrecht London New York, 2010.

\bibitem{Blumenhagen:2009yv}
R.~Blumenhagen, T.~W. Grimm, B.~Jurke, and T.~Weigand, ``{Global F-theory
  GUTs},'' \href{http://dx.doi.org/10.1016/j.nuclphysb.2009.12.013}{{\em
  Nucl.Phys.} {\bf B829} (2010)  325--369},
\href{http://arxiv.org/abs/0908.1784}{{\tt arXiv:0908.1784 [hep-th]}}.

\bibitem{Candelas:1996su}
P.~Candelas and A.~Font, ``{Duality between the webs of heterotic and type II
  vacua},'' \href{http://dx.doi.org/10.1016/S0550-3213(96)00410-5}{{\em
  Nucl.Phys.} {\bf B511} (1998)  295--325},
\href{http://arxiv.org/abs/hep-th/9603170}{{\tt arXiv:hep-th/9603170
  [hep-th]}}.

\bibitem{Candelas:1997eh}
P.~Candelas, E.~Perevalov, and G.~Rajesh, ``{Toric geometry and enhanced gauge
  symmetry of F theory / heterotic vacua},''
  \href{http://dx.doi.org/10.1016/S0550-3213(97)00563-4}{{\em Nucl.Phys.} {\bf
  B507} (1997)  445--474},
\href{http://arxiv.org/abs/hep-th/9704097}{{\tt arXiv:hep-th/9704097
  [hep-th]}}.

\bibitem{Perevalov:1997vw}
E.~Perevalov and H.~Skarke, ``{Enhanced gauged symmetry in type II and F theory
  compactifications: Dynkin diagrams from polyhedra},''
  \href{http://dx.doi.org/10.1016/S0550-3213(97)00477-X}{{\em Nucl.Phys.} {\bf
  B505} (1997)  679--700},
\href{http://arxiv.org/abs/hep-th/9704129}{{\tt arXiv:hep-th/9704129
  [hep-th]}}.

\bibitem{Andreas:1999ng}
B.~Andreas and G.~Curio, ``{On discrete twist and four flux in N=1 heterotic /
  F theory compactifications},'' {\em Adv.Theor.Math.Phys.} {\bf 3} (1999)
  1325--1413,
\href{http://arxiv.org/abs/hep-th/9908193}{{\tt arXiv:hep-th/9908193
  [hep-th]}}.

\bibitem{Hayashi:2008ba}
H.~Hayashi, R.~Tatar, Y.~Toda, T.~Watari, and M.~Yamazaki, ``{New Aspects of
  Heterotic--F Theory Duality},''
  \href{http://dx.doi.org/10.1016/j.nuclphysb.2008.07.031}{{\em Nucl. Phys.}
  {\bf B806} (2009)  224--299},
\href{http://arxiv.org/abs/0805.1057}{{\tt arXiv:0805.1057 [hep-th]}}.

\bibitem{Braun:2014pva}
A.~P. Braun, A.~Collinucci, and R.~Valandro, ``{Hypercharge flux in F-theory
  and the stable Sen limit},''
\href{http://arxiv.org/abs/1402.4096}{{\tt arXiv:1402.4096 [hep-th]}}.

\bibitem{Borisov93}
L.~Borisov, ``{Towards the Mirror Symmetry for Calabi-Yau Complete
  intersections in Gorenstein Toric Fano Varieties},''
  \href{http://arxiv.org/abs/alg-geom/9310001}{{\tt arXiv:alg-geom/9310001}}.

\bibitem{BatyrevBorisov}
V.~V. Batyrev and L.~Borisov, ``{On Calabi-Yau Complete Intersections in Toric
  Varieties},'' \href{http://arxiv.org/abs/alg-geom/9412017}{{\tt
  arXiv:alg-geom/9412017}}.

\bibitem{Friedman:1997yq}
R.~Friedman, J.~Morgan, and E.~Witten, ``{Vector bundles and F theory},''
  \href{http://dx.doi.org/10.1007/s002200050154}{{\em Commun.Math.Phys.} {\bf
  187} (1997)  679--743},
\href{http://arxiv.org/abs/hep-th/9701162}{{\tt arXiv:hep-th/9701162
  [hep-th]}}.

\bibitem{Curio:1998vu}
G.~Curio, ``{Chiral matter and transitions in heterotic string models},''
  \href{http://dx.doi.org/10.1016/S0370-2693(98)00713-8}{{\em Phys.Lett.} {\bf
  B435} (1998)  39--48},
\href{http://arxiv.org/abs/hep-th/9803224}{{\tt arXiv:hep-th/9803224
  [hep-th]}}.

\bibitem{Diaconescu:1998kg}
D.-E. Diaconescu and G.~Ionesei, ``{Spectral covers, charged matter and bundle
  cohomology},'' \href{http://dx.doi.org/10.1088/1126-6708/1998/12/001}{{\em
  JHEP} {\bf 9812} (1998)  001},
\href{http://arxiv.org/abs/hep-th/9811129}{{\tt arXiv:hep-th/9811129
  [hep-th]}}.

\bibitem{Gray:2013mja}
J.~Gray, A.~S. Haupt, and A.~Lukas, ``{All Complete Intersection Calabi-Yau
  Four-Folds},'' \href{http://dx.doi.org/10.1007/JHEP07(2013)070}{{\em JHEP}
  {\bf 1307} (2013)  070},
\href{http://arxiv.org/abs/1303.1832}{{\tt arXiv:1303.1832 [hep-th]}}.

\bibitem{DeWolfe:2004ns}
O.~DeWolfe, A.~Giryavets, S.~Kachru, and W.~Taylor, ``{Enumerating flux vacua
  with enhanced symmetries},''
  \href{http://dx.doi.org/10.1088/1126-6708/2005/02/037}{{\em JHEP} {\bf 0502}
  (2005)  037},
\href{http://arxiv.org/abs/hep-th/0411061}{{\tt arXiv:hep-th/0411061
  [hep-th]}}.

\bibitem{Hayashi:2009bt}
H.~Hayashi, T.~Kawano, Y.~Tsuchiya, and T.~Watari, ``{Flavor Structure in
  F-theory Compactifications},''
  \href{http://dx.doi.org/10.1007/JHEP08(2010)036}{{\em JHEP} {\bf 1008} (2010)
   036},
\href{http://arxiv.org/abs/0910.2762}{{\tt arXiv:0910.2762 [hep-th]}}.

\bibitem{Cordova:2009fg}
C.~Cordova, ``{Decoupling Gravity in F-Theory},''
  \href{http://dx.doi.org/10.4310/ATMP.2011.v15.n3.a2}{{\em
  Adv.Theor.Math.Phys.} {\bf 15} (2011)  689--740},
\href{http://arxiv.org/abs/0910.2955}{{\tt arXiv:0910.2955 [hep-th]}}.

\bibitem{Donagi:2004ia}
R.~Donagi, Y.-H. He, B.~A. Ovrut, and R.~Reinbacher, ``{The Particle spectrum
  of heterotic compactifications},''
  \href{http://dx.doi.org/10.1088/1126-6708/2004/12/054}{{\em JHEP} {\bf 0412}
  (2004)  054},
\href{http://arxiv.org/abs/hep-th/0405014}{{\tt arXiv:hep-th/0405014
  [hep-th]}}.

\bibitem{Curio:1998bva}
G.~Curio and R.~Y. Donagi, ``{Moduli in N=1 heterotic / F theory duality},''
  \href{http://dx.doi.org/10.1016/S0550-3213(98)00185-0}{{\em Nucl.Phys.} {\bf
  B518} (1998)  603--631},
\href{http://arxiv.org/abs/hep-th/9801057}{{\tt arXiv:hep-th/9801057
  [hep-th]}}.

\bibitem{Diaconescu:2003bm}
E.~Diaconescu, G.~W. Moore, and D.~S. Freed, ``{The M theory three form and
  E(8) gauge theory},''
\href{http://arxiv.org/abs/hep-th/0312069}{{\tt arXiv:hep-th/0312069
  [hep-th]}}.

\bibitem{Krause:2011xj}
S.~Krause, C.~Mayrhofer, and T.~Weigand, ``{$G_4$ flux, chiral matter and
  singularity resolution in F-theory compactifications},''
  \href{http://dx.doi.org/10.1016/j.nuclphysb.2011.12.013}{{\em Nucl.Phys.}
  {\bf B858} (2012)  1--47},
\href{http://arxiv.org/abs/1109.3454}{{\tt arXiv:1109.3454 [hep-th]}}.

\bibitem{Grimm:2009yu}
T.~W. Grimm, S.~Krause, and T.~Weigand, ``{F-Theory GUT Vacua on Compact
  Calabi-Yau Fourfolds},''
  \href{http://dx.doi.org/10.1007/JHEP07(2010)037}{{\em JHEP} {\bf 1007} (2010)
   037},
\href{http://arxiv.org/abs/0912.3524}{{\tt arXiv:0912.3524 [hep-th]}}.

\bibitem{Bershadsky:1996nh}
M.~Bershadsky, K.~A. Intriligator, S.~Kachru, D.~R. Morrison, V.~Sadov, {\em et
  al.}, ``{Geometric singularities and enhanced gauge symmetries},''
  \href{http://dx.doi.org/10.1016/S0550-3213(96)90131-5}{{\em Nucl.Phys.} {\bf
  B481} (1996)  215--252},
\href{http://arxiv.org/abs/hep-th/9605200}{{\tt arXiv:hep-th/9605200
  [hep-th]}}.
\end{thebibliography}
\end{document}